# sheldon

smart habitat
for the elderly

# State of the Art Report for Smart Habitat for Older Persons

April 2019

COST
EUROPEAN COOPERATION
IN SCIENCE & TECHNOLOGY

Funded by the Horizon 2020 Framework Programme
of the European Union

Old age is the loss of curiosity

José Martínez Ruiz "Azorín", Spanish writer

CeTEM

This publication is based upon work from COST Action CA16226,
supported by COST (European Cooperation in Science and
Technology).
COST is a funding agency for research and innovation networks. Our
Actions help connect research initiatives across Europe and enable
scientists to grow their ideas by sharing them with their peers. This
boosts their research, career and innovation.
www.cost.eu

cost
EUROPEAN COOPERATION
IN SCIENCE & TECHNOLOGY

Funded by the Horizon 2020 Framework Programme
of the European Union

# Executive summary

This document reports the State of the Art of science and practice on three topics related to smart and healthy ageing at home: furniture and habitats, Information and Communication Technologies (ICT), and healthcare. The reports were prepared by the working groups of COST Action CA16226, **Sheld-on**. **Sheld-on** is a network of researchers, user representatives, industry members, and other stakeholders. The three domains covered in this report were the areas of interest for three working groups from the COST Action. The aim of each working group was to assess the State of the Art for disciplinary understanding, identification of advances in smart furniture and habitat, products, industries and success stories. The findings on these topics of all working groups are compiled here. Due to the different backgrounds of the members of each of the working groups, the document is divided in three separate parts that can be considered as separate State of the Art reports. The goal of this document is to be used as input in the fourth working group of Sheld-on COST Action: Solutions for Ageing Well at Home, in the Community, and at Work, where experts from the three different domains converge to a single working group in order to achieve the action objectives.

The State of the Art for each WG was developed following a similar approach to its collation of information. Materials have been compiled from working group countries who share the most cutting edge and pivotal literature as well as other forms of research information and results (data). This includes research literature, industry literature, patents, project reports, white paper reports, and designs. The collection of this SoA report content has been built throughout this COST Action through WG meetings, conferences, outputs and networked discussions.

The report from Working Group 1 has focused on **Furniture and Habitat** (including architecture and services) that cover advancements made in improving lives of older persons through developments in technology, furniture design, and architectural environments.

The report provided by Working Group 2 contains examples and descriptions of the latest **ICT advancements** that are used for improving the lives of older adults. Since the research areas included in ICT developments are vast, the members of WG2 tried to cover as much as possible. The WG2 report itself reflects on this and contains several separate sections based on the more narrow research focuses that are covered.

Working Group 3 focused the SoA report on **needs of older persons** in relation to healthcare and smart living spaces. The report summarizes research done by WG3 members as well as results from others related to WG3 objectives and includes initial map of existing policies and practices regarding healthcare and smart living spaces.

In the three included reports on the state of the art, the members of each Sheld-on working group, performed a thorough survey that covered three distinct points of view on the same topic: Improving the lives of the older persons and allowing safe, comfortable, and healthy aging. Having separate working groups improved the quality of the presented work by providing diversity and allowing people of wide range of professional background to gain understanding about the used concepts, technologies and the state of the art research in the covered areas. What we find very interesting is that during the performed research, the three working groups had overlapping interests and research topics. For example, both WG1 and WG2 overlapped consistently with WG3 in the areas where the application of the research is concerned with the health and the products that are developed can be used in health care  environments. One can find many other examples of the interdisciplinarity of the topics covered in this COST Action within the reports. This further proves the benefit of bringing these three different research groups into the same Action to combine their knowledge and expertise. Creating this critical mass of knowledge and cooperation is actually one of the main objectives of the **Sheld-on** COST Action.

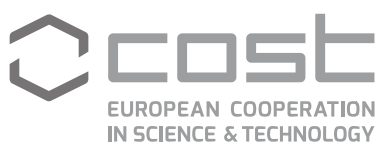

Funded by the Horizon 2020 Framework Programme of the European Union

# General Index

# Working Group 1
## Furniture and Habitat Industries

**Editors:**
Jake Kaner (UK)
Rafael Maestre (ES)

**Members:**
Ms Emilie Bossanne (FR)
Dr Michael Burnard (SI)
Prof Iana Codruta (RO)
Dr Levente Dénes (HU)
Ms Susanne Frank (NO)
Dr Josefina Garrido (ES)
Ms Valerie Gourvès (FR)
Mr Clément Grangé (FR)
Dr Preben Hansen (SE)
Prof Jasna Hrovatin (SI)
Dr Veronika Kotradyova (SK)
Dr Dimitrios Kraniotis (NO)
Dr Andreja Kutnar (SI)

Mr Dean Lipovac (SI)
Prof Jelena Matić (RS)
Mr Francisco Melero (ES)
Prof Georgios Ntalos (EL)
Mr Juan José Ortega (ES)
Prof Anders Q. Nyrud (NO)
Prof Alessandra Rinaldi (IT)
Ms Yifat Rom (IL)
Dr Anna Sandak (SI)
Ms Britt Vahter (EE)
Prof Joris Van Acker (BE)
Prof Joost van Hoof (NL)
Dr Martin Weigl (AT)

State of the Art Report for
Smart Habitat for Older Persons

# **Working Group 1**
## Furniture and Habitat Industries

**Acknowledgements:**
We would like to thank in particular, Blanca Puche and Guadalupe Santa for doing the thorough State of the Art search, where they provided a solid ground from where many parts of the final WG1 report were developed. Both Blanca and Guadalupe were guided by Rafael Maestre who also supplied helpful examples of patents, projects and literature.

Clement Grange and Valerie Gourves for great examples of live projects to improve lives through adaptable living structures.

Anna Sandak for providing material on smart furniture and sensorization in the home environment

Joris van Hoof for advice on medical structures and how medical products are selected in the domestic and institutional habitats.

Yifat Rom for providing work on the architecture solutions of older persons' designed homes.

Denes Levente for wearable devices used in the domestic habitat.

Michael Burnard for examples of projects and academic papers relating to specified furniture materials and their testing.

Georgios Ntalos for literature on furniture and the built environment.

Other working group members who provided literature and other contributions of useful material are also to be thanked.

We also acknowledge the support from the Sheld-on COST Action for organizing the meetings and supporting the work and all other members who participated in the STSMs and other members and non-members who gave their valuable contribution towards achieving the goals of the action.

This publication is based upon work from COST Action CA16226 - Indoor Living Space Improvement: Smart Habitat for the Elderly (Sheld-on), supported by COST (European Cooperation in Science and Technology).

# Index

Funded by the Horizon 2020 Framework Programme
of the European Union

# ¹ Introduction

The purpose of this report is to present a State of the Art search and analysis on **smart furniture and smart home for older adults**, under the concept of interior, architecture and services. The search has been focused on the **last 7 years**, except for the case of very relevant results. **Only in-force patents** were considered, even if applied for more than 5 years ago. First, an analysis of the results is presented in section 2. Section 3 describes the search profile or how it was carried out. The actual results are presented in section 4. Finally, the main conclusions of this report are presented in section 5.

Both the analysis/summary of section 2 and the actual results in section 4 follow the same structure, divided in four main categories:

- Commercial solutions
- Patents and Designs (only in force)
- Research projects
- Scientific publications

Within each of these categories, the results are further classified as:

- **Smart furniture and accessories for furniture**: This section will show the results related to furniture or accessories that can be applied to furniture and that have incorporated some type of electronics such as sensors, artificial intelligence, internet of things...
- **Sensorization in the home environment:** This section will show the results related to a smart home and a sensorization in the home environment to facilitate and improve the life of the elderly
- **Wearable devices**: This section will show the results of wearable devices such as bracelets or clothing that have incorporated technology and make life easier for the elderly
- **Architectural solutions**: This section will show the results of architectural solutions that have been applied to the work and domestic habitat of the older person nature and natural systems.

sheldon
smart habitat
for the elderly

# **2** State of the Art Summary and Analysis

This section provides a summarised analysis of the results that are later presented in detail within section 4. Both the analysis/summary of section 2 and the actual results in section 4 follow the same structure, divided in four main categories:

## **2.1** Commercial Solutions

**Smart Furniture**

A search of commercial solutions was carried out in order to determine what is already in the market and what has been announced and soon to be ready. By looking at the results it is evident that **furniture manufacturers have not yet widely adopted the integration of ICT** and intelligence in its product. Actually, there is almost not one that could be called "smart furniture", the only exception being **smart mattresses**.

For example, **Sleep Number** manufactures a smart mattress with large number of sensors and voice recognition. It estimates sleep quality and well-being based on bed movement, heart rate and respiratory rate. It can also give massages and help with snoring.

On the other hand, there are many **smart furniture accessories** that could be added to traditional furniture in order to achieve some kind of functionality that could make it "smart". The most common example is **occupancy sensors** typically connected to an alarm system that warns about users abandoning their bed, which is especially useful during the night. Some examples are Miray, Aidacare, Health and Care, Asistae and SafeBase.

A related but different approach is **Asistae**, an **advanced IoT solution** that is made up of a sensor for bed or armchair that is capable of sending information through regular mobile networks (e.g. 3G). A smartphone app can be used to provide alarms remotely to preapproved caretakers following some easily configurable time-based rules. This app can also provide recent historical occupancy.

Although not specifically developed with the older adult in mind, other examples that may help in AAL and that are related to smart furniture, are Nest Lock and Mirror. **Nest Lock** is a keyless lock that can be unlocked with a numerical code, with your phone remotely, or even with voice through Google Assistant. **Mirror** is an interactive mirror that shows an image of different trainings next to its user reflexion for technique improvement. This mirror could help keep older people active in the home.

**Interteam** is a Slovenian company that offers specialized products to furnish dwellings of the elderly.

## Sensorization In Home Environment

Most of the sensorization at home is done by systems that are made up of sensors distributed throughout the home. The data from these sensors are sent to a central processing hub where it can be stored and processed. In some cases, data are automatically analysed and some decision making is carried out (e.g. alert relatives). Some examples of these systems, though with some differences, are: **SentinelCARE, Just Checking, TruSense, GrandCare Systems.**

Other products are just made of one single device that tracks certain variables. **AbiBird** and **Canary Care** are just a simple wireless movement sensor (no internet connection needed). A smartphone app can be used by caretakers to monitor the movement remotely and receive notifications if no activity is detected after a given amount of time.

Similarly, **Elsi Smart Floor** is a sensing floor that monitors pressure (e.g. steps) and can non-invasively detect falls.

## Wearable Devices

During recent years the field of wearable devices has experienced an explosion of available solutions in the market. This also applies to the sector of AAL. Most of these devices are made of a smartwatch or smart-wristband that wirelessly transmits activity (and sometimes even health) data. This information is used to identify patterns and/or detect potential problems such as falls. A smartphone app provides direct contact to caretakers. Some examples of this powerful approach are **Senior Protection, Nectarine** and **CarePredict**. A simpler solution is provided by **Zembro**, a bracelet with an alarm button for sending a message to preselected contacts. Another one is **Contact**, a smartwatch with fall detection that includes GPS.

Other wearable devices include **E-vone** (a smart shoe with automatic fall detection and location SMS), **Pill Clip** (a small pill holder with LEDs that blink when it's time for medication), (a smart wristband wirelessly connected to a belt with airbags that deploy if fall is detected before the impact), and **QTUG** (can provide a fall risk score based on the data gathered by wearable sensors in the legs).

A unique wearable solution is provided by **Bioservo Technologies**, which has created a glove that improves **grip strength**. It detects the intended action with integrated pressure sensors and activates motors to support it.

## Architectural Solutions

- New product: **Illuminated Access Door** Manufacturer: Righini and SFL, lighting specialists.

## 2.2 Patents & Designs

**Smart Furniture**

Some patents are related to **bed occupancy for mattresses or beds**. These systems are usually designed to detect and warn caretakers when a person leaves the bed at night and takes more than expected to come back (e.g. ES1218994). A more advanced example is described in ES120 612U, a mattress that includes a variety of **temperature, humidity and pressure** sensors and a control circuit that sends that information wirelessly to a smartphone app.

A more general **sensorization of different pieces of furniture** (bed, armchair, nightstand…) with the goal of monitoring the status of a person is described in ES2536826 B1.

Some other patents focus on a variety of goals like **monitoring eating habits** (CN108721030), **monitoring bathroom usage** based on duration (CN108399713), and a **chair that helps its user to stand up** by inclining forward while monitoring and adapting to its user status (CN106667116). Finally, an elderly-friendly virtual **touch-type bedside multimedia** device is described in KR101687150.

• Prototype in book **Armchair "Up Down",** mentor: Jelena Matić, students: Marija Prelević, Ivana Ninković

**Sensorization In Home Environment**

Most of the patents in this domain focused on monitoring the older adult in order to detect potential problems. In this regard, a complete **Ambient Intelligence system** that uses a variety of sensors and can be connected to social networks is described in ES2462566.

Other patents focus in narrower areas. For example, in ES2421168 a system for people with hearing problems (including older adults) is described. A **bracelet vibrates** when the phone or doorbell is ringing, or the main door opens. CN108961710 presents an **electric carpet** for monitoring elderly users on top of it (walk, falls…).

Surprisingly, CN108961710 is the only patent that uses **video monitoring** and applies it to bed occupancy detection.

**Wearable Devices**

Activity monitoring is something very common nowadays within the area of wearable devices. However, there are not that many patents with a focus on elderly care. Only two patents of **activity trackers** (ES2394842 and WO2016075344) were found.

Another patent (WO2019036884) describes a wearable device that detects the status of a user (older person) and if certain values are exceeded a medicine is recommended.

**Architectural Solutions**

• Spatial installation OXYGEN for the Living room for the elderly in Belgrade.

**2.3** Research Projects

**Smart Furniture**

Projects in smart furniture do not abound. Here are some projects in completely different areas such as **smart kitchen, smart lamp, and robotic furniture**.

The goal of the "**Smart kitchen** for Ambient Assisted Living" project was to help elderly and/or disabled people to carry out their daily activities in an easier and simpler way. It uses a network of sensors to gather relevant data and obtain high level information such as fall or location detection.

The **Aladin Project** developed an innovative intelligent lamp that prevents falls through a combination of State of the Art analogue sensors capable of detecting the slowest movement and a proprietary software, PrediCare, that analyses behavioural patterns to predict future falls.

The **Baltse@nio**r project aims to support and help furniture/interior manufacturers across the Baltic Sea Region with knowledge, insights, tools and methodology to develop and market products with a focus on seniors. However, it is not smart furniture/home specific.

Roombot is a unique project that explores the design and control of modular robots, called Roombots or **furniture that moves, self-assembles, and self-reconfigures**. One of its applications is assistive furniture for the elderly or the disabled.

**Habitat seniors**, the first full adaptable accommodation designed by and for FCBA, (French: Forest, Cellulose Wood-Furniture Construction).

• New Product : **Entry & Ironing Space** Manufacturer: Optimum.
• New product : **Wall Cabinet With Integrated Door & Floor** Manufacturer: Righini and Optimum.

**Sensorization In Home Environment**

Some projects (**REAAL, LUCIA, CARE_HOME16**) developed AAL systems for older people monitoring through **non-invasive sensors**, and using **Artificial Intelligence** techniques, the system automatically learns the behaviour of the inhabitant, thereby establishing certain patterns that may represent an emergency, and when detected a warning is generated.

The objective of the **Activage** project is to build the first European IoT ecosystem and test it in 9 Deployment Sites across 7 European countries. It plans to reuse and scale up existing technol-ogy and make in interoperable with the aim of creating solutions and services that **promote active and healthy aging**.

The central objective of the **Listen** project is to design and implement a complete system (software and hardware), enabling robust hands-free large-vocabulary voice-based access to Internet applications in smart homes essential for the disabled and the **elderly**. Listen will provide natural voice control of the smart-home web-enabled functionalities, and access to specific Internet applications.

**Wearable Devices**

Recent research in wearable devices for older adults target **sensor integration in textiles, smart watches and smart insole**. Most of them detect or prevent falls, and some also extract well-being data.

The **Instinto** project developed techniques to **integrate sensors in textiles**, for their application in the field of health, specifically in prevention, detection and protection in falls in elderly people.

The **Invis Care** project tries to add connectivity and smart functionalities to any watch through a connected strap. The leading company is evolving the solution to address specific needs of the elderly.

The **Kytera Companion** consists of a wearable wristband, a mesh-network sensor system and a cloud-based software with a mobile app. Fall detection, well-being (hygiene, nutrition, sleep) and location are provided.

**Wiisel** project developed a wireless sensor built into an insole. Foot movement data is transferred to a mobile or computer, so that caretakers can evaluate the patient progress. It can detect falls and send an alert if it happens.

## 2.4 Scientific Publications

**Smart Furniture**

Different examples of application of technology to furniture have been found. From an AmI system that integrates sensors into different pieces of furniture (Bleda et al.), to a nutrition monitoring system (Zhou et al.), and even robotic furniture (Aguiar et al.). Furthermore, furniture for older people is discussed in terms of design basis (J. Zheng), design and building process (Brenes-Mora et al.), as well as smart furniture functions (Liangye Yu). An individual summary of each article follows below.

Bleda et al. describe an **AmI system for older adults** that **uses furniture as a key monitoring element** for the integration of wireless sensor nodes. It analyses which sensors are suitable for integration in the different pieces of furniture, their advantages and how they could be integrated.

**Dalila is an aerial kitchen furniture** for simplifying everyday activities of older users by using technology to create a safe environment with alerts perceived by all senses. The article (Brenes-Mora et al.) shows the **design and building process for smart furniture**.

Jiannan Zheng discusses the **design principle of intelligent furniture for the elderly** based on big data and intelligent technology.

An analysis of the smart home and related technologies is presented in a paper by Liangye Yu, **considering China**'s national conditions. The report studies the characteristics and needs of the elderly. Then it puts forward some **functions of intelligent furniture suitable for the elderly**.

A system for monitoring **nutrition related actions** is presented by Zhou et al. It is made of a smart table cloth with a fine-grained pressure textile matrix and a weight sensitive tablet. It detects actions such as cutting, scooping, stirring, etc., the identification of the plate/container on which the action is executed, and the tracking of the weight change in the containers.

A novel, networked and interoperative suite of **robotic furniture** is presented by Aguiar et al., part of home+, an assistive technology environment aimed at supporting aging in place. It describes the design and construction process for three robotic furniture pieces: a chair, featuring gesture-controlled assistive lift; a morphing side table; and an adaptive screen.

An **intent-based smart lift chair** is proposed by Hang Lu et al. It aims to analyse the user's physiological condition through pattern recognition. The paper introduces the idea of assistance-as-needed to encourage the elderly to improve their own motor function.

Stasidia (Church Stalls) of the Greek orthodox church – **a standing seat for elderly**. Ioannis Barboutis, Vasileios Vasileiou Faculty of Forestry and Natural Environment, Aristotle University of Thessaloniki. Proceedings of the XXVIth International Conference Research for Furniture Industry.

- **Critical points in the construction of aged people furniture**. Vasiliki Kamperidou.
- **Market potential and determinants for eco-smart furniture attending consumers of the third age**. Competitiveness Review, Vol. 26 Issue: 5, pp.559-574, Ioannis Papadopoulos, et al.
- **Ergonomic suitability of kitchen furniture regarding height accessibility**. Jasna Hrovatin, Silvana Prekrat, Leon Oblak, & David Ravnik.
- **Adaptability of kitchen furniture for elderly people in terms of safety**. Jasna Hrovatin, Kaja Širok, Simona Jevšnik, Leon Oblak, & Jordan Berginc.
- **Ambient Assisted Living,** Italian Forum (2017) Cavallo, F., Marletta, V., Monteriù, A., Siciliano, P. https://www.springer.com/it/book/9783319542829 - Book

## Sensorization In Home Environment

A wide variety of **AAL systems that carry out older people monitoring** (Rghioui et al.; Valero et al.; Pham et al.; A. Dasios et al.; Hernández-Peñaloza et al.) is presented within this subsection. Two different surveys are presented. One of them focuses on **sensor for monitoring** the elderly (S. Majumder et al.), whereas the other one covers smart home-based **remote healthcare** technologies (A. Dasios et al.).

IoT (Internet of Things) is applied to human **activity tracking for AAL** and e-health in a paper by Rghioui et al. It describes a IoT architecture and protocol for heterogeneous AAL and e-health scenarios where an IoT network is the most suitable option to interconnect all elements.

Valero et al. detail an integration and validation of multiple heterogeneous sensors with hybrid reasoners that support decision making in order to monitor personal and environmental data at a smart home in a private way.

**CoSHE** (Cloud-based Smart Home Environment) is a monitoring system for home healthcare (M. Pham et al.). It includes a smart home, a wearable unit, a private cloud, and a robot assistant. CoSHE collects physiological, motion and audio signals through non-invasive wearable sensors. It identifies daily activities, behavioural changes and monitors rehabilitation and recovery processes.

A **survey of different types of ambient-sensor-based elderly monitoring technologies** in the home is presented by Z. Uddin. It includes technologies, research works, and their outcomes. Ambient sensor-based monitoring technologies, multicomponent technologies, and even sensors in robot-based elderly care works were identified.

A comprehensive review on the State of the Art research and development in **smart home-based remote healthcare technologies** is described in a paper by S. Majumder et al.

A paper (A. Dasios et al.) reports hands-on experiences in designing, implementing and operating a **WSN-based prototype system for elderly care monitoring** in home environments. Environmental parameters like temperature, humidity and light intensity as well as micro-level incidents allow to infer daily activities like moving, sitting, sleeping, usage of electricity appliances and plumbing components.

A **radar and a PIR sensor for monitoring older people** is presented in a thesis by C. Mor. A previously developed system is improved by modifying its functioning, and by adding a more secure data storing.

A multi-sensor modular scheme for indoor (older) people monitoring is presented by Hernández-Peñaloza et al. for detection of abnormal events in PD patients.

- Ambient Intelligence Environments with Wireless Sensor Networks from the Point of View of Big Data and Smart & Sustainable Cities. 2018 ISSN:

2169-3536 DOI: 10.1109/ACCESS.2018.2849226 M. Espinilla(1); L. Martinez(1); J. Medina(1); C. Nugent(2). (1) Department of Computer Science, University of Jaén, Spain. (2) School of Computing, University of Ulster at Jordanstown, U.K.

- Radar Sensor System for Monitoring Elderly People At Home. UPCommons. Portal de acceso abierto al conocimiento de la Universidad Politécnica de Cataluña, España. Trabajo final de grado 2018-01. Catalán Mor, Joan Bru http://hdl.handle.net/2117/117778
- Smart Sensory Furniture Based on Wsn for Ambient Assisted Living. IEEE Sensors Journal ( Volume: 17 , Issue: 17 , Sept.1, 1 2017)pp. 5626 – 5636. (29 June 2017). Andres L. Bleda; Francisco J. Fernández-Luque; Antonio Rosa; Juan Zapata; Rafael Maestre.
- A Smart Kitchen for Ambient Assisted Living. Sensors 14, 1 (2014), 1629-1653; doi:10.3390/s140101629. Rubén Blasco (1), Álvaro Marco (1), Roberto Casas (1), Diego Cirujano (1) and Richard Picking (2).
- Integration of Multisensor Hybrid Reasoners To Support Personal Autonomy In The Smart Home. Sensors. 2014; 14(9):17313-17330. Miguel Ángel Valero (1), José Bravo (2), Juan Manuel Garcia Chamizo (3) and Diego López-de-Ipiña (4)
- Ami and Deployment Considerations In AAL Services Provision for Elderly Independent Living: The Monami Project. SENSORS 13 (2013), 8950-76. Falcó, Jorge; Vaquerizo, Esteban; Lain, Luis; Artigas Maestre, José Ignacio; Ibarz, Alejandro
- Wireless Sensor Network Deployment for Remote Elderly Care Monitoring. Athanasios Dasios, Damianos Gavalas, Grammati Pantziou, Charalampos Konstantopoulos
- Moving House and Housing Preferences in Older Age in Slovenia. Maša Filipovič Hrast, Richard Sendi, Valentina Hlebec, & Boštjan Kerbler, (2018) Taylor Francis.

## Wearable Devices

The design and evaluation of a smart insole is described in a work by Y. Charlon et al. for continuous monitoring of frail people at home. It measures number of steps, distance covered and gait speed.

A fall detection system is presented by P. Pierleoni et al. It consists of an inertial unit that includes triaxial accelerometer, gyroscope, and magnetometer with efficient data fusion and fall detection algorithms

## Architectural Solutions

- State of Art Senior and Habitat (2018) Valerie Gourves and Clement Grange, Report FCBA Institute, France. This report covers 27 studies, experiments and product developments through the work of the FCBA.
- The importance of assessing the environment to determine the most suitable location for seniors' residences so as to adjust the supply to the needs. et Psychologie Neuropsychiatrie du Vieillissement, 16(1), 31-38, 2018. Pierre-Marie Chapon - University of Lyon System, Guillaume Petit - University of Lyon System, Kévin Phalippon - Efferve'sens, Amberieu-en-Bugey, France.
- Environment and Living Conditions of Older Home Owners Gérontologie et société, 2017/1 (vol. 39 / n° 152) S Renaut, J Ogg, A Chamahian, S Petite.

- The "Mono-Polar" Young Retirees and Reorganisation of Domestic Settings Gérontologie et société, 2017/1 (vol. 39 / n° 152). Melissa petit.
- Choosing to Live in Sheltered Housing: Between Individual Trajectories and Public Policies. Gérontologie et société, 2017/1 (vol. 39 / n° 152). AB Simzac
- Déprise from the Prism of Intermediary Housing for the Elderly. Gérontologie et société 2018/1 (vol. 40 / n° 155). Laurent Nowik
- The daily sociability of elderly people in a shopping center: a peculiar leisure. Bulletin de l'association de géographes français, 95-1 | 2018, 79-96. Thibaut Besozzi
- State of the Art report on Design Methodology (2015) David Andrews and Stein Ove Erikstad, 12th International Marine design Conference Tokyo, Japan
- A novel accessibility assessment framework for the elderly: evaluation in a case study on office design. Panagiotis. Moschonas, Ioannis Paliokas, Dimitrios Tzovaras.
- The relationship of the elderly toward their home and living environment. Boštjan Kerbler, Richard Sendi, & Maša Filipovič Hrast,
- Healthy ageing at home: Ergonomics adaptations of interior design and self-assesed quality of life of older adults of the municipality of Ljubljana. Jasna Hrovatin, Saša Pišot, & Matej Plevnik.
- Preferences of Polish and Slovenian seniors concerning kitchen interior design. Beata Fabisiak & Jasna Hrovatin.
- Physical barriers and the use of assistive devices in senior citizens' everyday life. Mateja Berčan, Majda Pajnkihar, Jože Ramovš, & Zmago Turk.

# ³ Search Profile

Working group members have contributed to the report representing the span of European states within the network including international countries such as Israel.

Information has been included from those contributions in addition to the search conducted by the main proposer (CETEM).

This information has been discussed at working group workshops and round table meetings where SoA has been shared, discussed and evaluated.

## 3.1 Patent Databases

**Invenes**

Database of inventions of the Spanish Patent and Trademark Office. In order to disseminate the technological information contained in the patent documents and a national coverage, INVENES contains information on Spanish and Latin American patents and utility models, as well as on Spanish industrial designs.

**Espacenet**

Database with a high geographical coverage managed by the European Patent Office (EPO). It has fast, advanced or patent number search options and allows the use of Boolean and / or truncation operators.

## 3.2 International Classification of Patents

- G08B 21/00 Alarms responsive to a single specified undesired or abnormal condition and not otherwise provided for
- G08B 21/04 … responsive to non-activity, e.g. of elderly persons
- A47 Furniture; articles or appliances for domestic use
- A61B 5/00 Measures aimed at establishing a diagnosis

## 3.3 Commercial News and Solutions

- https://www.esmartcity.es
- https://www.casadomo.com/
- https://domotizados.co/
- http://geriatricarea.com/
- http://www.innovaticias.com
- https://es.digitaltrends.com
- https://computerhoy.com/

### 3.4 Database of Magazines and Scientific Publications:

**Redalyc**
Redalyc is an academic project for the diffusion in Open Access of the scientific publishing activity that takes place in and about Ibero-American States

**ScienceDirect**:
Access tool to the books and magazines of the Elsevier group (physical sciences and engineering, life sciences, health sciences and social sciences and humanities). There is the free version and the paid version.

**IEEE Xplore**
IEEE Xplore is a research database for discovery and access to journal articles, conference proceedings, technical standards, and related materials on computer science, electrical engineering and electronics, and allied fields.

### 3.5 Thesis Databases and Research Memories:

**Teseo**
Teseo is a database of doctoral theses that allows to retrieve information about doctoral theses defended in Spanish universities since 1976

**DART-Europe**
DART-Europe is a partnership of research libraries and library consortia who are working together to improve global access to European research theses. DART-Europe is endorsed by LIBER (Ligue des Bibliothèques Européennes de Recherche), and it is the European Working Group of the Networked Digital Library of Theses and Dissertations (NDLTD).

**Openthesis**
OpenThesis is a free repository of theses, dissertations, and other academic documents, coupled with powerful search, organization, and collaboration tools.

**Open Access**

Theses and Dissertations (OATD): OATD aims to be the best possible resource for finding open access graduate theses and dissertations published around the world. Metadata (information about the theses) comes from over 1100 colleges, universities, and research institutions. OATD currently indexes 4,814,919 theses and dissertations.

**3.6** Research Projects

**Cordis**

Cordis provides information on all EU-supported R&D activities, including programs (H2020, FP7 and older), projects, results, publications

**Keep.eu:**

Keep is a source of aggregated data regarding projects and beneficiaries of European Union cross-border, transnational and interregional cooperation programmes among the member States, and between member States and neighbouring countries.

**CDTI**

Center for Industrial Technological Development that finances technological innovation, technology aids and grants for technological development projects R + D + I

**Keywords**

Smart furniture, smart habitat, Elderly (Senior, Older persons)

# ⁴ Search Results

## **4.1** Commercial Solutions

This section details the commercial solutions launched during the last five years related to the project objective.

**Smart Furniture**

MIRAY Bed sensor

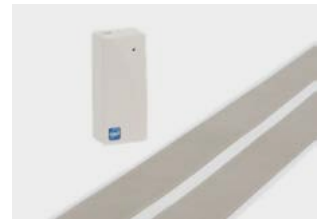

During the night shift in the residences and care centres it is necessary to control the activity of the patients who sleep in the centre. The bed sensor is a high-tech device that controls whether the patient gets out of bed

AIDACARE Mat with sensor

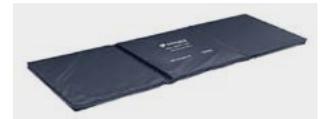

A mat next to the bed alerts caregivers in case a patient leaves the bed. The Aidacare sensitive mats for the elderly are made of safety foam and include a reflective strip for greater visibility.

HEALTH AND CARE Bed and chair occupancy alarms

There are a full range of chair leaving alarms and bed leaving alarms that enable a carer to monitor the activity of a person when getting out of bed. The bed and chair exit alarm monitors alert a carer when a person gets up out of their chair or gets out of bed. These bed sensor monitors and chair sensor monitors are especially useful when monitoring elderly patients who are prone to wandering, or those at risk of falls.

ASISTAE

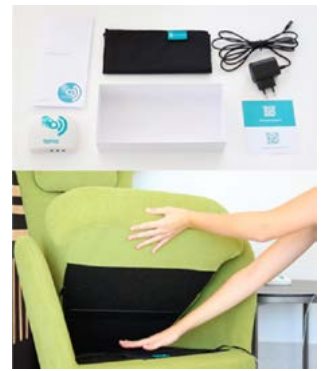

Asistae is positioned as an intelligent device that allows children to be informed 24 hours if their parents have had a problem, and provides the peace of mind to know if something abnormal has arisen.

Assist is a device equipped with a sensor that warns the children of the facts that can give us peace of mind that everything is going well informing the adult children of the elder if their mother or father have gotten out of bed or if they have gone already to sleep.

## SAFEBASE

For night supervision, it is interesting to quickly obtain information that a client is out of bed longer than usual at night with the risk of injury from falls, or if the client wakes up many times each night, which may indicate worries or prostate problems. A simple and discrete presence sensor is placed under the legs of the bed and collects data that is analysed in real time to identify abnormal patterns that must be addressed. Urgent alarms are sent by SMS. SafeBase also provides historical statistics, such as the development of sleep periods over time.

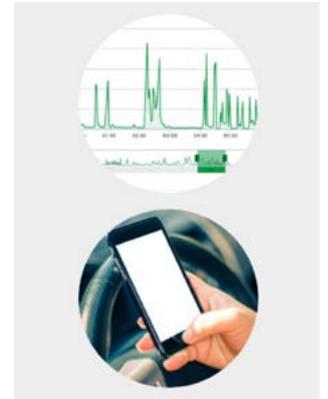

## SLEEP NUMBER

Sleep Number has created a smart mattress with a large number of sensors and voice recognition functions that offer the user many features. It can tell you how well you sleep, help you snoring, give you a relaxing massage, answer your voice commands and much more. It uses SleepIQ and ActiveComfort technology to monitor your bed movement, heart rate and respiratory rate to discover your well-being. It will help you detect a disease so you can act as soon as possible. SleepIQ, along with an application, can give you details about your nutrition, fitness, diet and other essential biometric information. Finally, it will give you a score based on your sleep patterns and will also give you recommendations to improve your sleep.

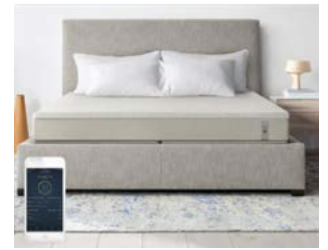

## NEST LOCK

Yale knows safe locks. Nest knows the connected home. Together, they have created a tamper-proof keyless lock that connects to the Nest app. Now you can lock and unlock a door from anywhere. Give people you trust an access code, instead of a password. Block and verify the status of your door with your voice using the Google Assistant. And always know who comes and goes. This solution can provide security for the elderly at home

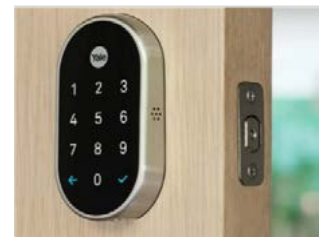

## MIRROR

Mirror is an interactive mirror in which different trainings of different disciplines are projected and which allows us to see our reflection at the same time to control our technique. It works through an app in which we can select a wide variety of workouts that are renewed weekly: from Yoga to boxing, through strength training, cardio or Pilates. This solution can help keep older people active in the home

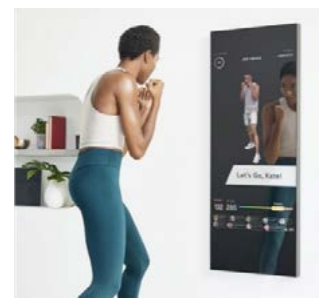

INTETEAM (workplace economy)

Today we are witnessing an ever-longer life expectancy. Over the years, fear and uncertainty in the movement can have a great impact on day-to-day activities. In order to provide the elderly with an adequate quality of life even at a higher age, the high requirements imposed by health care must be taken into account when designing and equipping the living and care facilities of the elderly.

**Smart Furniture**

CONCORDIA SYSTEMS

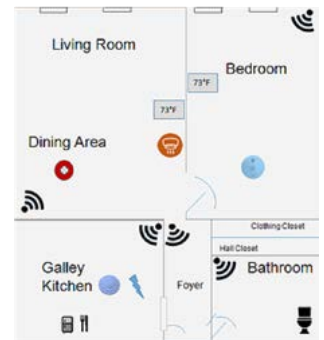

SentinelCARE is a service that connects seniors living independently with their adult children, caregivers and family members.

Motion sensors, door and specialty sensors are strategically placed and professionally installed in critical living spaces.

Typically, these include bedrooms, kitchen, bathrooms, foyer, walk-in closets, living and dining areas. Small and micro-sized contact sensors are placed on the refrigerator, freezer, cutlery drawer, medicine cabinets and entry/exit doors signalling when activity takes place.

Sensors connect to a communications hub that is installed in a closet or other unobtrusive location. All that is needed is a typical electrical outlet.

JUST CHECKING

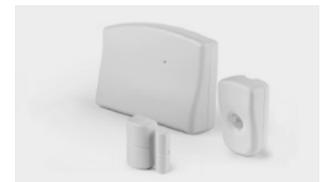

Small wireless movement sensors are placed around a property. They are easy to attach to walls, skirting boards or furniture using the sticky back Velcro strips provided. Using these sensors, Just Checking creates a chart of activity online, where you can see which rooms have been visited, when, and how long for. Just Checking also shows when a door is used and how long it is open for.

The door sensors are made up of two parts, the contact, and the magnet. These are placed side-by-side, with one part on the edge of the door and the other on the door frame. They can be used on both internal and external doors. Working in combination with the movement sensors, the door sensors tell you when visits have been received, when an individual leaves or enters the property, and how long the property appears to be vacant for.

## TRUSENSE

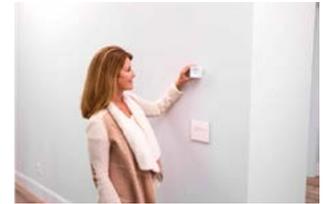

TruSense sensors help keep seniors safe at home. TruSense takes care of analyzing your user's daily habits such as time spent sleeping, in the kitchen or when leaving the house. If you change any of your routines, TruSense notifies family members, the user, and those who are registered to obtain the information. If necessary, the system can also notify TruSense's 24-hour emergency monitoring center via Amazon Echo Dot voice control.

## ABIBIRD

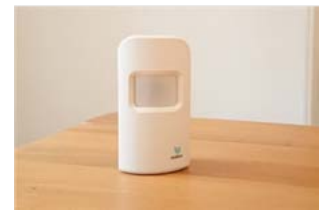

The AbiBird device understands and counts "normal" daytime and night-time activity inside a home. If a person stops moving for a period of time or the activity exceeds normal routines, AbiBird will immediately send an alert to family members or a caregiver's smartphone. The sensor is placed on a shelf or wall and has no camera or audio recording, allowing people to perform their daily activities without being observed, heard, or using a portable device.

## GRANDCARE SYSTEMS

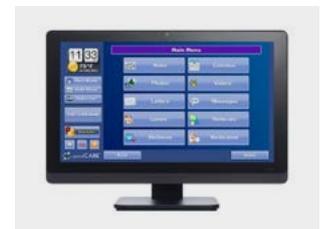

Caregivers or family members access the system by logging into the Online Care Portal. Wireless activity sensors, environmental sensors, and optional digital health devices can be added to the system as needed. These devices can be used to notify designated caregivers by phone, email, or text message if something seems wrong or if wellness readings are out of reach.

## CANARY CARE

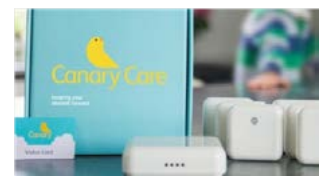

Canary Care uses small motion sensors that will monitor for activity and based upon the rules you create will let you know if no activity has been detected during specific times.

## MARICARE

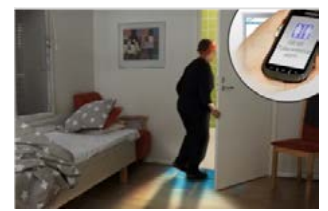

Elsi® Smart Floor is primarily an invisible nursing help-tool for fall-prevention and includes a range of fall detection, bed, toilet, toilet w. timer, dementia alerts, and burglar alarms, etc. It is a professional proactive care aid for Nursing Homes, Senior Homes, as well as Rehabilitation Centres and Hospitals. Furthermore, the system provides tools for tracking abnormal behaviour patterns 24/7 caused, for example, by infections. These early alerts reduce hospitalization time and costs.

## Wearable Devices

### SENIOR PROTECTION

Senior Protection has two appliances for the protection of the elderly in the home. On the one hand, there is a watch that has among its functionalities the physical activity indicator, LED signs, SOS button. On the other hand, it has a device for the home that offers voice chat to ask for help, SOS pushbutton and more.

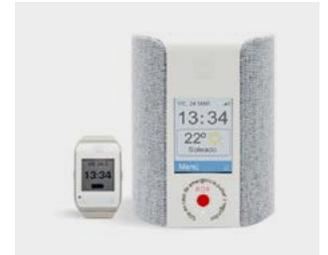

### E-VONE

E-vone is the first shoe connected with automatic fall detection and geolocation both indoors and outdoors. In the event of a fall, indoors or outdoors, your E-vone shoes send an independent alert to your family by SMS with the exact address where you are.

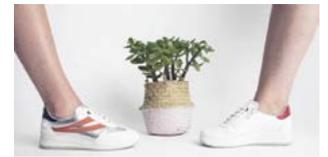

### NECTARINE HEALTH

Nectarine® is an artificial intelligence remote care solution. It has been developed to assist senior living and independent living facilities to deliver more efficient and higher quality care.

Nectarine® combines unobtrusive and lightweight hardware, artificial intelligence and an easy-to-use app to monitor the acute and subtle changes in a person's daily routine.

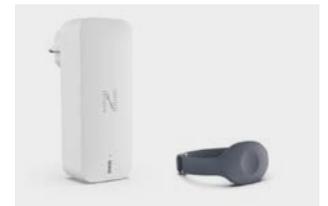

### CAREPREDICT

CarePredict provides useful information to help you provide better care for seniors. The end-to-end solution combines portable technology, intelligent indoor location tracking, deep automatic learning and sophisticated predictive analytics. Tempo is the portable device worn on the wrist and houses a sophisticated range of sensors capable of detecting the activities of a person's daily life. It provides a calling system and detects that ADLs such as eating, drinking, bathing, are being performed and where they take place. When the built-in assist button is pressed, an alert is sent to caregivers showing the exact location of the elderly person. Tempo also has two-way audio communication between the caregiver and the older adult for additional support.

This product has been patented by CarePredict since 2016. It currently has two patents related to health monitoring.

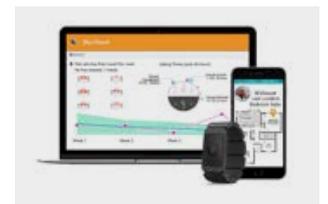

## PILL CLIP

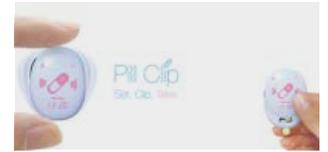

Korean designers Hoon Yoon and Chaemin Ahn created a novel pill-box with a built-in alarm. The product is called Pill Clip and was especially designed for older people who need to take their medication on a regular basis. The device is small and can be easily fastened, in the shirt pocket, like a pin, making it extremely practical and easy to use. The Pill Clip has an LED display that indicates the time and has buttons on the sides so people can program the times when they need to take their medications.

## HIP-HOP SMART WEARABLE

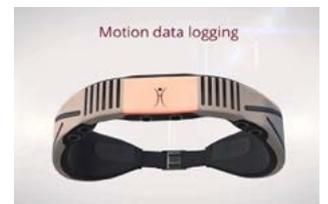

Hip-Hope is a smart wearable device, designed as a belt, worn around the user's waist. A proprietary multi-sensor system detects impending collision with the ground. Upon detection, two large-size airbags instantly inflate and protect the wearer's hips. Fall alert notifications are automatically sent to pre-defined destinations.

## KINESIS QTUG ™

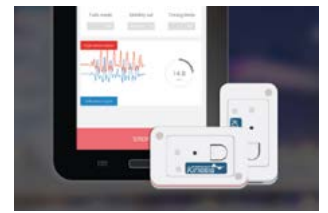

Quantitative Timed Up and Go (QTUG™) provides Falls risk score (a validated risk profile of a patients' future risk of having a fall). QTUG™ also provides an estimate of Frailty. The Falls risk and Frailty scores have been extensively validated by nine years of research. Proven technology - Improving quality of care for older adults.

## ZEMBRO

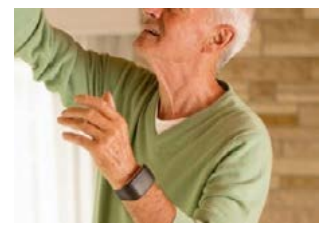

Zembro bracelet is the first intelligent bracelet specially designed for the elderly. It offers the elderly the freedom, independence and peace of mind.

If they press the button once, the clock will appear.

If they press the button twice, they'll see the battery status and other information.

If they hold the button down for four seconds, the bracelet will sound the alarm and inform all their Zembro connections. Zembro sends their connections a message, one by one. They can then telephone the bracelet. You they talk with each other, via the bracelet.

## CONTIN YOU

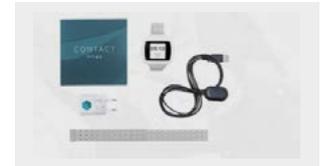

ContinYou developed Contact, a new smartwatch with built-in GSM, ultra-mobile automatic fall alarm and health monitor to help deal with some of the challenges of aging. It enables people to live more independent and active lives at home longer, despite illness or disabilities. Contact features: automatic real time alarms and alerts, predicts incidents, works both indoors and outdoors.

## BIOSERVO SEM ™

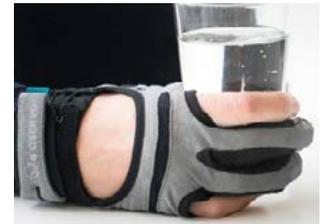

Bioservo Technologies is a world leading company in wearable muscle strengthening systems for people in need of extra strength and endurance. All products and systems are based on our award-winning SEM-technology and we are extremely proud to provide innovative solutions in order to keep people strong, healthy and motivated. The glove from Bioservo improves grip strength and reduces muscle strain for people with weak grip, this is achieved by sensors that detect the user's action and strengthens this action by activating the motors. The company Biosiervo Technology has a total of 10 patents related to the device.

## 4.2 Patents

**Smart furniture**

The following are the results closest to the object of the search, obtained in the Spanish patent database INVENES and the international patent databases Espacenet and Google Patents. The presented patents include only those that are currently "in force" at the time of preparing this report. Many designs and patents exist that provide solutions for the elderly in the domestic habitat and the working environment which have not been included here as they are beyond the scope of this current working group. It is also recognised that multiple interdisciplinary designs exist which cross the three working groups. It is anticipated that these will be further explored in the work of working group four (2019-21) that brings together the three core disciplines of the Action; Furniture and Habitat, IT and Health.

SYSTEM OF HELP AND SECURITY, FOR THE CARE OF OLDER PERSONS, SICK, CHILDREN AND BABIES

The present invention it can be used universally in a variety of **beds, cradles, chairs** and the like, located in homes, hospitals, nursing homes, residences and similar places. The system and method detect when the user has gotten out of bed, cradle, chair or similar and activate a courtesy lighting lamp to avoid falls due to lack of vision. Likewise, it allows to activate a plurality of alarms to warn caregivers and/or relatives when the user is detected to have been up for a long time. In this way, if the user falls when getting up and spends too much time away from the bed, crib, chair or similar, their relatives or caregivers receive a warning signal to go for it and avoid greater problems...

*Publication date: 16-10-2018 || Application number: ES1218994 || Palacios De La Olla, Ricardo*

MATTRESS

The present invention relates to a mattress, both for babies and adults, which has characteristics on the basis of which it allows obtaining data of the sleeping subject in order to be able to perform a remote follow-up, in order to know the comfort characteristics of the user or baby, such as if he has heat, is cold, sweats, or, in the specific case of a baby, if he has urinated. In the mattress there is at least one humidity sensor, a temperature sensor and a pressure sensor associated with a control circuit and this one to a wireless communication module, with means of communication with an application installed in a mobile phone, tablet or similar, through which to interpret said parameters obtained by the sensors...

*Publication date: 07-02-2018 || Application number: ES120 612U || ECUS SLEEP, S.L.U. (ES)*

SYSTEM FOR MONITORING THE
STATE OF A SUBJECT

A system for monitoring the state of a subject, specifically designed for installation in homes or rooms etc., comprising items of furniture with at least one built-in sensor capable of measuring a physical parameter of the subject; a processing unit linked to a piece of software for receiving an electrical signal from at least said sensor and a data communication module for wirelessly transmitting the parameters obtained by at least said sensor. As such, this processing unit generates a signal when it detects an anomalous value or pattern of behaviour linked to at least one parameter received by at least one sensor...

*Publication date: 09-12-2015 || Application number: ES2536826 B1 || CETEM*

MONITORING EQUIPMENT FOR EATING HABITS OF
ELDERLY PEOPLE IN GEROCOMIUM

The invention discloses monitoring equipment for eating habits of elderly people in a gerocomium. A wireless power supply device is arranged in a matching table, and comprises an energy transmitting coil, and the energy transmitting coil is connected with a relay; the relay is connected with a central processor; the central processor is connected with a reed switch and a GSM module; a wireless constant-temperature bowl is placed on the matching table, and comprises a bowl body and a power supply base; the bowl body comprises an inner bowl body and an outer bowl body; a wire placing groove is formed in the outer wall of the inner bowl body; heating wires are arranged in the wire placing groove; the lower ends of the heating wires are connected with an energy receiving coil through a temperature control switch; the temperature control switch and the energy receiving coil are arranged in the power supply base; a magnet is arranged in the power supply base; the central processor is connected with a pyroelectric infrared sensor...

*Publication date: 02-11-2018 || Application number: CN108721030 (A)*
*ZHU JIAYI; ZHOU JINGYA; HU YUQI; LI ANQI; DAI YAN (CN)*

INTELLIGENT MONITORING METHOD
FOR BATHROOM

The invention discloses an intelligent monitoring method and system for a bathroom. The method can comprise the steps of obtaining the distance between a **bathroom door** and a door frame according to a monitoring unit; obtaining a closing time of the bathroom door according to a counting unit; comparing a threshold value of a comparing unit with the closing time, and uploading a comparison result to a central unit; sending an **alarm** signal according to information of the central unit; cutting off the alarm signal through a control unit. The intelligent monitoring method and system for the bathroom have the advantages that through the cooperative use of the monitoring unit and the counting unit, the stay time of the elderly in the bathroom is monitored in real time, an alarm is given out in time when an emergency occurs, and the safety of the elderly living alone is improved…

*Publication date: 14-08-2018 ||Application number: CN108399713 (A)*
*WUHU TAILING INFORMATION TECH CO LTD (CN)*

### INTELLIGENT CHAIR FOR ELDERLY PEOPLE

The invention relates to an intelligent chair for **elderly people**, comprising a chair body convenient for elderly people to sit on and a monitoring **system for monitoring the body states of elderly people**. The intelligent chair for elderly people can be set to various shapes of an L-shaped chair manually by an operator, so that when the operator rises, the L-shaped chair inclines forward to be convenient for the operator to rise, thereby avoiding the phenomenon that elderly people tumble when rising due to insufficient strength of rear legs after long-time sitting. The intelligent chair is more comfortable to sit on when the operator reads book, reads papers or watches television, and makes the operator feel more comfortable during rest, such as doze, afternoon sleep and sleep. Simultaneously, the body state of the operator is monitored by a monitoring module to ensure safety of elderly people in all states. The intelligent chair for elderly people also can analyse and judge the state of the operator's need according to information collected by a control centre, thereby adjusting to a state fitting the operator in cooperation with the L-shaped chair and/or external intelligent household appliances…

*Publication date: 17-05-2017 || Application number: CN106667116 (A)*
*ANJI HUAQI FURNITURE CO LTD (CN)*

### AGED AND FEEBLE PERSON FRIENDLY MULTIMEDIA SYSTEM BASED ON VIRTUAL TOUCH SENSOR

The present invention relates to an elderly-friendly virtual **touch-type bedside multimedia** device and an implementation method for the same, and more specifically, to an image processing technology, which has a multifunctional bedside multimedia device such that an elderly person who is uncomfortable or battling performs emergency call, telephone use, Internet access, interior lighting control, and TV/radio control, has a virtual touch area created in a flat structure part such as a wall surface, and recognizes the motion of the elderly person and analyses depth information to control the bedside multimedia device…

*Publication date: 15-12-2015 || Application number: KR101687150 (B1)*
*DONG-EUI UNIV INDUSTRY-ACADEMIC COOP FOUND (KR)*

DESIGN OF FURNITURE
AND WOOD PRODUCTS

The categories of population that are socially or economically endangered due to different social circumstances, or have some specific inborn or acquired characteristics, are groups that are rarely or not sufficiently being considered when designing furniture. Reasons for this are numerous, and the inevitable impact of a market economy is sometimes crucial. In such circumstances, it is important for the designer to do best to include the needs of some of the vulnerable categories of the population as much as possible, through the design of products for the target groups he/she most often works with. Inclusive products exert far wider and more significant influences, while the designer shows through this a full responsibility for the society and high professional ethics.

The special feature of the **"Up Down" armchair** lies in the fact that its seat, when not bearing weight, takes a position at the height of the armrest, then gradually, with the help of an analogue mechanism  the car trunk supporting pistons, under the weight of the user reaches its standard use elevation. The described mechanism helps older or sick people to take a seat or get up more conveniently.

*MATIĆ J., UNIVERSITY OF BELGRADE, P. 102-105, ISBN 978-86-7299-276-2*

## Sensorization In Home Environment

WIRELESS WARNING THROUGH
ZIGBEE TECHNOLOGY

The present invention refers to a "wireless alarm" which is about a bracelet or anklet in charge of emitting a small vibration and warning by a small LED screen, when one of the following events occurs in the home: Call the doorbell or telephone, call home phone, opening of the main door.

Therefore, our invention consists of 5 main modules, 3 of them emitters (bell module, telephone module and door module), 1 transmitter / receiver (vibrating bracelet) and a data recording module in charge of communicating with a PC and storing the events that happen for a possible consultation, in which the date, the hour and the type of event happened will appear. All this will be done using a wireless technology called Zigbee.

The product is aimed essentially at people with hearing problems of different nature, (including elderly people and people with low levels of care)

*Publication date: 30-09-2014 || UNIVERSIDAD DE VALLADOLID (ES) Application number: ES2421168 B1*

## METHOD, DEVICES AND SYSTEM TO MANAGE DIGITAL ENVIRONMENTS

System consisting of one or several devices by an intelligent algorithm and different databases that is constituted in a digital environment that is managed automatically or through the intervention of the user and several devices. It is an alternative for the elderly who want to live independently and with greater security.

Gateway consisting of a physical device equipped with a series of hardware and software components that provide a new connectivity and social intelligence, capable of collecting information through: 1. A variety of sensors, 2. Learning algorithms installed on the catwalk. 3. Social networks, …

*Publication date: 25-05-2014 || GÓMEZ MAQUEDA, Ignacio (ES)*
*Application number: ES2462566 A1*

## ELECTRIC CARPET DATA ACQUISITION SYSTEM

The invention discloses an electric carpet data acquisition system, and belongs to the field of intelligent acquisition technology. The electric carpet data acquisition system comprises a carpet main body; an internal cavity of the carpet main body is provided with grooves; pressure sensors are uniformly arranged in the internal cavities of the grooves; a single acquisition module is connected with a single identification module through electrical output connection; the single identification module is connected with a single amplifier through electrical output connection; and the single amplifier is connected with an A/D converter through electrical output connection. The electric carpet data acquisition system is reasonable in design, and is convenient for installation; the pressure sensors are arranged in a household carpet, so that walking track of elders can be sensed at any time, the specific position information of elders can be acquired, when an elder stay at a position for a long time, alarming is sent, and information is sent to a service platform through wireless communication, so that a problem that severe results may be caused because of not in time find and help of elders fall at home is solved…

*Publication date: 07-12-2018 || TIANJIN TIANRUI CARPET CO LTD (CN)*
*Application number: CN108961710 (A)*

## METHOD FOR VIDEO MONITORING OF OUT-OF-BED AND ON-BED STATES OF ELDERLY

The invention provides a method for video monitoring of the out-of-bed and in-bed states of the elderly, comprising: a), arranging a camera; b), collecting images of the elderly getting out of the bed to form training and test samples; c), performing normalization and labelling of the images and forming label files; d), training a neural network; e), using the test samples to calculate the accuracy of the deep neural network, if the accuracy is lower than a set threshold, increasing the capacity of the training samples and performing retraining, and if the accuracy is not lower than the threshold, ending training; and f), placing real-time monitoring images into the neural network, and calculating the overlap ratio between the human body and the bed body to determine the elderly is in out-of-bed or in-bed state. The method utilizes the processing technology of the video image to monitor the out-of-bed and in-bed states of the elderly, and triggers alarming when abnormal conditions occur during the process of the elderly getting out of the bed, thereby preventing the elderly from having an accident when getting out of bed for a long time… Publication date: 06-11-2018 || SHANDONG UNIV OF FINANCE AND ECONOMICS; JINAN DONG-SHUO MICROELECTRONIC CO LTD (CN) || Application number: CN108764190 (A)

## Wearable Devices

PORTABLE AND ADAPTIVE MULTIMODAL MONITOR FOR HUMANS, BASED ON A BIOMECHANICAL-PHYSIOLOGICAL AVATAR, FOR DETECTING PHYSICAL RISK EVENTS

The object of the invention is a portable and adaptive multimodal monitoring system for humans, based on biomechanical-physiological avatar, able to detect and warn in real time of physical risk events, which include falls, but are not limited to them. The proposed system belongs to the electronics and new technologies sector, and proposes a solution capable of overcoming the existing problems in the current systems, relating to the reliability of impact detection and at its discretion. Its main application sector is in the social services sector for the **elderly**, people with movement limitations and chronic pathologies.

At least the triaxial accelerometric intelligent **sensor** can be worn without the need for clothes, either fastened to the skin by a hypoallergenic band-aid, on a **bracelet, or subject to an ornament of minimum dimensions**, and its position can be modified on the user's request to adapt to changes of plaster or ornament, adapting automatically and in an unguided way to the new conditions of fixation…

*Publication date: 06-02-2013 || UNIVERSIDAD DE SEVILLA (ES) Application number: ES2394842 A1*

ELDERLY PERSON PHYSICAL HEALTH MONITORING METHOD AND SMART TERMINAL

An elderly person physical health monitoring method and a smart terminal. The method comprises: a smart terminal worn on the body of an elderly person detects physical indicator information of the elderly person via a physical indicator sensor. The smart terminal worn on the old person detects the physical indicator information of the old man through the body indicator sensor. Preferably, the smart terminal pre-sets the medicine that the old person needs to take, and the time of taking the medicine. The smart terminal detects time information, and if the pre-set taking time of taking the medicine is reached, the corresponding medicine taking prompt information is output....

SYSTEM FOR MONITORING PHYSICAL
ACTIVITY IN ELDERLY PERSONS 

The invention relates to a system for monitoring physical activity in elderly persons, which is formed by: a portable monitoring device which allows the accelerations produced by the wearer's movements and activity to be measured at all times; at least one measurement hub which receives the measurements from the portable device; and at least one remote server which receives the measurements transmitted by the hub, records the measurements in a database and analyses them by means of a logic system therein for identifying the user's physical activity patterns...

## **4.3** Research Projects

### **Smart Furniture**

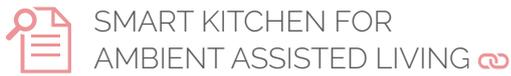

SMART KITCHEN FOR
AMBIENT ASSISTED LIVING

This work presents the design and development of an Intelligent Environment in the kitchen, which helps elderly and/or disabled people to carry out their daily activities in an easier and simpler way. The main fields to be dealt with are: methodological and technological. On the one hand, a systematic methodology is presented to extract the needs of specific groups in order to improve the information available by the design team of the product, service or system. From the technical point of view, the design of an intelligent assisted environment in the kitchen is approached, proposing and defining the architecture of the system. This architecture, based on OSGi (Open Services Gateway initiative), offers a modular system with high interoperability and scalability capabilities. In addition, a network of sensors distributed in the environment is designed and implemented in order to obtain as much information as possible about the context, presenting different algorithms to obtain high level information: fall detection or location. All the devices present in the environment have been modelled using the taxonomy proposed in OSGi4AmI, extending it to the most common household appliances in the kitchen.

*Date: 2013 || Author: Blasco Marín, Rubén (Universidad de Zaragoza)*

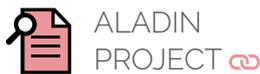

ALADIN
PROJECT

Domalys developed Aladin, an innovative intelligent lamp that prevents falls by monitoring and assisting the elderly. It does so through a combination of State of the Art analogue sensors capable of detecting the slowest movement and a proprietary software, PrediCare, that analyses behavioural patterns to predict future falls.

*Start Date: 1 May 2018. End Date: 31 August 2018*

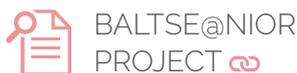

BALTSE@NIOR
PROJECT

Baltse@nior is an Interreg Baltic Sea Region project, with the purpose of supporting and helping furniture and interior manufacturers across the Baltic Sea Region with knowledge, insights, tools and methodology to develop and market products that meet the needs of the growing population of seniors.

*Start Date: May 2016. End Date: April 2019*

## ROOMBOT
## PROJECT

This project funded by the Swiss NCCR in Robotics explores the design and control of modular robots, called Roombots, to be used as building blocks for furniture that moves, self-assembles, and self-reconfigures. Modular robots are robots made of multiple simple robotic modules that can attach and detach connectors between units allow the creation of arbitrary and changing structures depending on the task to be solved. Compared to "monolithic" robots, modular robots offer higher versatility and robustness against failure, as well as the possibility of self-reconfiguration. The type of scenario that we envision is a group of Roombots that autonomously connect to each other to form different types of furniture, e.g. stools, chairs, sofas and tables, depending on user requirements. This furniture will change shape over time (e.g. a stool becoming a chair, a set of chairs becoming a sofa) as well as move using actuated joints to different locations depending on the users' needs. When not needed, the group of modules can create a static structure such as a wall or a box. Our dream is to provide multi-functional modules that are merged with the furniture and that lay users and engineers can combine for multiple applications.

The type of applications of such robots could be multiple: assistive furniture for the elderly or for people with a motor handicap, programmable conference rooms, interactive art, satellite or space station elements, etc. We are particularly interested in the first application, namely providing assistance to the elderly. This work will be done in collaboration with the CHILI lab and with DomoSafety.

Habitat seniors, the first full adaptable accommodation designed by and for FCBA, Sarah Couillaud, Yoann Montenot, Clément Grange. Habitat Seniors is the product of a consortium of French industrial groups, headed up by the FCBA. This is the first co-innovation project for adaptable habitats, designed from the ground up to adapt to seniors' residential needs. With this in mind, the developers, architects, manufacturers and users have all been involved throughout the design process, continually learning and improving in an effort to provide real solutions that meet the growing demand from seniors to stay in their own homes as long as possible. Several products were developed during this program.

New Product : ENTRY & IRONING SPACE Manufacturer: Optimum. Optimum has designed a cabinet which brings together all the functions needed in this space. Equipped with storage space and an ironing board, this keeps everything close at hand so there's no unnecessary movement or actions needed. Here, items "come to the seniors", so the residents don't need to bend down or work to access them - but nor are they ever in the way. A helper can perform these household tasks while still being able to talk to the senior and keep them company. They will also be out of the way if the senior is resting. What's more, it acts as a support to polish shoes, or even to just sit and have a rest thanks to the built-in fold-down seat

New product: WALL CABINET WITH INTEGRATED DOOR & FLOOR Manufacturer: Righini and Optimum. Optimum and Righini filed a joint patent for this wall cabinet, equipped with an integrated door and various different storage spaces. They drew on the skills and expertise of the Blum Company to design their interior solutions. This partition is mobile and can be moved by a professional, adapting the living space as and when necessary. Thanks to the built-in sound-proofing inside this unit, the residents can do two different things in two rooms without disturbing each other. What's more, Gerflor has provided a soft floor which provides protection against falls which is pleasant underfoot, soundless and resistant to punctures and dents from heavy objects, while Evidences Mobilier kitted out the living room with some armchairs featuring removable seats or casters and an integrated walking stick holder, an end table and a light, mobile table which can be moved easily.

The Research Centre of the Creative Furniture Industry (RC31) takes part in research and development projects, while promoting the use of wood as Slovenian sustainable raw material and encouraging development of high added value products, top design and high-quality production. Among other goals, it aims to develop furniture adapted to older adults. RC31 deals with research and development projects, resulting in the product samples, promoting the use of wood as Slovenian sustainable raw material and encouraging development of high added-value-products, top design and high-quality production. The centre will also provide the training and counselling services, including studies and analyses, as well as a variety of services to support the planning and implementation of development investments in enterprises. It will build on the new development approaches, new structures and new methods of production, using eco-technologies, eco-materials, eco transport logistics systems, as well as on education of demanding, sustainable consumer.

**Sensorization In Home Environment**

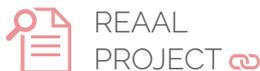

REAAL
PROJECT 🔗

The standard monitoring systems for the elderly in the home are effective in 70% of the cases, but there is a 30% of domestic accidents that still escape the control of the official buttons. For this reason and with the aim of offering a complete solution for the sector, **Ibermática**, a Spanish technology services company, has developed the REAAL project through its i3B Innovation Institute, based on the **Internet of Things and intelligent data analytics**. This initiative allows to monitor the activity of older people through **non-invasive sensors** and analyse the extracted data. Based on this information, and using **Artificial Intelligence** techniques, the system automatically learns the behaviour of the inhabitant, thereby establishing certain patterns that may represent an emergency.

*Date: 2017*

LUCIA
PROJECT

The objective of the "Lucia" project is to sensorize homes of elderly people and / or dependents who live alone so that a system based on artificial intelligence detects patterns of behaviour. It is about creating an AI system that is in continuous learning and is capable of analysing the behaviour patterns of its users, carrying out warning actions when significant alterations of these patterns are detected. Depending on the level of the alert they will be transferred to specialized personnel or to relatives and / or guardians. Initially the system uses 3 different types of sensors: movement, pressure and door opening. These three sensors are decided since the movement will inform us of the degree of activity in the house and the location of the service can be inferred by triangulating the information of these sensors; With the pressure we can detect when he is lying down or sitting in his favourite chair or bed, and with the door opening we can detect when he is away from home and therefore it is not necessary to activate alarms.

*Date: 2018*

DEVELOPMENT OF AN INTERACTIVE SYSTEM OF EVALUATION AND
TREATMENT OF IMPROVEMENT FOR DEPENDENT PEOPLE

Company: Tercera Edad Activa S. L. Programme: CDTI (Centro para el Desarrollo Tecnológico Industrial). Project financed by the FEDER funds of the Community of Madrid

*Date: 2014-2015*

AMI-IOT INTEGRATION MODEL FOR
THE CARE OF THE ELDERLY CARE_HOME16

Elderly people suffer physical and mental deterioration, which halt and/or difficult the control of the homework, loss of independency and autonomy, affecting their quality of life and well-ness. In this project, an integrated AmI-IoT layered model is designed that integrates functionalities of Internet of Thinks (IoT) and Environmental Intelligence (AmI) and provides a reference from monitoring and assistance of elderly people living alone. The model involves three segments responsible for automating housing, supervising the user, making decisions, monito-ring events, identifying habits and accessing AmI and IoT services.

*Date: 2017*

ACTIVAGE PROJECT. ACTIVATING INNOVATIVE IOT
SMART LIVING ENVIRONMENTS FOR AGEING WELL

The main objective is to build the first European IoT ecosystem across 9 Deployment Sites (DS)

in seven European countries, reusing and scaling up underlying open and proprietary IoT platforms, technologies and standards, and integrating new interfaces needed to provide interoperability across these heterogeneous platforms, that will enable the deployment and operation at large scale of Active & Healthy Ageing IoT based solutions and services, supporting and extending the independent living of older adults in their living environments, and responding to real needs of caregivers, service providers and public authorities...

*Start date 1 January 2016 End date 30 Jun 2020 || Coordinator: MEDTRONIC IBERICA SA (Spain)*

LISTENT PROJECT. HANDS-FREE VOICE-ENABLED INTERFACE TO WEB APPLICATIONS FOR SMART HOME ENVIRONMENTS

Nowadays, it is becoming increasingly affordable to enhance the home environment with several automation schemes, allowing remote control of e.g., heating/cooling, communication, lighting, media, etc. Such smart home functionalities are essential for people with disabilities and the elderly, as they not only provide assistive control of important everyday functionalities, but may prove to be life-saving in case of emergency. However, smart home functionalities may become useless for people who need those most, if they cannot be accessed via a natural, easy to use, interface. The central objective of LISTEN is to design and implement a complete system, including both the software and hardware components, enabling robust hands-free large-vocabulary voice-based access to Internet applications in smart homes. This would allow the users to have natural control (i.e., using their voice) of the smart-home web-enabled functionalities (e.g., turning on/off web-enabled "smart" appliances), but also to access specific Internet applications (e.g., web search, email dictation, access to social networks).

*Start date 1 June 2015 End date 31 May 2019*
*Coordinator: FOUNDATION FOR RESEARCH AND TECHNOLOGY HELLAS (Greece)*

**Wearable Devices**

INSTINTO
PROJECT

Research and development of an intelligent system based on integrated sensors and actuators in textiles applied to the prevention, detection and protection against falls of elderly people.

The main objective of the INSTINTO project is to develop integration methods and innovative design solutions, aimed at the integration of sensors in textiles, for their application in the field of health, specifically in prevention, detection and protection in falls in elderly people. The activities developed during the project are aimed at the generation of new knowledge through non-economic R & D activities, to improve the competitiveness of the centres involved and promote the development and strengthening of the competitive and innovative capacity of companies, by applying the knowledge generated during the investigation.

*Date: 2016 || Coordinator: AITEX and IBV. Project financed by the FEDER funds of the Valencian Community*

## INVIS CARE PROJECT. BRINGING CONNECTED TO THE ACTIVE ELDERLY 

At Invis, we have developed an advanced wearable technology that brings connectivity to the classic watch industry, while leaving traditional design without any changes. With our connected strap end users can still keep their watches that have a sentimental value for them and use new technology. Invis is a connected implants system, which brings smart functionalities without the need to invest in a traditional production process - one of the key competitive advantages. Aiming to become a leading smart OEM supplier for classic watch manufacturers we are currently evolving our solution addressing specific needs of the elderly and targeting the potential market at our reach worth over €90M

*Start date 1 June 2018 end date 31 August 2018 || Coordinator: INVIS SP ZOO (Poland)*

## KYTERA COMPANION. DISRUPTING THE ELDERLY HOME CARE WITH CONTEXTUAL ANALYSIS SOFTWARE 

KYTERA COMPANION, is a pioneering monitoring solution, consists of a wearable wristband, a mesh-network sensor system and a cloud-based software with a mobile app. Our technology can detect up to 95% of both soft and hard falls, through analysing the elders' activity patterns and deducing the activity context. Moreover, our solution provides accurate well-being (hygiene, nutrition, sleep) and location (based on full indoors positioning technology and advanced posture analysis) information, achieving almost zero false alarms. On top of that, we enable collateral financial savings compared to Residential Care Facilities for Elderly (can be up to €1,600/month in countries like the uk or France).

Kytera Technologies is an expert technology SME, consisting of experts with long-term experience from wireless communication industry.

*Start date 1 February 2018 end date 31 May 2018 || Coordinator: KYTERA TECHNOLOGIES LTD (Israel)*

## WISSEL PROJECT. WIRELESS INSOLE FOR INDEPENDENT AND SAFE ELDERLY LIVING 

WIISEL's main goal is to develop an unobtrusive, self-learning and wearable prevention and warning system to decrease the incidence of falls in the elderly population. The idea is that elderly people put a specifically designed insole in their shoes, which monitors their walking pattern. This way the system may detect changes in gait and balance in the daily environment of the older person, which helps predict the likelihood of falling.

*Start date 1 November 2011 end date 31 March 2015 || Coordinator: Eurecat (Spain)*

## Architectural Solutions

New product: ILLUMINATED ACCESS DOOR Manufacturer: Righini and SFL, lighting specialists. Righini and SFL, lighting specialists, put their expertise to work for the Habitat Seniors project and joined forces to develop a door integrated into the removable wall-cabinet, co-patented by Optimum and Righini and fitted with a low-energy lighting system. The LED lights show the senior where the door is, reducing the risk of falls and knocks. It runs off its own battery which recharges automatically when the door is closed, creating a lighting solution in the event of a power cut. This core interior design feature is both reassuring and decorative helps preserve a senior's independence - even if they are visually impaired.

International project HELPS, with the goal of supporting the elderly in regard to their habitats, resulted in a  brochure explaining the effective adaptations of indoor environments to the older adults.

Staff and students of the Slovenian Faculty of Design visited older adults at their homes to assess which adaptations would be the most useful to them. It was discovered that most adaptations are needed in bathrooms and kitchens. Computer visualizations of both minor (but simple and inexpensive) and substantial indoor adaptations were prepared for older adults, who were satisfied with the project and wished for these services to be available to them in the future. Following the project, a manual was prepared for the older adults, which presents common problems and adaptations of living environments

### 4.4 Scientific Publications

Next, the scientific publications related to the Project objective.

**Smart Furniture**

SMART SENSORY FURNITURE BASED ON WSN
FOR AMBIENT ASSISTED LIVING 

Ubiquitous computing has been defined as "machines that fit the human environment instead of forcing humans to enter theirs." An example of this type of approach is "Smart Sensory Furniture" (SSF) Project. SSF is an ambient assisted living system that allows inferring a potential dangerous action of an elderly person living alone at home. This inference is obtained by a specific sensory layer with sensor nodes fixed into furniture and a reasoning layer embedded in a PC that learns from the users' behavioural patterns and advices when the system detects unusual patterns. This paper aims to explain the SSF sensory layer, which is a distributed signal processing system in a network of sensing objects massively distributed, physically coupled, wirelessly networked, and energy limited. A complete set of experimental tests has been carried out. The results show the level of accuracy for each type of sensors and potential use. Finally, the power consumption was experimentally measured and the results show the low maintenance requirements of this solution. The complete system design is described and discussed, including the node mesh details, as well as the type of sensors and actuators and other aspects, such as integration issues and solutions. Start date 1 November 2011 end date 31 March 2015 || Coordinator: Eurecat (Spain).

TECHNOLOGY AVAILABLE TO THE ELDERLY: INTELLIGENT FURNITURE
DALILA MANUFACTURING PROCESS FOR THE ELDER USER 

Dalila is an aerial kitchen furniture which target is to make storage and everyday activities easier for an elder user. This product takes advantage of technology to create a safe environment where the user will be able to develop her or himself, using alerts perceived by all senses. The technology used in this product is simple and barely perceptive, therefore the product is quite intuitive. This article shows the design and building process for a smart furniture that makes interaction between the user and the kitchen easier, a product that won't limit the user's possibilities, unlike it will create a safe and danger free environment.

## ANALYSIS OF INTELLIGENT FURNITURE DESIGN FOR THE ELDERLY IN INTERNET OF THINGS ERA

This paper puts forward the design principle of intelligent furniture for the elderly based on big data and intelligent technology. The present situation and demand of intelligent furniture for elderly people are analyzed. Summed up the design method of intelligent furniture for elderly people.

## RESEARCH ON SMART HOME FOR THE ELDERLY

With the development of Internet of Things technology, smart home has become a hot topic in the society today. However, there are few smart homes designed for the elderly. Based on China's national conditions, this report analyzes the development of related technologies at home and abroad. Then, the report studies the characteristics and needs of the elderly. Finally, based on this, this report summarizes and puts forward some functions that intelligent furniture suitable for the elderly should have.

## SMART TABLE SURFACE: A NOVEL APPROACH TO PERVASIVE DINING MONITORING

We present a novel sensor system for the support of nutrition monitoring. The system is based on smart table cloth equipped with a fine grained pressure textile matrix and a weight sensitive tablet. Unlike many other nutrition monitoring approaches, our system is unobtrusive, non-privacy invasive and easily deployable in everyday life. It allows the spotting and recognition of food intake related actions, such as cutting, scooping, stirring, etc., the identification of the plate/container on which the action is executed, and the tracking of the weight change in the containers. In other words, we can determine how many pieces are cut on the main dish plate, how many are taken from the side dish, how many sips are taken from the drink, how fast the food is being consumed and how much weight is taken overall. In addition, the distinction between different eating actions, such as cutting, scooping, poking, provides clues to the type of food taken and the way the meal is consumed. We have evaluated our system on 40 meals (5 subjects) in a real-life living environment: for seven eating related actions (cutting, scooping, stirring, etc.), resulting in above 90% average recognition rate for person dependent cases, and spotting each action out of continuous data streams (average F1 score 87%).

## THE NETWORKED, ROBOTIC HOME+ FURNITURE SUITE: A DISTRIBUTED, ASSISTIVE TECHNOLOGY FACILITATING AGING IN PLACE.

We introduce and detail a novel, networked and interoperative suite of robotic furniture. This suite forms a key part of our development of home+, an assistive technology environment aimed at supporting aging in place. This paper elaborates the design and construction process for the three robotic furniture core elements of home+: a chair, featuring gesture-controlled assistive lift; a morphing side table; and an adaptive screen. The sensor suite, networking, and user interface for the system is described and discussed. We report on initial experiments with senior citizens using the system.

## THE DEVELOPMENT OF A SMART CHAIR TO ASSIST SIT-TO-STAND TRANSFERRING PROCESS.G

Standing up from a seated position, known as sit-to-stand (STS) movement, is one of the activities of daily living (ADLs) on a daily basis. As people age, physiological changes occur including reduced muscle strength and mass as well as sensory capacity. This may lead to difficulties in STS transferring process, with which the elderly may encounter sedentary lifestyle and contracted social space. There exist market available assistive lift devices with performance far from satisfaction, for the reason being that they fail to provide appropriate assistance. Thus, an intent-based smart lift chair is proposed and partially developed aiming to analyse user's physiological condition through pattern recognition.

## STASIDIA (CHURCH STALLS) OF THE GREEK ORTHODOX CHURCH – A STANDING SEAT FOR ELDERLY

In Greek Orthodox Church and the liturgy taking place there, the worshipers tend to stay up standing most of the time. Especially in monasteries, the elderly people (monks) need to be facilitated in attending the liturgies of long duration and therefore, a series of chairs of special type called stasidia or stalls, are placed in the church, between the columns of the aisles. The design of this ecclesiastical furniture is unique and appears only in Orthodox Church. Stasidia are made of several wood species and constructed in such a way, that the elderly people are supported by the backrest and arms of the furniture. Initially, the stasidi was stable, whereas subsequently a moveable wooden element (misericord) was added in the structure, in order to offer a slight support to standing worshipers. The aim of this furniture construction is more to facilitate and support the people standing, rather than people seating during the worship or prayer. As the decades passed, this kind of ecclesiastical chair was transferred and spread from the monasteries to the rest of churches and the stasidi from a furniture of simple design and construction, of plain assembly of wooden elements, transformed into a furniture of more complicated design, more ornamental, that now brings several furnishings and decorative carvings in the surfaces of back and arms. The main goal of this work is to examine the evolution of architecture and designs of stasidia through the centuries and the study of designs and manufacturing processes of stasidia today.

*Ioannis BARBOUTIS, Vasileios VASILEIOU Faculty of Forestry and Natural Environment, Aristotle University of Thessaloniki. Proceedings of the XXVIth International Conference Research for Furniture Industry22*

## CRITICAL POINTS IN THE CONSTRUCTION OF AGED PEOPLE FURNITURE

Even though the economic crisis has influenced the shopping behaviour of all people, elderly between 65 and 69 still consume and spend more money on furniture than younger people. Despite of this, furniture manufacturers seem to lack knowledge about the diverse needs of aged people, referring to furniture, mainly because of a poor communication between furniture industry and its end-users. In this study an attempt to approach and categorize the requirements and needs of aged people using furniture in their everyday life is implemented. The study used approximately 100 interviews with elderly people that were carried out on a continual process that lasted around 1.5 years in several furniture stores of Thessaloniki (North Greece), gathering data and information in a frame of conversations, trying to reach a deeper understanding of their habits, needs and requirements related to furniture. The results, were also correlated to corresponding literature, and revealed common wishes and needs for furniture that provide comfort, safety, functionality, pleasure and independence, since furniture could help elderly people continue to be active and self autonomous.

*Vasiliki KAMPERIDOU. ISSN 2069-7430 ISSN-L 1841-4737*

## MARKET POTENTIAL AND DETERMINANTS FOR ECO-SMART FURNITURE ATTENDING CONSUMERS OF THE THIRD AGE 🔗

This paper aims to discuss some of the findings of an ongoing "Green & Smart Furniture" (GSF) research project. It actually focuses on third-age consumers' behavior and interest in purchasing smart and eco-friendly wooden furniture, providing some critical implications for the successful design and production of GSF products in the framework of innovation and differentiation.

## ERGONOMIC SUITABILITY OF KITCHEN FURNITURE REGARDING HEIGHT ACCESSIBILITY

The study, carried out using a computer simulation model, analyzed the products of three Slovenian kitchen manufacturers. The cross section of accessibility in the wall cabinets was determined for different age groups of men and women. The results show that the efficacy of the volume in wall cabinets higher than 600 mm, in comparison to places where objects are easily reachable, is 30% lower for women, thus indicating the inefficiency of storage space in wall cabinets. In terms of accessibility, existing kitchens are not optimal for the elderly, and a model with a deeper worktop and wall cabinets lowered onto the worktop is proposed. Accessibility in such wall cabinets is increased by up to 70% if the body is moved forward by 30°.

*Jasna Hrovatin, Silvana Prekrat, Leon Oblak, & David Ravnik, (2015)*

## ADAPTABILITY OF KITCHEN FURNITURE FOR ELDERLY PEOPLE IN TERMS OF SAFETY

This research was carried out via individual surveys at the respondents' homes. 204 respondents participated in the research. The results show that most users do not realize that, with more appropriate kitchen equipment, they could perform daily tasks faster, safer, and with less effort. Common shortcomings include insufficient lighting (32 %), inappropriate sequential composition of work surfaces (56 %), ease of hygiene maintenance (68 %), inappropriately - shaped furniture (72 %), and tasks that become troublesome because of declining memory (75 %). It is necessary to design kitchen equipment specifically adjusted for the needs of the elderly.

*Jasna Hrovatin, Kaja Širok, Simona Jevšnik, Leon Oblak, & Jordan Berginc, (2012)*

### Sensorization In Home Environment

## AMBIENT INTELLIGENCE ENVIRONMENTS WITH WIRELESS SENSOR NETWORKS FROM THE POINT OF VIEW OF BIG DATA AND SMART & SUSTAINABLE CITIES

In the context of ambient-assisted living and monitoring activities and behaviours, the concept of smart homes/environments has emerged as a research and development field with many examples appearing in a range of different contexts (universities, research centres, hospitals, residences, etc.). Recently, a Smart Lab called UJAmI based on Ambient Intelligence was created at the University of Jaén in order to provide an environment where solutions in assistive technologies could be developed. This paper presents the experience of developing this smart lab within its first year. The initial space assigned to the smart lab is described through the creation of the smart lab in five perspectives: the layout of the smart lab, the selection of the devices, the deployment of the middleware, the technical infrastructures and the furniture. Finally, following our experience of creating the lab we reflect upon our experience and provide a set of guidelines and recommendations. Ongoing projects within this Smart lab are also in-

## RADAR SENSOR SYSTEM FOR MONITORING ELDERLY PEOPLE AT HOME 🔗

Considering the aging society tendency that the whole world is experiencing, the percentage of elderly people will incrementally represent a larger share of the total population. Older adults are more prone to develop health problems. For them to be successfully treated, early detection is crucial. So as to facilitate this, monitoring them will make for a remarkable improvement. A radar sensor which enables this possibility is what this thesis is about. Starting from an already developed project, several parts have been modified to make the system more efficient. The contributions implemented in the practical work of the thesis are about the modification of the functioning of the system, adding a more secure way of storing the data by introducing a Raspberry Pi for such purposes. Moreover, a PIR sensor, which is used to detect movement, is also added to the structure providing several benefits to improve the performance of the whole system. Last but not least, since this project affects the privacy of its users, an ethical discussion has been carried out to assure its viability.

*UPCommons. Portal de acceso abierto al conocimiento de la Universidad Politécnica de Cataluña, España. Trabajo final de grado 2018-01. Catalán Mor, Joan Bru*

## SMART SENSORY FURNITURE BASED ON WSN FOR AMBIENT ASSISTED LIVING 🔗

Ubiquitous computing has been defined as "machines that fit the human environment instead of forcing humans to enter theirs." An example of this type of approach is "Smart Sensory Furniture" (SSF) Project. SSF is an ambient assisted living system that allows inferring a potential dangerous action of an elderly person living alone at home. This inference is obtained by a specific sensory layer with sensor nodes fixed into furniture and a reasoning layer embedded in a PC that learns from the users' behavioural patterns and advices when the system detects unusual patterns. This paper aims to explain the SSF sensory layer, which is a distributed signal processing system in a network of sensing objects massively distributed, physically coupled, wirelessly networked, and energy limited. A complete set of experimental test has been carried out. The results show the level of accuracy for each type of sensors and potential use. Finally, the power consumption was experimentally measured and the results show the low maintenance requirements of this solution. The complete system design is described and discussed, including the node mesh details, as well as the type of sensors and actuators and other aspects, such as integration issues and solutions.

*IEEE Sensors Journal ( Volume: 17 , Issue: 17 , Sept.1, 1 2017)pp. 5626 – 5636. (29 June 2017). Andres L. Bleda; Francisco J. Fernández-Luque; Antonio Rosa; Juan Zapata; Rafael Maestre.*

## A SMART KITCHEN FOR AMBIENT ASSISTED LIVING 

The kitchen environment is one of the scenarios in the home where users can benefit from Ambient Assisted Living (AAL) applications. Moreover, it is the place where old people suffer from most domestic injuries. This paper presents a novel design, implementation and assessment of a Smart Kitchen which provides Ambient Assisted Living services; a smart environment that increases elderly and disabled people's autonomy in their kitchen-related activities through context and user awareness, appropriate user interaction and artificial intelligence. It is based on a modular architecture which integrates a wide variety of home technology (household appliances, sensors, user interfaces, etc.) and associated communication standards and media (power line, radio frequency, infrared and cabled). Its software architecture is based on the Open Services Gateway initiative (OSGi), which allows building a complex system composed of small modules, each one providing the specific functionalities required, and can be easily scaled to meet our needs. The system has been evaluated by a large number of real users (63) and carers (31) in two living labs in Spain and UK. Results show a large potential of system functionalities combined with good usability and physical, sensory and cognitive accessibility.

## INTEGRATION OF MULTISENSOR HYBRID REASONERS TO SUPPORT PERSONAL AUTONOMY IN THE SMART HOME

The deployment of the Ambient Intelligence (AmI) paradigm requires designing and integrating user-centered smart environments to assist people in their daily life activities. This research paper details an integration and validation of multiple heterogeneous sensors with hybrid reasoners that support decision making in order to monitor personal and environmental data at a smart home in a private way. The results innovate on knowledge-based platforms, distributed sensors, connected objects, accessibility and authentication methods to promote independent living for elderly people. TALISMAN+, the AmI framework deployed, integrates four subsystems in the smart home: (i) a mobile biomedical telemonitoring platform to provide elderly patients with continuous disease management; (ii) an integration middleware that allows context capture from heterogeneous sensors to program environment's reaction; (iii) a vision system for intelligent monitoring of daily activities in the home; and (iv) an ontology-based integrated reasoning platform to trigger local actions and manage private information in the smart home. The framework was integrated in two real running environments, the UPM Accessible Digital Home and MetalTIC house, and successfully validated by five experts in home care, elderly people and personal autonomy.

### AMI AND DEPLOYMENT CONSIDERATIONS IN AAL SERVICES PROVISION FOR ELDERLY INDEPENDENT LIVING: THE MonAMI PROJECT

We present a novel sensor system for the support of nutrition monitoring. The system is based on scooping, stirring, etc.), resulting in above 90% average recognition rate for person dependent cases, and spotting each action out of continuous data streams (average F1 score 87%).

### WIRELESS SENSOR NETWORK DEPLOYMENT FOR REMOTE ELDERLY CARE MONITORING

This paper reports hands-on experiences in designing, implementing and operating a wireless sensor network (WSN)-based prototype system for elderly care monitoring in home environments. The monitoring is based on the recording of environmental parameters like temperature, humidity and light intensity as well as micro-level incidents which allow to infer daily activities like moving, sitting, sleeping, usage of electricity appliances and plumbing components. The prototype is built upon inexpensive, of-the-shelf hardware (e.g. various sensors, Arduino micro-controller platforms, ZigBee-compatible wireless communication modules) and license-free software, thereby ensuring low system deployment cost. Upon detecting significant deviations from the ordinary activity pattern of individuals and/or sudden falls, the system issues automated alarms which may be forwarded to authorized persons via a variety of communication channels. Furthermore, measured environmental parameters and activity incidents may be monitored through web interfaces.

### AMBIENT ASSISTED LIVING, ITALIAN FORUM (2017)

This book documents the state of the art in the field of ambient assisted living (AAL), highlighting the impressive potential of novel methodologies and technologies to enhance well-being and promote active ageing. The coverage is wide ranging, with sections on assistive devices, elderly people monitoring, home rehabilitation, ICT solutions for AAL, living with chronic conditions, robotic assistance for the elderly, sensing technologies for AAL, and smart housing. The book comprises a selection of the best papers presented at the 7th Italian Forum on Ambient Assisted Living (ForitAAL 2016), which was held in Pisa, Italy, in June 2016 and brought together end users, technology teams, and policy makers to develop a consensus on how to improve provision for elderly and impaired people. Readers will find that the expert contributions offer clear insights into the ways in which the most recent exciti ng advances may be expected to assist in addressing the needs of the elderly and those with chronic conditions.

## INTERNET OF THINGS FOR MEASURING HUMAN ACTIVITIES IN AMBIENT ASSISTED LIVING AND E-HEALTH 

Internet of Things (IoT) is a new paradigm that combines several technologies such as computers, Internet, sensor networks, radio frequency identification (RFID), communication technology and embedded systems to form a system that links the real world with digital world. Currently, a large number of smart objects and different type of devices are interconnected and more and more they are being used in Ambient Assisted Living (AAL) scenarios for improving the daily tasks of elderly and disabled people. This paper presents an IoT architecture and protocol for Ambient Assisted Living and e-health. It is designed for heterogeneous AAL and e-health scenarios where an IoT network is the most suitable option to interconnect all elements. Finally, we simulate a medium-size network with four protocols especially designed for networks with important energy constraints in order to show their performance...

## INTEGRATION OF MULTISENSOR HYBRID REASONERS TO SUPPORT PERSONAL AUTONOMY IN THE SMART HOME 

The deployment of the Ambient Intelligence (AmI) paradigm requires designing and integrating user-centered smart environments to assist people in their daily life activities. This research paper details an integration and validation of multiple heterogeneous sensors with hybrid reasoners that support decision making in order to monitor personal and environmental data at a smart home in a private way. The results innovate on knowledge-based platforms, distributed sensors, connected objects, accessibility and authentication methods to promote independent living for elderly people. TALISMAN+, the AmI framework deployed, integrates four subsystems in the smart home. The framework was integrated in two real running environments, the UPM Accessible Digital Home and MetalTIC house, and successfully validated by five experts in home care, elderly people and personal autonomy...

## DELIVERING HOME HEALTHCARE THROUGH A CLOUD-BASED SMART HOME ENVIRONMENT (COSHE) 

The dramatic increase of senior population worldwide is challenging the existing healthcare and support systems. Recently, smart home environments are utilized for ubiquitous health monitoring, allowing patients to stay at the comfort of their homes. In this paper we presented a Cloud-Based Smart Home Environment (CoSHE) for home healthcare. CoSHE collects physiological, motion and audio signals through non-invasive wearable sensors and provides contextual information in terms of the resident's daily activity and location in the home. This enables healthcare professionals to study daily activities, behavioral changes and monitor rehabilitation and recovery processes. A smart home environment is set up with environmental sensors to provide contextual information. The sensor data are processed in a smart home gateway and sent to a private cloud, which provides real time data access for remote caregivers. Our case study shows that we can successfully integrate contextual information to health data and this comprehensive information can help better understand caretaker's health status...

### AMBIENT SENSORS FOR ELDERLY CARE AND INDEPENDENT LIVING: A SURVEY 🔗

Elderly care at home is a matter of great concern if the elderly live alone, since unforeseen circumstances might occur that affect their well-being. Technologies that assist the elderly in independent living are essential for enhancing care in a cost-effective and reliable manner. Elderly care applications often demand real-time observation of the environment and the resident's activities using an event-driven system. As an emerging area of research and development, it is necessary to explore the approaches of the elderly care system in the literature to identify current practices for future research directions. Therefore, this work is aimed at a comprehensive survey of non-wearable (i.e., ambient) sensors for various elderly care systems. This research work is an effort to obtain insight into different types of ambient-sensor-based elderly monitoring technologies in the home. With the aim of adopting these technologies, research works, and their outcomes are reported. Publications have been included in this survey if they reported mostly ambient sensor-based monitoring technologies that detect elderly events (e.g., activities of daily living and falls) with the aim of facilitating independent living...

### SMART HOMES FOR ELDERLY HEALTHCARE 🔗

Advancements in medical science and technology, medicine and public health coupled with increased consciousness about nutrition and environmental and personal hygiene have paved the way for the dramatic increase in life expectancy globally in the past several decades. However, increased life expectancy has given rise to an increasing aging population, thus jeopardizing the socio-economic structure of many countries in terms of costs associated with elderly healthcare and wellbeing. Smart homes, which incorporate environmental and wearable medical sensors, actuators, and modern communication and information technologies, can enable continuous and remote monitoring of elderly health and wellbeing at a low cost. Smart homes may allow the elderly to stay in their comfortable home environments instead of expensive and limited healthcare facilities. Healthcare personnel can also keep track of the overall health condition of the elderly in real-time and provide feedback and support from distant facilities...

## WIRELESS SENSOR NETWORK DEPLOYMENT FOR REMOTE ELDERLY CARE MONITORING 

This paper reports hands-on experiences in designing, implementing and operating a wireless sensor network (WSN)-based prototype system for elderly care monitoring in home environments. The monitoring is based on the recording of environmental parameters like temperature, humidity and light intensity as well as micro-level incidents which allow to infer daily activities like moving, sitting, sleeping, usage of electricity appliances and plumbing components. The prototype is built upon inexpensive, of-the-shelf hardware (e.g. various sensors, Arduino micro-controller platforms, ZigBee-compatible wireless communication modules) and license-free software, thereby ensuring low system deployment cost. Upon detecting significant deviations from the ordinary activity pattern of individuals and/or sudden falls, the system issues automated alarms which may be forwarded to authorized persons via a variety of communication channels. Furthermore, measured environmental parameters and activity incidents may be monitored through web interfaces...

## A MULTI-SENSOR FUSION SCHEME TO INCREASE LIFE AUTONOMY OF ELDERLY PEOPLE WITH COGNITIVE PROBLEMs 

Elderly people care is a major challenge for the smart-cities of future. This represents a valuable opportunity to develop scalable applications to cover the special needs in terms of health monitoring and accessibility for people with cognitive impairments. In this paper, a complete system to support daily activities of elderly people based on a multi-sensor scheme is presented. This system is intended to be deployed not only at home, but also at crowded places, such as daily care centers. A multi-layer architecture is drawn to ensure system modularity and interoperability of heterogeneous data with concurrent services. The proposed system includes a set of algorithms for data gathering and processing to detect abnormal events in the considered scenarios. The experiments performed in real scenarios have led to a good performance of the algorithms proposed as well as high accuracy in event detection for both environments...

## RADAR SENSOR SYSTEM FOR MONITORING ELDERLY PEOPLE AT HOME 🔗

Considering the aging society tendency that the whole world is experiencing, the percentage of elderly people will incrementally represent a larger share of the total population. Older adults are more prone to develop health problems. For them to be successfully treated, early detection is crucial. So as to facilitate this, monitoring them will make for a remarkable improvement. A radar sensor which enables this possibility is what this thesis is about. Starting from an already developed project, several parts have been modified to make the system more efficient. The contributions implemented in the practical work of the thesis are about the modification of the functioning of the system, adding a more secure way of storing the data by introducing a Raspberry Pi for such purposes. Moreover, a PIR sensor, which is used to detect movement, is also added to the structure providing several benefits to improve the performance of the whole system. Last but not least, since this project affects the privacy of its users, an ethical discussion has been carried out to assure its viability. Monitoring the well-being of elderly people at home by means of body and wall sensors.

## Wearable Devices

## DESIGN AND EVALUATION OF A SMART INSOLE: APPLICATION FOR CONTINUOUS MONITORING OF FRAIL PEOPLE AT HOME 🔗

The objectives of this work are to develop a technological solution designed to support active aging of frail older individuals and to conduct a first evaluation of the devices. We wish to bring a reflection in the field of connected health by setting up a remote medical follow-up. In this context, the connected object presented in this article aims at implementation a longitudinal follow-up of the walk by a health professional. Continuous remote data analysis applies behavior learning methods by modelling walking habits and allows the detection of deviations by application of thresholds defined by the expert. We propose an instrumented shoe insole to provide such monitoring (number of steps, distance covered and gait speed)…

## A HIGH RELIABILITY WEARABLE DEVICE FOR ELDERLY FALL DETECTION

Falls are critical events among elderly people that requires timely rescue. In this paper, we propose a fall detection system consisting of an inertial unit that includes triaxial accelerometer, gyroscope, and magnetometer with efficient data fusion and fall detection algorithms. Starting from the raw data, the implemented orientation filter provides the correct orientation of the subject in terms of yaw, pitch, and roll angles. The system is tested according to experimental protocols, engaging volunteers who performed simulated falls, simulated falls with recovery, and activities of daily living. By placing our wearable sensor on the waist of the subject, the unit is able to achieve fall detection performance above those of similar systems proposed in literature. The results obtained through commonly adopted protocols show excellent accuracy, sensitivity and specificity, improving the results of other techniques proposed in the literature.

## Architectural Solutions

### THE IMPORTANCE OF ASSESSING THE ENVIRONMENT TO DETERMINE THE MOST SUITABLE LOCATION FOR SENIORS' RESIDENCES SO AS TO ADJUST THE SUPPLY TO THE NEEDS

Seniors' residences must be located in areas that foster the well-being of both residents and staff. This study is unprecedented in France. It uses the multi-criteria decision-making method to classify the environmental targets in priority order, according to the importance they are given by the people living or working on the premises as related to their proximity to the residence. The results are then integrated into a mapping of areas in which the targets are geolocalized, thus highlighting the most suitable zones. The data collected from interviews and from the mapping vary from one residence to another. Nevertheless they all clearly point to the importance of a territorial approach before planning the building of such residences.

### ENVIRONMENT AND LIVING CONDITIONS OF OLDER HOME OWNER

The aim of this paper is to present a general descriptive account of the environment and living conditions of home owners aged 55 years and above using data on the opinions of households in the French Housing Survey, 2013 (Enquête Logement 2013, Insee). Four groups of home owners are identified according to the degree of satisfaction expressed with the quality of their environment and housing: the first group (46%) are largely satisfied with the quality of their neighbourhood and housing (G1. High quality of environment and housing); the second group (35%) are satisfied with their environment but much less so with their housing (G2. Calmness and rusticity); in the third group (10%), the convenience of housing is mitigated by a low environmental quality (G3. Interior comfort, poor quality of environment); finally, the home owners in the fourth group (9%) are much less satisfied with their housing and their environment (G4. Deteriorated living conditions). The intersection of the expressed opinions with resources reveal strong social inequalities between home owners: on the one hand, the higher social class households in Groups 1 and 3 and on the other hand, low incomes for the households of farmers, manual workers and employees in Groups 2 and 4. The paper concludes with a discussion on home adaptations for older home owners on lower incomes living in rural areas (G2) or in urban areas (G4).

*Gérontologie et société, 2017/1 (vol. 39 / n° 152) S Renaut, J Ogg, A Chamahian, S Petite.*

### THE "MONO-POLAR" YOUNG RETIREES AND REORGANISATION OF DOMESTIC SETTINGS

The transition to retirement can result in a new rapport with domestic space, as is particularly the case with the retirees in this study who were engaged in voluntary work or in work. Indeed, retirement deprives the individual of a public space outside of the home confines him to the private domestic sphere that he will progressively need to redefine. This article helps us to understand the evolution of refurbishments of domestic spaces at the time of retirement: what types of space younger retirees are deprived of? What spaces do they re-invest? What are the processes involved in the redistribution of spaces within the domestic setting, as well as with the exterior environment? We hypothesise that reorganisation of domestic spaces are influenced by reorganisation of lifestyle of recently retired people. Thus, in this paper two areas of activity are privileged: voluntary work and remunerated work of retired people, which can affect positively or negatively accommodation of domestic settings.

*Gérontologie et société, 2017/1 (vol. 39 / n° 152). Melissa petit.*

### CHOOSING TO LIVE IN SHELTERED HOUSING: BETWEEN INDIVIDUAL TRAJECTORIES AND PUBLIC POLICIES

This article proposes creating a connection between the residential process and public policies regarding collective housing such as sheltered housing. It examines the individual choices related to accessing collective housing for elderly people. This allows analysing the potential impacts of public policies regarding sheltered housing on elderly people's trajectories. After a short introduction of sheltered housing and related public policies, the article defines the reasons for joining these establishments and then the limits of that offer characterised by blockages of access. These elements are seen in the context of regulatory obligations regarding public policies on sheltered housing and the recent evolution of French legislation. The results are based on a collection of qualitative data gathered during a PhD research. The article is mainly based on interviews conducted with elderly people and managers in the establishments as well as providing an analysis of legal and operational documentation.

*Gérontologie et société, 2017/1 (vol. 39 / n° 152). AB Simzac*

### DÉPRISE FROM THE PRISM OF INTERMEDIARY HOUSING FOR THE ELDERLY

For elderly people, residential mobility can trigger considerable consequences, especially if it happens in tandem with events linked to individual ageing. This can be it is observed through the process of moving into intermediary housing (service-apartments cum residences for senior citizens, sheltered housing accommodation and other forms) which often concern elderly people in a situation of déprise. Based on research amongst people living in intermediary housing (IH), this article locates the different reasons which account for this residential choice and argues that these forms of habitat enable re-securing these persons. Depending upon the reasons that propel older people to move into IH and the interplay with other variables such as health, forms of sociability, etc., it appears that individuals do not benefit in a similar way from the services which are proposed within the premises of these diverse properties, and that the reconstruction of one's home and sense of place identical to the former residence remains an incomplete process for a some of the older movers. These observations lead us to identify of the notion of a residential bifurcation. To summarize the diversity of situations, we suggest three paths of ageing in IH which alternately constitute three illustrations of the notions of emprise, reprise et déprise.

*Gérontologie et société 2018/1 (vol. 40 / n° 155). Laurent Nowik*

### THE DAILY SOCIABILITY OF ELDERLY PEOPLE IN A SHOPPING CENTER: A PECULIAR LEISURE

Studying groups of elderly people who meet every day in a shopping center located in the city center, this article shows how people create their own place in the shopping mall and divert the conventional uses and signification to transform it in a meeting place. Their sociability appears as leisure, a playful form of socialization, which contains important identity issues, but creates problems with the economic logic of the shopping center because they never consume. In that sense, we can question this kind of leisure and the normalization of space process which aims to control this diversion of space.

*Bulletin de l'association de géographes français, 95-1 | 2018, 79-96 Thibaut Besozzi*

### STATE OF THE ART REPORT ON DESIGN METHODOLOGY

Although this paper is from another discipline it does serve as an example of how design methodology is central to and design project covering design for risk, literature review and handling uncertainty in future operating context. This may be helpful for the designers of **Sheld-on**.

*(2015) David Andrews and Stein Ove Erikstad, 12th International Marine design Conference Tokyo, Japan*

A NOVEL ACCESSIBILITY ASSESSMENT FRAMEWORK FOR THE ELDERLY:
EVALUATION IN A CASE STUDY ON OFFICE DESIGN

Elderly and impaired persons constitute an important part of our societies. Existing practices to test accessibility features on forthcoming consumer products and services rely on tests with real impaired users on the industrial prototypes. Our approach comes to automate the evaluation process and introduce it in early phases of the product design. The proposed accessibility assessment framework is based on the Virtual User Models (VUMs) concept. VUMs are models containing several parameters used for the emulation of the behavioural characteristics of impaired and elderly populations. In this paper, the simulation framework and a number of VUMs corresponding to real persons are evaluated using two variations of a workplace office design. Results indicated that VUMs are efficient predictors of the corresponding end user's behaviour and thus, their simulated performance can lead into decision making during the product test-and-redesign cycles.

*Panagiotis Moschonas, Ioannis Paliokas, Dimitrios Tzovaras. ISBN: 978-1-63190-011-2*

MOVING HOUSE AND HOUSING PREFERENCES
IN OLDER AGE IN SLOVENIA 

In situations where older adults require extensive care, more than 70% of them find old people's home "acceptable" or "perfectly acceptable". Interestingly, sheltered housing is perceived as more negative, although it offers some advantages compared to old people's homes. Among the reasons for this may be unfamiliarity with the sheltered housing and large financial burden it poses.

*Maša Filipović Hrast, Richard Sendi, Valentina Hlebec, & Boštjan Kerbler, (2018) Taylor Francis.*

THE RELATIONSHIP OF THE ELDERLY TOWARD THEIR
HOME AND LIVING ENVIRONMENT 

The analysis confirmed the assumption that the Slovenian elderly are also very attached to their homes or home environments and are satisfied with living there. In addition, the analysis showed some differences among the elderly in this regard depending on their age, where they live, and how long they have been living in their current homes.

*Boštjan Kerbler, Richard Sendi, & Maša Filipović Hrast, (2017)*

### HEALTHY AGEING AT HOME: ERGONOMICS ADAPTATIONS OF INTERIOR DESIGN AND SELF-ASSESSED QUALITY OF LIFE OF OLDER ADULTS OF THE MUNICIPALITY OF LJUBLJANA

The results showed that there is a significant difference between the subjective evaluation of the participants and the evaluation by the professionals regarding the estimation of appropriate lighting. The participants who estimated their quality of life as better are more likely to have sufficient or adequate lighting and more adaptations in the kitchen, which makes daily kitchen work easy and safe. Further investigation dealt with the ergonomic adaptation of the bathrooms, where we found out that less than 15 % of the participants installed the handrails in the shower or bath tub, which can significantly contribute to safety. Additionally, physical capability as part of quality of life negatively correlates with the number of adaptations made in bathroom ($r$ = -0.149; $p$ = 0.039), which refers to the fact that the adaptation of accessories (handrails) were only installed when the need for them appeared. With minor changes in the living environment and taking care for maintaining psycho-physical capabilities, older adults can easily and safely perform everyday tasks, which prolongs an individual's autonomy and independence – the fact we are still not sufficiently aware of.

*Jasna Hrovatin, Saša Pišot, & Matej Plevnik, (2016)*

### PREFERENCES OF POLISH AND SLOVENIAN SENIORS CONCERNING KITCHEN INTERIOR DESIGN

The preferences are related to the comfort and evoking positive feelings, sense of security and relaxation, closeness to nature, and childhood memories of home. Older adults prefer cosy spaces that are large and bright, full of plants, equipped with furniture, and elements made of natural materials, especially solid wood. Respondents from both countries agreed that style and colours of kitchen should have a relaxing and optimistic impact on their mental and physical state. Investigations on preferred colour range revealed that the elderly choose warm colours with pastel tones more likely for their ideal kitchen interior design.

*Beata Fabisiak & Jasna Hrovatin, (2014)*

### PHYSICAL BARRIERS AND THE USE OF ASSISTIVE DEVICES IN SENIOR CITIZENS' EVERYDAY LIFE

The age-related problems often necessitate the use of assistive devices to secure safety, autonomy and independent living of the elderly, especially the availability of telephone communication (landline, 100 %) and the assistive devices for people with movement disorders (walking canes, 27 %). The majority of the respondents also expressed the need for adaptations to their internal and external home environment (installing handles, 69 %) and for physical barrier removal at home and in public areas.

*Mateja Berčan, Majda Pajnkihar, Jože Ramovš, & Zmago Turk, (2010)*

# ⁵ Conclusions and future work

## 5.1 General conclusions

Even though **Smart Furniture** is consolidating as viable solutions for the care of older adults, its market is still in its infancy and very few products could be properly referred to as such, the only exception being **smart mattresses**. However, there are many **smart furniture accessories** to make traditional furniture "smart", such as **occupancy sensors** (Miray, Aidacare, Health and Care, Asistae and SafeBase).

Some recent market solutions (e.g. Asistae) are starting to use a much more flexible IoT approach, where the sensor is connected through long-range wireless communications (like 3G) to a cloud server that analyses the data, and an app allows access and control to the sensor data.

Patents also reflect the fact that most solutions are smart accessories for regular furniture, except for mattresses. Some protected functionalities are: bed occupancy, monitoring eating habits, and a chair to help its user to stand up. Some patents can get quite complex by including a variety of sensors in a mattress or even different pieces of furniture. Sensors remotely work with a smartphone app to detect anomalies and warn caretakers.

Research in smart furniture does not abound. One publication describes the use of **smart furniture as a key sensing element** within a complete AAL system. Other projects are very varied such as **smart kitchen, smart lamp and robotic furniture**. The former two include sensors to extract contextual information and detect events such as falls. The latter is quite unique since it creates furniture that moves, self-assembles, and self-reconfigures. Actually there are a few publications in the area of robotic furniture.

Other scientific publications in smart furniture for older adults discuss **design aspects**, and can even be very specific such a system for monitoring **nutrition related actions**.

**Sensorization at home**

Sensorization at home is done by **AmI** (Ambient Intelligence) systems with sensors distributed throughout the home, whose data is sent to a central processing hub. Some automatically analyse the data and perform decision making (e.g. alert relatives). Some market examples: SentinelCARE, Just Checking, TruSense, GrandCare Systems.

Patents also cover that AmI approach, but some others can get pretty specific such as a bracelet that vibrates to warn about the main door or phone, or sensing carpets. Video monitoring for bed occupancy is also protected.

Other products are just made of **one single sensor** device that tracks certain variables. AbiBird and Canary Care are made of a wireless movement sensor that works with a smartphone app for caretakers. Similarly, Elsi Smart Floor is a sensing floor that can non-invasively detect falls.

Research projects are mainly focused on AAL systems that use non-invasive sensors, and Artificial Intelligence techniques to automatically learn the behaviour of the inhabitant and find patterns that may represent an emergency, and when detected a warning is generated.

Similarly, scientific publications describe a wide variety of **AAL systems for older people monitoring in regards to activities and health**. Two different **surveys** are respectively focused on **sensors for activity monitoring**, and smart home-based **remote healthcare** technologies.

**Wearable devices**

Wearable devices have experienced an enormous general market growth even within the AAL sector. This can be seen throughout this report. However, few patent protections have been found, related with the technology or research projects aimed at this.

Most marketed products have the shape of a **smartwatch** that wirelessly transmits activity and health data, which are used to identify patterns or problems like falls. Caretakers are notified through a smartphone app. Some examples are Senior Protection, Nectarine and CarePredict. However, not many patents focus on elderly care.

Wearable devices can also be implemented in **other shapes** like insole (E-vone), pill holder (Pill Clip), or even a wristband connected to a belt with airbags (Hip-Hop). Even a robotic glove to add strength to its users hands (Bioservo Technologies).

Recent research in wearable devices for older adults target **sensor integration in textiles, smart watches and smart insole**. Most of them detect or prevent falls, and some also extract well-being data.

**5.2** Future Work

During the first year of the Action the working group membership from the COST Action 16226 Indoor living space improvement: Smart habitat for the elderly, met two times (WG1 minutes, may be found here; http://www.sheld-on.eu/working-groups/).

Future work will build upon the theoretical framework that has developed from the network of researchers, the STSMs and the international conference, as well as the meeting held to discuss research progress at the round table events held adjacent to the conference in Riga (Latvia, 2018) and the meetings in Alicante (Spain, 2018) and Porto (Portugal, 2019). The results from this work, formalised here in the State of the Art report, will allow the next phase of the work to explore solutions as stated in the Action MoU, working through the structure of the newly formed WG4 with its three sub groups.

# ⁶ Appendices

Further sources of information which have informative value and may be of benefit to the Action which may have currency and act as a catalyst to generate questions and ideas for development is that research which has been delivered as presentations at conferences, some of which has been published. Additional sources may be found through academic societies and websites which are mentioned below.

## 6.1 Conferences

SISTEMA INTELIGENTE DE ALERTA Y MONITORIZACIÓN DE PERSONAS MAYORES Y/O DEPENDIENTES QUE VIVEN SOLAS (PROYECTO "LUCÍA"). IV CONGRESO CIUDADES INTELIGENTES ⌗

The objective of the "Lucia" project is to sensorize homes of elderly people and / or dependents who live alone so that a system based on artificial intelligence detects patterns of behavior. The aim is to create an AI system that is constantly learning and is capable of analyzing the behavior patterns of its users, carrying out warning actions when significant alterations of these patterns are detected. Depending on the level of the alert they will be transferred to specialized personnel or to relatives and / or tutors. In 2017, the Island Council of Tenerife carried out with its own means a proof of concept (PoC) of the "Lucia" project, through which the homes of three elderly people living alone were sensed, obtaining positive results that encourage us to continue advancing in their developing.

*Félix Fariña Rodríguez, Consejero con Delegación Especial TIC y Sociedad de la Información, Cabildo de Tenerife.David Pérez Rodríguez, Director Técnico del Proyecto Tenerife Smart Island, Cabildo Insular de Tenerife Madrid, 30-31 Mayo 2018*

BUILDING THE GREEN-SMART WAY: EXPLORING CONDITIONS FOR GREEN AND SMART FURNITURE MANUFACTURING FOR PEOPLE IN THE THIRD AGE FOR THE ELDERLY. INTERNATIONAL CONFERENCE

The present paper discusses some of the findings of an ongoing "Green & Smart Furniture" (GSF) research project. It actually focuses on third-age consumers' behavior and interest in purchasing smart and eco-friendly wooden furniture. The research reveals the critical elements for a successful design and production of GSF products that will enhance both needs and expectations of the consumers. Thus, 399 specially constructed questionnaires were gathered during 2013, which were further elaborated and statistically analyzed with SPSS ver. 17 Results are quite encouraging for the enterprises that would decide to design and produce smart and eco-friendly wooden products. Consumers show a significant interest in purchasing GSF at a percentage of 70%. They are willing to pay an extra amount of 9% in average than regular prices of conventional furniture. The most important factors in purchasing GSF have been found to be price, quality, functionality, safety and ergonomics, as well as the type of raw materials.

*7th Annual EuroMed Conference of the EuroMed Academy of Business. Trigkas Marios et al.*

### BIOCLIMATIC DESIGN OF A RESIDENTIAL COMPLEX FOR THE ELDERLY. INTERNATIONAL CONFERENCE

Traditionally, a home for the elderly aimed exclusively at the accommodation and care of people of the 3rd and 4th age. In a modern and continuously evolving society, how this can remain stagnant? This design, which was presented as a diploma design work at the Architecture Department of the University of Patras, constitutes a new idea, a revision and redefinition of the meaning of a 'home for the elderly'. A strong emphasis is placed on modern way of life and how this can influence the needs of old people and be influenced by them. The basic concept is based on an open plan system, with bioclimatic features. The aim is to create a secure, pleasant and modern environment fulfilling the requirements of a contemporary building while keeping some elements of the past so as to achieve a balanced aesthetic result.

*"Passive and Low Energy Cooling 737 for the Built Environment", V Simantria et al. May 2005, Santorini, Greece*

### DESIGN AND ERGONOMICS OF MONITORING SYSTEM FOR ELDERLY

*5th International Conference, DHM 2014. Andreoni et al. Springer International Publishing Switzerland*

### INNOVATION IN FURNITURE COMBINING ECOLOGY AND ICT: THE GREEN AND SMART FURNITURE RESEARCH PROJECT (2015)

*I Papadopoulos et al. Proceedings of 7th Annual American Business Research Conference ISBN: 978-1-922069-79-5*

### INTEGRATION OF SMART HOME TECHNOLOGIES IN A HEALTH MONITORING SYSTEM FOR THE ELDERLY (2018)

*A. Arceleus et al. 21st International Conference on Advanced Information Networking and Applications Workshops. 0-7695-2847-3/07*

### INTELLIGENT FURNITURE DESIGN IN THE ELDERLY BASED ON THE COGNITIVE SITUATION (2017)

*X. Lu et al. MATEC Web of Conferences. 10.1051/matecconf/2017104030*

REFINEMENT OF TEMPERATURE
SENSING YARNS. (2018) 🔗

**6.2** Learned Societies

There are many academically based societies that are relevant to this working group and a sample of these is included here, that may be of interest to researchers and students as useful points of reference.

THE DESIGN RESEARCH
SOCIETY 🔗

The Design Research Society provides a special interest group on Wellbeing which includes a network of researchers to engage and discuss design related topics. This group positions itself as a community whereby those with expertise in design, psychology, engineering, architecture, human factors, technology development, HCI, healthcare and other disciplines can come together to nourish the interdisciplinary of the domain.

DESIGN FOR AN AGEING
POPULATION 🔗

The demographic landscape of our cities is changing fast, as our cities grow and the population ages. But how do architects respond to the challenge? How do we go about creating more 'age-inclusive' spaces? And are there ways we can cultivate a design sensibility more sensitive to the desires and needs of an ageing population?

THE DESIGN FOR AGEING
WELL PROJECT 🔗

The design for ageing well project will focus on bringing emerging wearable technologies to active members of older age groups who do not suffer from restrictive medical conditions. The proposal will address "Ageing well across the lifecourses: autonomy and independence" with a multi-disciplinary team incorporating researchers from technical textiles, wearable electronics and Information and Communication Technologies (ICT), and social and care sectors, with active participation from the users.

TRANSFORM AGEING AND SOCIAL
CARE – THE FILO PROJECT

We challenged social ventures to find solutions and create lasting change through the initiative, breaking down barriers and supporting ageing societies.

## 6.3 Non-reviewed literature

This list of literature which has not been reviewed is here for reference purposes if the reader should require a wider set of material. The references in these works may provide a link to further related reading.

- *Adaptability of Kitchen Furniture for Elderly People in Terms of Safety (2012) J. Hrovatin et al.*
- *CIRDO: Smart companion for helping elderly to live at home for longer (2014) S. Bouakaz et al.*
- *Cyber–physical cloud-oriented multi-sensory smart home framework for elderly people: An energy efficiency perspective (2017) M. S. Hossain et al..*
- *Delivering home healthcare through a Cloud-based Smart Home Environment (CoSHE) (2018) M. Pham et al.*
- *Design and evaluation of a smart insole: Application for continuous monitoring of frail people at home (2018) Y. Charlon et al.*
- *Design for AAL Integrated Furniture for the Care and Support of Elderly and Disabled People (2017) P. Beer at al.*
- *Evaluation of Three State of the Art Classifiers for Recognition of Activities of Daily Living from Smart Home Ambient Data (2015) T. Nef et al. 10.3390/s150511725*
- *Furniture for Later Life. O Johnsson PhD thesis. Department of Design Sciences, Division of Industrial Design. ISBN 978-91-7473-706-6*
- *Smart Home Assistant for Ambient Assisted Living of Elderly People with Dementia (2015) E. Demir et al.*
- *Smart homes and home health monitoring technologies for older adults: A systematic review (2016). L.Lieu et al.*
- *Smart Homes for Elderly Healthcare—Recent Advances and Research Challenges (2017) S Majumder. Sensors. 10.3390/s17112496*
- *The Elderly's Independent Living in Smart Homes: A Characterization of Activities and Sensing Infrastructure Survey to Facilitate Services Development (2015) Sensors. Q. Ni et al. 10.3390/s150511312*
- *Exergame technology and interactive interventions for elderly fall prevention: A systematic literature review (2017) Choi. Et al.*
- *Photodiodes embedded within electronic textiles. (2018) Scientific reports 8 (1): 16205. SATHARASINGHE, A., HUGHES-RILEY, T. and DIAS, T.. ISSN 2045-2322*
- *Developing novel temperature sensing garments for health monitoring applications. (2018) Fibres, 6 (3) : 46 LUGODA, P., HUGHES-RILEY, T., OLIVEIRA, C., MORRIS, R. and DIAS, T. ISSN 2079-6439*
- *A historical review of the development of electronic textiles. (2018) HUGHES-RILEY, T., DIAS, T. and CORK, C.R., Fibers, 6 (2): 34. ISSN 2079-6439*
- *Design for the environment in UK product design consultancies and in-house design teams: an explorative case study on current practices and opinions. (2013) RADLOVIC, P.F.M., LEMON, M. and FORD, P., The International Journal of Design Management and Professional Practice, 6 (2), pp. 73-83. ISSN 2325-162X*

**6.4** Websites

There are many websites that provide useful information on the topics under review, however this report has recommended a small number here as a starting point for students.

- *https://www.esmartcity.es/comunicaciones/comunicacion-sistema-inteligente-aler-ta-monitorizacion-personas-mayores-dependientes-viven-solas-proyecto-lucia*
- *http://www.newdynamics.group.shef.ac.uk/design-for-ageing.html*
- *https://www.architecture.com/knowledge-and-resources/resources-landing-page/design-for-an-ageing-population*
- *https://www.dezeen.com/tag/design-ageing/*
- *https://gtr.ukri.org/projects?ref=AH%2FM005607%2F1*

# Working Group 2
## ICT developments

**Editors:**
Petre Lameski (MK),
Michal Isaacson (IL)
Kuldar Taveter (EE)

**Members:**
Mr Aliaksei Andrushevich (CH)
Dr Carmelo Ardito (IT)
Dr Frantisek Babic (SK)
Mr. Alfonso Bahillo (ES)
Prof Sabina Baraković (BA)
Prof Jasmina Barakovic Husic (BA)
Mr Miguel Ángel Beteta (ES)
Mr Andres Lorenzo Bleda (ES)
Mr Gregorio Cañavate (ES)
Dr Claude Chaudet (CH)
Dr Ivan Chorbev (MK)
Dr Radu Ioan Ciobanu (RO)
Ms Elena Coroian (DE)
Dr Ciprian DOBRE (RO)
Prof Hazim Kemal Ekenel (TR)
Dr Moriah Ellen (IL)
Dr Francisco Florez-Revuelta (ES)
Prof Ivan Ganchev (IE)
Prof Nuno Garcia (PT)
Dr Andrej Grguric (HR)
Prof Andrei Gurtov (SE)
Mr Paul Held (DE)
Prof Darko Huljenic (HR)

Prof Florin Ioras (UK)
Dr Aleksandar Jevremovic (RS)
Dr Saso Koceski (MK)
Dr Artur Krukowski (EL)
Dr Marcel Kyas (IS)
Dr Agnieszka Landowska (PL)
Ms Anne-Marie Lipphardt (DE)
Dr Madhusanka Liyanage (FI)
Dr Rafael Maestre (ES)
Prof Constandinos Mavromoustakis (CY)
Prof Michael Mrissa (SI)
Prof Nuno Pombo (PT)
Prof Lilia Raycheva (BG)
Dr Ioannis Refanidis (HE)
Mr Kasper Rodil (DK)
Mr Jose-Luis Romero-Gazquez (ES)
Dr Susanna Spinsante (IT)
Dr Juan Suardiaz Muro (ES)
Prof Carlos Valderrama (BE)
Ms Willeke Van Staalduinen (NL)
Prof Trajkovik Vladimir (MK)
Dr Eftim Zdravevski (MK)

# Working Group 2
## ICT developments

**Acknowledgements:**

We acknowledge the contribution from each of the working group members towards making this report:

We thank Eftim Zdravevski, Vladimir Trajkovik and Ivan Chorbev for contributing towards the quantitative analysis of the studies;

Frantisek Babic and Kasper Rodil for their contribution in the Participatory Design/Co-design section;

Agnieszka Landowska, Jasmina Baraković Husić, Sabina Baraković, Claude Chaudet, Enida Cero, Francisco Florez-Revuelta, Andrushevich Alexey for their contribution in the Adoption, accessibility, acceptance and legal section;

Radu-Ioan Ciobanu, Ciprian Dobre, Pau Climent-Perez and Francisco Florez-Revuelta for the Ambient living, wearables and sensors section;

Hazim Kamal Ekenel and Lilia Raycheva for the Supporting emotional well-being, people with cognitive decline and ADL section and bibliography contribution;

Frantisek Babic for Neural settings medical data analytics section;

Ace Dimitrievski, Carlos López and Rafael Maestre, Haiyang Zhang, Ivan Ganchev and Andrei Gurtov for Solutions for caregivers, security and privacy section.

We also acknowledge the support from the Sheld-on COST Action CA16226 for organizing the meetings and supporting the work and all other members who participated in the STSMs and other members and non-members who gave their valuable contribution towards achieving the goals of the action.

# Index

Funded by the Horizon 2020 Framework Programme of the European Union

cost
EUROPEAN COOPERATION
IN SCIENCE & TECHNOLOGY

Funded by the Horizon 2020 Framework Programme
of the European Union

# ¹ Introduction

This document is prepared as part of the COST Action **Sheld-on** CA16226. The goal of this document is to identify and report on the state of the art in ICT developments targeted at enhancing the wellbeing of older adults and of those who care for them. The document is divided in 7 sections, each identifying different aspects of these technologies and reporting on different areas of research and development. The focusing groups were selected based on the WG2 meetings discussions, each targeting specific aspects of the ICT developments. The focusing groups are as follows:

- Co-design/participatory design
- Adoption/accessibility/acceptance/legal
- Ambient living/wearables/sensors
- Supporting emotional well-being/ Supporting people with cognitive decline/supporting ADL
- Natural Settings Medical data analytics
- Solutions for caregivers, security and privacy

In the following sections the document describes the state of the art in each of the focusing groups. The document also gives brief analysis of the publishing trends in each focus groups based on relevant keywords. The analysis was performed using a semi-automated Natural Language Processing (NLP) Toolkit¹² and used the IEEEXplore, PubMed and Springer digital libraries for searching and selecting publications. The following keywords were used for the automated search for each of the focusing groups:

"AAL", "Ambient assisted living", "active aging", "elderly care", "ELE", "human activity recognition", "enhanced living environment", "smart home", "smart habitat", "tele-care", "smart furniture".

Based on the found publications, additional selection was performed using properties that are specific for the focusing groups. The initial findings based on the keywords are depicted in Figure 1.1 :

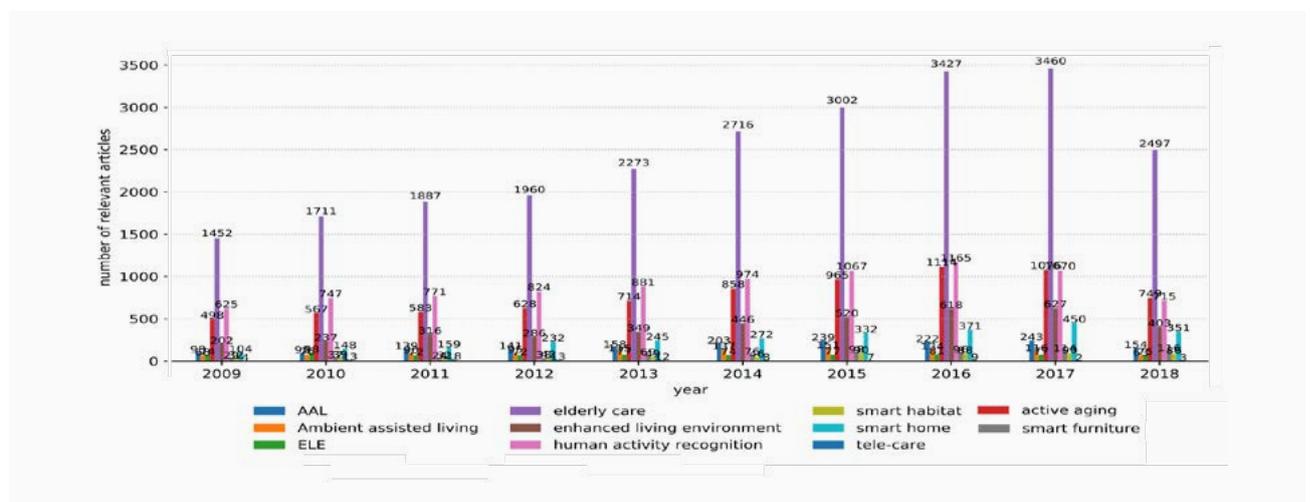

*Figure 1.1: Number of publications per keyword per year*

# ² Co-design/participatory design

## 2.1 Quantitative analysis of publications

The publications that contained the globally selected keywords where filtered by the following properties:

"**Robotics**": [["social robotics", "social robot"],
["prototype", "field test"],
["ideation", "idea", "concept"],
["shopping robot"],
["guiding robot", "assistive robot"],
["camera robot"]].

"**co-design**": [["participatory design", "PD", "co-design", "cooperative design", "co-operative design", "bottom up"],
["top down", "off-the-shelf", "generic", "generalized"],
["involvement", "include", "inclusive", "inclusiveness"],
["inclusive design", "user sensitive"],
["stakeholders", "users", "patients"],

["feedback"],
["diversity", "diverse"],
["early involvement", "initial involvement"],
["context"],
["workflow", "habit", "routine"]].

"**sensor position**": [["wearables", "smartwatch", "smart watch", "bracelet", "smartphone", "smart phone", "wristband"],
["environmental", "ambient"],
["body", "hand", "hip", "ankle", "pocket", "arm", "neck", "ear"],
["furniture", "chair", "table", "bed", "mattress"],
["clothes", "apparel", "garment", "shirt", "pants", "shoes", "knit", "sawn", "jacket", "pajamas", "gloves", "glasses", "socks"]]]

Based on the properties the NLP Toolkit did automated selection of relevant papers. The results of the survey are shown on the following figures:

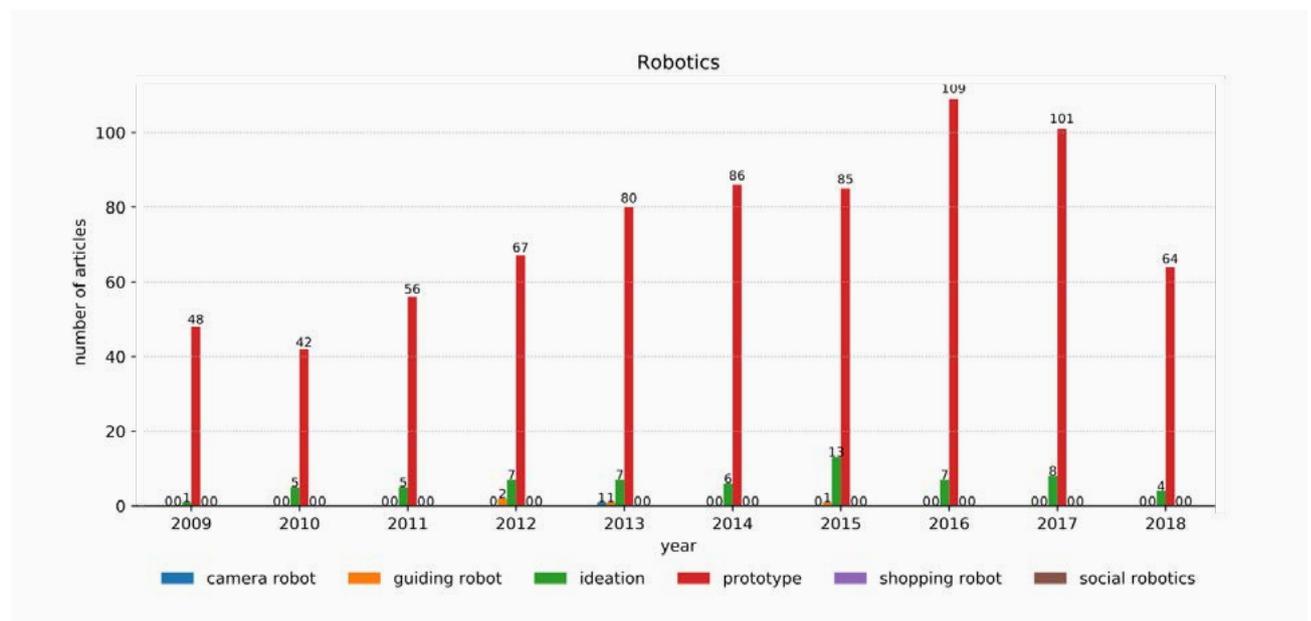

*Figure 2.1 Publications that contain the "Robotics" properties per year*

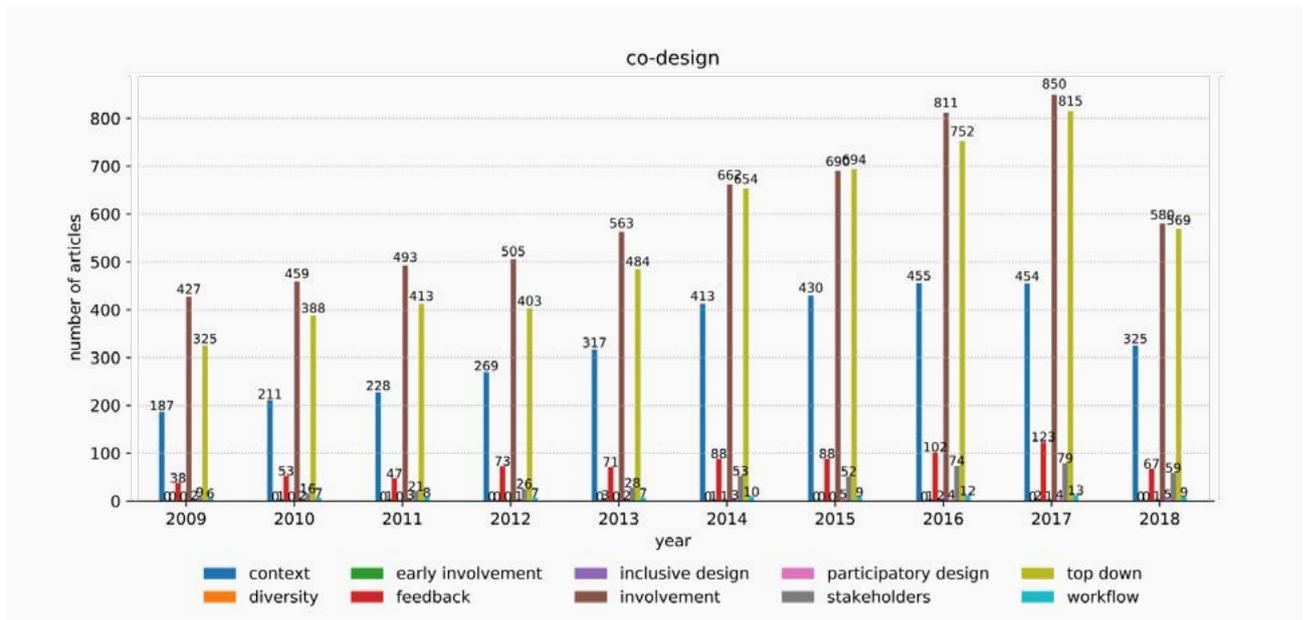

*Figure 2.2 Publication that contain the "Co-design" properties per year*

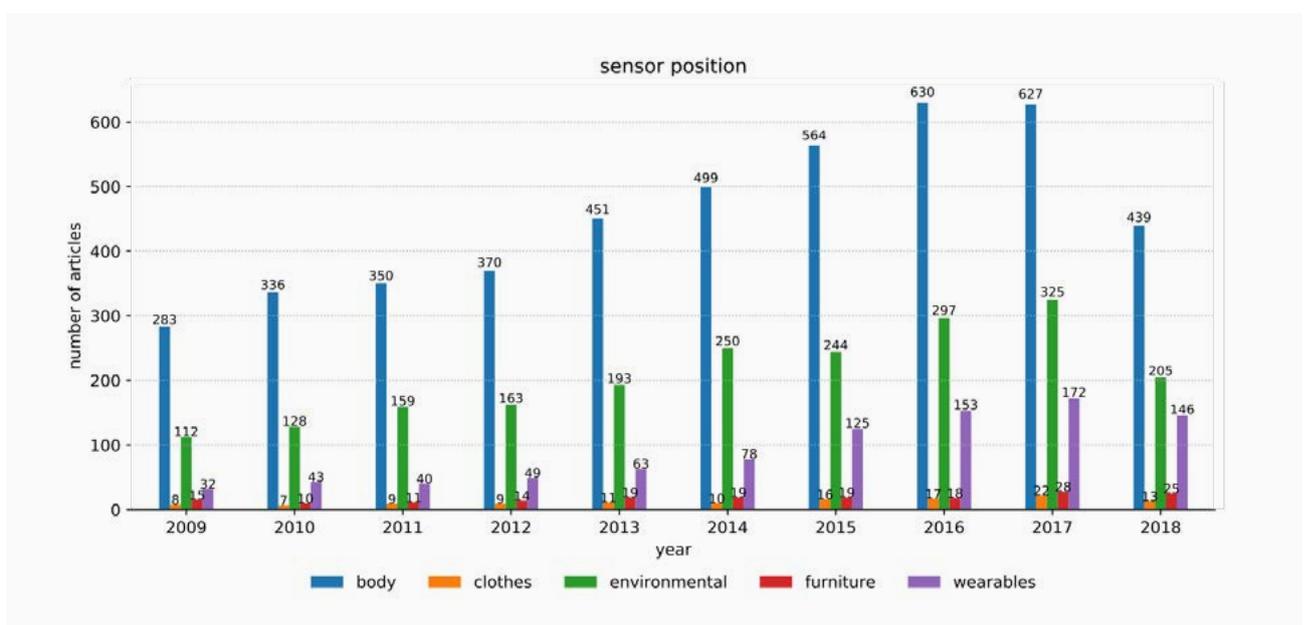

*Figure 2.3 Publications based on mentioned sensor position property per year*

As it can be observed from Figures 3-1 to 3-3, there is an increasing trend in publications for co-design. Furthermore, there is very small number of publications for guiding robots and social robots which gives us insight that there is much work to be done in that area. Also, most of the publications consider wearable sensors and there is not enough work done in the research community on furniture-based sensors which is also a field that needs to be considered.

## 2.2 Participatory design definition and state of the art

PD (originally co-operative design, now often co-design) is an approach to design attempting to actively involve all stakeholders in the design process to help ensure the result meets their needs. It means that users are understood as experts on relevant tasks, activities, and processes[1]. We expect their detailed view and constructive feedback on the presented ideas and proposals.

In the case of older adults, the management of PD is affected by the diversity of health conditions, contextual factors and daily activities[2]. Aarhus et al.[3] exemplify how similar-aged older adults have different attitudes and perspectives towards technology. Grönvall and Kyng[4] explain how moving the PD process into a home environment affects the organization, participation and recruitment, especially when working with ill and fragile participants. Gregor and Newe,ll[5] propose a methodology called User sensitive inclusive design allowing designers to accommodate diversity within the group of older adults. Geurts et al.[18] explore the relation between ethic in information and communication technologies and the participatory involvement of older people in making Gerontechnologies accessible.

The UTOPIA project is focused on developing effective methods for the early involvement of older people in the development of information technology-related products through a diverse user base[6]. Ekdahl et al[7] discuss older patients' needs and preferences when making decisions about care and medical treatment, discussing the limitations involved when engaging older participants.

Huldtgren et al.[8] present a concept for community-based co-creation where the goal is to facilitate long-term collaboration with various stakeholders in a community-based environment. Scandurra and Sjölinder[9] emphasize the importance of understanding the context around older users. The danger of "overselling" the outcome of PD activities to such participants is also stressed. Lindsay et al.[10] argue for the importance of empathy when engaging older adults in order to provide meaningful engagements. Massime et al. emphasizes the need to respect individual's contributions to the design process[11].

The mentioned papers focus mainly on the heterogeneous character of the target group of older people. The authors tried to find the key points how to overcome these barriers or to propose the new methodologies for this purpose. We should focus on these aspects:

- Keeping older participants focused on the topics and giving them clear opportunities to present their ideas during the meetings.
- Respecting the differences in the target group and do not try to sell our idea of the solution at all costs.
- Supporting their creative thinking on new technologies and things which they cannot imagine or do not meet yet.
- Understanding the context and workflows of their daily activities as an important input to the design process.

**2.3** Social robotics and participatory design survey based on publications from the Participatory Design Conference

**Motivation**

One of the current trends of ICT is the domain of robots. In the context of the CA16226 the field of social robotics holds promise as potential extension of in-home caregiving to be a personal assistant which can assist with solving various tasks.

In general, the topic of social robotics in an aging context has the following coloration – much development is limited to industrial off-the-shelf products or highly generalized solutions where the application context is only assumed after the development. In contrast to this top-down approach, this review will look at the state of robotics (also historically) in a design methodological field known as Participatory Design (PD). PD is historically a bottom-up, emancipatory design approach where technological systems are co-designed with end-users.

A multidisciplinary gerontechnology[18] must therefore resort to an analysis of the needs and characterisics of older adults with regard to technology. Indeed, each has his own values, wishes, desires and individuality greatly influence the idea of what is useful. The idea of the participation of the aging population being involved in the design of the prototypes for these new technologies, in both meeting the objectives sought and in the developed form, is widely acclaimed but little used in practice. However, from an ethical point of view, it appears essential to base an approach on the famous slogan 'nothing about us, without us'.

The first part is to identify papers which research surrounds robotics and the second part is to investigate the concept of living conditions and satisfaction of older adults in these technological contexts provided by the found papers.

**Search scope and strategy**

The following literature was identified by an automated string search on all proceedings counting full- and short -research papers from the biannual Participatory Design Conference (PDC) from 1990 to 2018. Total volume for the search was 531 research papers. Workshops, posters, installations etc. were removed as these do not by format carry enough information to properly gauge the contribution(s).

The search string used was a truncated [robot] allowing variations and partial congruence such as [anth*robot*ic], [robot*ic], [robot*ic] [robot*s] etc. thus in this manner also identifying hyphenation associated with the search string, such as [social-robotics].

The return of the search showed a total amount of 177 hits in papers having any variation of [robot] in the text – here also counting bibliography, copyright statements etc. After filtering these for duplicates – 29 unique papers were identified.

## Limitation

The review is limited by scoping only literature from PDC.

## Manual sorting of identified papers

Identified papers have been manually inspected. Only papers which clearly address the reflection on, design of and/or construction of robots are included.

After the manual inspection leading to Table 1 - a reverse search on [age], [elder*], [senior] was conducted to scope out oversights in the manual inspection.

## Results

Table 1 below shows the included literature and a breakdown on contributions to the Robot topic.

The first-time robots are mentioned directly in the literature is in the proceedings of the PDC in 2002. Before this point in time the mentioning of robots is limited to examples of potential technology in line with all other types of technology. But looking across the publications added to Table 1 it is clear that the promise of finding contributions to the combination of robot and older adults is unfulfilled. While robot is on the agenda in PD it only surpasses stages of ideation when off-the-shelf robots have been used in workshops (see Walker et al., 2012[16]; Iivari and Kinnula, 2018[14]). One paper both demonstrates use of a robotic prototyping system and in a context of a Community. The main finding is that these robots, which the design partners develop become ways for themselves as Community members to become critically engaged towards their own life-worlds (DiSalvo et al., 2008[13]). Critical engagement as an outcome of designing robots is also backed up by (Brown and Choi, 2018 [12] ), also in a personal life context.

While it is not uncommon to find contributions mostly relating to values, criticality and reflection on technology, very little addresses the concept of robots. Although Leong and Robertson (2016)[16] present interesting insights from discussing technology with senior citizens and gives insights into values and older adults participation in design - the robotic outcomes are not identifiable.

The demographic and epidemiological transition to a hyper-connected society pose new challenges. This has led in particular by a growing interest in technologies. Indeed, these new technologies will be able to offer additional support to older adults in the fulfilment of their needs and the management of their everyday life's. The socioeconomic innovation is presented as an opportunity to improve both the quality of life of the aging population, offering services to them and creates new opportunities for financial investments. However, to do this it is important not to limit innovation to the development of advanced technologies, but to ask the question of the usefulness and accessibility of the devices. The question of ethics must be at the heart of the debate dealing with the pursuit of technical excellence and socio-economic sustainability.

In conclusion, the field of PD (limited by the PDC proceedings and search strategy) does not lead the way for how to approach the co-design of robots in the context of healthy ageing. Robot solutions are primarily limited to ideation and very little tangible robotic development is described.

The scope should be opened up towards more technical domains and a more open study on designing ICT (not limited to a robotic context) with senior citizens (this is the most fitting label).

| Reference (APA) | | | | | | |
| --- | --- | --- | --- | --- | --- | --- |
| Design partner | Technological construction and type of robot | Off-the-shelf robot | Primary content in relation to robots | Level of Robot content | Context | Aging/ elderly |
| Farber, A., Druin, A., Chipman, G., Julian, D., & Somashekher, S. (2002). How Young Can Our Technology Design Partners Be?. In PDC 2002 (pp. 272-277). | | | | | | |
| Children | Minimal prototype Unclear | No | Ideation | None | Unclear | No |
| Taxén, G. (2004). Introducing participatory design in museums. In Proceedings of the eighth conference on Participatory design: Artful integration: interweaving media, materials and practices-Volume 1 (pp. 204-213). ACM. | | | | | | |
| High school students | No Guiding robot | No | ideation | Idea | Museum | No |
| Reddy, M., Hughes, N., & Turnbull, N. (2004). Anthrobotics: Science-by-Doing in Higher Education. In PDC 2004 (pp. 53-56). | | | | | | |
| High school students | No | No | Curriculum development | Applied | Higher education | No |
| Holmquist, L. E. (2008, October). Bootlegging: Multidisciplinary brainstorming with cut-ups. In Proceedings of PDC 2008 (pp. 158-161). Indiana University. | | | | | | |
| Industry and academia | No Various conceived ideas | No | Design workshops | Idea | Unclear | No |
| Weiss, A., Wurhofer, D., Bernhaupt, R., Beck, E., & Tscheligi, M. (2008, October). This is a flying shopping trolley: A case study of participatory design with children in a shopping context. In Proceedings PDC 2008 (pp. 254-257). Indiana University. | | | | | | |
| Children | Shopping robot | No | Design workshops | Idea | Shopping | No |
| DiSalvo, C., Nourbakhsh, I., Holstius, D., Akin, A., & Louw, M. (2008, October). The Neighborhood Networks project: a case study of critical engagement and creative expression through participatory design. In Proceedings of PDC 2008 (pp. 41-50). Indiana University. | | | | | | |
| Community members | Working prototypes Camera robot | No Project developed Canary robotic prototyping platform | Design workshops Critical engagement with technology | Full cycle | Community | Unclear |
| Kilbourn, K., & Bay, M. (2010, November). Foresight and forecasts: participation in a welfare technology innovation project. In Proceedings of PDC 2010 (pp. 255-258). ACM. | | | | | | |
| staff | No | No | Reflections on robots and current practice | reflections | hospital | minimal |
| Walker, S., Bell, S., Jackson, A., & Heery, D. (2012, August). Imagine real avatars and flying shepherds: involvement and engagement in innovative ICT. In Proceedings of the PDC 2012. (pp. 101-108). ACM. | | | | | | |
| Public | No aerial | Yes Giraff telecare robot | Engagement and Reflections on novel applications for improved bandwidth | Applied | bandwidth | No |
| Leong, T. W., & Robertson, T. (2016, August). Voicing values: laying foundations for ageing people to participate in design. In Proceedings of PDC 2016 (pp. 31-40). ACM. | | | | | | |
| seniors | None | No | Workshops on values in ICT for the elderly | None | None | yes |

| Reference (APA) | | | | | | |
|---|---|---|---|---|---|---|
| Design partner | Technological construction and type of robot | Off-the-shelf robot | Primary content in relation to robots | Level of Robot content | Context | Aging/elderly |
| Brown, A. V., & Choi, J. H. J. (2018, August). Refugee and the post-trauma journeys in the fuzzy front end of co-creative practices. In Proceedings of the PDC 2018 (p. 15). ACM. | | | | | | |
| Refugees | No Social-robot (for feeling less-lonely) | No | Design workshops | Idea | Unclear | No |
| Iivari, N., & Kinnula, M. (2018, August). Empowering children through design and making: towards protagonist role adoption. In Proceedings of PDC 2018 (p. 16). ACM. | | | | | | |
| Children | No | Yes Lego robots | Learning | Applied | learning | No |

## **2.4** Bibliography

# ³Adoption, accessibility, acceptance and legal

## 3.1 Quantitative analysis of publications

The properties used for selection of publications are the following:

"**Adoption, Acceptance, Accessibility**": [["acceptance"],

["accessibility"],

["adoption"],

["usability", "utilization", "utilize"],

["assistive technologies", "assistive", "support"],

["comfort"],

["well being"],

["quality of life", "QoL"],

["psychology"]],

"**End-users**": [["family", "next in kin"],

["doctors"],

["facility", "nursing home", "hospital"],

["care-providers", "nurse", "staff"],

["elderly", "patients", "older", "generation", "age", "old", "aged"],

["disabled"],

["impaired"],

["verified", "verifications"]],

"**Legal issues**": [["legal", "law", "bylaw", "licit", "lawful", "legalisation", "legalization", "legalise", "legalize"],

["liability", "liable", "responsible"],

["protocol"],

["patent"],

["procedure"],

["regulation"],

["procurement"],

["GDPR"]]]

The results of the quantitative analysis of publications is depicted in the following figures:

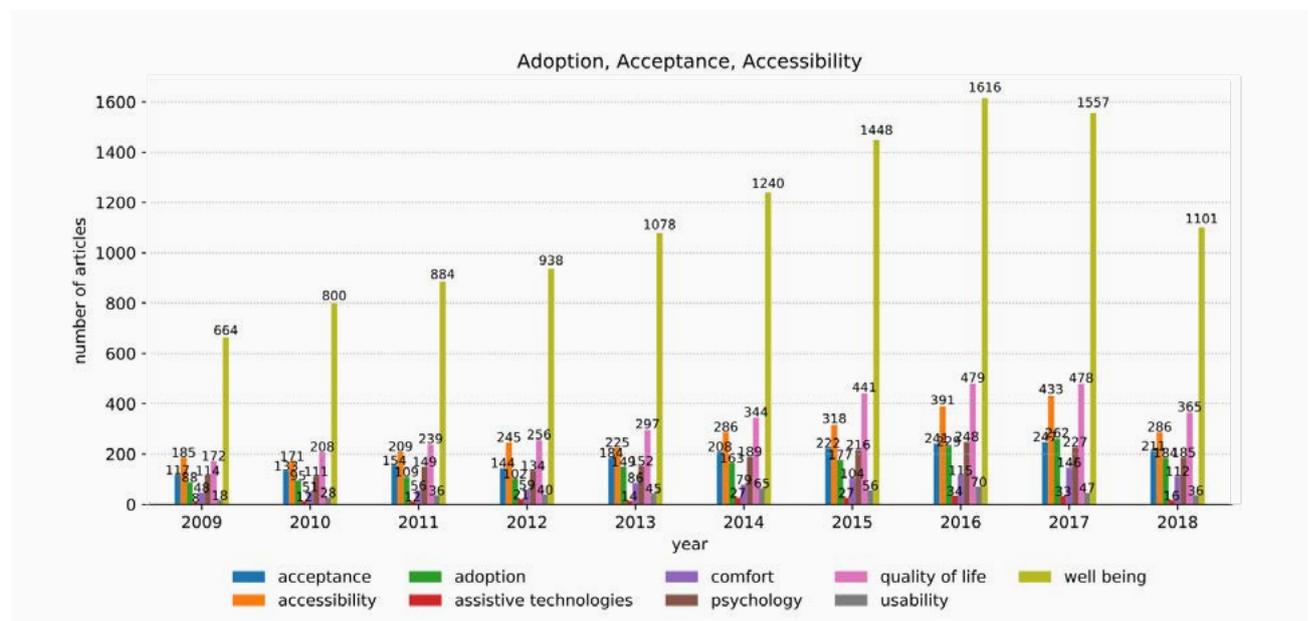

*Figure 3.1 Adoption, Acceptance, Accessibility selected publications per year*

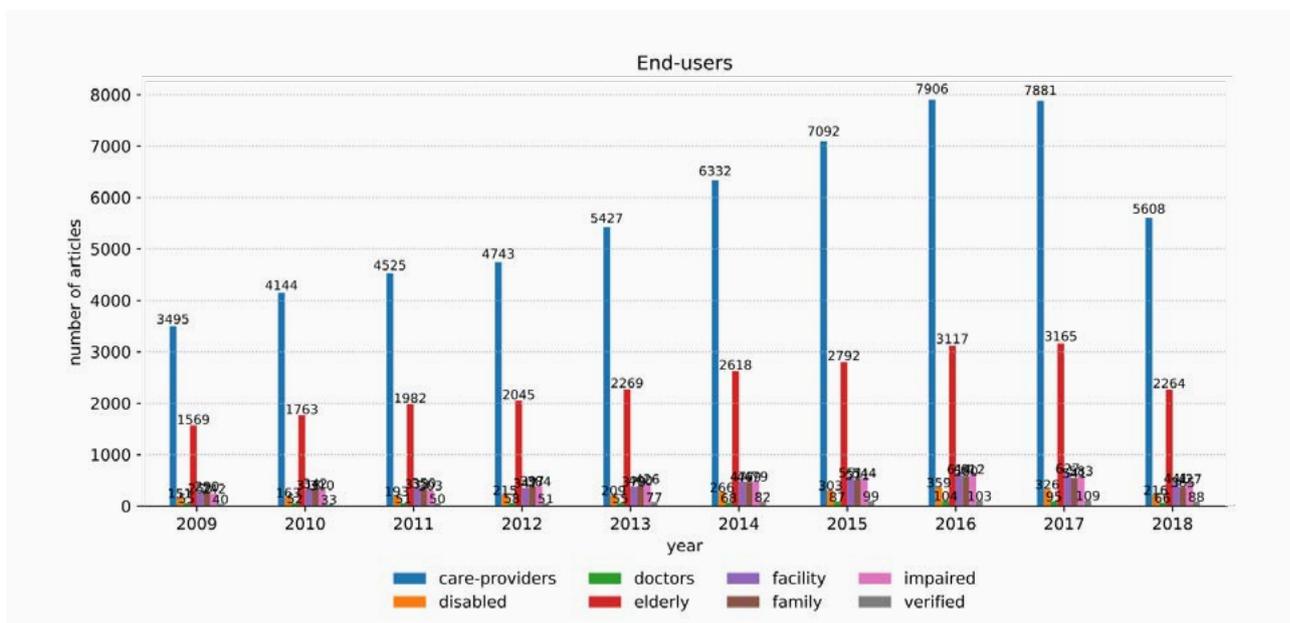

*Figure 3.2 End-users properties selected publications per year*

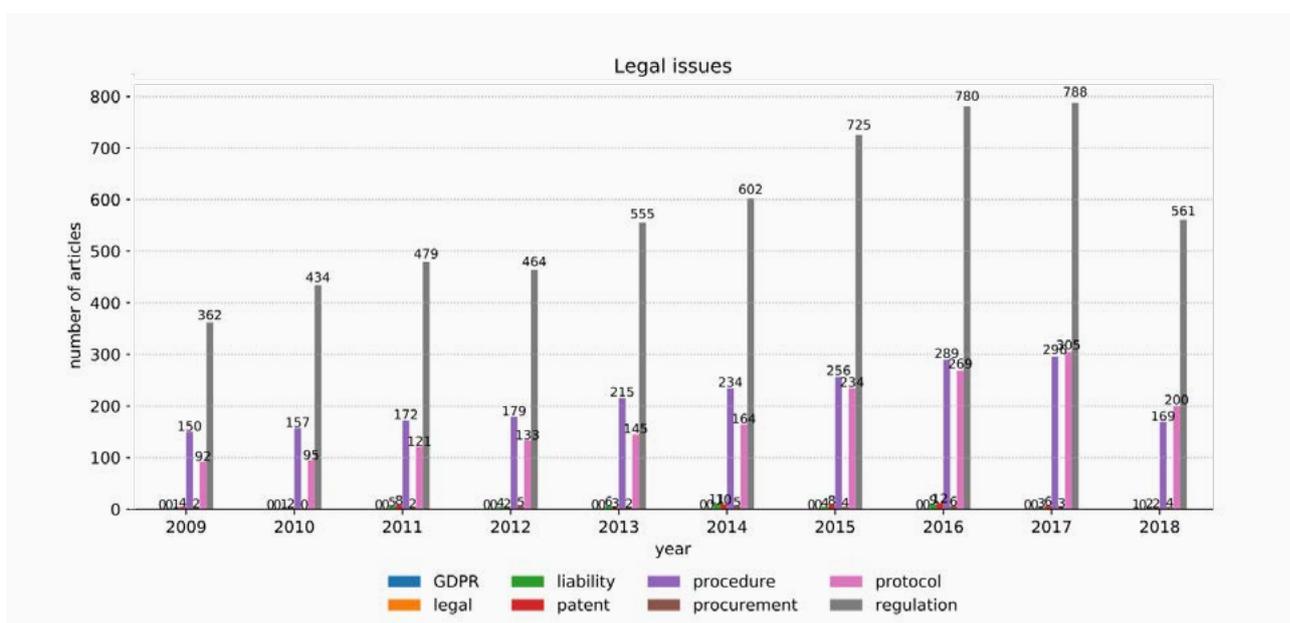

*Figure 3.3 Legal issues properties selected publications per year*

The increasing publications number shows the increasing interest in the field of Adoption/Acceptance/Accessibility in the scientific community. The legal issues are also increasing slightly in interest and receive attention too. The following text gives a more thorough analysis of the Adoption/Acceptance/Accessibility of assistive technologies.

## 3.2 Adoption/Acceptance/Accessibility of assistive technologies for older adults

**Introduction - method**

In the following years, all life spheres will be flooded with smart things and systems. These systems will completely change everyday activities by creating opportunities for development and innovation, which in turn will bring in countless benefits. Smart systems will connect homes, cars, governments, health, etc. This concept will also change the way people interact with the society and things around them and try to simplify our lives. Generally, the ultimate goal of smart concept should be to improve our Quality of Life (QoL).

QoL is a multidimensional concept, which emphasizes the self-perceptions of an individual's current state of mind affected in a complex way by the person's physical health, psychological state, personal beliefs, social relationships, and their relationship to salient features of their environment. Simplification and facilitation of everyday activities and improvement of QoL is especially important for older adults as their number is significantly increasing according to available statistics. Authors in[15] analyses the interest of seniors regarding technology and healthcare according to fundamental terms such as adaptation and acceptability of gerontechnologies. Their work articulates the shared challenges between the technological and social gerontological domains in order to drive the commercial logic and technological innovation according to the complex need expressed by the elders and their families. They discuss the limits to overcome and the ethical stakes to pursue in order to develop participatory approaches ensuring both social innovation and economic development. Their study showed several aspects to be considered in order to build acceptance models. Those are not really organized as a methodology but presented as advices helping to build the right gerontechnology approach. In this context, one particularly important aspect is the generation effect. In sociology, a generation is described as a group of people born during the same period. Therefore, these people have a shared context and have lived through similar events. The gap between generations was previously related to the age difference between people, but its nature has now changed. Indeed, our society is prone to develop new technologies which are modifying significantly our surroundings. This fast evolution increases the gap between young people and non-technologically savvy seniors. A second important aspect when building acceptability and adoption is limited to understanding the health objectives. Indeed, the popularity of geriatrics tends to convey a predominantly medical view of the concept. However, as stated by the World Health Organization (1946), health is not just the absence of disease. It is better defined as 'a state of complete physical, cognitive and social well-being.'

The number of people aged 60 years or older will rise from 900 million to 2 billion by 2050, and the population ageing is happening more quickly than in the past. The World Population Aging Report shows that older population growth rate is more rapid in developing countries than developed countries. Doing a simple math leads to the conclusion that assistive technologies should be devoted to a large extent to improvement of QoL of older adults given that they are expected to compromise 22% of entire world population. Indeed, the concept of Gerontechnology[15] is at the crossroads between the world of research and technology. Therefore, it needs a greater knowledge of the multiple characteristics of the ageing process. In this way, for example, the aging population, despite being heterogenic, is the evidence of an increased propensity to develop sensory and cognitive losses. This information is essential for the creation of a prototype. However, it should not be overshadowed that these limitations and their perceptions are intrinsically affected by other variables like age, gender, life trajectory, social capital and education that are likely to affect or not the recourse to the proposed device and its modalities.

This document refers the State of the Art study in topics related to adoption of assistive technologies for older adults. Adoption of technologies was perceived in a broad sense, including acceptance and accessibility issues. By assistive technologies we mean any kind of supporting electronic device or a software program, that was designed and developed to support health and well-being of older adults. Terms of adoption, acceptance and accessibility are treated here as equivalent, although there are differences between them. Practically, papers use those three words to express research related to how older adults react to new technologies.

In order to perform state of the art two separate studies were conducted:

- Meta-analysis of smart ageing solutions described in the literature.
  The meta-analysis followed the procedures suggested by[2].

- Systematic literature review of selected bibliographic repository.
  The systematic review was based on methodology proposed by Kitchenham[1].

Based on the state of the art, the main gaps in this research domain were also identified.

**State of the Art Review**

META-ANALYSIS OF SMART AGEING SOLUTIONS

Smart ageing is a wide concept defined as technology and innovation usage in both the public and private sectors to produce products, services, solutions, and systems to improve QoL of people ages 50 and over. Healthy ageing is another term used to describe the concept of enabling older people to enjoy a good QoL. Term mentioned in World Health Organization (WHO) is active ageing and is defined as the process of optimizing opportunities for health, participation, and security in order to enhance QoL as people age.

Smart ageing ecosystem includes key determinants of healthy ageing as described in EuroHealthNet, and covers access to services, employment and volunteering, physical activity, social inclusion and participation, new technologies, diet and nutrition, long-term care, environment and accessibility, education and life-long learning. On the other hand, QoL has eight dimensions as suggested by Eurostat: material living conditions, health, education, productive and valued activities, governance and basic rights, leisure and social interactions, natural and living environment, and economic and physical safety.

Each smart ageing determinant should be contained at least in one QoL dimension. For example, New Technologies (determinant) contributes to Leisure and Social Interactions (dimension) of older adults given that it provides new ways of entertainment and communication options. However, it contributes negatively to Personal insecurity given that usually people over 50 are insecure when using new devices, applications, etc., resulting in the withdrawal and abstinence from new technology products.

This mapping provides a connection between smart ageing determinants and QoL dimensions, and results in QoL indicators for older adults[10]. Knowing which features of smart aging products and services affect which QoL dimension of older adults allows better targeting and effectively achieving the ultimate goal – better QoL of older adults. Research and industry communities should consider them when developing their products and services.

Today we are witnessing multiple smart ageing solutions being developed and produced, with many more in the announcement, but the QoL of older adults is not noticeably improved yet. Motivated by this, we seek to give an answer to the following question: are the existing smart ageing solutions succeeding in direct improvement of QoL of older adults?

We have analyzed 35 existing smart ageing solutions (Figure 3.1a): 44.4% dealing with Smart Home, 27.8% with Smart Health Monitoring, 22.2% with Smart Fall Detection, and 2.8% with Smart Application and with Service Robot. Regardless of context, the addressed solutions are associated to smart ageing determinants with the following percentages (Figure 3.1b): Long-term Care 85.7%, Access to Services 54.3%, Physical Activity 40%, Environment and Accessibility 11.4%, New Technologies and Social inclusion and Participation each 5.7%, Diet and Nutrition 2.8%, and Employment and Volunteering and Education and Life Long Learning 0%.

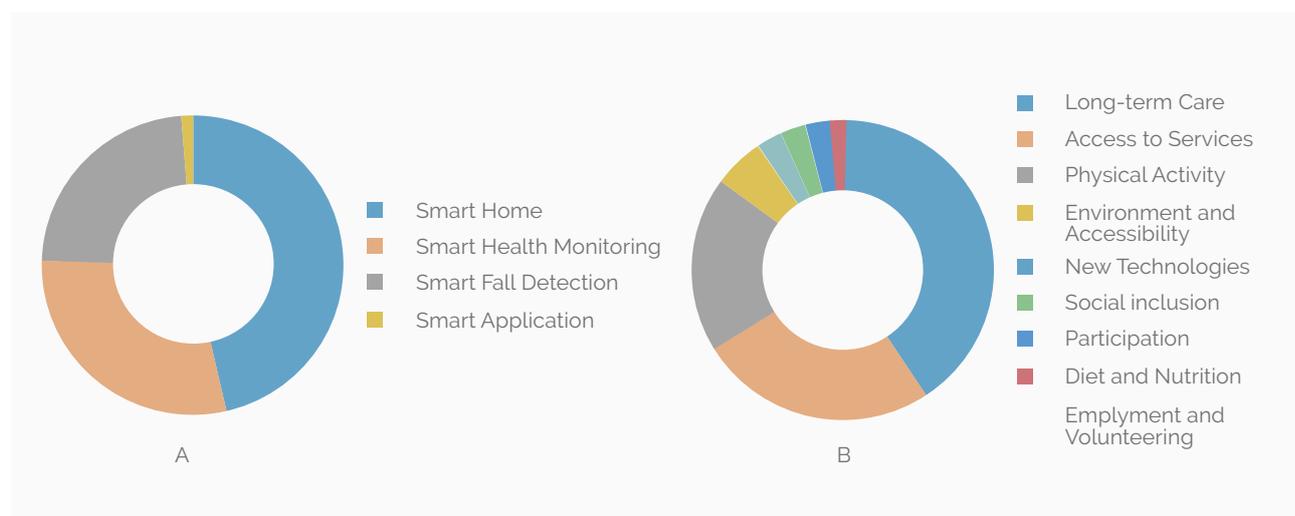

*Figure 3.4 Meta-Analytical Review of Smart Ageing Solutions: (a) Smart Context; (b) Smart Ageing Determinants*

Further on, only 30% of addressed solutions consider older adults as end service users, while the remaining ones are developed or produced for people around older adults such as family, doctors, care providers, etc. (Figure 3.2a). This means that the proposed solutions may help and simplify activities for people that take care of older adults. Older people may benefit from that indirect help, but not necessarily. In fact, in most cases they have not been asked does the solution help given that only 8.3% of analyzed approaches are verified by older adults, leaving the rest unverified by the ones for whom they are allegedly developed (Figure 3.2b).

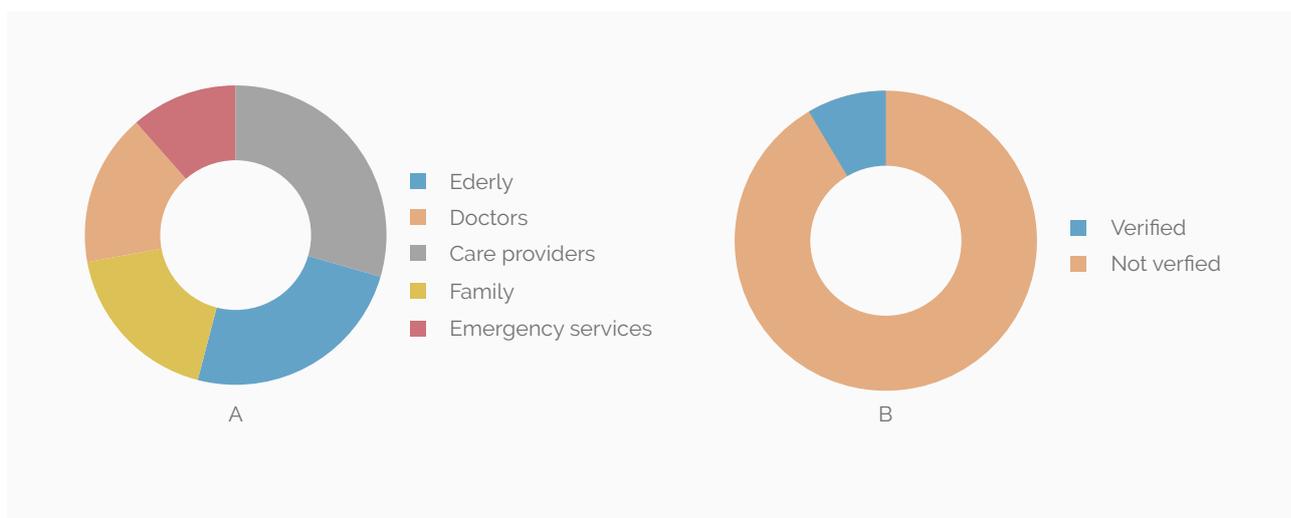

*Figure 3.5 Meta-Analytical Review of Smart Ageing Solutions: (a) Service Users; (b) Service Verification with older adults*

Meta-analytical review of the existing smart ageing solutions (i.e., different assistive technologies) shows that they do not necessarily directly contribute to QoL of older adults due to several reasons. Firstly, the existing solutions have not addressed various dimensions of QoL of older adults, nor have they included different smart ageing determinants, meaning that they are not multidimensional and comprehensive, i.e., not in line with QoL nature. Furthermore, the majority of solutions have not been developed for direct usage by older adults at all, but for people around them. Last and most important, most solutions were never properly verified by older adults so one cannot say that they have been beneficial to them[10].

SYSTEMATIC REVIEW OF LITERATURE - QUANTITATIVE ANALYSIS

The systematic review was based on methodology proposed by Kitchenham[1]. There was only one repository explored, i.e. Web of Science. The search keywords were grouped into following groups:

- Adoption / Accessibility / Acceptance / Utilisation

- Technolog* / ICT / sensor* / device* / program*/wearable*
- Elder* / age / generation* / old

Only articles written in English were included. There was no limit on publication dates.

From the quantitative perspective, the results are quite interesting. Figure 3.6 shows both exact queries, as well as the size of output paper sets. While search was performed in topic field (TS), the output set was quite large, therefore a restriction to title only (TI) was applied. A significant observation could be made – when searching without condition (3) regarding older adults, one might obtain thousands of papers, while adding this condition results in a limited set of 79 papers, which indicates, that only 2,5% of research papers on adoption and acceptance of technology refers to older adults people.

**Search History:**

| Set | Results | | | | Edit Sets | Combine Sets ○ AND ○ OR Combine | Delete Sets Select All ✖ Delete |
|-----|---------|---|---|---|-----------|------|------|
| | | Save History / Create Alert | | Open Saved History | | | |
| #4 | 3,219 | (TI=((adoption OR acceptance OR accessibility) AND (ICT OR technolog* OR sensor* OR device* OR wearable*))) AND LANGUAGE: (English) AND DOCUMENT TYPES: (Article) *Indexes=SCI-EXPANDED, SSCI, A&HCI, CPCI-S, CPCI-SSH, BKCI-S, BKCI-SSH, ESCI, CCR-EXPANDED, IC Timespan=All years* | | | Edit | ☐ | ☐ |
| #3 | 79 | (TI=((adoption OR acceptance OR accessibility) AND (ICT OR technolog* OR sensor* OR device* OR wearable*) AND (elder* OR age OR generation* OR old))) AND LANGUAGE: (English) AND DOCUMENT TYPES: (Article) *Indexes=SCI-EXPANDED, SSCI, A&HCI, CPCI-S, CPCI-SSH, BKCI-S, BKCI-SSH, ESCI, CCR-EXPANDED, IC Timespan=All years* | | | Edit | ☐ | ☐ |
| #2 | 48,968 | (TS=((adoption OR acceptance OR accessibility) AND (ICT OR technolog* OR sensor* OR device* OR wearable*))) AND LANGUAGE: (English) AND DOCUMENT TYPES: (Article) *Indexes=SCI-EXPANDED, SSCI, A&HCI, CPCI-S, CPCI-SSH, BKCI-S, BKCI-SSH, ESCI, CCR-EXPANDED, IC Timespan=All years* | | | Edit | ☐ | ☐ |
| #1 | 6,531 | (TS=((adoption OR acceptance OR accessibility) AND (ICT OR technolog* OR sensor* OR device* OR wearable*) AND (elder* OR age OR generation* OR old))) AND LANGUAGE: (English) AND DOCUMENT TYPES: (Article) *Indexes=SCI-EXPANDED, SSCI, A&HCI, CPCI-S, CPCI-SSH, BKCI-S, BKCI-SSH, ESCI, CCR-EXPANDED, IC Timespan=All years* | | | Edit | ☐ | ☐ |

*Figure 3.6 Numbers of papers found in adoption/acceptance/accessibility search*

Another observation reflects on the growth in the interest in the topic. Figure 3.7 presents citations per year charts in the domain of acceptance/adoption by older adults. From less than 10 citations yearly in 2010, the citations count raised to more than 200 yearly in 2017. This shows a significant raise in interest of this area of research in the last decade, however the number of papers and citations in the domain still seems low.

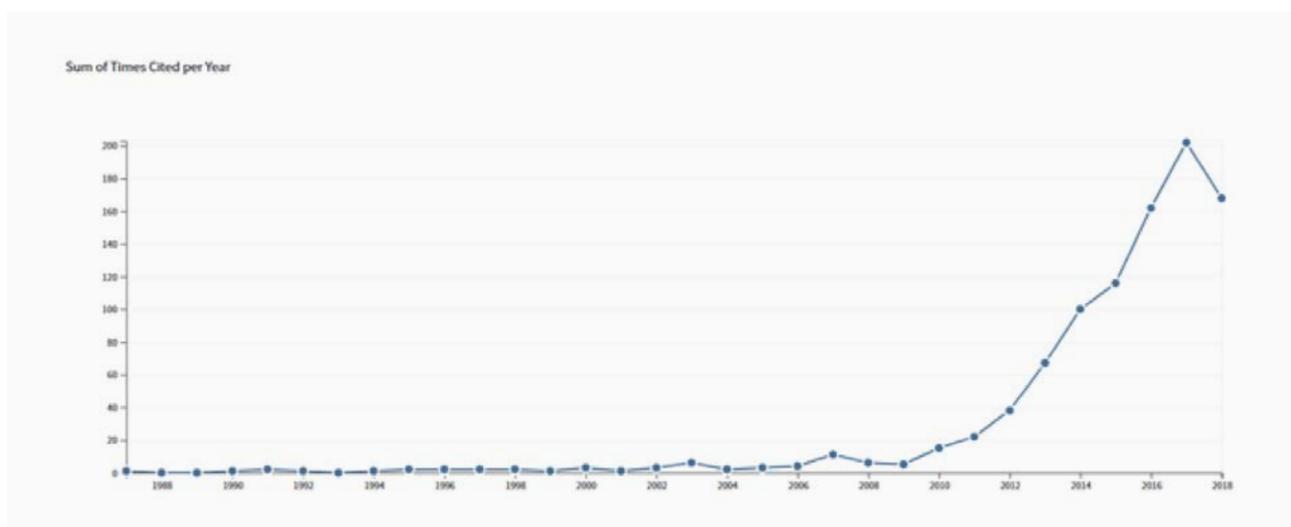

*Figure 3.7 Citations per year on adoption/acceptance by older adults.*

Another observation results from analysis of domains, that are represented by the chosen set of papers. Figure 3.8 shows number of papers in specific Web of Science domains. Please pay attention, that this area of research is multidisciplinary and requires combination of ap-proaches from different disciplines, including: ergonomics, health care, computer science, psychology and more.

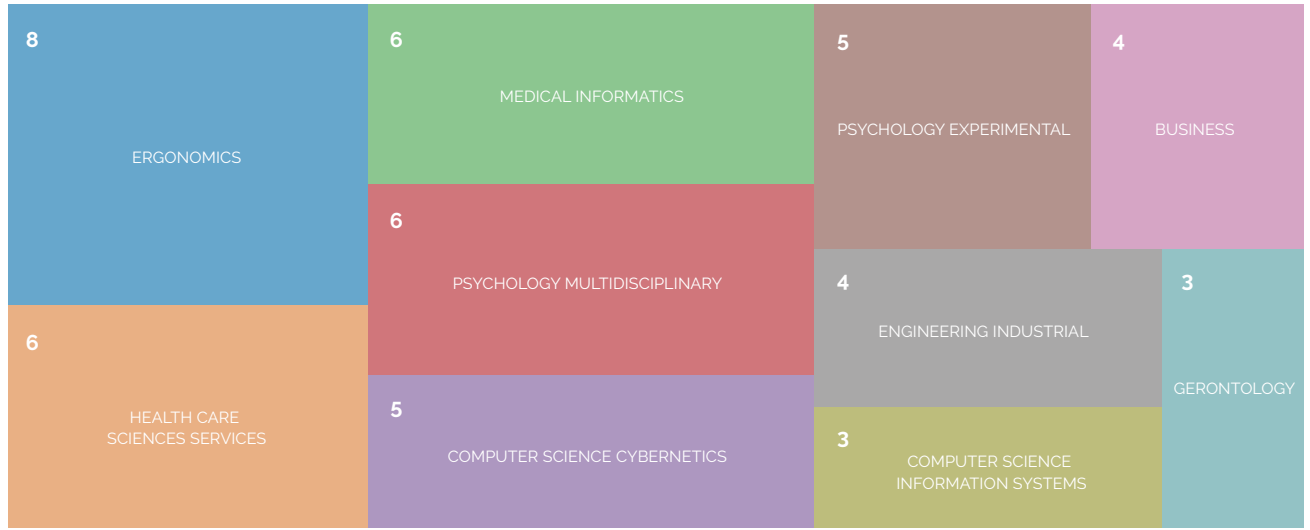

*Figure 3.8 Papers per discipline*

Another observation results from analysis of domains, that are represented by the chosen set of papers. Figure 3.8 shows number of papers in specific Web of Science domains. Please pay attention, that this area of research is multidisciplinary and requires combination of approaches from different disciplines, including: ergonomics, health care, computer science, psychology and more.

**Basic technology acceptance models**

One might find some basic models of technology acceptance[3-9]:

- Technology acceptance model TAM,
- Extended technology acceptance model TAM2,
- Extended (contextual) technology acceptance model TAM3,
- Unified Theory of Acceptance and use of technology UTAUT,
- Model of adoption of technology in households MATH.

TECHNOLOGY ACCEPTANCE MODEL TAM

Technology acceptance model (TAM) was first proposed by Davis, Bagozzi et Warshaw in 1989[3]. It was the first model of technology acceptance proposed and was extended multiple times in following research. The TAM model is based on 6 concepts: (1) External variables, (2) Perceived usefulness, (3) Perceived Ease of use, (4) Attitude towards using, (5) Behavioral intention to use, (6) Actual system use. Scheme of dependencies between the concepts is provided in Figure 3.9. The TAM model is quite general – any external variables could be included. Please note, that the attitude towards using (4) is distinguished from behavioral intention to use (5) and the actual system use (6). TAM model in the basic or extended form is used for explaining technology adoption among older adults.

*Figure 3.9 Technology Acceptance Model[3]*

EXTENDED TECHNOLOGY ACCEPTANCE MODEL TAM2

The basic TAM model defines "external variables" as general concept, therefore more acceptance models were proposed to explore what they are. TAM2 model was proposed by Venkatesh and Davis[4][5]. The model consists of concepts derived from TAM: perceived usefulness, perceived ease of use, intention to use and usage behaviour. Apart from the basic concepts the following variables were introduced:

- Voluntariness
- Experience
- Subjective norm

- Image
- Job relevance
- Output quality

- Result demonstrability

TAM 2 model is presented in Figure 3.10.

*Figure 3.10 Extended Technology Acceptance Model TAM2[4]*

EXTENDED (CONTEXTUAL) TECHNOLOGY ACCEPTANCE MODEL TAM3

Another extension of the basic TAM model was proposed by McFarland and Hamilton[6]. It adds contextual information to variables influencing the acceptance of technology. Apart from the concepts derived from basic TAM model (Perceived ease of use, perceived usefulness, system usage), the TAM3 model introduces following variables:

- Other's use
- System quality
- Organizational support
- Prior experience
- Anxiety
- Task structure
- Computer efficacy

The interrelations between the concepts in TAM 3 model are presented in Figure 3.11.

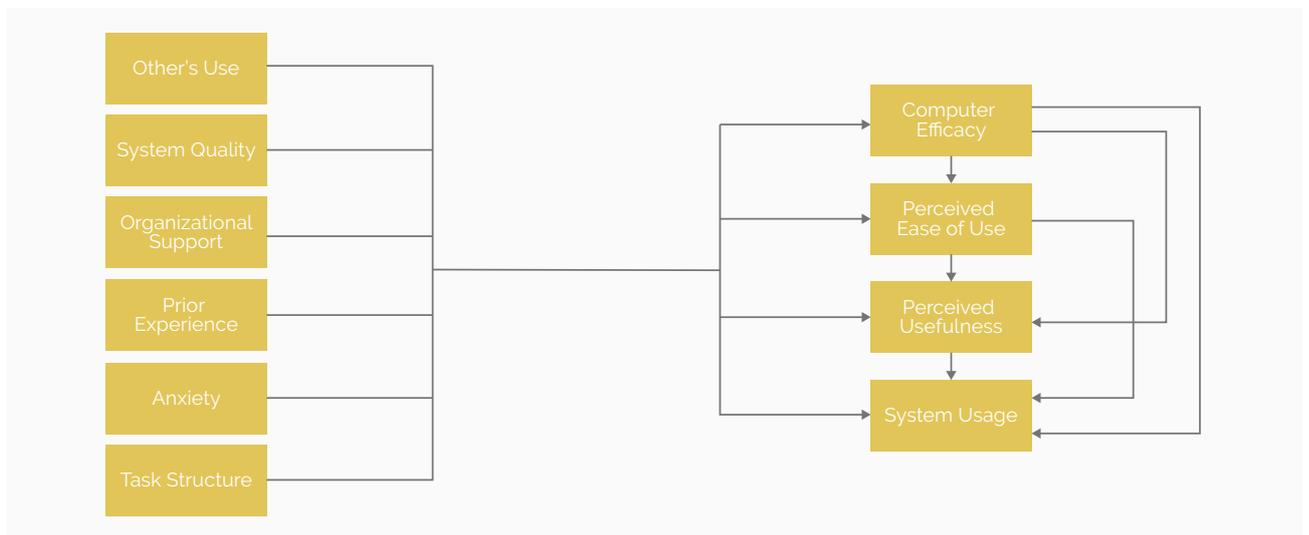

*Figure 3.11 Extended (Contextual) Technology Acceptance Model TAM3[6]*

UNIFIED THEORY OF ACCEPTANCE AND USE OF TECHNOLOGY UTAUT

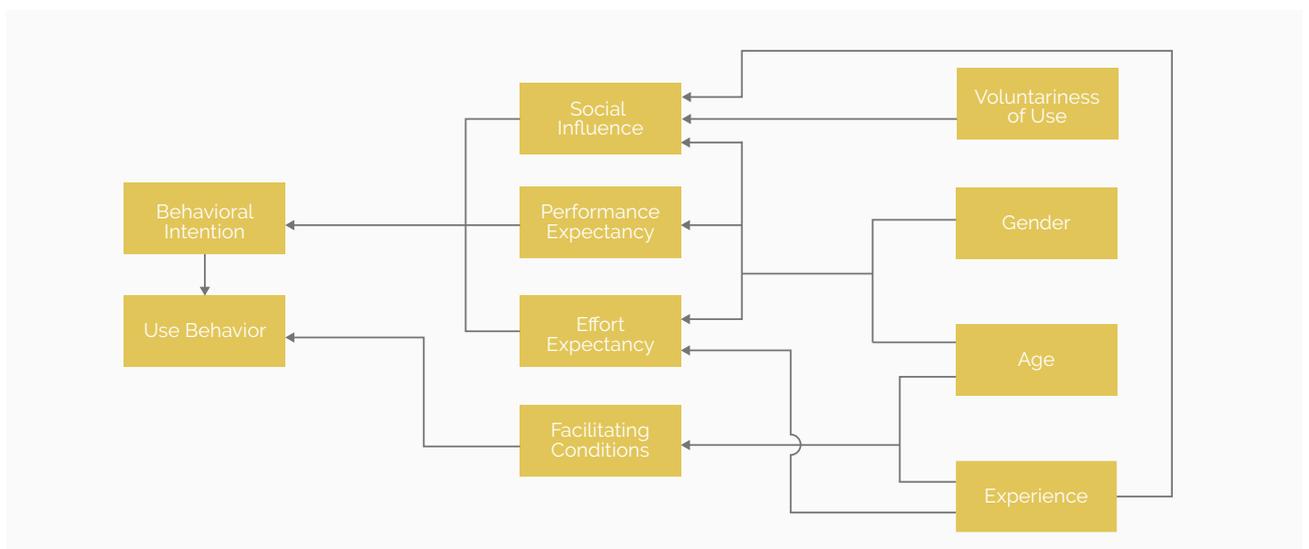

*Figure 3.12 Unified Theory of Acceptance and use of technology model UTAUT[7]*

After examination of eight technology acceptance models available so far (in 2003), Venkatesh et al proposed a unified acceptance model. Concepts in the model include: performance expectancy, effort expectancy, social influence, facilitating conditions. Among variables, that influence those gender, age, experience and voluntariness of use were proposed. Behavioural intention and Use behaviour are adapted from TAM model in UTAUT. Figure 3.12 shows interrelations between the concepts and variables in the UTAUT model.

MODEL OF ADOPTION OF TECHNOLOGY IN HOUSEHOLDS MATH

There is another model of adoption of technology in households, which have been used in analysis of the acceptance concepts and variables. Behavioural intention, as a basic concept in the model, is preceded by 13 concepts divided into 3 categories. Attitudinal Beliefs category contains 5 concepts regarding app types: Application for Personal Use, Utility for Children, Utility for Work, Applications for Fun, Status Gains. Nominative beliefs category holds 3 social concepts: Friends and Family influences, Secondary Sources' influences, Workplace referents' influences. Control beliefs category contains the following 5 concepts: Fear of technological advances, declining cost, cost, perceived ease of use, self-efficacy. The model is presented in Figure 3.13.

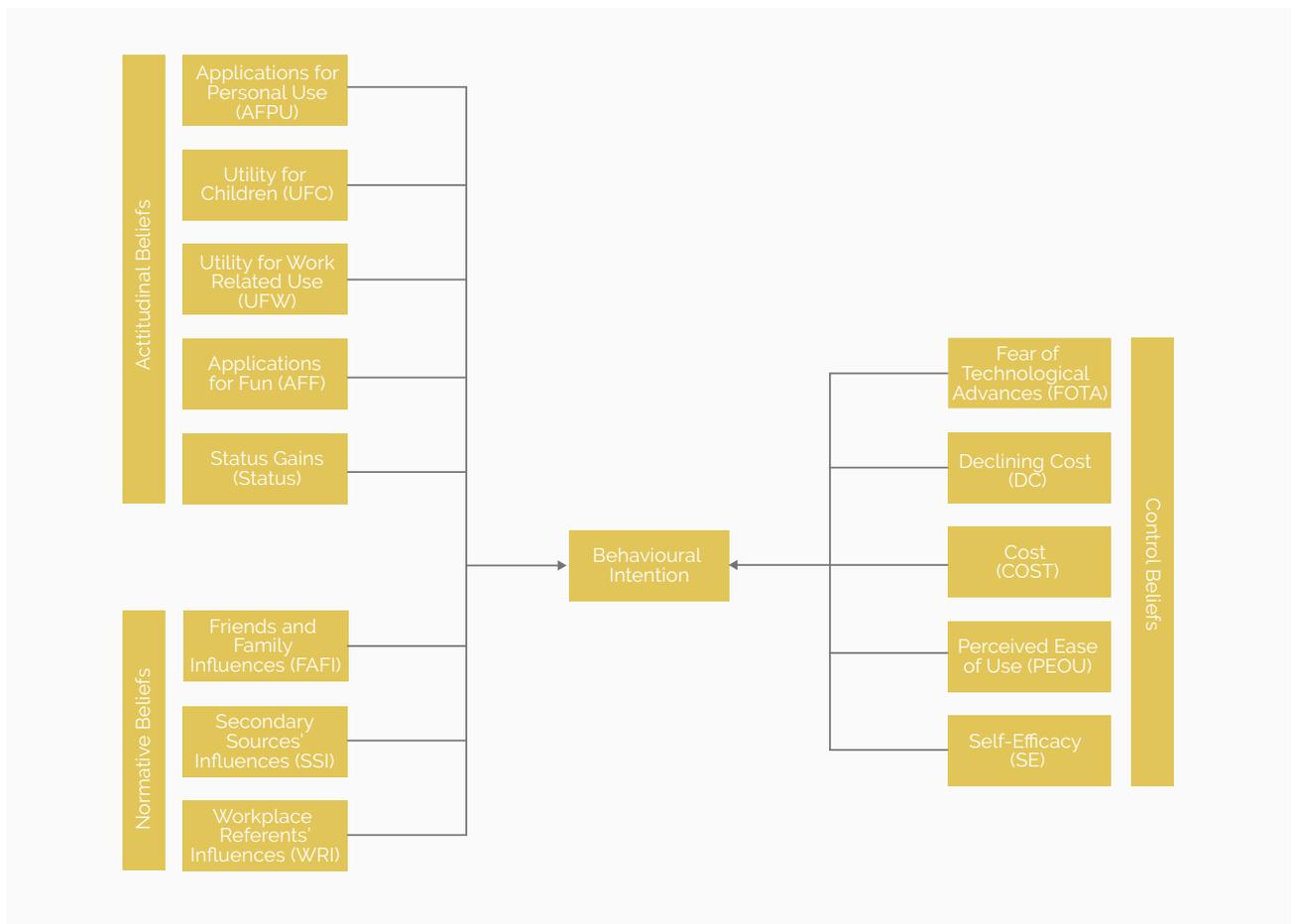

*Figure 3.13 Model of adoption of technology in households MATH*[8,9]

Each one of the general technology adoption models, described above, were used in research regarding older adults' acceptance of technology.

## Specific technology acceptance models for older adults

Literature reports three adoption/acceptance models, that are proposed specifically for older adults as users:

- Senior technology acceptance model STAM
- Extended Technology Acceptance Model for the Elderly ETAME
- Model of Technology Adoption by Older Adults MATOA
- Elderadopt

### SENIOR TECHNOLOGY ACCEPTANCE MODEL STAM

STAM[11] Questionnaire and interviews in 6 elderly care structures in Hong Kong  Explores several factors (age, gender, education, income, health satisfaction, movement ability, social) Basically it confirms intuition and was based on a small sample.

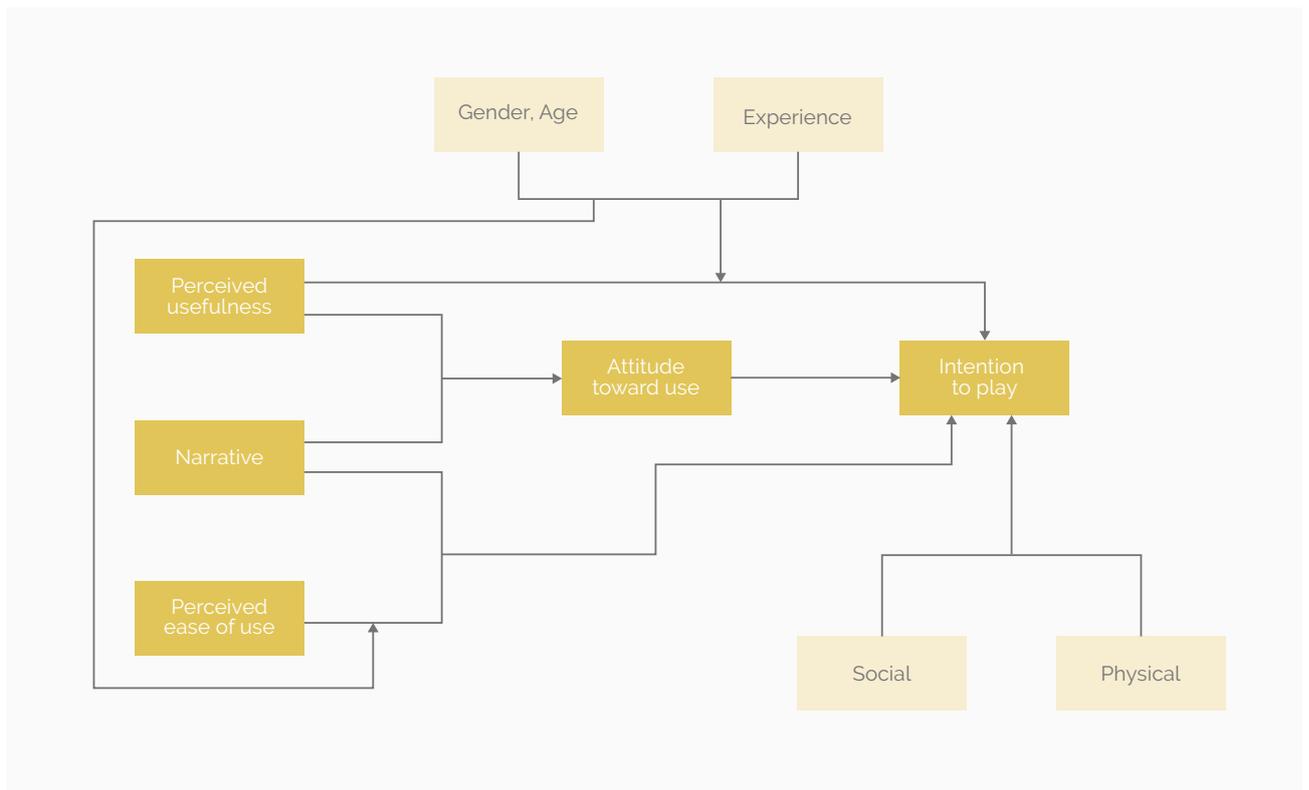

*Figure 3.14 Extended Technology Acceptance Model for the Elderly (ETAME)[12]*

## MODEL OF TECHNOLOGY ADOPTION BY OLDER ADULTS MATOA

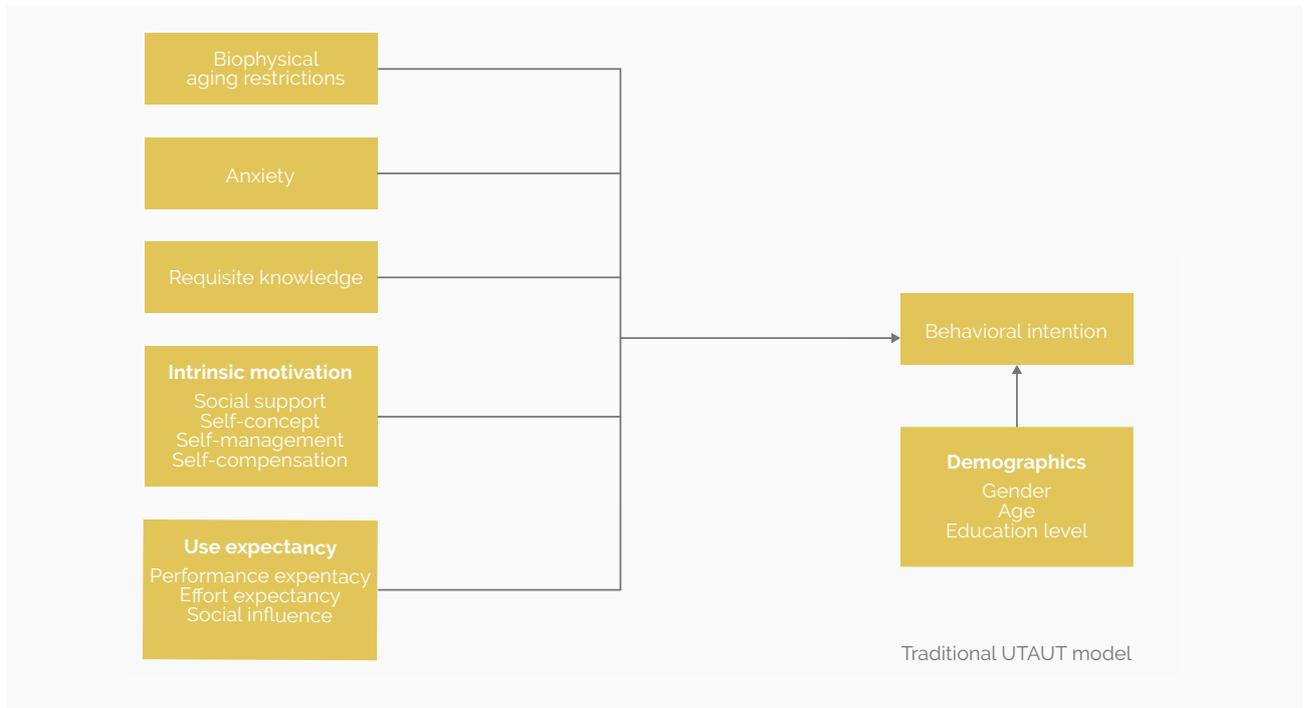

*Figure 3.15 Model of Technology Adoption by Older Adults[13] 3.2.4.3*

## ELDERADOPT

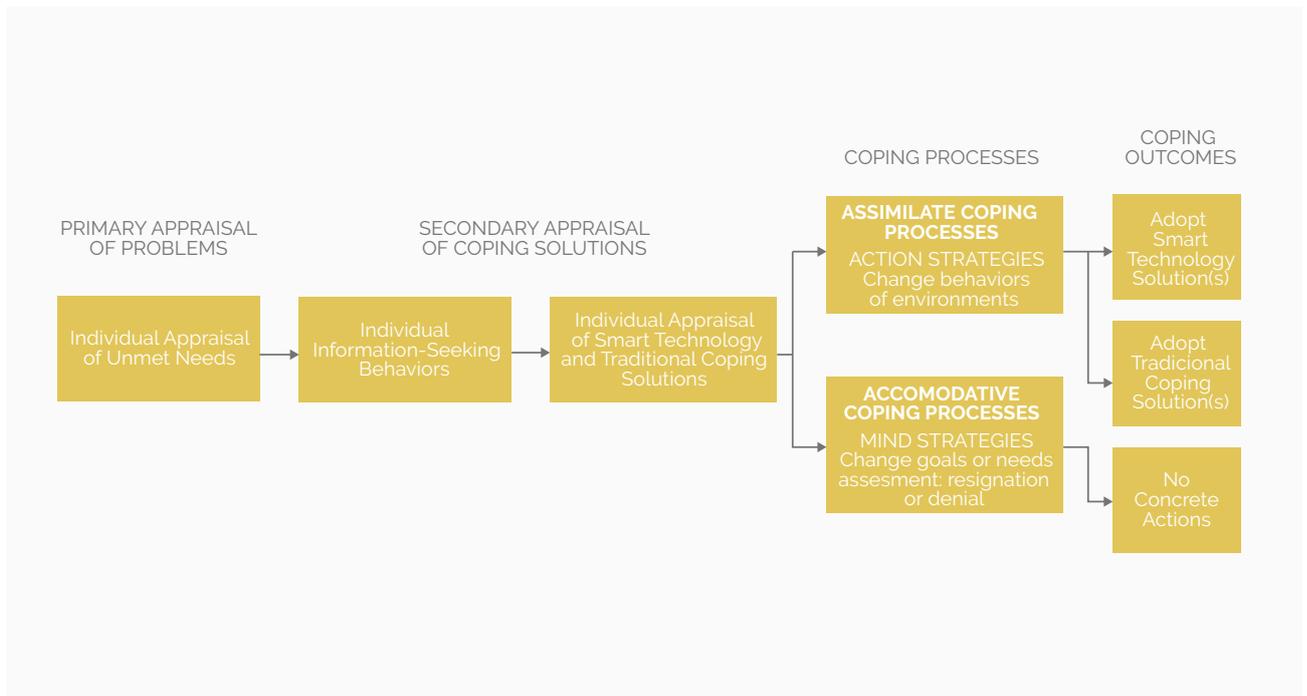

*Figure 3.16 Elderadopt model – appraisal problems[14]*

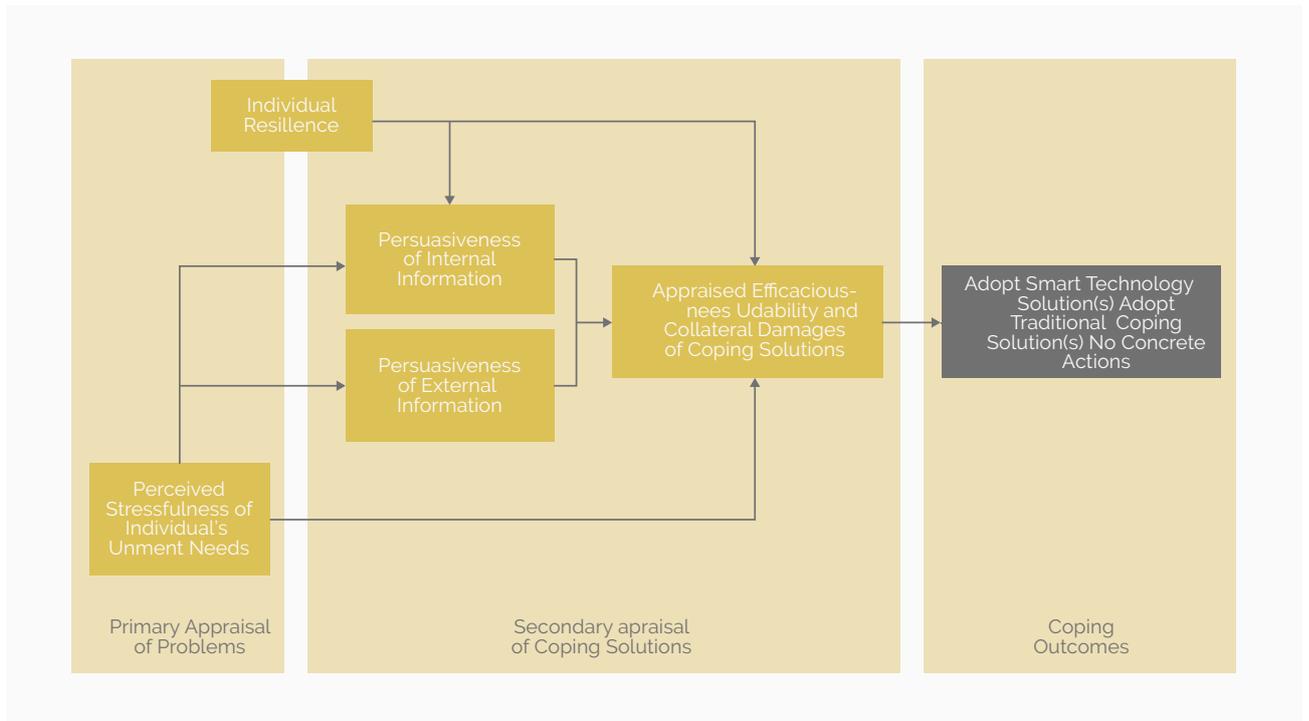

*Figure 3.17 Elderadopt model – basic concepts[14] 3.2.5*

**Relevant literature from systematic review**

Table 2 holds the most relevant papers retrieved by the systematic review:
Table 2 Relevant research for systematic review regarding Adoption, Accessibility and Acceptance

| No | Reference | Method | Main findings |
|----|-----------|--------|---------------|
| 1 | Steele, R, Lo, A, Secombe, C, Wong, YK, Elderly persons' perception and acceptance of using wireless sensor networks to assist healthcare, INTERNATIONAL JOURNAL OF MEDICAL INFORMATICS, 2009, 10.1016/j.ijmedinf.2009.08.001 | Focus groups-based analysis of factors influencing adoption and acceptance of sensors for health monitoring | Main findings: <br> · Independence of living and prolongation of independence is a critical value for older adults. <br> · The main concern was not privacy, but affordability. <br> · For female older adults, the social perception of visible sensor/device matters. <br> · Embedded solutions are preferred, as they are less visible and require less interaction and control. |
| 2 | Mercer, K, Giangregorio, L, Schneider, E, Chilana, P, Li, M, Grindrod, K, Acceptance of Commercially Available Wearable Activity Trackers Among Adults Aged Over 50 and With Chronic Illness: A Mixed-Methods Evaluation, JMIR MHEALTH AND UHEALTH, 2016, 10.2196/mhealth.4225 | Semi-Experiment. A group of 32 elderly tested wearable activity trackers in a random order. Acceptance criteria were developed from TAM model. | The wearable sensors were highly accepted, with. Preference was given to the ones that were: <br> · Of acceptable accuracy <br> · Compatible with cheap PC Or Android devices <br> · Well – documented (easy and extensive instructions provided in paper form) <br> · Fitbit zip device the Best to use, learn, etc. although another one was the most precise |
| 3 | Hanson, VL, Influencing technology adoption by older adults, INTERACTING WITH COMPUTERS, 2010, 10.1016/j.intcom.2010.09.001 | Investigated factors influencing use of technology. Experiments based on web search problems | Main findings: <br> · Older adults are equipped with older devices, not up to date operating systems versions <br> · They are rather neglecting the need to upgrade <br> · They are less robust to changes in software, upgrades etc. If it requires to learn new operational procedures <br> · Older adults were slower, but managed to complete at the same rate as the younger participants <br> · For older adults a technology must clearly address a need and must be perceived usable. <br> · In the second experiment eye tracker was used and revealed some differences |

| No | Reference | Method | Main findings |
|----|-----------|--------|---------------|
| 4 | Chen, K, Chan, AHS, Gerontechnology acceptance by elderly Hong Kong Chinese: a senior technology acceptance model (STAM), ERGONOMICS, 2014, 10.1080/00140139.2014.895855 | Questionnaire and interviews in 6 elderly care structures in Hong Kong | Explores several factors (age, gender, education, income, health satisfaction, movement ability, social). Basically, confirms intuition |
| 5 | Lee, C, Coughlin, JF, PERSPECTIVE: Older Adults' Adoption of Technology: An Integrated Approach to Identifying Determinants and Barriers, JOURNAL OF PRODUCT INNOVATION MANAGEMENT, 2015, 10.1111/jpim.12176 | | The 10 factors identified as the facilitators or determinants of older adults' adoption of technology provide a holistic framework that covers social contexts of use and delivery and communication channels as well as individual characteristics and technical features: value, usability, affordability, accessibility, technical support, social support, emotion, independence, experience, and confidence. |
| 6 | Peek et al. - Factors influencing acceptance of technology for aging in place: A systematic review - IJMI 2014 | Meta-analysis | Explores literature for criteria that influence acceptance. Distinguish two stages: pre-implementation and post-implementation. End by grouping influencing factors in 6 groups (core characteristics (age), concerns at technology, expected benefits, need, existence of alternatives, social influence). Questions the TAM/UTAUT models for acceptance. |
| 7 | Peek, Sebastian P. M. - Understanding technology acceptance by older adults who are aging in place: a dynamic perspective - Ph.D Thesis, 2017 | Study groups and meta-analysis | Ph.D Thesis that tries to understand better the factors that influence technology adoption by older adults. First chapters explore the difference in perception from different stakeholders (older adults, social workers, technology designers, policy makers, etc.). Second chapters look at specific issues (acceptance criteria, reasons for adoption, influence of the family,...). |
| 8 | Di Lecce, V., et al.: Smart Postural Monitor for Elderly People. In: IMEKO TC4, Spain (2017). | | A low-cost Smartphone application that is able to transform a bed or an armchair into an intelligent device is proposed for unskilled users. |

| No | Reference | Method | Main findings |
|----|-----------|--------|---------------|
| 9 | 11. Wang, L, Rau, PLP, Salvendy, G, OLDER ADULTS' ACCEPTANCE OF INFORMATION TECHNOLOGY, EDUCATIONAL GERONTOLOGY, 2011, 10.1080/03601277.2010.500588\ | Survey using paper-based questionnaires (two parts: (1) collect demographic information; (2) investigate variables contributing to older adult's technology use). 233 seniors included in the study (72 male, 157 females, 4 unknown) with average age 68 and four level of education. Data analysis performed using descriptive statistics od subject's current use of information technology, exploratory factor analysis (needs satisfaction, public acceptance, perceived usability, support availability), multiple regression analysis. | Four factors are identified as ones that impact older adults information technology acceptance (needs satisfaction, public acceptance, perceived usability, support availability). Future research activities should be directed to perform this study on certain information technology products and investigate the impact of additional factors such as adult's experience, interest, and cognitive abilities. |
| 10 | 12. Ehmen, H, Haesner, M, Steinke, I, Dorn, M, Govercin, M, Steinhagen-Thiessen, E, Comparison of four different mobile devices for measuring heart rate and ECG with respect to aspects of usability and acceptance by older people, APPLIED ERGONOMICS, 2012, DOI10.1016/j.apergo.2011.09.003 | Qualitative subjective user ratings and structured interviews (along with evaluation by expert and use of questionnaire) of 12 older participants (4 female, 8 male) with average age 71. Data analysis was both qualitative and quantitative. Descriptive statistics were used to analyze user ratings. Statistical significance was tested along with fisher's exact test, Cramers' V as coefficients of contingency. | Usability and acceptability of 4 different devices to be worn for the measurement of heart rate or ECG were analyzed. There were a relatively high acceptance concerning the belts but none of them were completely usable. Several important deficiencies in the design were identified: locking mechanisms, buttons/clasps, sensors positioning, belt length adjusting, etc. Further study should consider effect of long term wearing and randomized assignment of each device with larger group of participants and taking into account gender specific preferences for wearing. In addition, additional variables in the understanding of the psychological factors that equate to acceptance/rejection of device should be considered. |
| 11 | 13. Nayak, LUS, Priest, L, White, AP, An application of the technology acceptance model to the level of Internet usage by older adults, UNIVERSAL ACCESS IN THE INFORMATION SOCIETY, 2010, DOI10.1007/s10209-009-0178-8 | Cross-sectional study with 592 older adults (236 males and 365 females) that completed a postal questionnaire. Multiple regression analysis. | Shih's TAM model was validated, emphasizing the need to include demographic variables because of direct/indirect influence on Internet usage. Further studies should consider: potential differences between Internet users, attitude and perceived usefulness from website design point of view, Internet marketing strategies, training that accommodates their needs. |

| No | Reference | Method | Main findings |
|----|-----------|--------|---------------|
| 12 | 15. Ehrenhard, M, Kijl, B, Nieuwenhuis, L, Market adoption barriers of multi-stakeholder technology: Smart homes for the aging population, TECHNOLOGICAL FORECASTING AND SOCIAL CHANGE, 2014, DOI10.1016/j.techfore.2014.08.002 | Two-step qualitative research approach: (1) to determine specific market and organizational barriers in the adoption of Smart Homes, (2) to compare our findings in the case of study to other service domains. 14 interviews were conducted. | End-user requirements should be considered when designing and implementing Smart Home platform. Platform management is crucial. Price has strong effect on user behavior. Role of government can improve regulating the market, enforcing standards, safeguarding privacy, subsidizing pilot projects or initial investment costs. |
| 13 | 18. Young, R, Willis, E, Cameron, G, Geana, M, Willing but Unwilling": Attitudinal barriers to adoption of home- based health information technology among older adults, HEALTH INFORMATICS JOURNAL, 2014, DOI10.1177/1460458213486906 | Qualitative study - interviews with 35 American adults aged 46-72 years. | Home-based health information systems should incorporate familiar computer applications, alleviate privacy and security concerns, and align with older adults' active and engaged self-image. |
| 14 | 19. Hong, SJ, Lui, CSM, Hahn, J, Moon, JY, Kim, TG, How old are you really? Cognitive age in technology acceptance, DECISION SUPPORT SYSTEMS, 2013, DOI10.1016/j.dss.2013.05.008 | Empirical study comparing to groups of participants (cognitive age = chronological age vs cognitive age < chronological age) in the context of mobile data services. Research model was tested using partial least squares (PLS). | To empirically explore the moderating effect of cognitive age on theoretical relationships in other IT acceptance frameworks. Investigate whether different conceptualizations of self-perceived age to yield other interesting patterns |
| 15 | 26. Hsiao, CH, Tang, KY, Examining a Model of Mobile Healthcare Technology Acceptance by the Elderly in Taiwan, JOURNAL OF GLOBAL INFORMATION TECHNOLOGY MANAGEMENT, 2015, DOI10.1080/1097197 8X.2015.1108099 | 390 elderly (Taiwan), provided with a connected watch and a questionnaire. | Taiwan, study of the factors that would influence or affect the use of connected (mobile) health. Based on TAM and extends it to the specific case of mobile health watch. <br><br> In the scope and solid analysis, but maybe quite specific focus (mobile watch). The model still deserves to be mentioned and compared to state of the art. |
| 16 | 36. Chiu, CJ, Liu, CW, Understanding Older Adult's Technology Adoption and Withdrawal for Elderly Care and Education: Mixed Method Analysis from National Survey, JOURNAL OF MEDICAL INTERNET RESEARCH, 2017, DOI10.2196/jmir.7401 | 16 older adults taking in-depth interviews | Technology adoption behavior can be divided into 3 stages and 8 factors: preadoption (self-management, self-compensation, self-image, negative perception of technology), adoption (technology adoption barriers and adoption and usage), and postadoption (refusal to change to new technology and technology value seeking) |

| No | Reference | Method | Main findings |
|---|---|---|---|
| 17 | 37. Raymundo, TM, Santana, CD, Factors Influencing the Acceptance of Technology by Older People How the elderly in Brazil feel about using electronics, IEEE CONSUMER ELECTRONICS MAGAZINE, 2014, DOI10.1109/MCE.2014.2340071 | 100 seniors taking socioeconomic survey, the rating scale of IADL, a scale for the acceptance of technology based on the principles of the scale of attitudes toward computer use and questionnaire about the factors that influence the used of technologies by older people. | A significant number of older adults are afraid of technology and have difficulty using it. Gender, as well as cultural issues are associated with acceptance, while on the other hand age does not influence the acceptance of technology. In addition to age and gender, the authors identified the variables that affect the use and acceptance of technology such as: motivation, importance and usefulness of technology, difficulties, and feelings. |
| 18 | 38. Golant, SM, A theoretical model to explain the smart technology adoption behaviors of elder consumers (Elderadopt), JOURNAL OF AGING STUDIES, 2017, DOI10.1016/j.jaging.2017.07.003 | Survey | Theoretical model consisting of nine major and nine minor constructs and 21 propositions that articulate their relationships suggest 10 action strategies that would increase adoption of new technologies among older adults. |
| 19 | 41. Hauk, N, Huffmeier, J, Krumm, S, Ready to be a Silver Surfer? A Meta-analysis on the Relationship Between Chronological Age and Technology Acceptance, COMPUTERS IN HUMAN BEHAVIOR, 2018, DOI10.1016/j.chb.2018.01.020 | Meta-Analysis of 144 studies | A meta-analysis revealed that links between age to perceived usefulness and intention to use were both fully mediated through perceived ease of use. The results were moderated by type of technology, such that age-effects were only evident for technologies that do not address the prevailing needs of older adults, which led to conclusion that the age is only related to specific technology perceptions and only for specific technologies. This is a basis for start of age-sensitive design of specific technologies. |
| 20 | 45. Theng, CS, Sagadevan, S, Malim, NHAH, Leisure Technology for the Elderly: A Survey, User Acceptance Testing and Conceptual Design, INTERNATIONAL JOURNAL OF ADVANCED COMPUTER SCIENCE AND APPLICATIONS, 2017, WOS:000423921400014 | Survey and interviews | Improvements in cognitive abilities after using leisure technologies are reported. Different types of leisure technologies benefit older adults in different ways. The enhances version of leisure technology conceptual design by combining puzzle, music and art is proposed based. In addition the proposal contains features such as user friendly interface, in-game rewards, reminder system and social networking to encourage older adults to play. |
| 21 | 46. Wang, KH, Chen, G, Chen, HG, A MODEL OF TECHNOLOGY ADOPTION BY OLDER ADULTS, SOCIAL BEHAVIOR AND PERSONALITY, 2017, DOI10.2224/sbp.5778 | survey with 286 participants who were all aged over 46 years | proposes MODEL OF TECHNOLOGY ADOPTION BY OLDER ADULTS MATOA |

| No | Reference | Method | Main findings |
|----|-----------|--------|---------------|
| 22 | 48. Meyer, J, Older Workers and the Adoption of New Technologies in ICT-Intensive Services, LABOUR MARKETS AND DEMOGRAPHIC CHANGE, 2009, DOI10.1007/978-3-531-91478-7_5, 10.1007/978-3-531-91478-7 | Literature survey | Article concludes that the older the workforce, less likely is the adoption of new technologies or software. However, the research results show that better results in terms of adoption have been achieved in case when the companies reorganized their work regardless of the age of workforce. Finally, the analysis shows that there are further factors affecting the adoption of new or significantly improved technologies and software such as the change of market or customer requirements and the introduction of product innovation. |
| 23 | Baraković, S., Skorin-Kapov, L. Survey of research on Quality of Experience modelling for web browsing. Quality and User Experience, 2017. DOI: 10.1007/s41233-017-0009-2 | Experiment with 77 participants | Age per se does not affect user perception related to technology, but other user and context factors that differ across different age groups, such as previous experience with a given technology, visual difficulties, users' character, culture, etc. |

## Gaps in the research

A detailed literature review has identified both: the general adoption models used in evaluation of the older adults' acceptance of technologies and adoption models that were specifically designed to match the older adults' acceptance of technologies.

There are well-established areas of research, such as privacy issues, that are addressed extensively in the recent years.

However, one might point out several gaps in the research:

- Apart from multiple models, there was no study on how the concepts in diverse models overlap.
- Framework for user experience testing with the older adults is missing.
- A challenge was identified on how to reach people, who are not willing to participate in the studies. It seems research is addressing a non-representative population. Perhaps some insight reasons why older adults do not participate should be investigated (from the relatives?).

- Various QoL dimensions and smart ageing determinants should be considered. No study was found on how QoL dimensions map to concepts used in acceptance/adoption models.
- The solutions are developed for direct usage by older adults or surrounding stakeholders. For the second category privacy issues must be investigated in detail.
- Technologies for direct use by older adults are frequently not adapted enough to deficits of older adults– by design or by implementation. Especially varying abilities of older adults should be taken into account.

- The solutions should be verified by older adults, while universal questionnaires and methods are missing.

- Personal, socioeconomic and cultural contexts are rarely explored in combination in acceptability/adoption/legal issues research. Context is of a particular importance – a diverse factors might influence acceptance at home, at elderly care center or in a medical context.

# 4 Ambient living, wearables and sensors

## 4.1 Quantitative analysis of publications

The properties used for selection of publications are the following:

"cloud platforms": [["azure"], ["google cloud"], ["aws", "amazon"], ["IBM cloud"]],

"communication": [["smart gateway"],
["edge"],
["cloud"],
["fog"],
["IoT"],
["NB-IoT"],
["bluetooth"],
["zigbee"],
["universaal"],
["fiware"],
["wifi", "WLAN", "wireless"],
["LORA", "low range comm", "low range communication"],
["4G", "LTE"],
["5G"],
["broadband"],
["fiber", "optics"]],

"deployment scenarios": [["smart home"],
["smart habitat"],
["facility"],
["retirement village", "retirement home"],
["AAL", "ambient assisted"],
["ELE", "enhanced living environment"],
["outdoor", "outside"]],

"measured signals": [["physiological", "physiologic"],
["heart rate", "heartbeat", "heart beat", "pulse"],
["temperature", "thermometer"],
["blood pressure", "Hg", "hydrargyrum", "auscultation", "sphygmomanometers", "mercury", "aneroid"],
["location", "gps", "position", "tracking", "motion"],
["brain activity", "eeg", "electroencephalogram", "electro encephalogram"],
["insulin", "glucose", "BGM", "HbA1c"],
["sleep"],
["oxygen", "so2", "oximetry"]],

"sensor types": [["kinect", "rgbd", "3d camera"],
["audio", "microphone", "voice"],
["video", "camera"],
["pressure", "bed", "mattress", "piezoelectric"],
["intertial", "accelerometer", "accelerometry", "gyro", "gyroscope"],
["ECG", "cardiogram", "electrocardiogram"],
["EEG", "electroencephalogram", "encephalogram"],
["EMG", "electromyogram"],
["GPS"]]]

The results of the NLP toolkit using the keywords and performing the selection given the properties are presented in the following figures:

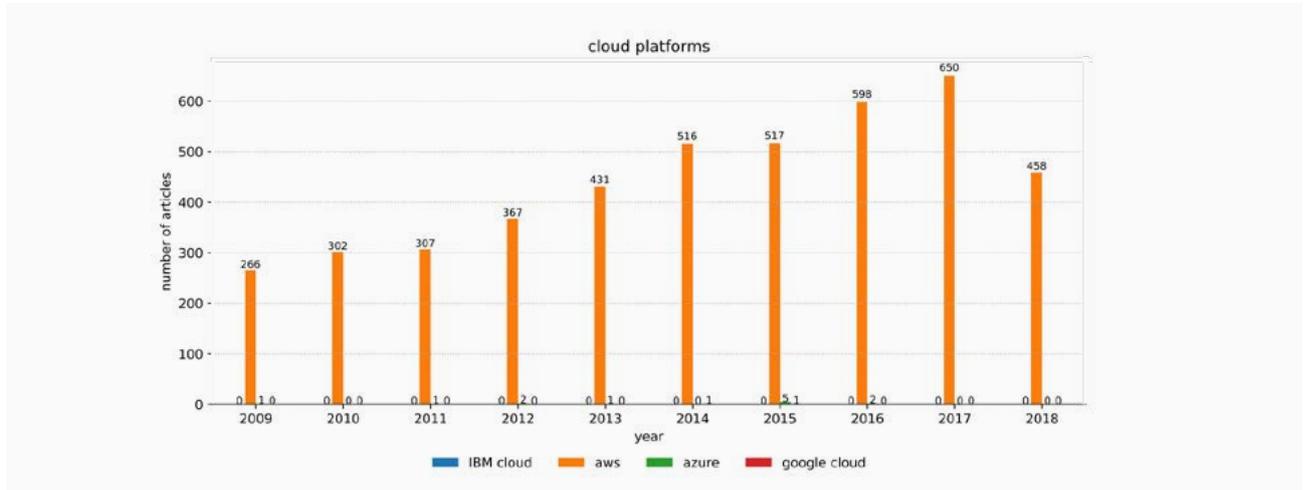

*Figure 4.1 Cloud platforms analysis per year*

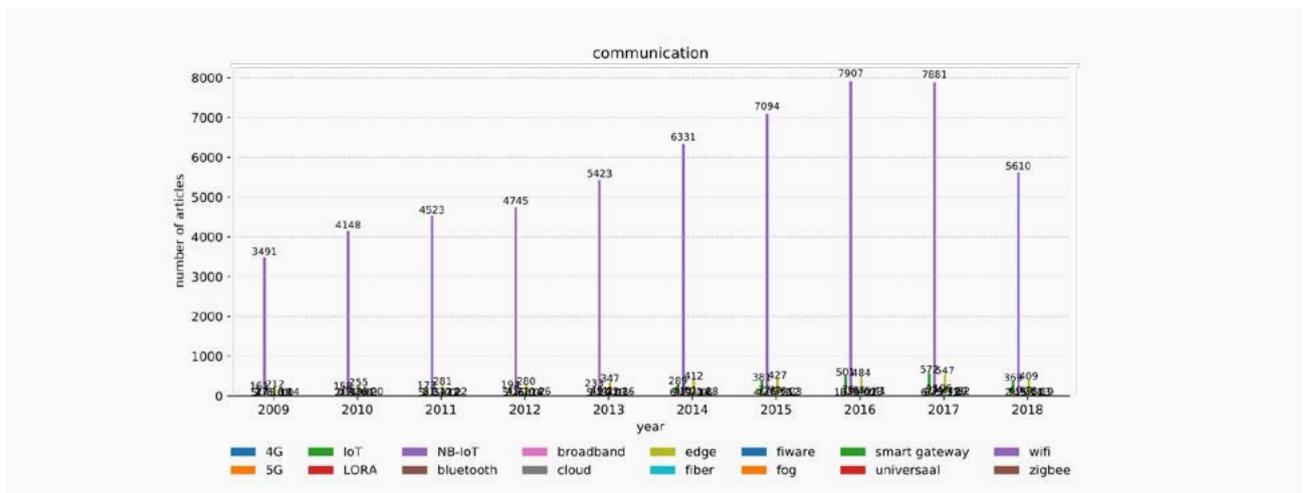

*Figure 4.2 Communication properties analysis per year*

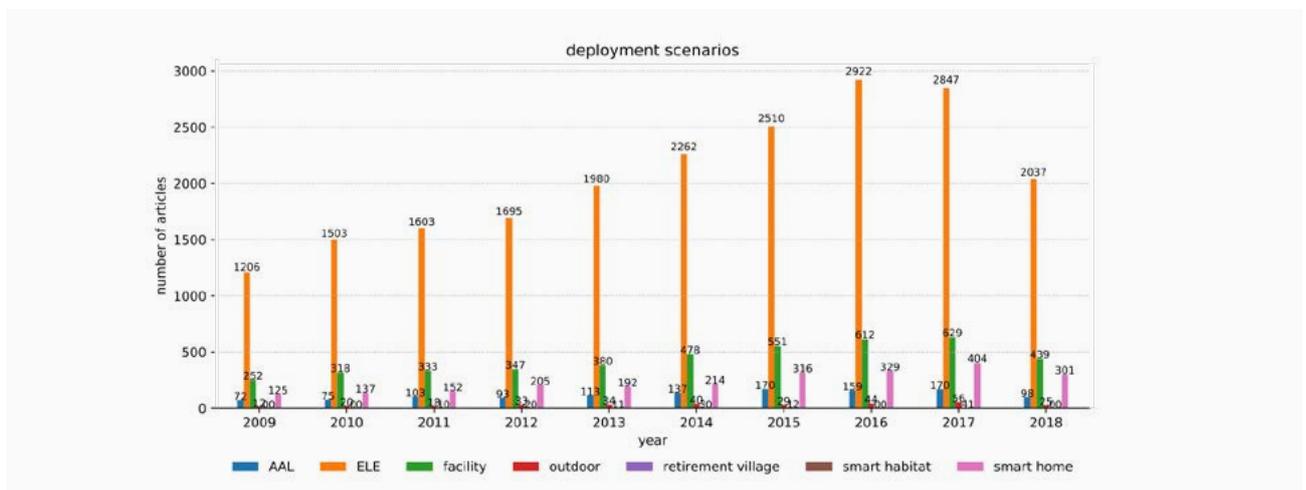

*4.3 Deployment properties analysis per year*

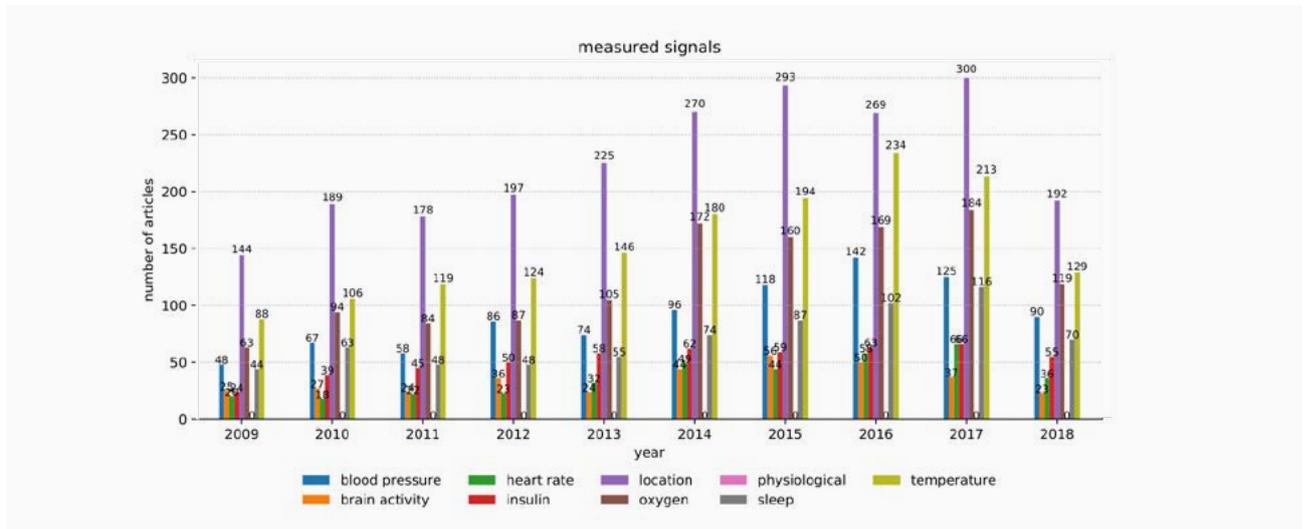

*Figure 4.4 Measured signals property analysis per year*

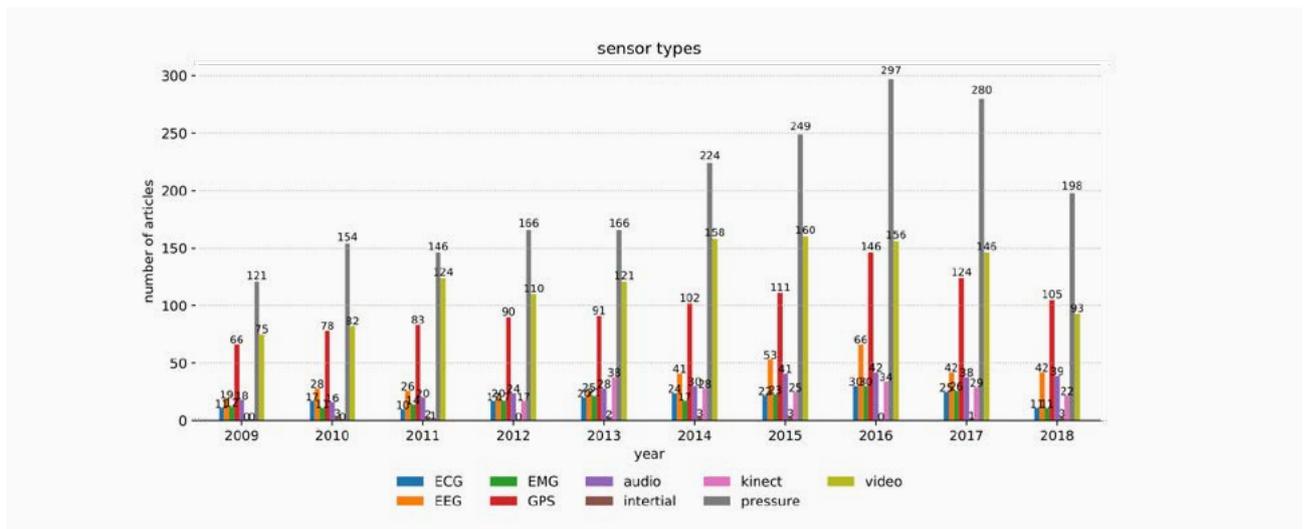

*Figure 4.5 Sensor types analysis per year*

The above figures give a valuable insight in the developments of the sensors for healthcare and health monitoring of the aging population. In the following text an analysis of the current state of the art and some conclusions are presented.

### **4.2** Communication technologies for AAL

In this section, we present a comparative study of existing wireless and mobile support technologies for AAL, extracted from[99]. The technologies chosen for this analysis are Bluetooth and Bluetooth Smart (BLE), NFC, ZigBee, Wi-Fi and Wi-Fi Direct, LoRa, 3G and 4G.

**Bluetooth**  is a low-power and low-cost short-range communication technology for fixed and mobile devices, invented by the Swedish company Ericsson in 1994. It uses short wavelength UHF radio waves in the ISM band from 2.4 and 2.485 GHz, and has become ubiquitous in most mobile devices nowadays. The latest version is 4.2, which offers support for IoT by enabling IP Connectivity and delivering new privacy methods and increased speed. In 2010, a new version of Bluetooth was added to the standard by the Bluetooth Special Interest Group, called **Bluetooth Smart**  (or Bluetooth Low Energy). Its purpose was to provide a considerably lower power consumption, while keeping a similar communication range, and it was created to be used in devices powered by small, coin-cell batteries (like watches or toys). Moreover, small devices such as sports and fitness, health care, keyboards and mice, beacons, wearables and entertainment devices benefit from Bluetooth Smart, since it makes their operation possible for more than a year without recharging.

**NFC**  is a short-range wireless connectivity technology that enables smartphones and other devices to establish radio communications with each other by bringing them in close proximity. It is mostly used for making transactions, exchanging digital content, and connecting electronic devices, being compatible with many existing contactless cards and readers. An important characteristic of NFC, which distinguishes it from similar solutions and has ensured its success, is that devices that use it are often connected to a cloud, so NFC enabled smartphones can come with dedicated apps such as "ticket" readers, while a third-party NFC device can act as a server for any action. From a hardware standpoint, it uses electromagnetic induction between two loop antennae when the operating devices are in proximity.

**ZigBee**  is a specification for a suite of protocols destined for IoT devices with small, low power digital radios, standardized in 2003 by the ZigBee Alliance. It is based on an IEEE 802.15.4 standard and can be used to transmit data over long distances by passing it through a mesh network of intermediate devices, in an opportunistic fashion. Similar to NFC, ZigBee is generally employed by applications with low data rates that need to have high autonomy, such as wireless light switches, electrical meters with in-home-displays, traffic management systems, etc. The latest version, ZigBee 3.0, was released in 2015 and offers a unified protocol specification for all levels of the network, especially the application level. Thus, according to the ZigBee Alliance specifications, everything (from joining a network to device operations like on and off) is defined, so devices from different vendors can work together seamlessly.

**Wi-Fi**  is a networking technology that uses the 2.4 GHz UHF and 5 GHz SHF ISM radio bands for offering wireless connection between devices and access points. It follows the IEEE 802.11 standards, and is completely ubiquitous nowadays, helping connect devices such as personal computers, video consoles, smartphones, digital cameras, TVs, tablets, digital audio players, etc. However, for situations where two devices in close range wish to communicate but have no access point to connect to, there is also Wi-Fi Direct , which is a related technology that allows devices to communicate through the wireless interface without needing a fixed AP.

LoRa  is a proprietary technology for Low-Power Wide-Area Networking (LPWAN), which offers a long-range protocol for public and private networks with a low power consumption. It uses the LoRaWAN protocol to perform communication between remote sensors and gateways connected to the network and stands at the basis of the Internet of Things. Its main goal is to be used for collecting data from small sensors and devices towards static gateways that are able to process and aggregate the data.

3G  and 4G  are the third and fourth generations of mobile broadband technology, according to a set of standards specified by the International Telecommunication Union. There are two 4G candidates available commercially, namely Mobile WiMAX and LTE. 3G and 4G are employed for wireless voice telephony, mobile Internet access, fixed wireless Internet access, video calls, or mobile TV, and at least one of them is present on most of the smartphones available nowadays. They are mainly used as a way of connecting to the Internet when no Wi Fi connection is active, and each mobile carrier provides 3G/4G support for a certain cost.

A comparison between the technologies presented above, based on various criteria, is shown in the table below. The first row in the table shows which of them require an infrastructure, and which do not. The ones that do not require the existence of a prior infrastructure (i.e., Bluetooth and Bluetooth Smart, NFC, ZigBee, and Wi-Fi Direct) can be used for device to device communication between nodes in range, as an alternative to the solutions that require an infrastructure.

| Metric | Bluetooth | BLE | NFC | ZigBee | WFD | Wi-Fi | LoRa | 3G | 4G |
|---|---|---|---|---|---|---|---|---|---|
| Infrastructure | No | No | No | No | No | Yes | Yes | Yes | Yes |
| Max range (m) | 100 | 50 | 0.2 | 100 | 200 | 100 | 2200 | 1000+ | 1000+ |
| Speed (Mbps) | 2.1 | 1 | 0.4 | 0.25 | 250 | 600 | 0.05 | 28 | 300 |
| Power | 1 | 0.05 | 0.05 | 0.33 | 33 | 33 | 0.05 | 16 | 20 |
| Security | WPA2 | AES | N/A | AES | WPA2 | WPA2 | AES | KASUMI | SNOW |

The table also shows, in the second row, the maximum range in meters for each of the analyzed technologies. It can easily be seen that NFC has by far the worse range, since it cannot go farther than 20 cm. This happens because communication is done through electromagnetic induction, which cannot be performed if the two communicating devices are farther away. Other than NFC, the other technologies offer similar values, with the mention that Bluetooth Smart has a lower range than regular Bluetooth, since it was designed with power saving in mind, thus reducing the maximum range. 3G and 4G technologies connect to cell towers which offer very strong signals, so the values are in the range of kilometers (LoRa also connects to a gateway).

The third row of the table presents the transfer speed (in Megabits per second) of each of the analyzed protocols. Wi-Fi clearly has the highest speed, which can be even higher (up to 1300 Mbps) if an 802.11ac router is used . 4G can reach about half the speed of Wi-Fi, and from the no-infrastructure protocols, Wi-Fi Direct has the highest speed. Since NFC and ZigBee were created for short communications (e.g., between various sensors in a smart home), they have very low speeds. The speeds obtained by Bluetooth and Bluetooth Smart are higher with one order, and Bluetooth even supports speeds up to 24 Mbps if used in tandem with Wi-Fi .

However, power consumption should also be taken into consideration when choosing a protocol, since ALL nodes are generally battery-powered mobile devices. Thus, we show in the fourth row of the table above the power consumed by each of the analyzed technologies. We have taken Bluetooth as a gauge, which is why its power is shown as 1. The Bluetooth Smart low-power protocol 4.3

**4.3** Sensors for AAL

**State of the art survey on multimedia sensors for AAL**

The field of computer vision has been growing steadily and attracting the interest of both researchers and the industry. Especially in the last decade, enormous advances have been made with regard to automated and reliable recognition of image or video content. Additionally, recent advances in 3D scene acquisition, such as the Microsoft Kinect depth sensor, represent a huge leap forward, enabling 3D modelling and body pose estimation in real time with low cost and mostly simple setup solutions. Such advances are very relevant to the field of active and assisted living (AAL).

Traditionally binary sensors have been employed to provide the infrastructure of a smart home upon which services can then be provided to assist people and provide comfort, safety and e-health services, among others. However, binary sensors reach their limits when complex scenarios have to be taken into account that require a broader knowledge of what is happening and what a person is doing. One or more visual sensors can provide very detailed information about the state of the environment and its inhabitants when combined with the aforementioned pattern recognition and machine learning techniques.

Even though the idea of using cameras to monitor older or impaired people easily raises privacy concerns, computer vision has been considered widely for AAL[14][18]. This is due to the multiple types of AAL scenarios in which the use of cameras would still be acceptable, such as in public facilities, nursing centers and hospitals, and during specific activities or events, such as tele-rehabilitation or safety assessment. Since image and video can provide very rich data about a person's activity, research has also been carried out to enhance monitoring systems with security[65] and privacy protection techniques[3][17]. In this sense, cameras can provide rich sensor data for human monitoring, not only complementing systems with networks of binary sensors, but potentially replacing them in a near future.

Methods applied in computer vision for activity recognition in AAL vary greatly in complexity, with some only aiming at basic active versus sedentary identification, or very basic pose estimation (e.g. for fall detection), only consid- ering standing, sitting or laying poses as part of the proposed methods. Other works focus on more complex actions, such as primitives involved in the development of an activity at home, mainly with an interest on activities of daily living (ADLs) such as grooming, cooking, exercise, etc. Only few works have managed to capture knowledge from very long-term data capture, in what is identified as behaviours. Furthermore, researchers use the words action, activity and behaviour interchangeably, thus adding to the confusion when comparing the literature. This makes taxonomies necessary to identify different levels of complexity, time span, and abstraction[18].

Occlusions are one great impediment for methods identifying activities in indoor environments. Therefore, alternative top-view[20] and/or multi-camera[15] methods have been proposed to overcome the problem, since some views might show less occlusions, and additional information can reveal hidden portions of the scene from a different angle. Fusion of information from different cameras poses an additional level of complexity as it requires synchronisation and picking a fusion scheme. That is, fusion could happen at different levels: at the input by averaging or concatenation (feature fusion), during learning (e.g. view independent model training), at the decision (parallel models per view with a voting/weighting scheme), or as data fusion before feeding to the model (e.g. multiple 2D data fused into a 3D representation)[15].

Up until recently, most activity recognition methods were based on non-neural machine learning techniques. Very popular were bag-of-visual-words a visual extension of a common natural language processing (NLP) method: bag-of-words[16]. Similarly, also Fisher Vectors (FV)[62]. Random forests[70], kernel SVMs[61], spatio-temporal manifolds[8], and hidden Markov models (HMMs)[47 68] were also common.

The advent (or rather rebirth) of neural networks and deep learning (DL) have marked the start of a "new era" in many fields including computer vision. As a related field, AAL does not scape this trend either. Several reviews show that human action recognition are the application most used in this aspect, mostly because is one of the most complex (e.g. as compared to fall detection or other applications of AAL) [13 33 34 37 66].

Human action recognition is mostly viewed as a classification task. Most used networks for HAR are spatio-temporal or multiple stream networks. Spatio-temporal networks include extensions to convolutional neural networks (CNNs) that take into account temporal information: 3D-CNNs in which convolutional blocks have been augmented to work with 3D blocks of XYT pixel colour information[35] with or without motion information as an additional input channel at the input layer, such as an optical flow[34 48]. Extensions to CNNs also include temporal pooling[74]. Spatio-temporal networks also include recurrent neural networks (RNNs) using long short-term memory (LSTM) blocks, as well as hybrid CNN-LSTM networks.

Multiple stream networks include those that train colour (RGB) and motion (e.g. optical flow) information in parallel 'sub-networks' that are connected at the decision-making fully-connected layers via their softmax scores[53], or earlier, which is shown beneficial[30].

Most recently published methods tend more towards the use of DL-based methods. Among these, convolutional neural networks (CNNs) are still very widely used[22–24 38 40 45 54 58 59 63], which is also confirmed by recent reviews[29]. However, as stated in[34] with 3D spatio-temporal extensions of CNNs it is difficult to determine which number of frames should be ideal during training. In this sense, hybrid spatio-temporal networks with CNNs connected to RNNs using LSTM blocks seem to be gaining momentum as more methods appear that rely on this or similar approaches[1 2 39 43 51 64 75] another option are dynamic images a type of spatio-temporal template from a video, similar to motion history and energy images (MHI/MEI), but using rank pooling within a CNN[7].

Another interesting trend in machine learning, and specifically in DL-based techniques is that of using synthetic data[9 48 59], since neural networks generally require datasets with enough variability to generalise well. Synthetic data also helps with unbalanced classes (e.g. falls tend to be a minority case[36]), and data collection efforts. Furthermore, ground truth is known beforehand.

Human behaviour analysis has typically been performed from fixed cameras, that is, cameras located in the environment. However, a more recent trend is to use cameras worn by the individuals to monitor, thus identifying the activities performed by the information given by where and what the person is looking at. This receives the name of egocentric vision and the reader is referred to the review by Nguyen et al.[42] for further reading. This modality has some advantages, as fixed cameras have a limited field of view and are often-times impeded by occlusions produced in cluttered environments (e. g. by columns, furniture, or other people). Furthermore, manipulated objects tend to be prominent in these so called first-person or egocentric vision systems, thus facilitating ADL recognition. Indeed, current literature suggests that presence of objects is extensively used for the recognition of ADLs in AAL contexts. This is motivated by the idea that objects provide essential information about a person's activity. Constraining the scenario to an area of interest is often useful for improved object detection and classification (avoids irrelevant objects being detected). This is performed either by foreground detection[28] [49], gaze estimation[60 73], or saliency estimation[10 19 60 69]. Object-based approaches exploit features from objects (e.g. via feature point descriptors such as SIFT[11], SURF[4], and HOG[21]), hands, and relationships between them to model activities[26 27 32 41 46]. Once detected, objects can be classified. Support vector machines (SVMs) have been shown to be the most used tool for training and recognizing objects[12 27 28 31 41 46 49 50]. However, although object-based approaches have proven popular, object recognition remains a challenge due to the intra-class variations found in unconstrained scenarios. As a consequence, the performance of current systems is far from satisfactory. To alleviate this, algorithms to constrain the scenario to an area of interest have been presented, either by foreground detection[28 49], gaze estimation[60 73], or saliency estimation[10. 19. 60. 69.]. Flow or motion-based systems on the contrary can identify activities based on the motion of eyes, head, and (upper) limbs, as well as sometimes on the apparent motion of the environment with respect to the camera wearer (for instance, frequency of jitter may determine differences between walking, running, or performing other activities)[44 52]. Hybrid systems can benefit from both information sources to make better informed decisions on the type of activity performed by the wearer. To detect actions (primitives) and activities, several methods are proposed. The Bag-of-Words (BoW) approach has shown to be a popular method to model actions and has been extended in different ways: bag of active objects[25 56 73], bag of object (and wrist) interactions[6], and bag of oriented pairwise relations[5]. Temporally-aware methods are also popular, including extensions to BoW with spatio-temporal pyramids[41 46 57], temporal templates[55], or hidden Markov models (HMMs)[72 73], and related extensions, such as Dynamic Bayesian Networks (DBNs)[55 67 71].

**State of the art survey on activity recognition for AAL**

In the light of recent advancements in technology, especially in domains like artificial intelligence and machine learning, most of the gadgets people use in their day-to-day life appear not to be used to their full potential, and mobile devices make no exception. Activity recognition is one of the many sub-domains of mobile technology that escalated quickly over the past few years[103]. Its areas of use cover smart homes[101], fitness tracking, healthcare monitoring solutions[89][98], security and surveillance applications[105], tele-immersion applications[106], etc. Nowadays, with the boost of power the smartphones have been given lately, most of this technology is being transferred to the mobile world. Currently, the simplest and most common usage of activity recognition on phones is represented by fitness applications, especially running tracking, a simple search on the Play Store offering tens of choices. Recently, given the whole uncertainty surrounding the security and privacy of user data, steps have been taken towards using activity recognition for user authentication.

As mentioned, multiple ways of recognizing human activity have already been suggested. Human physical activity recognition based on computer vision[100] is one of them. As an example, one solution[93] used Kinect sensors to detect skeleton data from the human body and identify the activity based on the information extracted. The authors achieved good results, correctly identifying the performed activity in approximately four fifths of the attempts. However, the main drawback of this method is the necessity of a visual sensor, which, although it has a good accuracy, does not allow for too much mobility.

Other research in the area of activity recognition takes into consideration the possibility of using RFIDs to identify the action a person is executing[91]. This is done by exploiting passive RFID technology to localize objects in real-time and then infer their movements and interactions. The authors are able to recognize actions such as making coffee or tea, preparing a sandwich, or getting a bowl of cereal, with probabilities higher than 90%.

During the last decade, with the new discoveries in the world of artificial intelligence and machine learning, activity recognition has received a significant share of attention. Throughout the course of these last 10 years, constant improvements have been made in this domain, early studies starting from gesture spotting with body-worn inertial sensors[95] and state of the art solutions offering complex human activity recognition using smartphone and wrist-worn motion sensors[106].

One of the early attempts to recognize human activities using sensors was through body-worn sensors such as accelerometers and gyroscopes[97], in order to detect a vast range of user activities, some of these being pressing the light button, doing a handshake, picking the phone up, putting the phone down, opening a door, drinking, using a spoon or eating hand-held food (e.g., chocolate bars). For this, a set of sensors was placed on the arm of the tester, having one end on the wrist and one end on the upper arm, near the shoulder. The data collected consisted of relative orientation information, such as the angle the hand is being held at and the movement of the hand.

Another solution experimented with human activity recognition using cell phone accelerometers[97]. The authors chose to use only the accelerometer sensor because, at that time, it was the only significant motion sensor used by most mobile devices. In the meantime, several other sensors have reached the vast majority of smartphones. The activities under observation are walking, jogging, sitting, standing, climbing stairs and descending stairs.

A survey of similar approaches has been performed in 2015[107], where the authors analyze activity recognition with smartphone sensors. They categorize the type of sensors existing in smartphones (accelerometer, ambient temperature sensor, gravity sensor, gyroscope, light sensor, linear acceleration, magnetometer, barometer, proximity sensor, humidity sensor, etc.), as well as the types of activities that can be recognized, ranging from simple activities like walking, sitting, or standing, to more complex ones such as shopping, taking a bus, or driving a car. Furthermore, activities can also be split into living activities (eating, cooking, brushing teeth, etc.), working activities (cleaning, meeting, taking a break, working, etc.), or health activities (exercising, falling, following routines, etc.). Very importantly, the authors also extract some challenges related to smartphone-based activity recognition, among which are subject sensitivity, location sensitivity, activity complexity, energy and resource constraints, as well as insufficient training sets.

One of the more recent and complex studies proposed a solution based on both smartphones and wrist-worn motion sensors[106]. The main idea of this approach is that the way smartphones are held by their users (e.g., in the trouser pocket) is not suitable for recognizing human activities that involve hand gestures. That is why additional sensors are used besides the ones from the device. Both sets of sensors include accelerometer, gyroscope and linear acceleration.

An interesting project that also uses multiple sensors from Android devices is SociAAL, where virtual living labs are created using a framework specifically designed for this purpose, called PHAT[87]. The devices in the project use Android and are able to take advantage of all sensors that can be found on such a device, including camera, microphone, accelerometer, user input, etc.[88]

One area that has resonated a lot with activity recognition lately is sport, especially fitness and running. There are countless examples of applications which use human activity recognition to help users track their training sessions. Samsung Health offers such a tool beside many more, but there are other apps that focus entirely on running and walking, like Nike+ and Endomondo, or even cycling and swimming like Strava. Although these apps recognize a very limited number of activities, they have excellent results, being extremely accurate in detecting the type of activity performed, offering their clients benefits like auto pause when they detect the user is not running anymore and personalized training patterns.

Furthermore, in recent years there has been a significant growth in the amount of smart watches and fitness bands such as Fitbit and Apple Watch, that are able to track the number of steps, sleep patterns, passive periods, etc. These kinds of devices have started being used as components of more complex systems that employ many sensors to perform in-depth activity recognition for scenarios such as healthcare[90], healthy aging[102], persuasive technology for healthy behavior[92], etc.

People's tendency and pleasure to play never disappears and there have been a lot of improvements in the gaming area. In the past few years, activity recognition has become an active part of playing, with the creation of technologies like Kinect, Playstation Move and Nintendo Wii. While some of these recognize activity only by using visual computing (like Kinect), the others also rely on sensors. The Nintendo Wii has a remote which has a motion sensor that makes the activity recognition possible. All these are used in different ways to play games and even to create a healthy habit. By recognizing the activities that a user is performing, the computer is able to understand the user and give a response based on human reactions. This way, the human-computer interaction is possible.

Recent studies have shown that user activity and behavior can be used to indicate the health status of humans. Investigations have led medical experts to the conclusion that there is a strong correlation between the amount of physical activity and the different diseases related to obesity and metabolism. Considering the great quantity of data that can be collected regarding a person's activity, the idea of using these data to gather information about the human medical condition has grown rapidly. Although it is believed that this may be a better solution than a time-limited medical appointment, it is not regarded as a replacement, but more like an additional tool.

One of the earliest attempts to tackle this problem[96] was made using a Bluetooth-based accelerometer worn in 3 positions, and RFID modules used to tag objects used daily. The paper concludes that most people stay in one of the following states: sitting, standing, lying, walking and running. Also, the described solution includes a wrist-worn sensor used to detect hand activity in order to improve the accuracy of human activity recognition. This sensor is also fitted with an RFID reader to detect tags placed on certain objects. When the sensor reads a nearby tag, it offers data describing the hand motion when using that object.

## **4.4** State of the art conclusions on AAL

The encompassing area of Ambient Assisted Living (AAL) represents the group of technological solutions that help older adults to maintain their independence (or perceived self-esteem, life quality expectations in terms of social inclusion, depression or simply frailty) for longer periods of time than expected under no-support assumptions[76]. AAL is, in this sense, combining innovation from multi-disciplinary domains, from geriatrics and social sciences to ICT (e.g., in the form of technologies designed for monitoring and enhancing living environments). AAL aims to support older adults in their homes (through health, safety and security), at work (enabling access), in the community (offering social inclusion, as well as entertainment and leisure) and when they are on the move (through activity management and mobile services). Healthy aging is a natural continuation of the idea of ambient assisted living, and recent research has focused on finding communication and inter-connection solutions that are able to offer these benefits for older adults.

At the advent of AAL, conventional wireless sensor networks were enhanced with a distributed database approach[76], where each sensing node would send collected data to a database, which would in turn be accessed in a conventional manner. However, there were plenty of disadvantages to this approach, not least of them being inefficient storage and high bandwidth consumption. An alternative is to take a distributed processing approach, where data handling and decision making is delegated to each node, which would require intelligent agents running in a multi-agent system (where an agent would be responsible of one or more sensing nodes). O'Grady et al.[76] present a thorough review of the state of the art at the beginnings of AAL in 2010, and the research directions that were envisioned then, such as agent-based autonomic systems or component and service-based systems.

In this area of research, sensors need to be unobtrusive and pervasive, as shown by Buzzi et al.[77]. There, the authors propose a conceptual model focused on data-driven technologies and methods. This model focuses on four levels of human functioning: individual (which is ensured through prevention and rehabilitation), close relationships (solved through assistance), community (addressed through understanding) and society (where the older adults need to be successfully integrated). A pervasive ICT solution that attempts to ensure the functioning of these four categories would need to include predictive analytics, remote monitoring, interaction activities, and user online platforms.

AAL technologies are now moving towards the IoT era, where devices are interconnected between each other and through a server or a cloud, and information collected from them is aggregated and analyzed, in order to infer situations and even make decisions. Thus, researchers in the area of communication for AAL and healthy aging focus on creating unobtrusive IoT frameworks[78][79] and better methods of communication using the available low-power protocols that connect "things" in the AAL system possess (LoRaWAN, NB-IoT, Wi-Fi Direct, Bluetooth Low Energy, etc.). Furthermore, intelligent devices such as smartphones are being employed to further enhance existing systems. As an example, Marques and Pitarma propose an indoor monitoring system for AAL based on IoT[80], where devices located in a sensor layer collect data and send it to a coordinator in a separate layer, which connects to a backend where data is collected and analyzed, and where decisions are made. However, in such a situation, if there are many sensor devices in the first layer, the gateway might become extremely crowded. Thus, alternative solutions propose the use of opportunistic communication[81], where nodes at the bottom layer are able to communicate directly with each other and make decisions locally. Furthermore, they can aggregate their data and only send the aggregated values to the gateway layer or to the backend, decongesting the communication channel and increasing throughput.

Some researchers, such as Yacchirema and Palau[83], focus on building the IoT gateway technology, by proposing a solution that is able to inter-connect devices using multiple protocols (such as ZigBee, Bluetooth, Wi-Fi) by translating the sensor data into a uniform format, while also providing data storage and other services. More interestingly, solutions nowadays have also begun adding the AI component[84]. For example, Bonaccorsi et al.[82] propose a cloud robotic system for provisioning the assistive services for promoting healthy aging. Their system is composed of a service robot connected to a cloud platform, that offers localization-based services to target users in multiple environments.

Another important aspect for AAL systems is the AI component, especially for recognition of activities of daily living. The number of utilized sensors is ever increasing and the need for automated approaches for sensor and feature selection is becoming more and more present. In[87] the authors propose an automated approach for the feature and sensor selection, particularly suitable for activity recognition.

Based on the discussion above, we believe that there are several directions that the areas of AAL and healthy aging need to focus on in the near future:

- Improving the communication methods (between sensors, or from sensors to gateways or to the backend) – this implies proposing new protocols or new methods of lowering the load on the network, by finding ways of decreasing the amount of data that needs to reach the backend and performing more decisions locally.
- Adding the AI component – since the amount of data produced by IoT systems has increased significantly in recent years (veering extremely close to Big Data), novel techniques that are able to distinguish between useful and redundant data by learning from the past and adapting need to be employed.

- Building smart gateways – one of the main problems nowadays is that there is no unification of IoT devices and protocols; if someone wants to build a large IoT system for AAL and healthy aging, they need to either find only sensors created by the same producer, or they need to create additional components that are able to translate data between sensors; this is not feasible, which is why researchers nowadays are focusing on proposing smart gateways and frameworks that significantly lower the effort of adding new sensors and applications to an IoT infrastructure[85].

# ⁵Supporting emotional well-being, people with cognitive decline and ADL

## 5.1 Quantitative analysis of publications

The properties used for selection of the publications for the analysis are:

"Application": [["rehabilitation"],
　　["fall prevention", "fall detection"],
　　["ADLs", "ADL", "activities of daily living", "routine activities",
　　"routine activity", "regular activity"]],
"Basic ADLs": [["eating", "eat", "feeding", "feed", "food", "drink"],
　　["bathing", "bath", "shower"],
　　["dressing", "dress"],
　　["toileting", "toilette", "hygine", "brushing", "combing"],
　　["transferring", "transfer", "mobility", "locomotion"],
　　["continence"]],
"Healthy ageing": [["employment"],
　　["access to services", "access", "services"],
　　["phisical activity", "activities", "leisure"],
　　["social inclusion", "inclusion", "participation"],
　　["education", "learning"],
　　["living conditions"],
　　["health", "care"],
　　["safety"],
　　["long-term"],
　　["diet", "nutrition", "food"]].

"Medical conditions": [["cognitive impairment", "cognitive
　　decline"],
　　["early degeneration"],
　　["chronic disease"],
　　["diabetes", "insulin"],
　　["mental health"],
　　["happyness", "happy", "smile", "smiling"],
　　["Alzheimer's", "alzheimers", "alzheimer"],
　　["dementia"],
　　["Schizophrenia"],
　　["depression", "depressed", "sad", "sadness"]],
"Quality of Life": [["physical health", "physical"],
　　["psychological state", "physiological", "psychology"],
　　["personal beliefs", "personalized", "personalization", "prefer-
　　ence"],
　　["social aspect", "social", "social relationships"],
　　["environment", "surroundings"],
　　["Quality of Life", "QoL"]]]

Based on these properties, the obtained quantitative results are presented in the following figures:

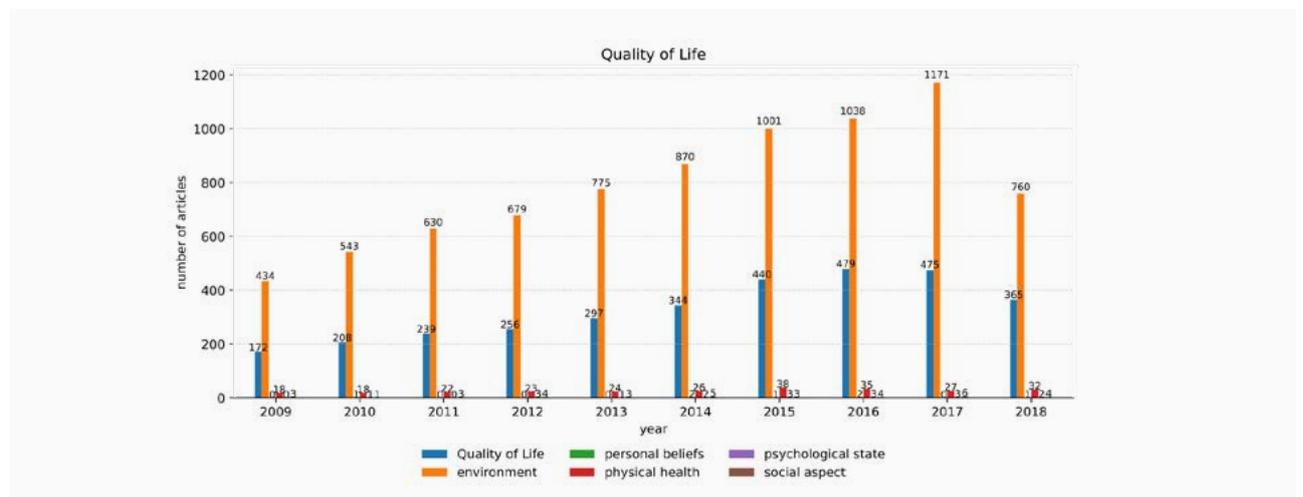

*Figure 5.1 QoL properties publications per year*

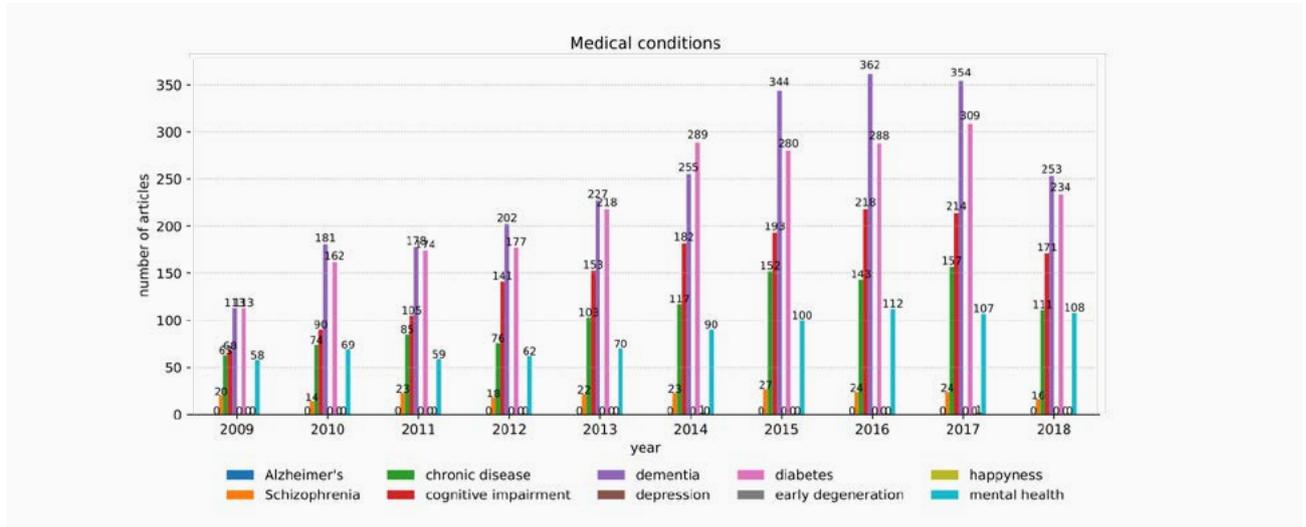

*Figure 5.2 Medical conditions properties publications per year*

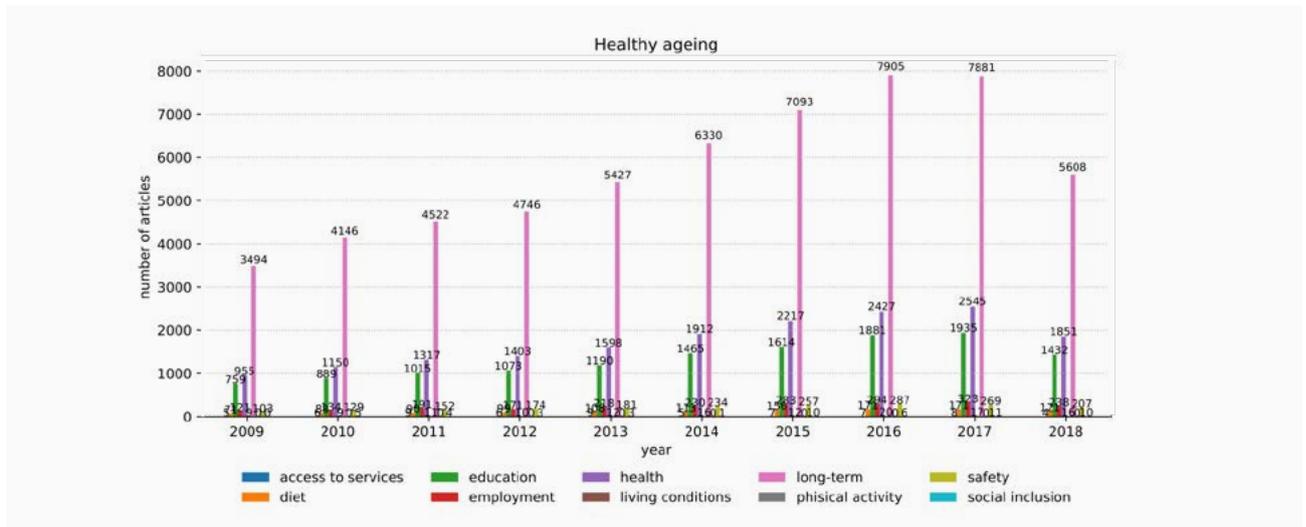

*Figure 5.3 Healthy aging properties publications per year*

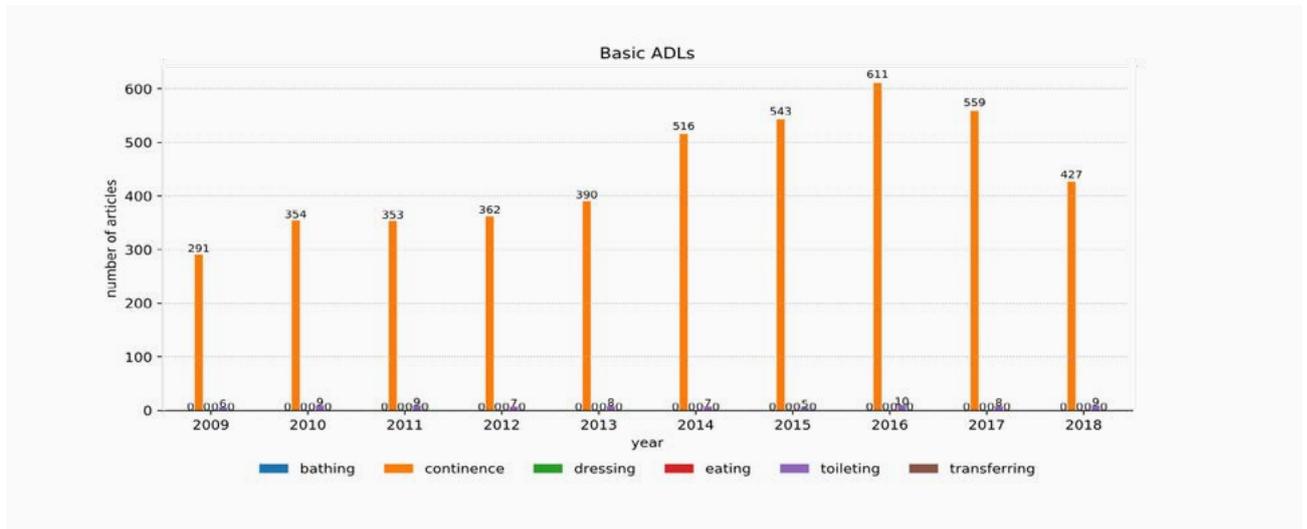

*Figure 5.4 Application properties publications per year*

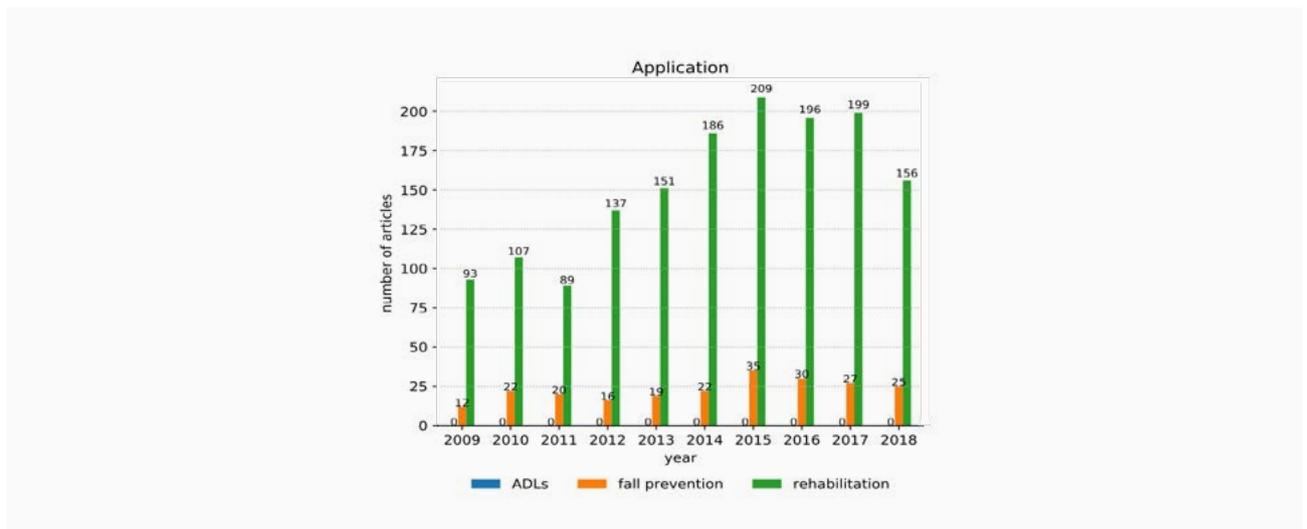

*Figure 5.5 Application properties publications per year*

Being an important issue, the increasing number of publications covering the research areas are quite logical. As part of the research promoted by this COST Action a separate review of technological solutions for people with Alzheimer diseases was published (Maresova et al. 2018).

## 5.2 Brief analysis of the state of the art on systems supporting emotional wellbeing

In the **SenseCare**[2] model has been proposed to monitor and care for the emotional well-being of older people and also people with dementia in the smart home environment. The SenseCare model utilizes multiple modalities as a sensor data which are video, audio, electrodermal activity and photoplethysmography. These modalities are combined to feature extraction and facial expression recognition by using machine learning algorithms. Video/audio features can be used outside the cloud due to privacy concern. Shortly, emotions can be stored according to diary of activities and these activities can be recorded with their expression such as "happiness" or "unhappiness".

A universal model for the home of the well-being of older adults has been explained in[2]. Several existing technologies have been suggested to use in order to improve older adult's lives by monitoring their activities, thanks to the communication of several electronic devices in their home environment. The proposed model has been defined with three dimensions: health, activity and emotional. This model includes three entities which are agents, devices, geography, and a Home Processing Unit (HPU) to manage all the data and take decision.

The Facial Action Coding System (FACS) is used for two purposes in emotional dimension. One of them is recognizing facial expression to measure the user mood. Applying the facial expression to the avatar to communicate with the user is defined as the second purpose. Happiness is defined as more important than the other emotions. This model concentrates on smiling to recognize happiness. Due to the fact that smiling is not continuous process, detection of the number of smiles per day is used rather than measuring the amount of time for smiling. In addition, 10 actions can be defined as a trigger to make people happy such as "Sunny weather" and "Looking back at old photographs".

Heart rate activity during watching different movies has been estimated by using facial videos and associated with mental health in this study. In the proposed approach in[3], 300-frame window at a time is extracted from a video sequence. Facial landmarks are tracked for each frame and the plethysmograph (PG) signals are generated. Heart rate (HR) is estimated from PG signals entire sequence by sliding the window by 10 frames. Heart rates are used for training. In summary, facial videos are classified as depressed or healthy by using the SVM classifier. According to experimental results, estimated heart rate activity can be used effectively for analysis of a person's mental health to diagnose depression.

A hierarchical learning framework has been proposed to estimate the Visual Analogue Scale (VAS) scores from facial images. VAS is defined as one of the pain evaluation scales. According to[4], this is first approach to predict pain intensity from face images. This proposed framework has two stage: (i) a bidirectional long short-term memory recurrent neural net (LSTM-RNN) is used for the estimation of Prkachin and Solomon Pain Intensity (PSPI) levels from facial landmarks for each face image and (ii) Hidden Conditional Random Fields (HCRFs) is utilized PSPI scores to estimate the personal VAS score. Individual Facial Expressiveness Score (I-FES) is suggested to personalize the model for each person. UNBC-MacMaster Shoulder Pain Expression Archive

Database[5] has been used to evaluate proposed method that performs better than existing nonpersonalized approaches.

In[6], two main contributions are pointed to estimate symptoms of Schizophrenia from facial behavior analysis. The first contribution is to record symptom assessment interviews in the patient's home and mental health service in the UK which is more similar to real-life conditions than previous studies. In addition, 91 out-patients were used in this study and this is almost 3 times more than the previous dataset. The second contribution is to propose Deep Neural Network (DNN) in order to analyze the videos for facial expressions recognition and diagnosing the symptoms of schizophrenia according to two assessments which are Positive and Negative Syndrome Scale (PANSS), and Clinical Assessment Interview for Negative Symptoms (CAINS). This proposed network, called SchiNet, includes two main stage: (i) different DNNs for facial expression recognition (e.g. smiles and activations of facial muscles for each frame), and (ii) a DNN with two-parts: Gaussian Mixture Model (GMM) and Fisher Vector (FV) layers to extract high layer features from video interview and regression layer to estimate symptom. Experimental results demonstrate that facial expression is associated with symptoms of schizophrenia and SchiNet can be used to estimate symptom severity effectively.

**5.3** Brief analysis of the state of the art on improvement
of everyday life of older adults

Population ageing threatens the social exclusion of older people. That is why it is important to improve accessibility, functionality, and safety at home, at work and in society.

The collected papers represent the efforts, to find solutions that allow the senior citizens to live safely, comfortably, and healthily in their everyday life. The titles are grouped in several strands, such as: habitat, furniture, healthcare, ICT, policy papers.

**Habitat**

The papers, collected in this strand, describe the smart home technologies, which are also often referred to as home automation or domotics (from the Latin "domus" meaning home). Among the most popular smart home technologies are the remotely controlled programmable electrical appliances such as heaters, ovens, laundry machines, etc. The smart homes provide their inhabitants security and energy efficiency through the usage of modern communication technologies. This is very important especially for those older adults who live alone, if they are skillful enough at using smartphone, tablet or computer. Thus, the digitally literate older adults can control the utilities in their homes by using smart devices, such as smart home app on their smartphone or other networked device.

The selected papers give examples of good practices in some institutional nursing homes for the elderly, as well as in some small municipalities with programs for support of the older adults, who live alone. However, these practices are rather sporadic than applied on a regular basis. Besides, the digital literacy deficits hamper the effective usage of the smart technologies by the elderly.

**Furniture**

Another group of papers focuses on the usage of specialized furniture tailored for the specific needs of senior citizens. These devices are often referred to as "everyday assistance solutions" and include applications such as smart locks, electrically adjustable seats and beds, robot vacuums and carpet cleaners, a large variety of kitchen devices, etc. Although they greatly facilitate the day-to-day lives of the older adults living alone, these devices are not very popular in Bulgaria and other countries in the region of south eastern Europe due to two major challenges: the high cost associated with the uniqueness of the products (if designed properly to the needs of the user) and the insufficient digital literacy of older adults, who need to use them.

**Healthcare**

For the senior citizens, living alone, probably the most useful remote-controlled devices are those, related to their health care. However, the publications in the media on that matter are comparatively scarce. Although there is a general knowledge about the lifesaving emergency opportunities of the medical alert pendant, the usage of this comparatively simple device to summon help in short order is not popular. This is also valid for the motion wearable sensors that can help the activities of the older adults who live alone to be detected by their relatives or trusted people. Monitoring of these activities, as well as detecting whether the necessary medications are taken in a timely manner, is perceived rather than future services than day-to-day practices. High prices and deficits of digital and mental sustainability and reliability are among the main challenges to the broad application of these remote smart services in everyday life of the older adults living alone.

## ICT

Population ageing and the development of modern communication environment are two inter-linked processes. The trends of demographic crisis in the country determine the need for urgent prevention of digital generation divide, i.e. of the vulnerability and the social exclusion of older people from the contemporary ICT world, which is necessary and important for their everyday well-being at home and in society. In the selected publications the information and communication technologies are represented as pervasive assistive tools for the daily life of the senior citizens. It is underlined that in contemporary times all improvements related to providing a more comfortable, secure, healthy and socially meaningful life for the older population are technologically driven. The need for digital literacy is described as urgent for supporting the older adults to better manage their daily living. The challenges of providing suitable IT equipment and maintenance, as well as the difficulties of their usage by older adults people with physical or sensory disabilities are also pointed out.

## Policy papers

The amount of attention devoted to older people is still not proportionate to the challenges they face in the modern world. For instance, in the United Nations Universal Declaration of Human Rights (UDHR), adopted in 1948, as well as in the International Covenant on Economic, Social and Cultural Rights (1966) and the International Covenant on Civil and Political Rights (1966), which lay the basis of the International Bill of Human Rights (adopted in 1976), age discrimination is not explicitly referred to.

The Madrid International Plan of Action on Ageing (MIPAA) and the Political Declaration, adopted by the Second World Assembly on Ageing in April 2002, nowadays are still among the global guiding documents that have a priority focus in the areas of the rights of older adults and their well-being in a supportive environment (United Nations, 2014). Certain provisions on the equal and respectful treatment of old people are present in the Charter of Fundamental Rights of the European Union, drafted in 2000, which entered into force after the Treaty of Lisbon on December 1, 2009.

Since the start of the second decade of the 21st century, there have been active efforts to promote the adoption of a special Convention on the Rights of Older Persons by the UN. Alas, with no success, so far.

The Active Ageing Index (AAI), jointly developed in 2012 by the United Nations Economic Commission for Europe and the European Commission as a key monitoring tool for policy makers to enable them to devise evidence-informed strategies in dealing with the challenges of population ageing and its impacts on society. The Index is built on four domains: employment; participation in society; independent healthy and secure living; capacity and enabling environment for active ageing. Among the six factors of the fourth domain are use of ICT, social contacts and educational attainment.

Merit deserves the publications related to the institutional policies on ageing. In the selection are presented policy papers of the United Nations, the Council of Europe, the European Commission, etc. Some of the activities of the government of Bulgaria are also included in the selection. The National Strategy for Long-Term Care, adopted by the Council of Ministers in 2018, outlines the vision of a country with fast ageing population how to tackle the variety of problems. For that reason, a comprehensive Action Plan for the Period 2018-2021 for the Implementation of the National Strategy for Long-Term Care has been developed. It outlines the main measures to be undertaken as well as the allocated budget of close to 1 mln. Euro.

The concept of "active aging" in the National Strategy is based on three main interrelated aspects: work, in tune with the focus of the Organization for Economic Cooperation and Development in Europe on the ability of older adults to realize a real economic and social contribution to society; health, connected with the EU Communication Europe for All Ages, according to which better healthy living conditions should be ensured to the older adults; and social, related to the definition of "active aging" of the World Health Organization regarding the necessity of complex participation of the senior citizens in the social, economic, cultural, spiritual and civil life.

In order to deal with the challenges to population ageing, it is important to analyze how these issues are presented in the media ecosystem (traditional media – press, radio and television; social media - blogosphere, and social networks), and to seek information on the current good practices and deficits regarding digital technologies in the everyday life of the older adults.

# ⁶ Natural settings medical data analytics

## 6.1 Quantitative analysis of publications

The properties used for selection of the publications for the analysis are:

"**Data Analytics**": [["Recommencation systems", "Recommenca-
    tion engine"],
    ["Expert systems"],
    ["Bio-markers", "cut-off values"],
    ["Diagnostics"],
    ["Data mining", "mine data"],
    ["AI", "artificial intelligence"],
    ["Image processing"],
    ["Recognition"],
    ["Risk estimation", "Risk estimate"],
    ["Prediction"],
    ["Machine learning",
    "probability",
    "ML",
    "DL",
    "Deep learning",

    "Logistic",
    "Linear",
    "ANN",
    "NN",
    "Neural nets",
    "Neural networks",
    "Gaussian",
    "Bayes",
    "Decision Tree"]],
"**Medical Databases**": [["electronic health records", "ehr", "medi-
    cal records", "health records", "medical data", "health data"],
    ["database", "data base", "data warehouse", "data marts"],
    ["digitalization", "digital"],
    ["integration", "harmonization", "compliance", "ETL"],
    ["information system"]]]

Based on these properties the following results were obtained and presented in the figures:

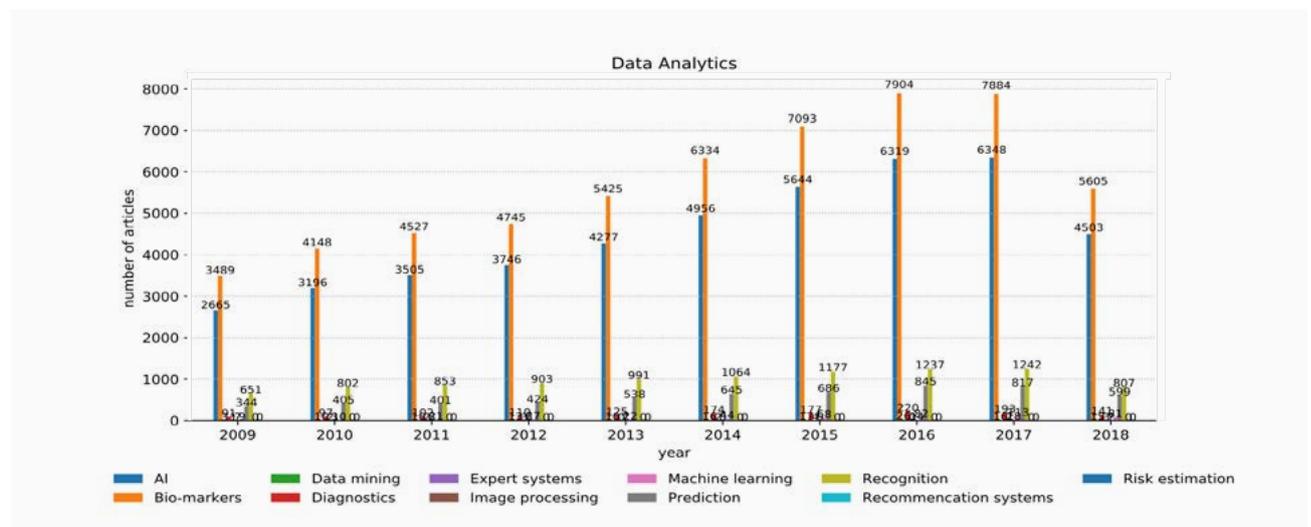

*Figure 6.1 Publications per year based on Data Analytics property group*

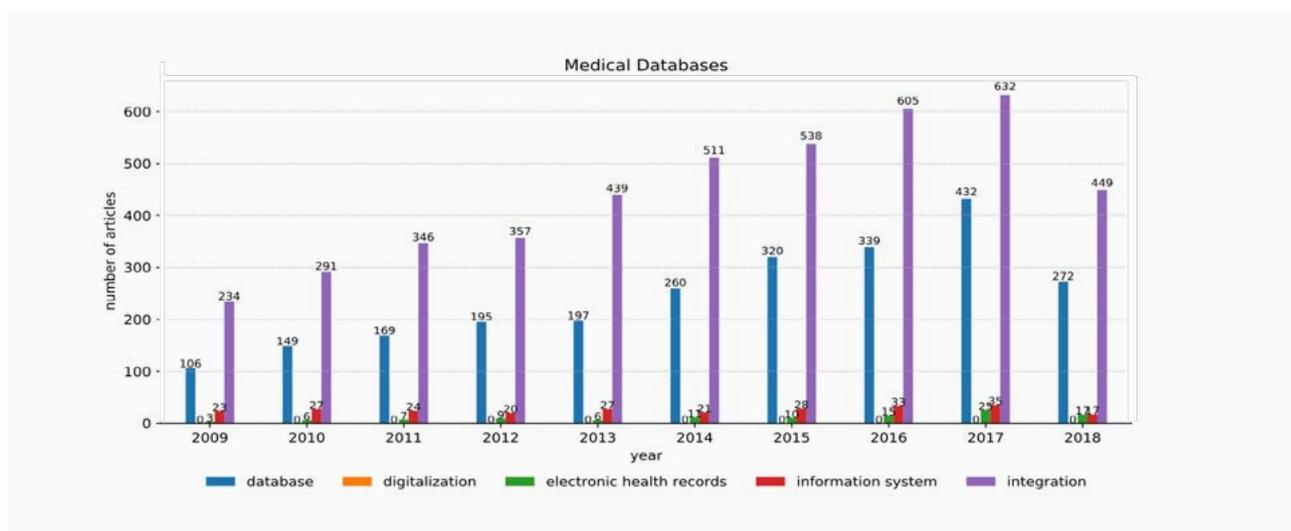

*Figure 6.2 Publications per year based on Medical Databases property group*

Based on the findings, the most papers are focused on integration and databases and for the data analytics the most papers are focused on using AI and Bio-markers for processing of the medical data. The trends of all properties are increasing each year which means that the interest in the scientific community for researching these topics is also increasing.

## 6.2 Brief qualitative survey and conclusions

The barriers for the use of data from electronic health/medical records include a limited scope of collected data, no systematically recording of the data, a lack of compatibility in variables coding and prescribing guidelines, different terminology, meanings and missing relations between other databases[1][2]. These barriers affect the process of data processing and analysis, such as missing values, a curse of dimensionality, bias control and own limitations of relevant observation study[3]. Existing heavy regulation and bureaucracy have slowed innovation in the process of digitizing healthcare. Azaria et al. proposed a decentralized record management system to handle electronic medical records using blockchain technology[4]. The similar approach is proposed by various researchers to enable the patient to own, control and share their own data easily and securely without violating privacy[5][6][7]. Dubovitskaya et al. present a prototype of a framework for managing and sharing data for cancer patient care through blockchain[10].

The blockchain is a distributed database that contains an ordered list of records linked together through a chain, on blocks. This approach is faced by various challenges, such as transparency and confidentiality as on a blockchain network everyone can see everything[8].

European health systems and databases are diverse and fragmented; it is a lack of harmonization of data formats, processing and transfer[9]. We need to improve existing methods, tools and legal frameworks to generate, collect and analyze data more effectively. For example, the EHR4CR project has designed a secure platform for eleven participating hospitals and ten pharmaceutical companies located in seven European countries [11].

When we will be able to provide sufficient quality of the electronic medical records and related services, we will be able to provide useful and correct analytical results supporting decisions during diagnostic and therapeutic procedures. Secondary use of this data for research purposes can bring a wide range of benefits, such as new diagnostic procedures, new key biomarkers, hidden relations between biomarkers or drugs, etc. New information systems and technologies should be a patient-oriented allowing an individual access to the data and services. Different stakeholders should be involved in the whole development cycle to ensure the result meets their needs. It is important to build an active awareness of new solutions or services to increase the number of users and also to help reduce a resistance to changes.

# ⁷ Solutions for caregivers, security and privacy

## ⁷·¹ Quantitative analysis

The properties selected for the further filtering of the publications are the following:

"**Caregiver solutions**": [["Monitoring"],
["Detection"],
["Tracking"],
["Recommendation"],
["Caregivers", "Hospitals", "nurses", "nursing home", "retirement home", "retirement village"]],
"**Privacy and security**": [["privacy", "privateness", "discrete", "discreteness"],
["intrusive", "intrusiveness"],
["security", "secure", "Cybersecurity"],
["risk-awareness", "threat awareness"],
["security protocol"],
["PAN", "Personal Area Network"],
["consent", "approval", "approve"],
["safety", "safe", "risk-free", "low-risk", "riskless", "impregnable", "attentive"],
["reliability", "reliable", "dependable", "dependability"]],
"**Security solutions**": [["PIN"],
["PAKE", "Password-Authenticated Key Exchange"],
["BAKE", "Biometric-Authenticated Key Exchange"],
["kyes"],

["SSL", "AES", "RSA", "private-public key"],
["multi-factor authenticattion", "2-factor authenticattion", "two-factor authenticattion"],
["authentication", "password"],
["access rules", "whitelist", "blacklist"],
["certificates"],
["encryption"],
["firewall"],
["biometric", "fongerprint", "retina", "face recognition"]],
"**Threats**": [["threats"],
["vulnerabilities"],
["malicious", "bad actors"],
["data leak"],
["breach"],
["eavesdropping", "eavesdrop"],
["interruption"],
["DOS", "DDOS"],
["man in the middle", "MITM"]]]

Based on these properties and property groups the following results were obtained as depicted in the figures:

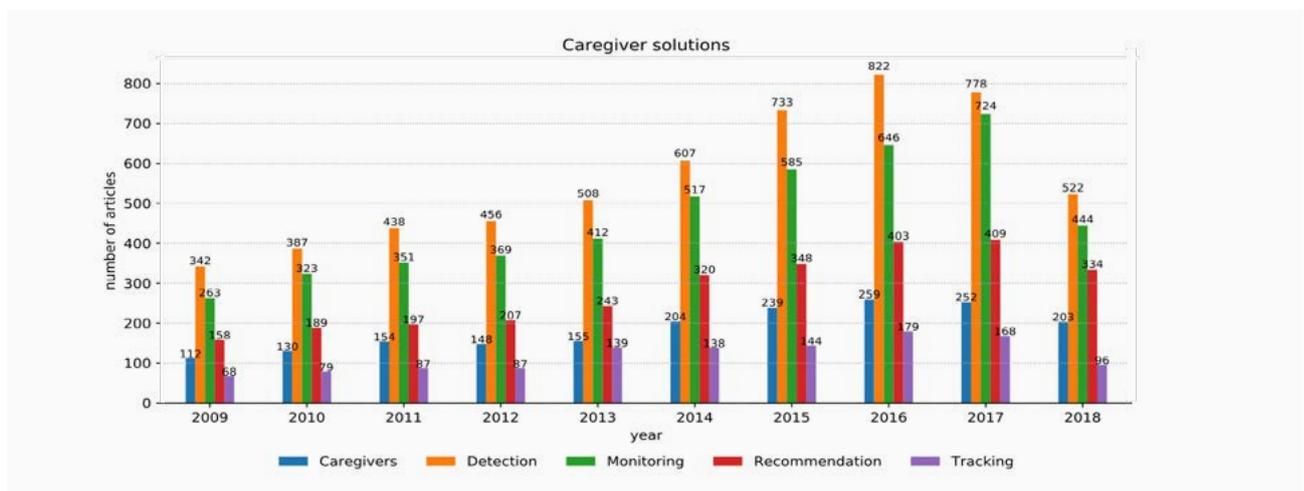

*Figure 7.1 Publications per year based on the Caregivers sollutions property goup*

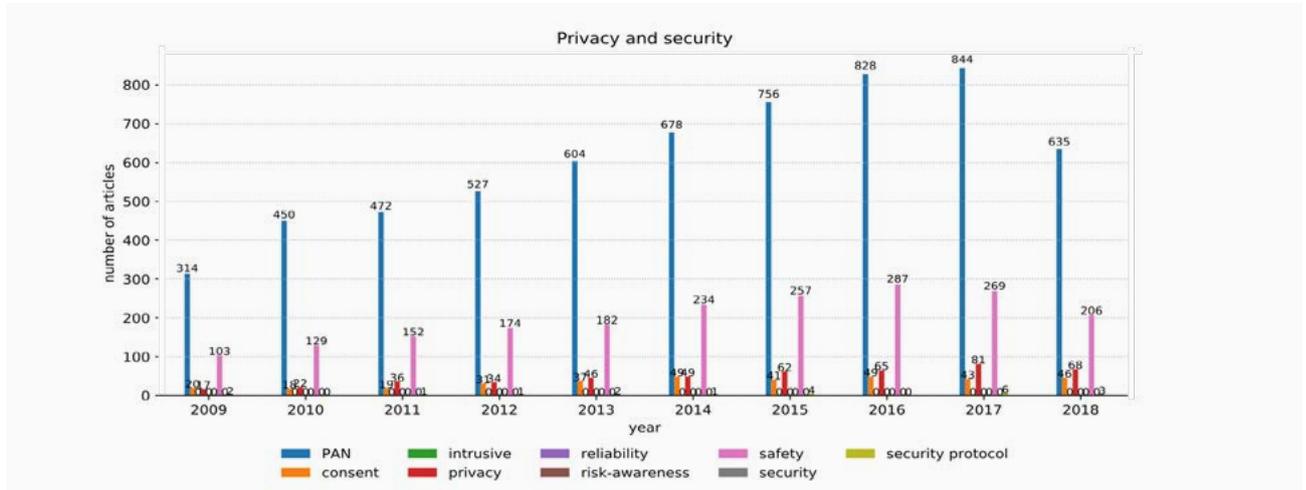

*Figure 7.2 Publications based on Privacy and Security property group*

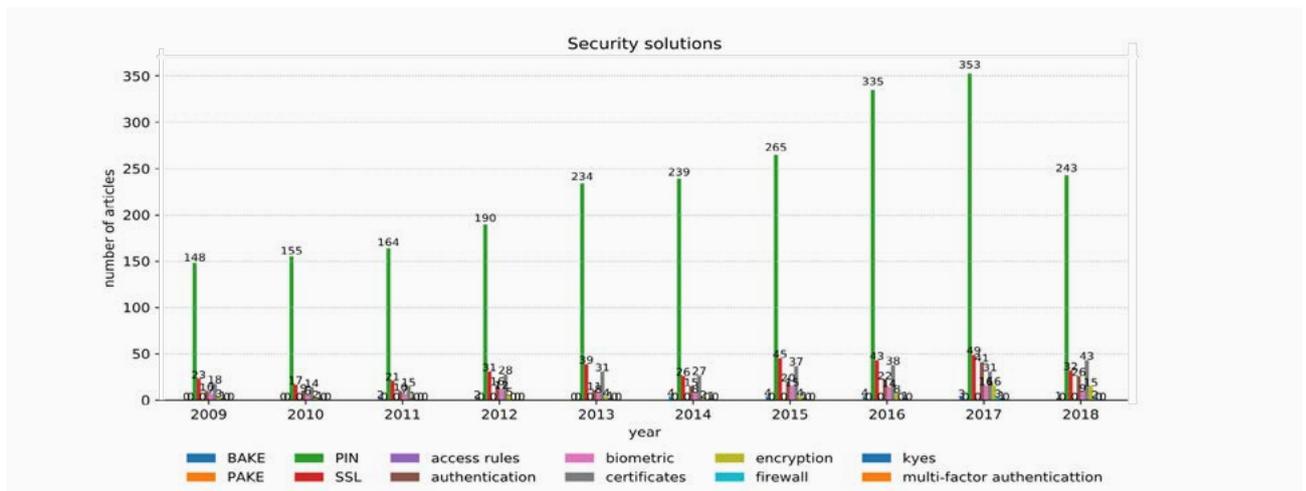

*Figure 7.3 Publications based on the Security sollutions property group*

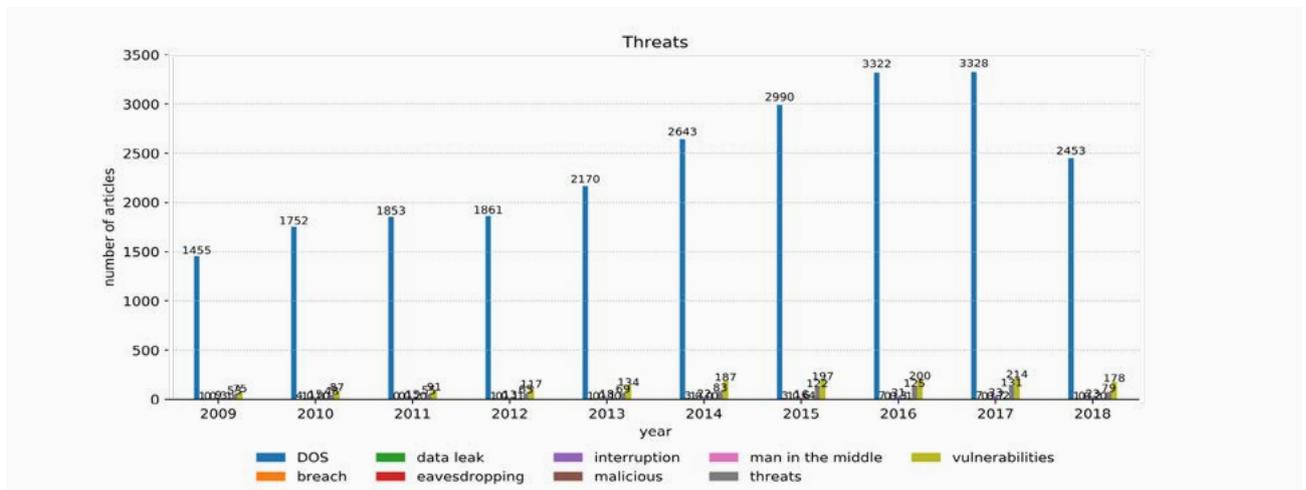

*Figure 7.4 Publications based on Threats property group*

The findings in the quantitative analysis show increasing trend of research interest for this important topic. The security and privacy are especially seriously treated by the European legal bodies, however there are still many challenges that need to be overcome.

## 7.2 Brief qualitative review and conclusions

As emerging field Ambient Assisted Living (AAL) benefits from and is often cross connected from other newer research fields such as Internet of Things (IoT). Security considerations are often more present in the later stages. As AAL field matures, more researchers tackle the issue of security and privacy. Cybersecurity is critically important not only because of the growing number of threats, vulnerabilities, and bad actors, but because the technology becomes intrinsically more sensitive with greater impact in many aspects of person's life[12].

A survey has shown that people care about the security and privacy of their data, especially in the case of medical information obtained by connected health. The attitude towards data security and privacy doesn't differ with age, experiences with disabilities, or care needs[13]. However, a survey conducted by Bellekens et al.[14] indicate that the vast majority of the participants trust the services they are using with regards to security and privacy. The users also demonstrate limited knowledge and awareness of current threats and consider themselves and the devices to be immune to a large majority of threats. Numerous participants also demonstrate an unclear understanding of risks and threats, further compromising their awareness.

Acceptance of technology may be hampered by fear of future security challenges and monitored by medical personnel[21]. To limit these effects, the older people are looking for valuable tools that do not take away their dignity and privacy[19]. They seek the development of discrete and light technologies that do not designate them as elderly or infirm, but allow them to integrate into society. As a result, they are more attracted to watch or mobile phone-based remote assistance devices. In the end, Lai et al. (2010) assessed the Chinese elder's perception of ICT development improving home security[22]. The results confirm that 91% of the sample believe that non-intrusive surveillance at home system is useful for meeting this purpose. However, while the majority of seniors ask for the device, few say they want to personally access this gerontology. They justify their way of thinking by saying that monitoring is not perceived as useful in their real daily life, for all their needs.

When technology needs to collect private and intimate (psychic, mental and/or social) to ensure the safety of the elder, several ethic rules must be respected[19]. Thus Leikas and Koivisto[20] argue that stakeholders have to:

- Collect only those user-related data essential to the achievement of the specific objectives set;
- Provide to the user with the ability to quickly and easily control and verify private information, while restricting third party access (strict confidentiality guidelines);
- Inform the user about the span of the storage of private information and respect fixed deadlines;

- Inform the user about the content of the data collected, the reasons and the way in which they will be exploited;

- Allow the user the ability to easily cancel the permission to access, collect and exploit personal data.

An important aspect of security and privacy is its intimate relation to social exclusion and discrimination[19]. The Satori project (2015) reflects these concerns in the context of the UE[20]. In particular, the project sets out the social exclusion that older people and people with disabilities can face in relation to information and communication technologies. Indeed, the inclusive design principle emphasizes the accessibility and needs of the weakest members of the target group when designing products and services. This principle promotes socially sustainable development because the concept of accessibility refers to obstacles which could exclude users with weaker skills or prevent the diversification of users through specialized designs. For older people and people with disabilities, the legislative pressures against discrimination will lead to the development of accessible systems needed to promote independent living, improve quality of life and social participation, and increase the potential of the market. However, this principle can hinder a rapid evolution of products when concepts such as customizable, derived and adaptable design are not involved.

A study of security with emphasis on data protection and privacy is done by Stutz et al. in[15]. Taking relevant provisions from the European data protection directives and national laws from EU member states, the authors advocate for data economy and data avoidance. Avoiding collection of data that is not related to the AAL system objectives and providing only parts of the data to that are relevant to the health care providers reduces the risks of data exposure. If data shall be processed by AAL systems and made available to third parties, an explicit consent from the data subject is needed. In line with these considerations, and also having in mind that besides the care receiver, the AAL sensors might collect information about visitors and third parties it is important to conduct a careful review of the planed setup of sensors, actors and data processing at a very early stage. Aside from medical data collected by various sensors AAL systems often collect non-medical data such as location information from GPS or bluetooth beacons. When designing the system this paper proposes separation between data that is absolutely required and data that is optional and can but does not have to be consented to. Starting from the default setting of most privacy the care receiver, trough simple and clear interface, can opt in for more data collection while reserving the right to revoke or modify the consent at any time. For the data that will be collected, the next steps are to create inventory (ex. personal data, medical data, location-based data). For each set of data, protection requirements are defined based on the impact of any data loss, damage or disclosure (ex. normal, high, very high).

Analysis of e-health and AAL security mechanisms, with emphasis on the architecture is conducted in[12]. The paper provides an IoT protocol architecture and examines security tools and techniques that can be leveraged as part of the deployment of IoT in e-health and assisted living applications. The authors provide the following considerations for the IoT security challenges: IoT technology is relatively new, less understood than traditional ICT systems; are widely scattered geographically; are widely scattered administratively, where multiple, often heterogeneous, environments, processes, technologies, and security mechanisms exist; they tend to be silos of their own and tend to be vendor-specific; IoT systems tend not to follow an accepted layered architecture; they have limited memory and computing capabilities and consume limited power; IoT endpoints may use different addressing models. To better understand what security measures are needed, as well as for defining what security functions must be implemented at various physical or logical points in the overall IoT ecosystem, a layered architecture approach is proposed. From the few architectures that are considered the paper focuses on Open Systems IoT.

Reference Model (OSiRM) which has 7 layers and is particularly focused on security. The authors conclude with several observations. The OSiRM definition may also be useful by defining what kind of IoTSec is needed for each instance of the layers/applications/things. A simple approach to security by attempting to segregate the IoT traffic away from the institution's Intranet will not provide reliably-strong security. Approach of just relying on the intrinsic security of the Personal Area Network (e.g., ZigBee) is not providing a complete an end-to-end solution. The attack surface with IoT is very large, with the multiple attack planes, exposing the device itself, the cloud services, the mobile apps, and the corporate network itself. The use of 3rd party software libraries by vendors developing IoT systems is a severe problem.

While AAL access gateways can communicate with the cloud utilizing advanced security measurements, at the level of sensors there are many restrictions in terms of memory, processing power and hardware capabilities. To address this, Al-Hamadi et al.[16] present a lightweight security protocol to secure the medical information transmitted from the biosensor to the gateway. The security protocol comes with acceptable low computation overhead and hence results in minimal processing delay. It is usually difficult for the adversary to intercept the communication between the biosensors and gateway due to low power and short distance of the signal. However, as the security risks for this type of data, including eavesdropping, interruption, modification of data, and unauthorized access, can be significant, it is important to apply proper protections. The proposed protocol uses pre-deployment key technique and the counter method. The protocol contains two parts, a key exchange process and secure communication process. The protocol performance was evaluated using MATHLAB based simulation and was shown that it doesn't add significant delays and can be used for real time data transmission. The protocol was formally analyzed using ProVerif tool in order to validate its correctness and absence of vulnerabilities.

Another secure system for ECG data transmission based on advanced encryption standard (AES) is proposed at[17]. The System-on-Chip (SoC) for the proposed algorithms was implemented by the authors on the Xilinx ZC702 prototyping board. The achieved hardware implementation results have shown that the proposed AES and ECG identification-based system met the real-time requirements. This approach offers greater communication security compared to the previous paper, and potentially can be used by system that transmit data on longer range, but the implementation is done in hardware. Results presented in the paper show that the proposed implementation only needs 30% of hardware resources and 107 mW to process an ECG sample in 10.71 ms.

A strong, multi-factor authentication for AAL environments is discussed in[18]. This paper addresses the challenges faced by the AAL care receivers including those who suffer from a cognitive or physical disorder and can have difficulty correctly entering a Personal Identification Number (PIN) on a keyboard or recalling the correct sequence of letters in a password. The solution is offered in biometric authentication using technologies such as fingerprints or speaker recognition. The author reviews both Password-Authenticated Key Exchange (PAKE) and Biometric-Authenticated Key Exchange (B-AKE) and concludes that both PAKE and B-AKE protocols ensure mutual authentication that helps protect AAL inhabitants from phishing and man-in-the-middle attacks without users needing to possess and manage digital certificates or understand the complexities of their proper use. Authentication methods can be coupled from easy to remember passwords and multiple biometric technology offerings.

The research trend is showing an increasing interest in providing adequate security for AAL. As AAL systems move toward utilizing IoT technologies, and as more standards are introduced, the security aspects are expected to be better addressed in the future.

Security and privacy are critical topic in handling health care data with sensor networks[77]. An intelligent system utilizing the concept of smart spaces was developed to assist elderly at home[76]. A video monitoring system with sensor video streaming is supporting privacy-preserving consent from patients[75]. Such systems should also be usable and energy-efficient to avoid the needs for frequent battery replacement[74].

## 7.3 State of the art of recommendation techniques

Intelligent recommendation techniques have been widely used in both online systems and mobile applications generally to assist users in finding the information they are potentially interested in or for suggesting the most suitable services (service instances) the users may avail at the moment. More specifically in the area of improving the quality of life (QoL) of elderly, these techniques could be used to send/push to users personalized recommendations/advices for actions or physical activities they need to perform in order to be safe, feel comfortable, and stay healthy in their environment, by preserving good health condition, social connections, etc.

Recommendation systems have been extensively studied for decades, as tools for the discovery of high-quality personalized recommendations from a large volume of choices. The goal of recommendation systems is to provide users with accurate recommendations as fast as possible[12]. Different recommendation techniques have their own pros and cons, due to different recommendation tasks and different types of data sources they rely on. Many approaches have been proposed in the literature aiming for more accurate recommendations, among which collaborative filtering (CF) is the most popular. CF has achieved great success in recommendation systems since it only requires user-item interactions to make recommendations[13]. CF approaches are typically favoured over Content-Based Filtering (CBF), which is another popularly deployed recommendation technique, due to their overall better performance in predicting common behavior patterns, and their ability to address data aspects that are usually difficult to profile using CBF[15], even though some researchers argue that CF is not so appropriate for use in systems that are working with high degree of confidentiality and thus prefer to use CBF for generating predictions and recommendations, e.g. in connected health[14]. However, the CF-based recommendation approaches are domain-free and usually only depend on the user-item interactions, i.e. they are easier to adapt to real-work scenarios than the CBF-based approaches. This makes CF the most successful and widely deployed recommendation technique as it only relies on past user's behavior (e.g. user's previous transactions, ratings, etc.) without requiring specific domain information[16 15 17].

CF provides recommendations based on user-item (user-service) ratings, and independently from the nature of users and of items/services. The CF goal is to discover items/services, which the target user would be interested in, by utilizing the information from the user-item (user-service) matrix. There are a few variations of CF. The CF could be oriented towards either rating prediction or item/service ranking[18 19 13]. Rating prediction estimates the ratings first, and then recommends the items/services with highest estimated ratings. Item/service ranking directly generates an ordered list of items/services that the user would be most interested in. These are two distinct recommendation tasks, and most algorithms are designed only for one of them[19]. They are different in some respects. First, their objective is different, where rating prediction aims to predict the ratings as accurate as possible, while item/service ranking aims to generate a ranked list of items/services and does not care about the possible ratings users may give. Second, in the model-based CF approach, the training process for rating prediction involves usually only positive entries in the user-item (user-service) rating matrix, while for the item/service ranking task, both positive and negative entries are considered. Third, in the evaluation process, rating prediction is evaluated by computing the difference between predicted and actual rating, while item/service ranking is evaluated as a classification task.

Every recommendation systems need to keep track of interaction between users and items/services, which is treated as evidence of user's tastes and interests. The interaction may result either in explicit feedback (i.e., integers within a certain range) or implicit feedback (i.e., binary data collected from user behavior) 18 20. Explicit feedback refers to numeric ratings of items/services explicitly given by users. Implicit feedback refers to the form of binary ratings gathered automatically in an unobtrusive manner from users' search history, usually including item/service click logs, purchase records, etc. Explicit feedback can be used for the both aforementioned recommendation tasks[2 6 21 22 23 24]. Implicit feedback is usually only utilized for item/service ranking [25, 26] since all implicit ratings are considered to be of the same value. Compared to explicit feedback, implicit feedback is closer to the real-industry perception of the problem and potential recommendation solutions, because users are usually reluctant to spend extra time or effort supplying explicit feedback, while implicit feedback can be collected automatically at a much larger and faster scale with no user efforts needed [25, 28]. Although implicit feedback is prevalent in real-world applications, explicit feedback is still valuable to exploit, as it typically contains valuable information that reflects users' preference.

A widely accepted taxonomy divides CF approaches into two categories[21]: (1) memory-based CF, where the recommendations are based on the assumption that users who share common interests have similar taste, or items/services with similar features have similar rating patterns; (2) model-based CF, which utilize various machine learning or data mining techniques to discover complex patterns in the training data in order to make predictions based on these, and compute recommendations by means of prediction models that have been trained from the user-item (user-service) matrix [13, 12]. These two types of CF recommendation approaches complement each other by tackling the same problem from two different perspectives[29].

Generally, memory-based algorithms are simpler and easy to implement, while they obtain reasonably accurate results [12, 27]. Two of the most widely deployed memory-based approaches are based on the k-nearest neighbor (KNN) heuristic – the so-called user-based KNN and item-based KNN[19]. User-based KNN[31] predicts the rating of item/service i by user u using the existing ratings given to i by the set of users that are most similar to u. Similarly, item-based KNN [22, 32] predicts the rating of item i by user u using the existing ratings given by user u to the set of items/services that are most similar to item i. Both user-based KNN and item-based KNN need to go through two steps to make a prediction: (1) find the k most similar neighbors to the target user/item, and then (2) aggregate the neighbors to generate the predictive score [12, 30]. In the first step, the most important part is to compute the similarity between users or items/services. The two most popular choices for similarity metrics are: (1) Pearson's correlation coefficient (PCC)[33], which measures the extent to which two vectors are linearly related to each other; and (2) cosine similarity (COS)[34], which measures the similarity between two vectors by computing the cosine of the angle between them[32]. In the second step, the predicted rating is calculated based on the ratings of the k-nearest neighbors selected in the first step, e.g. as the weighted sum of the ratings of the same item/service, provided/supplied by the neighbors of the target user (usually coupled with PCC), or as the average sum of other ratings (always coupled with COS for item-based KNN).

Due to its simplicity and flexibility[30], the nearest-neighbor based CF has been extensively studied, including different similarity measures [35, 36, 37], alternative strategies to select the neighbors[38], etc. Two drawbacks of the memory-based approach are: (1) low efficiency, since the computation of similarity between users/items is expensive (quadratic time complexity) as all users/items need to be examined to make a single prediction[27], and (2) the performance of recommendation heavily depends on the similarity measure[39].

Model-based approaches utilize various machine learning techniques for the CF task, by recognizing complex patterns in the training data, based on which intelligent prediction can be made[1]. Model-based approaches can partially solve the two drawbacks that the memory-based approaches suffer from (i.e. data-sparsity and scalability) and are able to achieve better recommendation accuracy and coverage than memory-based approaches[40], due to the fact that they train a model based on global rating data, while memory-based approaches only focus on the local rating information[40]. Various machine learning / data mining algorithms have been utilized for making recommendations in the past, such as Restricted Boltzmann Machines[41], regression-based models[42], latent factor models (i.e., latent semantic models[43], matrix factorization[4], etc.). Among them, the latent factor models are the most popular CF model-based ones. Most latent factor models are based on matrix factorization (MF). It has been experimentally demonstrated in the Netflix Prize competition that the MF models outperform the classic neighborhood models[44]. The MF technique was first proposed for use in recommendation systems by Funk, named Regularized SVD[15]. It aims at learning a low-rank approximation to a rating matrix, where both users and items/services are characterized by vectors of latent factors inferred from the rating matrix. The elements in these vectors measure the extent to which users/items possess those latent factors[15].

Most of the model-based recommendation approaches need to go through a model learning process depending on the recommendation model. In particular, this is also the case with MF-based models. A suitable learning method in the case of explicit user feedback relies on imputation to fill the missing elements in the rating matrix in order to make it denser. However, imputation is very expensive and may involve inaccurate information [15, 18]. A more recent and popular approach is to only factorize observed ratings instead of the whole matrix [44, 21, 15]. The idea is to learn the model by approximating the previously observed ratings.

In recent years, MF has also been applied for item/service ranking task in the case of implicit feedback, where positive-only feedback is available in the form of binary ratings in the user-item (user-service) matrix. One of the most representative works is[25], which is an adaption of the rating-based MF model. It still minimizes the square error, but by employing weights for increasing the impact of positive feedback[18]. Rendle et al. propose a more up-to-date generic approach for learning models for personalized ranking, namely Bayesian Personalized Ranking (BPR)[26], which has been experimentally proven to be more effective. The assumption behind BPR is that user u prefers an observed item/service i to an unobserved item/service j, with a higher predicted rating value. The assumption is modelled by using the Bayesian formulation to find the correct rankings for all items/services by maximizing the posterior probability [26, 45].

The reason for MF being favoured over memory-based approaches is not only in its outstanding performance and simplicity but also in its ability to incorporate additional information in the way the following models do:

- Biased MF – the standard MF does not consider user- and item/service bias information, which is an important factor for accurate prediction [46, 15]. For example, some users have overall higher rating behavior and some items/services are more popular than others. The bias MF considers more features that contribute to the rating value and thus demonstrates better performance compared to the standard MF [46, 15];
- SVD++ – this is a State of the Art recommendation model that considers both the explicit and implicit influence of user-item (user-service) interactions[21];
- Neighborhood-based MF – this approach, proposed by Guo et al.[29], incorporates item similarity into MF, with a predicted rating for items by users. Zheng et al.[47] propose a similar recommendation model, but instead of considering the top-N neighbors of items, they utilize users' neighborhoods, i.e. the set of top-N users similar to the target user;
- Social Trust Ensemble Recommendation (RSTE) – by leveraging social/trust networks to improve the recommendation performance, this model diffuses the user feature vector with the vectors of the user's trusted friends to model a specific rating[48]. Another model in this category is SocialMF[49], which is based on an assumption similar to the one made by RSTE but uses a different implementation strategy, where the interest of the user's trusted friend is incorporated by a regularization term in its objective function.

For CF approaches, the quality of recommendations depends on how well they address several severe problems they inherently suffer from, such as data-sparsity, cold-start, and scalability. These three problems relate to two issues: effectiveness (data-sprasity and cold-start) and efficiency (scalability). To some extent, these two issues are in conflict, and it becomes very important to consider them simultaneously[50].

Dimensionality reduction techniques can partially alleviate the data-sparsity and scalability problems, but they need to undergo expensive model training [12, 50, 27]. MF-based models are among the most popular dimensionality reduction approaches, which also perform well when processing large recommendation system databases[51]. Latent Semantic Index (LSI) is another popular dimensionality reduction method, which is usually combined with singular value decomposition (SVD) [52, 50]. SVD methods are able to provide good quality recommendations, however, they are computationally very expensive and thus can be only utilized in static off-line settings without real-time information changes.

The common strategy for addressing the data-sparsity and cold-start problems is to adopt additional information, which is incorporated into the set of ratings in order to make more accurate recommendations [14, 51]. In the literature, various types of additional information have been exploited for utilization in CF. These could be rich side information sources related to users and items/services[50], such as user demographics [64, 54], the social/trust relations between users [55, 56, 19], tags [57, 58], user reviews [59, 60], etc., or interaction-associated information related to the interplay of users and items/services[39], such as timestamps that record the time when a user interacted with an item/service [15, 61], location where the interaction was made [62, 63], etc. In particular, the cold-start problem is often being alleviated by employing hybrid recommendation approaches. Kaššák et al.[64] introduce a group recommender, where the predicted rating of an item recommended by CBF is recalculated by CF. An alternative approach is the one by De Campos et al. who propose a Bayesian network model for hybrid recommendation by combining CBF and CF[65]. Yu et al.[45] propose to model information related to users and items as a heterogeneous information network, the meta-paths in which are combined with user feedback for generating semantic-aware recommendations.

Although a large amount of work has been done, the problems CF suffers from still have not been well-addressed [21, 13]. A number of novel approaches to deal with aforementioned issues have been recently proposed in[67–72]. In[73], a novel recommendation approach specifically addresses the data-sparsity and scalability issues by exploiting the intrinsic information existing within the user-item (user-service) rating matrix. More specifically, it proposes to incorporate user representations, learned from users' rating history, into the MF framework. Based on it, two rating-based recommendation models are proposed – UserMF and UserReg –, which incorporate user feedback information into MF at a rating-model level and objective-function level, respectively. Based on the State of the Art SVD++ model, the assumption behind the UserMF model is that user's preferences can be reflected not only by the user global representations learned from rating data factorization but also by the set of items/services that the target user liked in the past. The second model, UserReg, takes a different implementation approach into the influence of user feedback by incorporating a regularization term to constrain the MF objective function. In UserReg, every user representation is regularized by the latent factors of the items/services that the user is interested in. Through extensive experiments, conducted over three large-scale real datasets, the performance of the two proposed models was evaluated against that of six baseline and State of the Art CF recommendation models. The results con-firmed that the proposed recommendation models can achieve as good or even better prediction accuracy compared with the existing MF-based models, with significant improvements of computational efficiency. In addition, the training time of the UserReg model is 200 times faster compared to the SVD++ model. Both proposed models are easy to scale up on larger datasets without dragging down the training efficiency.

For addressing the data-sparsity issue, another approach, proposed in[73], integrates variable item/service weights with a ranking-based recommendation model, where learning is driven by Bayesian Personalized Ranking (BPR). The result is a novel ranking-based model, Weighted-SLIM, which aims to provide users with high quality of top-N recommendations based on implicit feedback. WeightedSLIM is able to take different item/service weights into consideration when computing the predicted score for each unobserved user-item (user-service) pair, where item/service weights are calculated by ItemRank[66] and optimization is driven by BPR[26]. Extensive experiments, conducted on three benchmark datasets, confirmed that the proposed model outperforms the existing ranking-based models on a number of different evaluation metrics.

For tackling both the data-sparsity and cold-start issues,[73] proposes a novel MF model, FeatureMF, which takes into account item/service features. The assumption behind this model is that item-related information is available but limited. Unlike previous attempts to integrate item features in MF, FeatureMF diffuses the item latent factors with latent factors of item features. More specifically, it extends item latent vectors with item representation learned from rich side information associated with items. Experiments, conducted in both presence and absence of a cold-start item problem, demonstrated that FeatureMF achieves better prediction accuracy compared to other existing baseline and State of the Art MF-based models for CF. The conclusion is that the FeatureMF model can successfully alleviate the data-sparsity and cold-start item problems in recommendation systems.

## 7.4 Review of recent products and industrial achievements

An important aspect of research which is usually ignored is the existence of industry examples, achievements and products that use technologies for the improvement of the life of the aging population. We present a small overview of several companies and products that correspond to some of the research and technology areas covered by the **Sheld-on** COST Action:

## Products mainly focused on the aging population:

| Product | Major properties and characteristics |
|---|---|
| ASISTAE  <br> Device to monitor the activity of elderly <br><br> SPAIN | • Through a textile sensor, the device can detect when a person lies/ sits down or gets up. <br> • Possibility of checking if the person follows his routine. <br> • Focused for older adults people who live alone by themselves. |
| GREYMATTERS  <br> An interactive life storybook app for the tablet that aims to improve quality of life for people with dementia and their caregivers. <br><br> USA | • Through visual reminiscence, paired with music and games, the app helps patients and families preserve yesterday's memories, as well as share today's joyful moments. |
| CAREPREDICT TEMPO  <br> Smart bracelet designed for residences, hospitals, etc. <br><br> USA | • Tempo is a wrist-worn wearable that houses a sophisticated array of sensors able to detect an individual's activities of daily living (ADLs) and location, while providing a touch-button call system for real-time communication with caregivers. <br> • Another important part of understanding the ADL patterns of individuals is knowing where specific activities are occurring. <br> • CarePredict knows the exact indoor location of every person wearing a Tempo. |
| ACTIVE PROTECTIVE  <br> airbag belt <br><br> USA | • Belt that automatically deploys airbags over the hip when a fall is detected. <br> • Was founded on the simple belief that hip fractures among older adults, and their devastating consequences, could be prevented through the use of wearable technology |
| UNALIWEAR  <br> Voice-controlled smart watch <br><br> USA | • Voice-controlled device designed to keep you independent, active and safe. <br> • Reminder of medications, detection of falls, emergency assistance anywhere... <br> • Voice prompts. No buttons to press. <br> • It can help you find your way home if you become lost or get help for you. |
| BECLOSE  <br> Elderly monitoring <br><br> USA | • It consists of several sensors distributed around the house to perform a wireless and discreet follow-up of the routines of an older adult. <br> • Opening / closing of doors, activity levels, use of medicines... <br> • Sensor data is collected and sent to the cloud where it is stored and analysed. |
| LIVELY WEARABLE  <br> Urgent response button device <br><br> USA | • The locked pill dispensers are for those at greater risk for taking the wrong medication at the wrong time. All compartments are locked and only the specified compartment will unlock at the right time. <br> • With the push of a button the pill dispenser will open a two-way voice channel with the medical alert professionals at the UL certified monitoring center. <br> • No need for phone line or internet! |

| Product | Major properties and characteristics |
|---------|--------------------------------------|
| DRING <br>Smart cane for elderly<br><br>FRANCE | • A simple push of a single SOS button triggers a remote alarm and automatically calls your loved ones. Uses the GSM networks.<br>• The loss of verticality of the connected object triggers the alert and the assistance request. Built-in motion sensors (accelerometer and gyroscope) detect a possible fall and immediately warn your family.<br>• Smart sensors also prevent the triggering of false alarms.<br>• The cane always remains active. |
| DOMOALERT <br>Telecare system<br><br>SPAIN | • Connect at any time to see and communicate with the person you care.<br>• In a situation of danger or medical emergency, the person can pulse the panic button to send a warning of help.<br>• The system can include door opening sensors to detect if the person has left the house. |
| SENSOVIDA <br>Telecare system<br><br>SPAIN | • Sensovida is a telecare system for the elderly and / or with some type of disability.<br>• Automatic detection of risk situations (routines).<br>• Real-time tracking with mobile app.<br>• Emergency button (inside/outside home).<br>• Sensovida analyses parameters such as sleep, activity, etc... and generates a weekly wellness report. |
| FEARLESS <br>Sensory alarm to detect home accidents<br>Merchandising <br><br>AUSTRIA | • The Fearless Comfort System is a sensory alarm that detects accidents in the homes of some of society's most vulnerable people who wish to live in their preferred environment.<br>• A system that no longer requires a sensor to be worn on the body, no need to operate an alarm button, but is very easy to use despite complex technologies.<br>• Fearless combines several functions such as falling, getting up and detecting movement in the room and includes preventive measures such as automatic control of light during night time sleep. |
| HELICOPTER <br>Regular checking of health parameters for older people<br><br>DENMARK, SWEDEN, ITALY & ROMANIA | • The HELICOPTER proposal aims at exploiting ambient-assisted living techniques to provide older adults and their informal caregivers with support, motivation and guidance in pursuing a healthy and safe lifestyle.<br>• The system will gather data coming from a heterogeneous set of (mostly off-the-shelf) devices, including medical, environmental and wearable sensors, to provide a qualitative and quantitative assessment of the activities carried out. |
| AHEAD <br>Smart hearing aid & smart glasses<br><br>SPAIN, GERMANY, AUSTRIA & DENMARK. | • The AHEAD project aims at increasing the quality of life of older adults by assisting them for keeping an active and independent life.<br>• To improve quality of life of older adults, the project makes use of devices that older adults have already adopted: eyeglasses and hearing aid.<br>• The integration and combination of advanced and innovative sensing as well as ICT based modules will result in supporting system for improving daily life and health condition. |

| Product | Major properties and characteristics |
| --- | --- |
| TEXIBLE WISBI <br>Smart bed<br>Merchandising <br><br>AUSTRIA | • TEXIBLE Wisbi is a smart bed insert that will automatically send an alert when your supervised person is on a wet surface or leaves the bed.<br>• TEXIBLE Wisbi is hygienic, washable and very easy to use.<br>• Embedded, imperceptible sensors are detecting moisture or if someone is leaving the bed. No matter where you are, you will be reliably alerted via the app. |

## Products not only focused on the aging population:

| Product | Major properties and characteristics |
| --- | --- |
| BEDDIT 3 <br>Sleep monitor<br><br>USA | • Measures the quality of sleep through parameters such as sleep time, breathing, heart rate, the environment (temperature, humidity...) |
| SLEEP TRACKER <br>Smart bed<br><br>USA | • Firmer or softer? You both get to choose your ideal level or firmness, comfort and support. You can change it whenever you like.<br>• The Smart bed connects to your favorite apps, so you will know how life affects your sleep and how sleep affects your life. |
| ARIA2 <br>Smart wifi scale<br><br>USA | • Weight, percentage of body fat, BMI...<br>• Possibility of registering the user's nutrition to keep track of calories.<br>• Monitor evolution through diagrams and graphics. |
| BODYGUARDIAN HEART <br>Cardiac monitoring system<br><br>USA | • Is a discreet wireless remote monitor that records important physiological data such as heart rhythm, ECG, respiratory rate and activity.<br>• Allowing healthcare providers to monitor key biometrics outside the clinical setting, while patients go about their daily lives, thus safeguarding the health and safety of individuals every day. |
| LECHAL INSOLES <br>Smart insoles<br><br>INDIA | • Liftware's selection of stabilizing and levelling handles and attachments are designed to help people with hand tremor or limited hand and arm mobility retain dignity, confidence, and independence.<br>• It contains sensors that detect the movement of the hand and a small computer that directs two motors to move the accessory in the opposite direction to any detected tremor. |

# 8 Final remarks

In this report we present the state of the art as prepared by the members of Working Group 2 in the COST Action **Sheld-on**. The report clearly shows the trends in the ICT developments towards improving the life of the aging population. The report also gives some thorough analysis of the different aspects of the ICT developments presented in a form of separate reports from each focusing group. In the following section all of the referenced publications are presented as a single reference list. Most of the publications to be used as a reference list for the state of the art in ICT developments towards a safe, comfortable and healthy life of the aging population.

# 9 Future work

Based on the given state of the art analysis there are quite a few approaches that can be taken to furtherly improve the ICT technologies in order to aid the aging population. The work presented here is to be used as a baseline for further research that is to be performed within the **Sheld-on** COST Action. Based on this ICT technologies-oriented report and the reports of the other working groups, a wider multidisciplinary approach will be used to make foundations for collaboration between experts from scientific areas that are very different in nature, in order to give suggestions and directions for solutions for healthy aging that will give benefit to the society and meet the challenges of the future.

# ¹⁰ Full bibliography

(in alphabetic order)

# **Working Group 3**
## Healthcare

**Editors:**
Signe Tomsone (LV)
Petra Marešova (CZ)

**Members:**
Prof Sabina Baraković (BA)
Prof Jasmina Barakovic Husic (BA)
Mr Hrvoje Belani (HR)
Ms Paula Conte García (ES)
Ms Irene Coppola (MT)
Ms Carina Dantas (PT)
Dr Moriah Ellen (IL)
Ms María Fernández-Vigil (ES)
Prof Nuno Garcia (PT)
Dr Réka Geambasu (HU)
Dr Jonathan Gomez-Raja (ES)
Dr Lucía González López (ES)
Mr Marko Gošović (ME)
Dr Andjela Jaksic-Stojanovic (ME)
Dr Marcin Kautsch (PL)
Dr Veronika Kotradyova (SK)
Prof Ondřej Krejcar (CZ)
Prof Birgitta Langhammer (NO)
Mr Igor Ljubi (HR)
Dr Joana Madureira (PT)
Prof Pierre Mallia (MT)

Dr Oscar Martinez Mozos (ES)
Dr Ana Mendes (PT)
Dr Sarmite Mikulioniene (LT)
Prof Emanuele Naboni (DK)
Prof Anders Q. Nyrud (NO)
Ms Maria del Mar Olmo (ES)
Prof Dobrinka Peicheva (BG)
Prof Antti Peltokorpi (FI)
Mr Tomàs Puebla (ES)
Dr Jelena Ristic Trajkovic (RS)
Prof Heidi Salonen (FI)
Dr Lina Seduikyte (LT)
Dr Michalis Skarvelis (EL)
Dr Hilde Thygesen (NO)
Ms Willeke Van Staalduinen (NL)
Dr Martin Weigl (AT)

# **Working Group 3**
## Healthcare

**Acknowledgements:**
We acknowledge the contribution from each of the working group members towards making this report:

We thank Birgitta Langhammer for contributing to section on personal needs of elderly related to physical health;

Oscar Martinez Mozos for contribution to section on personal needs of elderly related to mental health;

Ana Mendes, Joana Madureira, Lina Seduikyte , Martin Weigl, Heidi Salonen for their contribution to section on indoor environmental factors for the elderly;

Veronika Kotradyová for contribution to section on principles of spatial design for well-being;

Ondrej Krejcar for contribution to section about smart furniture in patent and literature databases;

Sarmite Mikulioniene, Willeke van Staalduinen, Carina Dantas for their contribution to section on social context of improvement in indoor living for older people;

Petra Marešova, Willeke van Staalduinen, Carina Dantas for their contribution to section about economic consequences of investing in smart habitat;

Baraković Sabina and Baraković Husić Jasmina on their contribution to section about quality of life in context of smart ageing;

Jonathan Gomez-Raja for conribution to policy and practices section.

This publication is based upon work from COST Action CA16226 - Indoor Living Space Improvement: Smart Habitat for the Elderly (Sheld-on), supported by COST (European Cooperation in Science and Technology).

sheldon
smart habitat
for the elderly

# Index

cost
EUROPEAN COOPERATION
IN SCIENCE & TECHNOLOGY

Funded by the Horizon 2020 Framework Programme
of the European Union

# 1 Introduction

Developed countries are currently undergoing demographic changes which entail the rising number of senior citizens. This particular demographic group is prone to suffering from numerous chronic diseases. The link between old age and chronic disease e.g. is illustrated by USA population reference bureau[1], according to which up to 19 million people need to provide day-to-day primary assistance to their elderly family members.

There are already a number of technologies in use, including digital devices, smart sensors and intelligent applications that assist elderly people with their everyday needs in their own homes. Developing a strategy for an integrated technological solution would resolve many issues faced by elderly patients and would lead to improving their quality of life, health, and safety [2,3,4]

The objectives of COST Action CA 16226, Indoor living improvement: Smart Habitat for the Elderly (**Sheld-on**), Working Group 3 was:

- to explore the specific needs of elderly on physical and psychological level,
- to describe social, economic and technological consequences related to the care of elderly, to specify quality of live concept in context of smart environment,
- to review policies in different countries,
- to find good practices examples from different countries,

in order to promote safe, comfortable and healthy living at home.

WHO[5] describes healthy ageing as the process of developing and maintaining functional ability that enables well- being in older age. Functional ability comprises the health related attributes that enable people to be and to do what they have reason to value. It is made up of intrinsic capacity of the individual and relevant environmental characteristics and the interactions between individual and these characteristics. Intrinsic capacity is the composite of all the physical and mental capacities of individual.

Environment comprises all factors in the extrinsic world that form the context of an individual's life (from micro level to macro level). Within these environments are a range of factors, including the built environment, people and their relationships, attitudes and values, health and social policies, the systems that support them, and the services that they implement.

We took this explanation of functional capacity of older people as base for WG3 report on the state of the art, where we explored needs of older persons in relation to healthcare and smart living spaces and tried to map existing policies and practices regarding healthcare and smart living spaces in order to propose topics for future studies. Report summarizes research done by WG3 members as well as results of additional search of information related to WG3 objectives.

# ² Needs of elderly

From the point of view of professionals, needs of the elderly people could be analysed from different points of view such as:

- Health: maintenance of physical function, social activities
- Physical: independence in activities of daily living, mobility, falls, sight (light), hearing (socialization), teeth (nutrition)
- Psychological: anxiety, depression, worsened cognitive function (e.g., dementia)
- Social: safety, accessibility, participation

From the points of view of elderly people, needs may also be analysed from different perspectives such as:

- Dependence
- Loneliness
- Living conditions
- Economic issues
- Person's involvement in decision making about solutions in problematic situations

Needs of elderly are described below in three categories: from physical, mental and contextual point of view.

## 2.1 Personal needs of elderly related to physical health

**Biological changes**

Physical capacity of elderly relates with biological characteristics, such as VO2, muscle mass, sensory function (eye sight, hearing, fine motor function, etc.), memory, and neuro-motor function (speed, coordination, dual-task) and decrease with age[6]. These biological changes can be influenced by lifestyle from early to old age. Inactive and unhealthy lifestyle decreases biological reserves and leads to poor health and disability ,whereas an active healthy lifestyle maintains an independent lifestyle and leads to maintenance of reserves[8,9,10,11]. VO2max decreases approximately 5- 15 % per decade beginning at 25-30 years of age. This decline can be attributed to age; reductions are noticeable in maximal cardiac output and in maximal arteriovenous oxygen (a-v O2). However, older adults have a 10 to 30 percent higher VO2max if they were performing endurance training as young adults. The increase in VO2max in older adults is reflected in improvements in both maximal cardiac output and arterio-venous O2 difference. The magnitude of these VO2max adaptations in older adults depends on the training intensity[12,13,14]. VO2 may be viewed in terms of the Metabolic Equivalent of Task (MET), or simply metabolic equivalent, which expresses the energy (cost) of physical activities. At rest, 1MET is equivalent to 3.5ml/min/kg.

One can distinguish a "cut-point" between high and low levels of endurance capacity, which is 18.3ml/

min/kg[15]. A lower level <13 ml /kg/min indicates a need for help in activities of daily living[16].

Muscle mass is reduced for 40% in persons at 70 years of age. Strength, in terms of 1 repetition maximum, decreases from the age of 50 by 1-2 % every year. Power, that is, strength / time, decreases by 3.5 % per year[17]. This decline can be slowed down by exercise.

Disuse of skeletal muscle that is, inactivity, increases muscle loss resulting in more pronounced muscle atrophy. Other contributing factors affecting muscle mass in ageing are neuromuscular realignment (changes in motor units and innervation of fibres), reduction in growth factors, and changes in muscle protein turnover[18].

Below you can see possible outcomes in muscle mass and strength throughout a lifetime (Figure 2.1).

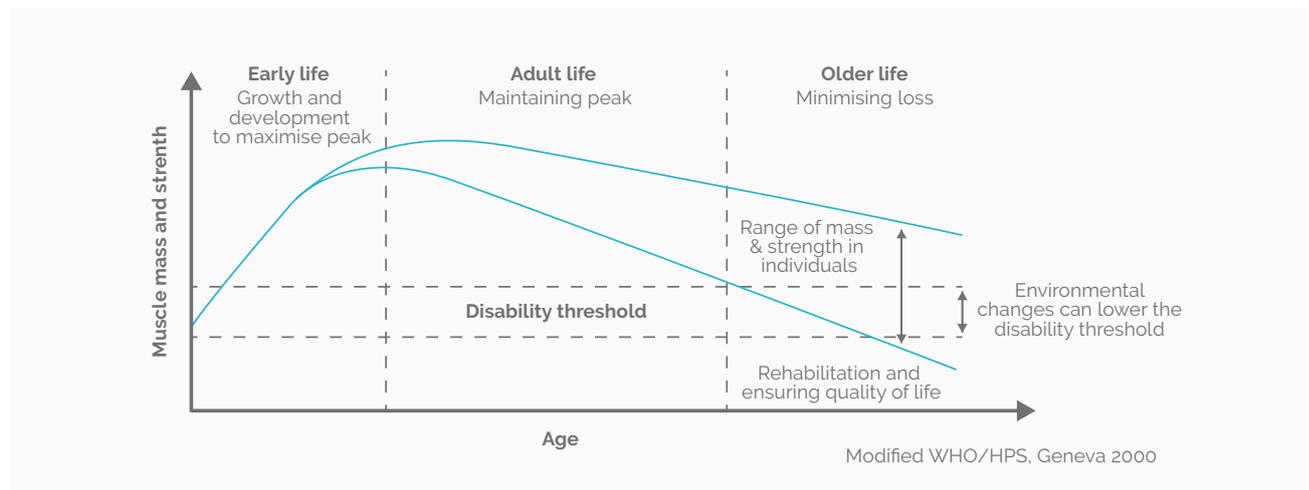

*Figure 2.1. Outcomes in muscle mass and strength throughout a lifetime*

To maintain physical capacity in balance, complex internal and external processes need to work in tandem. In ageing persons, balance is worsened due to one or several of these processes functioning inadequately [19] [20]. Balance is also affected by external context or environmental influences, such as physical surroundings, noise, stress, and other stimuli. Both the internal and the external processes are modifiable. LIVING AT HOME – PHYSICAL FACTORS

Falls can have serious consequences; about 1/3 of adults over the age of 65 falls each year. The risk of falls increases proportionately with age. At 80 years, over half of seniors fall annually. For a person, falls can be fatal. For the society, great costs are generated for the proper care and rehabilitation (Figure 2.2.).

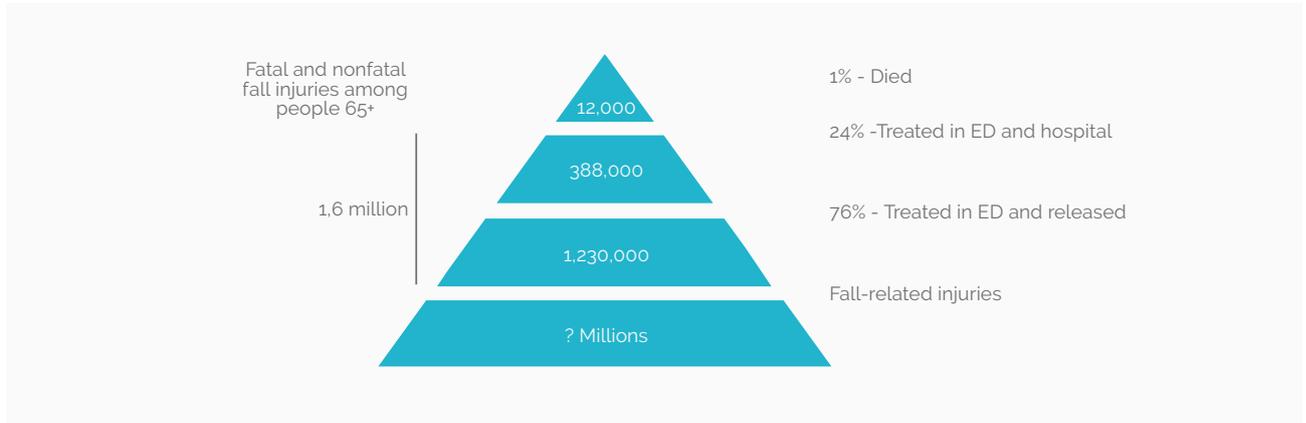

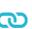
*Figure 2.2. Costs of falls*

The reasons why older people fall are complex (Figure 2.3.).

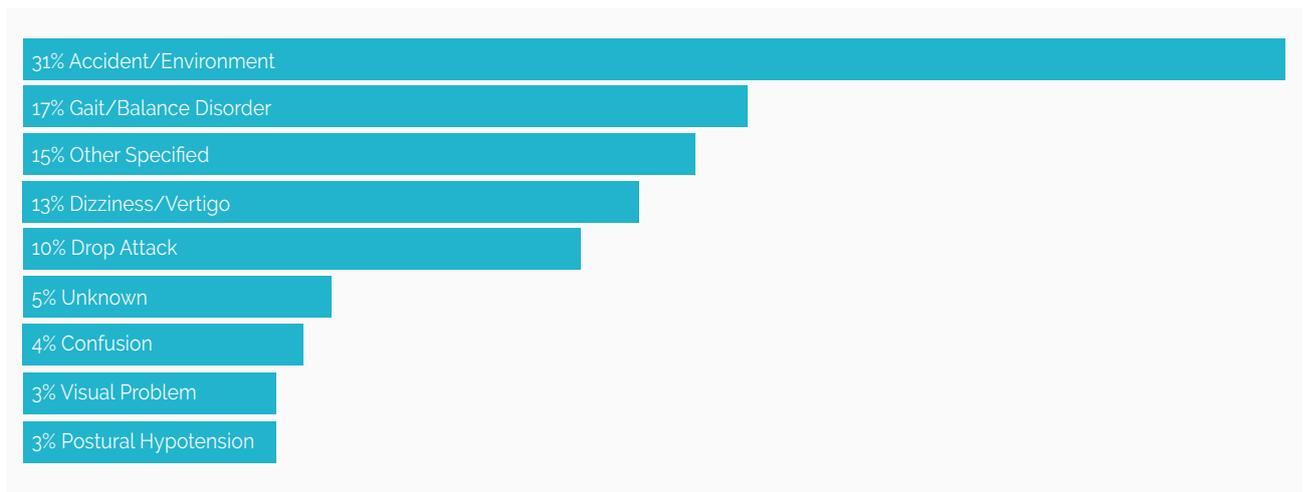

*Figure 2.3. Reasons for falls*

On the individual level, exercise and training are recommended to maintain endurance, strength, and balance. Equally important is to adapt the environment and make use of assistive devices in order to lower the disability threshold. The most favourable solution is to combine the two.

Frailty

A growing body of evidence indicates that biological aging or frailty influences many negative health-related outcomes, however, frailty is likely poorly understood and under-recognized by the public-at-large[22]. The conceptual model suggests that culture, knowledge about aging, and stereotypes influence adults' beliefs and perceptions. Ageing adults determine their health priorities, and then subconsciously or consciously determine which of these are controllable. If deemed controllable and important, the elderly may participate in health behaviours to mitigate the age-related decline. If deemed uncontrollable or less important, adults may focus their efforts on accepting them. Some findings suggest that frailty is a subjective term and that subjects often optimistically do not identify themselves as frail.

Due to the decline of immune functioning related to age and chronic disease, older adults are a well-recognized susceptible group. In this context, the aging phenomenon offers great opportunities as well as great challenges for all societies, namely to ensure that the increase in longevity is accompanied by good health and quality of life, while balancing socio-economic costs. Frailty is an age-related syndrome characterized by a state of increased vulnerability with reduced physiological reserves required to respond to stressors[23]. This syndrome is the most common condition in older adults leading to disability, institutionalization, and death. Higher prevalence of frailty in elderly community dwellers has been recorded in several studies. Frailty prevalence may range between 20% and 30% among adults aged 75 years or more. Several factors that may contribute to the condition have been suggested, however, a definitive explanation is still missing as well as evidence of why frailty develops in some individuals and not in others. Environmental factors, such as air pollution and other contaminants, have been associated with certain age-related disorders, such as Alzheimer's disease, for example. These factors may also play a role in the development of frailty syndrome, especially if considering the fact that numerous factors likely contribute to the onset of this condition. Therefore, environmental exposure should also be considered when studying the potential risk factors for the development of frailty. It is essential to understand the frailty status of people before the clinical symptoms develop. In this way, timely interventions can be organized that can sometimes reverse the symptoms. Furthermore, an early frailty assessment may lead to a more effective response of the healthcare system and thus reduce socio-economic costs of a long–term care. Different areas, such as environmental health and social sciences, should be considered in frailty aetiology. Bio-gerontology aims to define biological markers of aging (and frailty) –that are more revealing than chronological age - to recognize the biological age of populations, groups, and individuals. Currently, there are several gaps in the scientific literature regarding frailty syndrome, specifically in their definition, gold-standard models, causal-effects, prevention, and treatment.

In summary, with advanced age, biological and physical reserves are reduced and it is vital for the elderly to remain active to maintain their abilities and capacities. Illnesses, trauma and multimorbidity may speed up the decline in older adults and lead to their dependence in many activities.

Example 1- Interventions for elderly people in Norway (see in Attachment 1)

## 2.2 Personal needs of elderly related to mental health

We are currently experiencing unprecedented levels of ageing in the population in regions around the world, especially in the EU and Japan. By 2060 there will be around 151 million people aged over 65 years in the EU[24]. Increased longevity is a positive result of improved living conditions and healthcare, but it presents formidable challenges for public and private budgets and services as well as for elderly and their families. Older citizens in Europe and Japan wish to stay in their homes for as long as possible and enjoy an active and healthy ageing. The number of older adults living alone has increased to 13.14% in EU[25]. However, those in later stages of life are more at risk for age-related impairments, such as poor mental and physical health, frailty, and social exclusion that have considerable negative consequences for an independent life. Moreover, it will be difficult to cover all demands for home assistance in the near future due to shortages of available health workers and doctors to provide personal home care as a result of a high life expectancy (currently 79 years in EU) and low birth rate (1.6 in EU), leading to a projected ratio of older people against the rest of population of 50.2% in EU by 2060[26],[27]. Therefore, there is an urgent need for new and innovative forms of support and health care for older people wishing to stay at home. Compared to the rest of the population, older people are more likely to experience psychological distress due to bereavement, a drop in socio-economic status with retirement, or a disability[28]. Depression affects 12% of the elderly in EU (11.3 million) and it is the main illness in older people ahead of dementia[29]. Symptoms of depression in older adults are often overlooked and untreated because they coincide with other problems encountered by older adults, therefore, extra care should be taken. Symptoms include impairment of quality of life, inability to enjoy life, tiredness, decreased motivation, lack of concentration, and lowered functional capacity together with feelings worthlessness, helplessness, hopelessness, and guilt[30].. Depression has a direct impact on physical health and vice versa. Older adults with physical health conditions such as heart disease, diabetes arthritis, or kidney disease have higher rates of depression[31],[32],[33],[34]. Conversely, untreated depression in an older person negatively affects the outcome of other diseases, such as heart conditions, lupus, or AIDS[35],[36],[37],[38].

Finally, and most importantly, depression is the main cause of suicide in older people in western countries[39]. Depressed persons may require substantial care that may cause distress in caregivers or family members. Partners of depressed people have more difficulties in maintaining social and leisure activities, while the relationship with their partner may suffer as well. Relatives and spouses of depressed patients require attention as they are also dealing with the negative effects of depression[40],[41],[42]. The socio-economic costs of depression in the elderly include the loss of opportunities for social leisure, early hospitalisation and nursing home admission, more frequent (costly) professional help, up to 50% higher healthcare costs, and premature mortality[43],[44],[45].

Besides depression other conditions, like, anxiety and decreased cognitive function (e.g., dementia) can influence ability of older people to live an independent life at home.

## 2.3 Indoor environmental factors for the elderly

Indoor environments should safeguard and enhance occupants' health, comfort, and productivity, as people spend around 90% of their lives indoors[46]. There is still limited knowledge regarding the causes of symptoms observed in nonindustrial indoor settings, such as office buildings, recreational facilities, schools, and residences.

Indoor environmental problems are related to bone frailty, and skeletal and muscular structure degeneration. Exposure to indoor air pollution is an important stimulus for the development and exacerbation of respiratory diseases, such as asthma, chronic obstructive pulmonary disease, lung cancer, and cardiovascular disease. This situation is further exacerbated by the increased amount of time older adults spend indoors.

It has been estimated that older persons spend approximately 19–20 h/day indoors, and many spend all their time indoors in elderly care centres. Due to these conditions, older people are more susceptible to the effects of air pollution, therefore, monitoring indoor environmental quality should be a public health priority. Thermal comfort is one of the indoor environmental factors that affect health and human performance. Determined by temperature, humidity and air movement, it has a very significant impact on the general well-being and daily performance of building occupants. Poor thermal environment can also aggravate the impact of air pollutants on occupants' health. The ability to regulate body temperature tends to decrease with age. In general, elderly seem to perceive thermal comfort differently from the young due to a combination of physical aging and individual differences. These are too large to draw an unequivocal conclusion on the requirements of older adults regarding their preferred thermal environment. Nevertheless, self-reported poor health was significantly associated with poor TC. Exposure to cold has often been associated with increased incidence and severity of respiratory tract infections.

**Indoor air quality**

- The living environment is very important for overall human well-being. Indoor air quality (IAQ) is one of the most important factors influencing indoor microclimate. At the same time, airborne particulate matter (PM) is considered to be one of the key pollutants due to its complexity and adverse health effects. PM, especially its fine fraction, referred to as PM2.5 (particles with aerodynamic diameter lower than 2.5 mm) has been associated with various adverse health effects[47]. A report from WHO stated that on a global scale, 4-8% of premature deaths are related to the exposure to PM in the ambient and indoor environment.

- Major pollution sources in indoor environments include building materials, combustion processes, furniture, various household and personal care products, and plastic utensils. The concentration of an indoor air pollutant depends not only on its indoor emission rate, but also on the rate at which it is being transported from outdoors to indoors (if applicable), and the rates at which it is scavenged by indoor surfaces, consumed by indoor chemistry, and removed by ventilation or air cleaning[48]. Construction practices and indoor sources differ among countries depending on the socio-economic conditions of the occupants.

- Increasing requirements for the building energy efficiency (EE) raise new challenges for IAQ management. The countries within the European Union have assumed commitments to build low energy consumption buildings from 2016 to 2020. This usually means improving EE and air tightness of the building envelope. In the future, EE of existing buildings must also be improved. The main goal of the building refurbishment process is energy saving and improvement of building systems, but the improvement of occupants' wellbeing should also be considered as one of the most important refurbishment goals. From this perspective, IAQ research in low energy/refurbished buildings is of high importance. The modification of building systems, including structures (e.g. insulation of external walls) and heating, ventilation, and air conditioning (HVAC) systems, and new building materials, may have a significant influence on IAQ and subsequently, PM levels[49].

- Results[50] indicated that adequate ventilation must be assured in low energy buildings. While low occupancy of single family buildings may be beneficial to keep most of the pollutants within the recommended values, stronger emitting sources may be difficult to manage, especially immediately after installation. Checking of indoor air quality is recommended before occupancy to avoid exposure to high pollutants concentrations from the interior decoration and furnishing. Moreover, the concept of low energy buildings should be coupled with "healthy sustainable building" concept, aiming to avoid usage of highly emitting building and furniture materials.

- Energy efficiency, thermal comfort, and IAQ should be considered when designing heating, ventilation, and air conditioning (HVAC) systems for low energy and other buildings. Energy demand for space heating in such buildings is comparatively low, therefore ventilation systems can be used for heating purposes by introducing warm air even with the low air change rates[51].

- The stratification of air can have effects on the concentration of pollutants in the breathing zone and ventilation effectiveness may be significantly affected by air distribution method and the position of the pollution source. Contaminant stratification is possible at lower flow rates and higher density of contaminants.

- Combined air heating and ventilation systems are often used in low energy buildings. However, running these systems at the heating mode increases vertical air temperature gradient in rooms and can have a negative effect on indoor air quality.

- IAQ and thermal comfort is important for all generations, including elderly people. Sometimes older people need to learn how to use new tools appearing in their living environment or behave in different ways. This includes the regulation of HVAC system and thermal comfort devices in case of mechanical ventilation, or learning about the advantages of opening windows more frequently in insulated buildings where natural ventilation is present.

Example 2- Achievements and findings on Indoor Air Quality from Holzforschung, Austria (see in Attachment 1)

**2.4** Bibliography

# ³ Environment, design and smart furniture for elderly

## 3.1 Principles of spatial design for well-being

**Review of the literature**[52] about the built environment, as it impacts the health of older people, introduces health care professionals and researchers to the Our Voice framework for engaging older people as citizen scientists in order to empower them as agents of change in improving their local built environment and ultimately advancing community health.

**Eleven principles of spatial design for well-being are** also useful in creating healthy indoor environments. Human behaviour, outlook, general wellbeing and everyday social interactions are directly tied to the natural and built environment. Spaces and their structures influence our everyday lives and may have far-reaching consequences, potentially impacting our long-term mental and physical health. Recent multidisciplinary research of human centred design on platform of BCDlabis summarized into 11 features of supportive environment for contemporary humans – "cultural animals". Lack of a supportive environment contributes to civilisation diseases relevant for public health.

Needs of human beings are changing with time. In the senior age, the formulated 11 principles of spatial design for well-being are even more important.

FEELING OF SAFETY; PROSPECT AND REFUGE (1.+2)

First of all it is A feeling of safety, combined with the competence to manage risks and attractions, is connected to other important features - possibility to combine prospect with refuge, which is related to protecting one's back while having an overview of the surroundings. It is clearly reflected in space occupation, especially in public spaces. This need is related to seeing and being seen, where humans need to feel in control.

CONTACT WITH OUTDOOR ENVIRONMENT (3.)

A 3rd important feature of spatial design for well-being is enabling contact with the outdoors (at least visual) during the day and the possibility to control it – which is still strongly undervalued in many working environments.

Multiple scientific studies have pointed out on the benefits and importance of nature for people, and especially for older populations[53]. These benefits (increased participation in physical activities, improved mental health and cognitive function and an increase in social interaction) found through access to nature are key ingredients to well-being during ageing. The biophilia hypothesis suggests that there is an innate connection between humans and nature and that people tend to show a positive response when they experience a connection with nature.

Nature-Based Design[54] [55] is seen as a step in changing this kind of action, an alternative that is found in looking at nature as design inspiration and process knowledge. This implies creating a sensitive and responsive design for elderly that highlights a) visual connection to nature, b) pulpability and soundness of the nature and c) nurturing a sense of place, a community in which the role of aesthetics is crucial for behavioural change.

PERSONAL SPACE and INTIMACY vs SOCIALISATION (4.+5)

Another important issue is the need for personal and intimate space, one's own territory, and competence to occupy and control it. In the old age, controlling own space becomes especially important.

This is related to the need to switch between privacy and socialisation according to one's wishes. It depends on space arrangement supporting or postponing communication, about which a lot of knowledge is found in proxemics and anthropology. The elderly appreciate if they live in a mixed heterogeneous community in regard to age and social status.. This makes them feel included.

APPROPRIATE SCALE (6)

A sixth feature is the appropriate scale and harmonised proportions of buildings and their indoor living and working environment, where humans spend a lot of time.

ATTACHMENT (7)

A seventh important characteristic is the possibility (or competence) to be attached to a place or products, to have the competence to adapt and personalize them to mirror and extend the self into the space, and thus gain a state of self-identification. This is important for both cultural and evolutionary reasons. Attachment is strongly connected with the 8th feature- maintaining the cultural sustainability through giving local identity to the built environment and lifestyle.

LOCAL IDENTITY (8)

Securing identity in private and public spaces and preventing loss of local identity due to globalization.

BODY CONSCIOUS (9)

It is important to have the possibility to prevent pain and body deformations due to inappropriate products and environmental settings; freedom in choice of body position and using body conscious products are crucial.

APPROPRIATE ENVIRONMENTAL STIMULATION (10)

The tenth feature is the selection of adequate sensual stimuli- spaces where people spend a lot of time should not be too stimulating. It is important to provide the elderly the right amount of both stimulating and relaxing special arrangements.

MORE NATURAL MATERIALS (11)

Indoor materials should be carefully chosen. Natural materials should have a priority, since they can improve human well-being.

Example 3- Introduction to the topic "person-environment interrelation" (see in Attachment 1)

## 3.2 Smart Furniture in Patent and Literature Databases current state and results

Smart Devices and Smart Furniture as components of a Smart Home have attracted significant interests in the past decades. As a typical example, we can mention recent research for clocks with integrated wireless energy harvesting and sensors (Figure 3.1.)[56] and Google Home smart speaker.

Smart Furniture phrase is unfortunately used in various ways, connections, and meanings; from design of smart furniture to wall mounted electric sockets with internet connection. Here are some examples of existing definitions of smart furniture:

- Ito, Iwaya et al. in 2003[57] defined: „Smart Furniture is a platform for systems to realize Smart Hot-spot. By simply placing the Smart Furniture, we can turn legacy spaces into Smart Hot-spots. Smart Furniture is needed to be equipped with networked computer, I/O devices and sensors. Coordination with existing network infrastructure or user's devices are also required.".

- Vaida, Gherman et al. in 2014[58] provide a definition: "Smart Furniture is the furniture which brings added value, functionality, comfort and elegance to fit every personalized requirement issued by the user".

- Braun, Majewski et al. in 2016[59] defined: Smart Furniture is able to detect the presence, posture or even physiological parameters of its occupants" (Figure 3.2.).

- According to Technavio's smart furniture market research report[60]: "Smart furniture is powered by technological advances such as network connectivity via Bluetooth or Wi-Fi and others, which helps users enhance their furniture beyond its basic analog functions. Smart furniture helps consumers in browsing the Internet for news feeds, weather forecast updates, listen to music. It also offers wireless charging slots for smartphones and has features like distance operation and others".

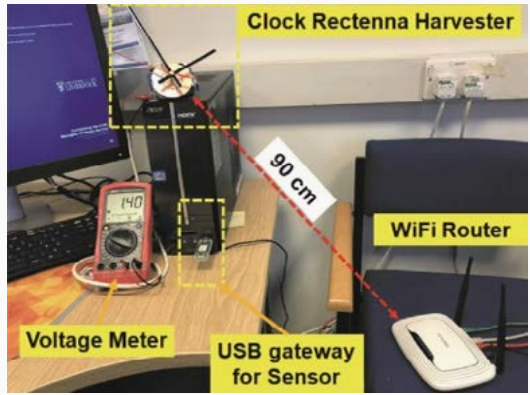

*Figure 3.1. Left: Smart Furniture in a real application - clock antenna harvests energy from a typical WiFi router at a distance of 0.9 m[61]. Right: The block diagram for energy-harvesting quartz clock and its application in energy storage and wireless sensing*

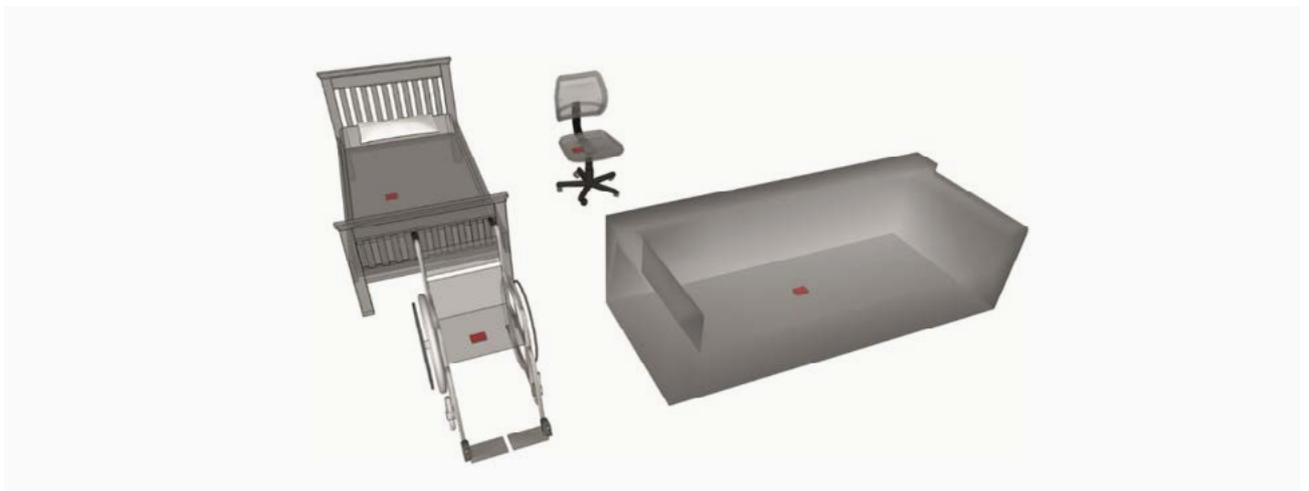

*Figure 3.2. Potential scenarios for wireless occupancy systems. Bed on top left, office chair on top right, wheelchair on bottom left, and couch on bottom right[62]*

Due to the overlap between industry, technology, and people,both basic and applied research needs to address the available literature on the topic as well as topics related to intellectual property (patents). To find results that match the definition of "Smart Furniture", titles, abstract and keywords will need to be searched. Number of search results according to the year is presented below (Figure 3.3.).

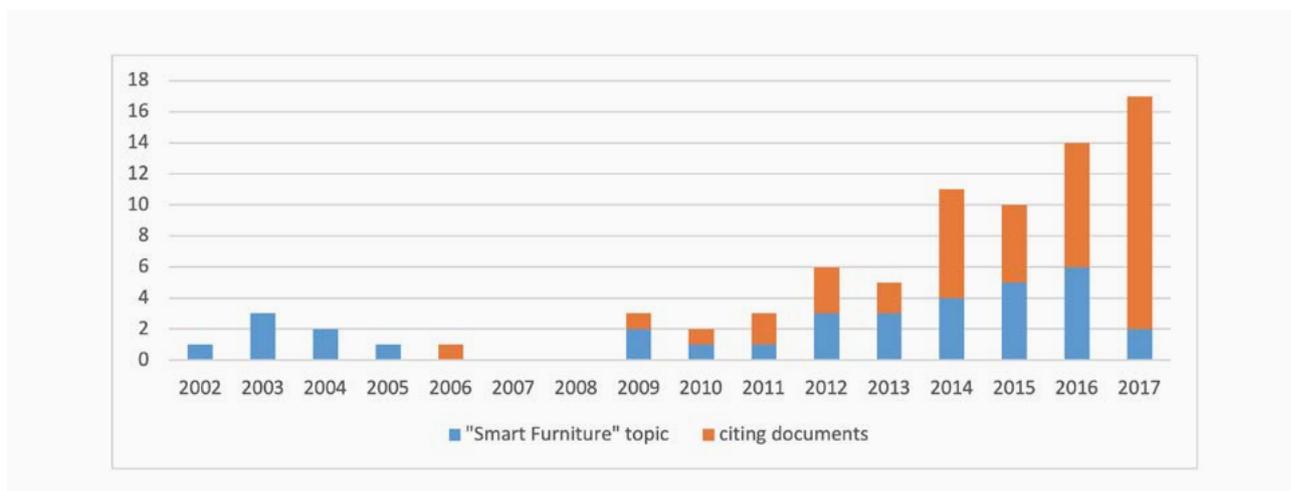

*Figure 3.3. Distribution of publications for the term "Smart Furniture" in ISI WOK database
(35 – blue colour) and citing documents of this set (53 – red colour) since 2002*

Trend showed by ISI WOK database (Figure 3.3.) covered all important journal articles and conference papers dealing with the term "Smart Furniture". Figure also demonstrates the increasing number of citations through the years, since the phrase was first presented and described by a Japanese professor[63].

There is one existing definition of "Smart" in the scientific book from Poslad 2009[64], where he defined and described Ubiquitous Computing as the umbrella term for three different areas: Smart Devices, Smart Environment, and Smart Interaction. He defined "Smart" as: "Concept smart simply means that the entity is active, digital, networked, can operate to some extent autonomously, is reconfigurable and has local control of the resources it needs such as energy, data storage". His book, with 257 citations at SCOPUS database, is the most cited book in the field of ubiquitous computing.

**Patent Databases**

Patent database ESPACENET returned 181 results based on the search phrase "Smart Furniture" (as topic search) for years between 1998 and2017 (there are no results for 2018 yet and older patents are not relevant for our criteria) (Figure 3.4).

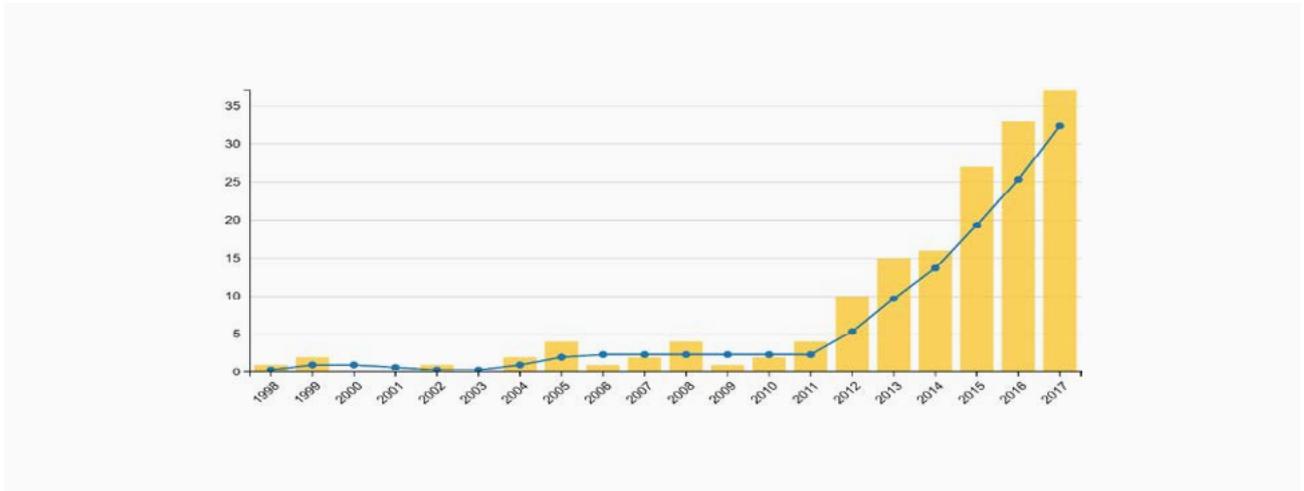

*Figure 3.4. Distribution of patents throughout the years for Smart Furniture topic at ESPACENET database (181 publications in total)*

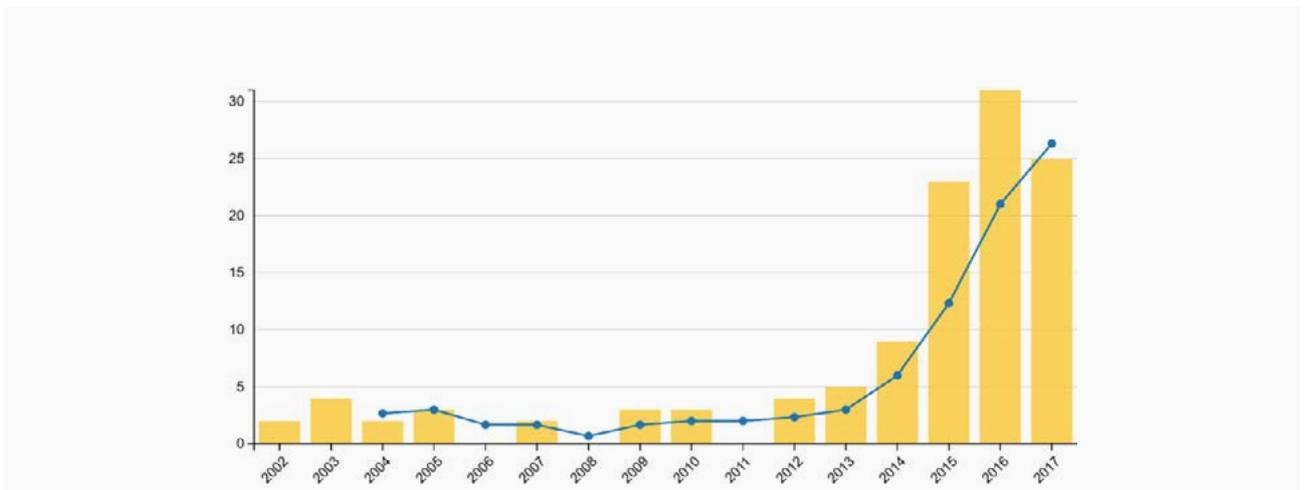

*Figure 3.5. Distribution of patents throughout the years for "Smart Furniture" phrase anywhere in the application text in the ESPACENET database ordered by application date (total of 117)*

The first relevant patent in the history containing the phrase "Smart Furniture" is probably "RFID smart office chair" by Hagale et al (2004) (Figure 3.6.) (4x in Abstract, 40x in Claims, 27x in Description). This patent application contains "Smart Furniture" phrase 71 times (4x in Abstract, 40x in Claims, 27x in Description). This patent is also the most cited (cited 71 times by other patents) from all patents covered by this search. Total number of citations is 123.

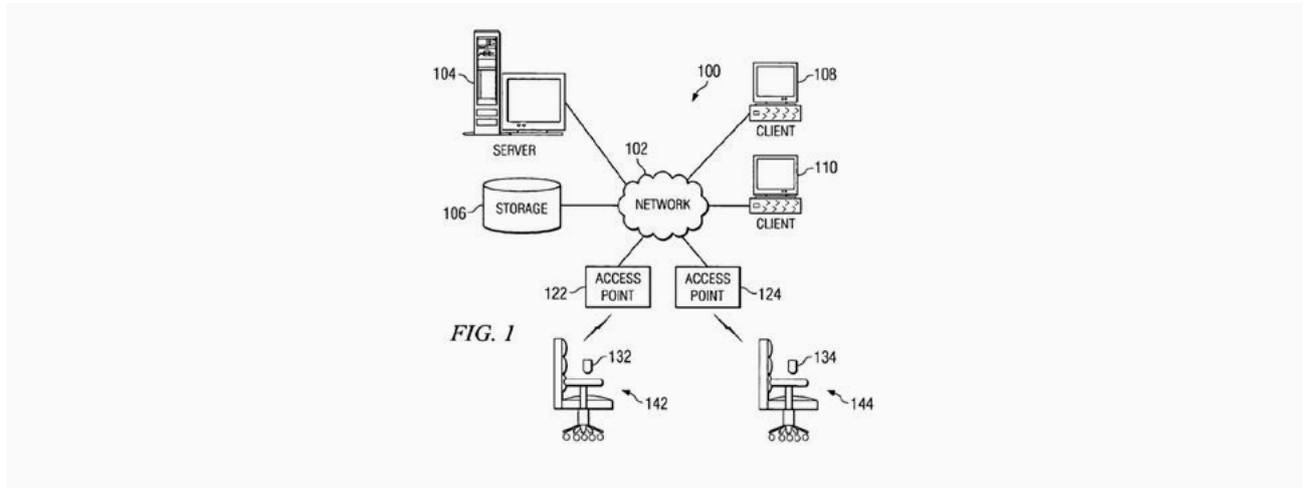

*Figure 3.6. Schema of Smart Furniture by Hagale et al. from IBM company in August 2004[65]*

Other patents from search unfortunately do not have any citations related to "Smart Furniture" with the exception of "Novel dining table capable of achieving combined and separate use of mahjong machine and dining table" from Chen Jinchen - China[66] with one cited patent.

One of the most recent (2018) patent applications comes from China and has spread worldwide[67].It deals with a personalized Smart Furniture which can be controlled with multiple options based on y gesture recognition and emotion recognition. They described three types of inputs that are analysed: electrical, audio, and video signals[68].

## **3.3** Bibliography

# ⁴ Social, economic and smart ageingcontext⁶⁹

## ⁴·¹ Social context of improvement in indoor living for older people

The impact of demographic ageing within the European Union (EU) is likely to be of major significance in the coming decades. Consistently low birth rates and higher life expectancy are transforming the shape of the EU-28's age pyramid; probably the most important change will be the marked transition towards a demographically mature population with a more even distribution of different age groups in society.

The population of the EU-28 on 1 January 2018 was estimated at 512.4 million. Young people (0 to 14 years old) made up 15.6 % of the EU-28's population, while persons considered to be of working age (15 to 64 years old) accounted for 64.7 % of the population. Older people (aged 65 or over) had a 19.7 % share (an increase of 0.3 % compared with the previous year and an increase of 2.9 % compared with 10 years earlier).

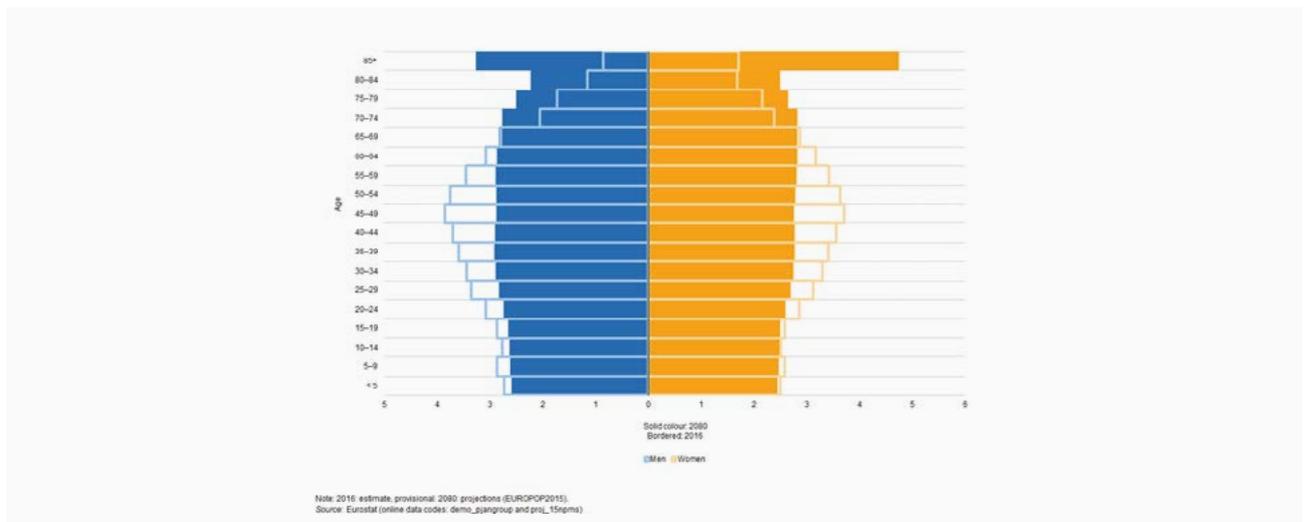

*Figure 4.1. Population pyramid EU-28, 2016 and 2018 (% of the total population)4*

The ongoing population ageing entails manifold challenges practically to all life domains– economic, education, healthcare and social security systems. It causes changes also in family structure and family relationships, construction of individual life courses, as well as perceiving of meaning of old age. The population aging in the context of limited public resources (during economic crises even shrinking) forces us to look for innovative ways to ensure the harmonious development of society as a whole, with particular emphasis on ensuring the sustainable quality of life and well-being of older people. Therefore the aging of the population, which is an essential achievement of the development of civilization, emerges also as a special stimulus for the creation of technological and social innovations. The spectrum of technological innovations is large – it can include such products and services as so called "calm technologies" (devices informs but doesn't demand our focus or attention[70]); "ambient intelligence" (devices work in concert to support people in carrying out their everyday life activities, tasks and rituals in an easy, natural way using information and intelligence that is hidden in the network connecting these devices ); and "smart technologies" (electronic gadgets that facilitate the provision of complex solutions to the user in the simplest way possible[72]).

Some of researchers introduced smart ageing concept[73,74,75]. The term is still in its infancy, therefore it isn't surprise that its perceptions vary. Some authors (De Keulenaer & Dome, 2005; Baraković Husić, et al 2019) give an accent to technology while explaining the concept – namely, for them it is a technology assisted living at home for individuals of advancing age. For others (Murata, 2015) smart ageing have two meanings: a) "we can become smarter as we age"; b) "we need smarter solutions to the challenges due to the ageing of society"[76]. In other words, smart ageing could be described as the phenomenon of technology and innovation usage to create products, services, solutions, and systems in order to improve quality of life of people leaving in an ageing society. However before we adopt this term it would be worth to rethink its possible negative effects for older persons as well as for smart-technology developers: on the one hand it means segregation of older persons on the base of their age only and description of them as stigmatized group and on the other it narrows the potential target group smart-technology-developers would be interested in. It seems, that a better way would be to use the term technology-for-smart-housing and an-active-lifestyle instead of smart ageing.

So among varied smart solutions developed to meet different needs of older persons the idea of smart homes [77] becomes an important one in ageing society because of these **prerequisites**:

1. As people age, homes can become an environment where the most time is spent. In this case, the requirements for home quality, safety, security, and ergonomic qualities greatly increase. There is a need for home gadgets and appliances that make it a great place to live in. It is important to note, that this type of smart habitat – improvement of general comfort – can be relevant aim to any person, no matter how old he/sheis;

2. If the frailty or health condition of the older person is progressing with the age, but housing is not adapted to social and /or health care functions, emerges the threat to institutionalization of hospitalization of older person, as a result of his/her autonomy loss / complex health care needs. In this case, different technological tools that provide the possibility for older adults with chronic conditions and complex needs to remain at home and maintain an acceptable quality of life (independent living and social participation) are very important[78]. The solution to remain in their own home for as long as possible often is the preferred solution for the majority of society. And again, it is important to point out, that this type of smart habitat – equipped with technologies which enhance one's quality of life, help monitor health and live safely in their homes[79]– can be relevant aim to any person with special needs, no matter how old he/she is;

3. Beneficiaries of smart habitat are not only frail older persons or older persons with chronic conditions but also their carers (be they informal or formal) and health care practitioners. Older persons are living in different types of households: single, one generation (usually with spouse or partner, but could be with siblings and the like), three or more generation households (usually with the family of the adult child). Some others inhabit different types of institutional households. In any case, the person's limited ability to perform basic daily activities requires involvement of carer (informal or formal, living under the same roof or in the near distance) and / or health care professionals, therefore smart housing plays an important role for care providers. Smart devices can release care takers from routine activities and instead allow them to focus on quality care.

It seems that the **demand** for smart solutions of complex problems rising in ageing societies will increase due to several factors:

1. Demographic drivers (increasing absolute numbers of older people; the increasing life expectancy at older ages due to improved quality of life, spread of healthy life style, and health services);

2. Social drivers (striving for better quality of life and social inclusionof older adults; changing ability of family members to take care for each other (due to changing family structure and functions – less children available for care of aged parents, and increased mobility of family members –often children are living at distance from parents in need for care);

3. Changes in human capital of older people – each next generation of older adults is more educated and more IT literate than previous, the increase in older persons living alone and preferring to stay at home; in general each next generation of older adults is more healthy, but we are "less tolerant" to such limitations of our functioning as inabilities – limited mobility, hearing impairments, or visual impairments, different degree difficulties within strumental activities such as shopping,transport, housekeeping;

4. Most of the existing dwellings are not adapted to different kind of needs, which are changing during the life course (for exsample, technical problems with difficult locks, luminous intensity, light switches, stairs, high doorsteps, narrow doorways, small bedrooms, etc.) In order to meet the challenges of people wishing to stay at home as long as possible we need to build new smart homes or renovate existing ones.

Despite the anticipated increase in demand, there are also **barriers** to developing smart homes[80],[81]:

1. It seems, that there still is hardly possible some kind of inventorization of enormous diversity of needs (often complex), services offered and technological solutions (variety of technological brands and (sometimes incompatible) systems);

2. Potential technology developers are not familiar with the problems experienced by end users, as well as potential users do not know what the market is offering;

3. Controversial technology acceptance. Also there is evidence, that older people are ready to accept new technologies, in public discourse there still circulates stereotypes about technology's expensiveness, complexity and insecurity, about ethical and privacy issues related to use of innovative technologies. Such a negative attitudes can slow development of smart homes.

4. Systematic review of articles on smart homes and home health monitoring technologies for older adults[82] revealed some weaknesses in the research on smart homes:

there wasn't provided evidence that smart homes or home health-monitoring technologies help to address the conditions ofdisability prediction and health-related quality of life or fall prevention; there were no studies about usability of home health technologies; there was a lack of economic assessment studies regarding the cost-effectiveness of smart homes and home health-monitoring technologies; it was pointed, that research was carried out exclusively in developed countries(developing countries not yet included); and finally, in the research there was little attention paid to the implementation of low-cost technologies (what would be of special interest).

Example 4- Social inclusion from the perspective of older adults in Lithuania and Portugal (see in Attachment 1)

**4.2** Economic consequences of investing in smart habitat

The demographic old-age dependency ratio (people aged 65 or above relative to those aged 15-64) is projected to increase significantly in the EU as a whole in the coming decades. Being about 25% in 2010, it has risen to 29.6% in 2016 and is projected to rise further, in particular up to 2050, and eventually reach 51.2% in 2070. This implies that the EU would move from four working-age people for every person aged over 65 years in 2010 to around two working-age people over the projection horizon.

As a result, the proportion of people at working age in the EU-28 is shrinking while the relative number of those retired is expanding. The share of older people in the total population will increase significantly in the coming decades, as a greater proportion of the post-war baby-boom generation reaches retirement. This will, in turn, lead to an increased burden on those at working age to provide for the health and social expenditure required by the ageing population for a range of related services.

Health care services represent a high and increasing share of government spending and total age-related expenditure. Furthermore, the ageing of the EU population may entail additional government expenditure. This makes public spending on health care an integral part of the debates on the long-term sustainability of public finances.

The projection for those aged 80 years and more will almost triple by 2060. This trend will cause an increase of social expenses in forms of pensions, healthcare and institutional or private care. Under this scenario, public spending on the older people will be a major problem in upcoming years.

The demographic change will have considerable consequences for the EU public finances. Based on current policies, it is estimated that 'exclusively' age-related (pensions, health, and long-term care) public expenditure will increase by 4.1 percentage points of GDP between 2010 and 2060, from 25% to 29%. Only expenditure on pensions is expected to increase from 11.3% to nearly 13% of GDP by 2060. However, there are significant differences between countries, de-pending largely on the progress made by each country in the reform of the pension system, which confirms the need for policy action to meet the challenges of an ageing population[83],[84].

Demographic change not only causes financial burden on society. Many older people, especially post-war baby boomers in Western European countries, were able to build up pensions, which they use to travel, hobbies and other pleasures for after working lives. This phenomenon is generally called Silver Economy. A study on the silver economy, initiated by the European Commission in 2018, estimates a baseline value of € 3.7 trillion (2015) for Europe's Silver Economy, primarily comprising private expenditure by older people (50 plus) on various goods and services, from housing to recreation[85].

The increasing reliance on healthcare and preventive measures results in higher expenses in health and social systems[86],[87],[88] that are intended as help for people in all age groups. From the financial point of view, it is not enough to use individual expense data to decide about the implementation of new interventions. The evaluation needs to be based on the comparison of effectiveness and success rates of a specific intervention, its impact on the quality of patients' lives, as well as on the financial cost.

The importance of the significant financial burden associated with the care for elderly was recognized by many studies[89] and international strategies (e.g., Innovation for Active & Healthy Ageing, Integrated care: health and social care become one)[90],[91], including the current study that was conducted by the team members. The study was able to show, on an example of people with neurodegenerative illnesses, that the expenses for their care are considerably higher than expected. (Table 4.1.). The need to find the solution to this problem is obvious[92].

| Product | Mean (billions EUR) | | | Range (billions EUR) | | |
|---|---|---|---|---|---|---|
| Year | 2010 | 2030 | 2050 | 2010 | 2030 | 2050 |
| Alzheimer's disease (AD) | 182,880 | 256,400 | 342,785 | 60,616- 316,97 | 84,985-444,402 | 113,618-594,130 |

*Table 4.1. Costs of the treatment and the care for patients with neurodegenerative illnesses. Source: (Maresova et al.b, 2016)[93]*

Integrated ICT solutions that would improve healthy and safe aging of elderly play an important role in this area[94],[95],[96]. The potential of these solutions is also seen in the possible reduction of health and social expenses.

An ongoing study compares five different population scenarios of people with Alzheimer's disease –it involves the administration of different kinds of medication compared with a scenario with no change in treatment up to the year 2080. The research uses computer simulations to formulate predictions. Changes in economic impacts are expected starting from the year 2023, when new medication is anticipated on the market. The results clearly illustrate that any intervention aimed to keep the patient in any stage of the disease for a longer period of time means the rise of treatment and care costs as well as the growth of the number of Alzheimer's disease pa-tients. Prolonging the life of Alzheimer's disease patients is meaningful in terms of their quality of life, so when new medication is introduced, the society must be ready to bear the increased economic burden.

This model is a base for other future variants based on the improvement of environmental conditions (smart habitat) and their impact on savings in care.

### 4.3 Quality of life in context of smart ageing

In the following years all life spheres will be flooded with smart things and systems. These systems will completely change everyday activities by creating opportunities for development and innovation, which in turn will bring in countless benefits. Smart systems will connect homes, cars, governments, health, etc. This concept will also change the way people interact with the society and things around them and try to simplify our lives. Generally, the ultimate goal of smart concepts should be to improve our Quality of Life (QoL).

QoL definition and meaning varies for people of different gender and age groups and their cultural, economic, and educational background. To simplify, QoL is related to an overall enjoyment of life. However, some authors use a term "satisfaction" to define QoL, some use "well-being", while others use "apperception" in defining it. The example of the first type is: "QoL is the degree of need and satisfaction within the physical, psychological, social, activity, material, and structural area". Second type defines QoL as "a state of well-being which is a composite of two components: 1) the ability to perform everyday activities which reflect physical psychological and social well-being, and 2) patient satisfaction with levels of functioning and the control of disease and/or treatment related symptoms". The third is used by the World Health Organization (WHO) and World Health Organization Quality of Life (WHOQOL) Group. They define QoL as "apperception of one's position in life in the context of the culture and value systems in which they live and in relation to their goals, expectations, standards and concerns". To sum, QoL is a multidimensional concept which emphasizes the self-perceptions of an individual's current state of mind affected in a complex way by the person's physical health, psychological state, personal beliefs, social relationships, and their relationship to salient features of their environment.

Simplification and facilitation of everyday activities and improvement of QoL is especially important for the elderly since their number is significantly increasing according to available statistics. The number of people aged 60 years or older will rise from 900 million to 2 billion by 2050, and the population ages even quicker than in the past. The World Population Aging Report indicates that the growth rate of the older population is more rapid in developing countries than in developed countries.

With this in mind, it is clear that smart things and systems should be largely devoted to improving QoL of elderly, given that they will represent22% of the entire world population. This concept is named smart ageing. Smart ageing is defined as technology and innovation usage in both the public and private sectors to create products, services, solutions, and systems to improve

QoL of people aged 50 and over. Healthy ageing is another term used to describe the concept of enabling older people to enjoy a high QoL. Term mentioned in WHO is active ageing and is defined as the process of optimizing opportunities for health, participation, and security in order to enhance QoL as people age.

**Smart Ageing Ecosystem**

Smart ageing ecosystem[97] includes key determinants of healthy ageing as described in Euro-HealthNet, and covers economic, social, and environment dimensions in objective and subjective conditions and all aspects of human QoL (see Figure 4.1). On the other hand, QoL has eight dimensions as suggested by Eurostat: material living conditions, health, education, productive and valued activities, governance and basic rights, leisure and social interactions, natural and living environment, and economic and physical safety.

Each smart ageing determinant should be contained at least in one QoL dimension (see Figure 4.2). For example, New Technologies (determinant) contributes to Leisure and Social Interactions (dimension) of elderly given that it provides new ways of entertainment and communication. However, it contributes negatively to Personal insecurity given that usually people over 50 are insecure when using new devices, applications, etc., resulting in the withdrawal and abstinence from new technology products. Using New Technologies allows elderly to increase their incomes by having jobs, which consequently affects their Material Living Conditions. New Technologies also have an impact on 1) Education, by expanding the options of older adults to gain knowledge and stay competitive, 2) Productive and Valued Activities, since they can perform contemporary tasks that are useful to themselves and the community, and 3) Health, by, at a minimum, better monitoring of their health conditions.

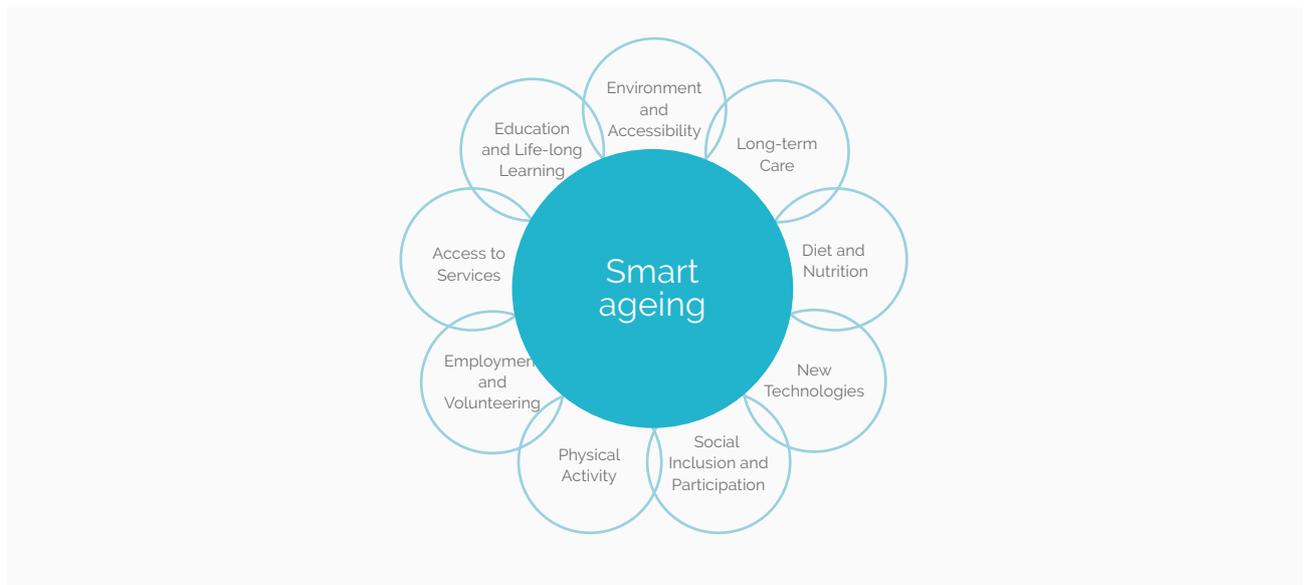

*Figure 4.2.Smart ageing determinants*

This mapping provides a connection between smart ageing determinants and QoL dimensions, and results in QoL indicators for elderly. Knowing which features of smart ageing products and services affect which QoL dimension of elderly enables better targeting and effectively achiev-ing the ultimate goal –improved QoL of elderly. Research and industry communities should consider this when developing their products and services.

**Research findings on smart ageing and QoL**

Today we are witnessing multiple smart ageing solutions being developed and produced, with many more in the announcement, but the QoL of elderly is not noticeably improved yet. Motivated by this, authors[98] seek to give an answer to the following question: are the existing smart ageing solutions succeeding in direct improvement of QoL of elderly?

The results of the conducted survey[99] give a clear answer to it – they are not. The authors could not conclude that the existing smart ageing solutions necessarily directly contribute to QoL of elderly due to several reasons. Firstly, the existing approaches have not addressed various dimensions of QoL of elderly, nor have they included different smart ageing determinants, meaning that they are not multidimensional and comprehensive, i.e., not in line with QoL nature. Furthermore, the majority of solutions have not been developed/produced for direct usage by elderly but for people around them. Last and most important, most solutions were never properly verified by the elderly so one cannot say that they have been beneficial to them.

This conclusion opens a wide area of research issues to be addressed and corrections to be applied in the future. However, in order to succeed in improving QoL of elderly, research and industry communities are recommended to develop smart ageing solutions that cover all elderly QoL dimensions and utilize smart ageing features to do so. An important recommendation is to ask the elderly what would help them and increase their QoL in a certain context, produce solutions directly for that population, and afterwards verify their products and services with them.

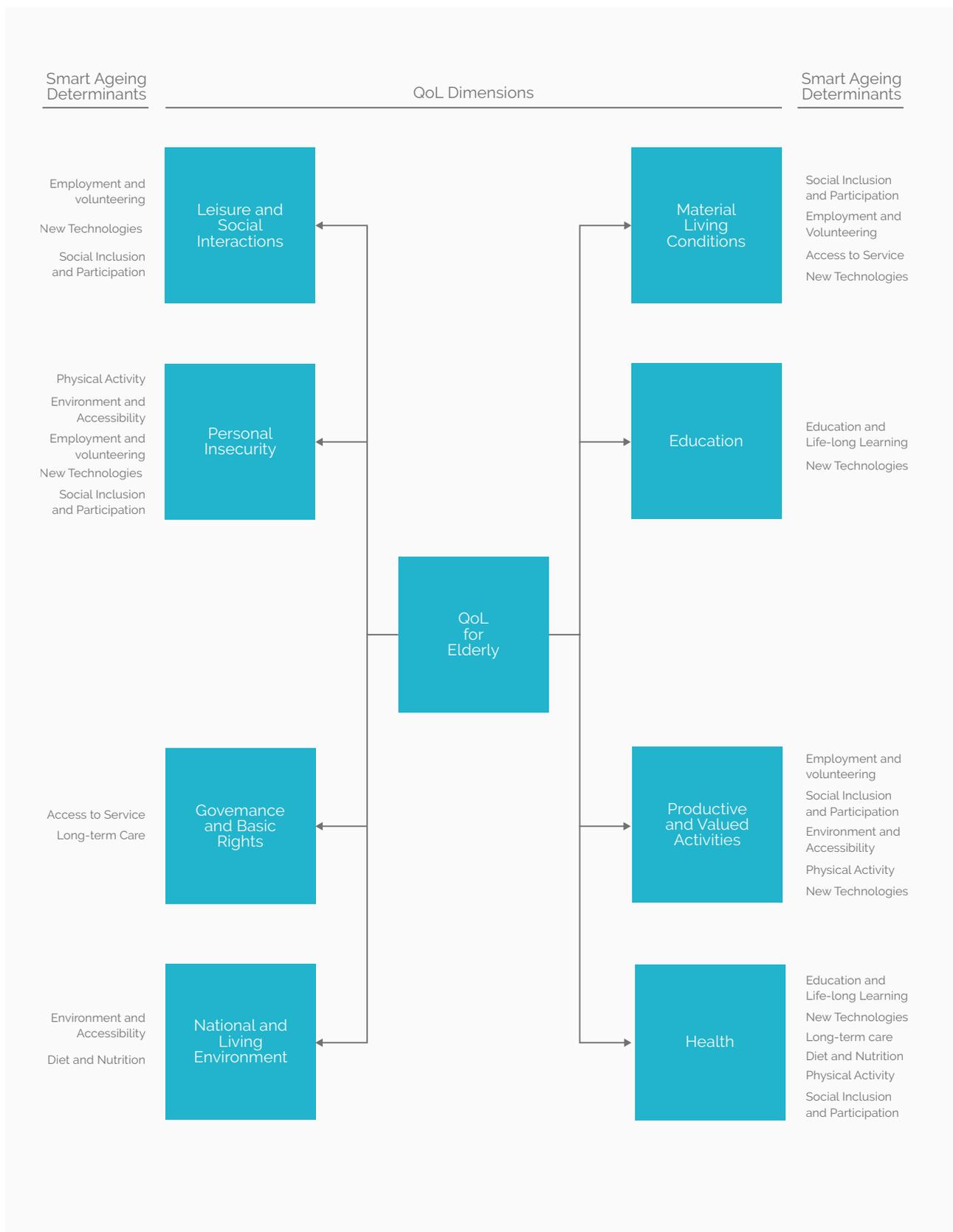

*Figure.4.3. Mapping between QoL dimensions and smart ageing determinants*

In summary, mapping the social, economic and smart ageing context we learn that demographic change opened up a new era of society in Europe. Social, economic and smart developments are taking place on every geographical and societal level. Quality of life and social inclusion are the main challenges for older people. How to achieve this, will be further explored in the next chapter: policy and practices.

## **4.4** Bibliography

# ⁵ Policy and practices

Active and healthy ageing (AHA) is one of the societal challenges we will be facing in the upcoming decades as demographic ageing heavily impacts on society and economy. Finding innovative and sustainable solutions to help individuals of advanced old age to continue living at home independently is critical in the context of both demographic change and budgetary constraints. Policies on ageing may perceive 'old age' as a status rather than the result of a process and change of paradigm is needed. In order to gain the ability to remain active, the individual needs to invest in this process throughout the life course.

Overview of existing policies and current practices gives insight about societal structures and awareness-raising actions that would enable persons to live an active life-style as long as possible.

The 2018 Ageing Report: Economic and Budgetary Projections for the EU Member States (2016-2070)[100] looks at the long-run economic and fiscal implications of Europe's ageing population.

Strategy and action plan for healthy ageing in Europe, 2012–2020[101] outlines synergies and complementarities in cooperation with partners and with European Commission initiatives. In implementing this strategy and action plan, the Regional Office will ensure that all countries in the WHO European Region are adequately covered, as population ageing is spreading fast in the Region, making the need to prepare health and social care systems for ageing populations particularly urgent.

WHO Age-friendly Environments Programme[102] is an international effort to address the environmental and social factors that contribute to active and healthy ageing

WHO Active Ageing A Policy Framework This Policy Framework[103] is intended to inform discussion and the formulation of action plans that promote healthy and active ageing.

Age friendly cities of WHO[104] is a network the encouraging the development of age-friendly cities, to tap the potential that older people represent for humanity. It outlines the challenge facing cities, and summarizes the research process that led to identifying the core features of an age-friendly city. Since the launch of the Global Network of Age-friendly Cities and Communities, 800 realties have joined the network and this number is still growing. The WHO concept addresses the optimisation of the physical and social environments of older persons and defines necessary provisions (such as ICT and healthcare).

European Scaling-up Strategy in Active and Healthy Ageing[105] states that health and care services in Europe are undergoing changes to adapt systems to a growing demand caused by ageing and the expansion of chronic diseases. This restructuring, which combines health and social care resources, involves developing and testing innovative solutions and eventually the large-scale implementation of the most successful practices. The multitude of good examples developed throughout the EU has led to a realization that a comprehensive scaling-up strat-egy is needed at European level. The European Innovation Partnership on Active and Healthy Ageing (EIP AHA or Partnership), which brings together key stakeholders in this policy area and supports the good practices and References Sites developed by its partners, can act as a catalyst to foster scaling-up across regions and countries.

A European Innovation Partnership on Active and Healthy Aging (EIPAHA)[106] platform is a communication and information hub for all actors involved in Active and Healthy Ageing throughout Europe. It contains an action group about "Age Friendly environments" whose main objective on "Innovation for age friendly buildings, cities, and environments" is to bring together partners from all over Europe who are committed to implementing strategies for the creation of age-friendly environments which support active and healthy ageing of the European population.

Artificial Intelligence for Health and Health Care[107] study centres on how computer-based decision procedures, under the broad umbrella of artificial intelligence (AI), can assist in improving health and health care. Although advanced statistics and machine learning provide the foundation for AI, there are currently revolutionary advances underway in the sub-field of neural networks. This has created tremendous excitement in many fields of science, including medicine and public health. First demonstrations have already emerged, showing that deep neural networks can perform as well as the best human clinicians in well-defined diagnostic tasks. In addition, AI-based tools are already appearing in health-oriented apps that can be employed on handheld networked devices such as smart phones.

Smart cities: digital solutions for a more liveable future[108] state that after a decade of experimentation, smart cities are entering a new phase. Although they are only one part of the full tool kit for making a city great, digital solutions are the most powerful and cost-effective recent additions to that tool kit. The report analyses dozens of current applications and finds that cities could use them to improve some quality-of-life indicators by 10–30 percent. It also finds that even the most cutting-edge smart cities on the planet are still at the beginning of their journey.

• Smart cities add digital intelligence to existing urban systems, making it possible to do more with less. Connected applications put real-time, transparent information into the hands of users to help them make better choices. These tools can save lives, prevent crime, and reduce the disease burden. They can save time, reduce waste, and even help boost social connectedness. When cities function more efficiently, they also become more productive places to do business.

• MGI assessed how dozens of current smart city applications could perform in three sample cities with varying legacy infrastructure systems and baseline starting points. We found that these tools could reduce fatalities by 8–10 percent, accelerate emergency response times by 20–35 percent, shave the average commute by 15–20 percent, lower the disease burden by 8–15 percent, and cut greenhouse gas emissions by 10–15 percent, among other positive outcomes.

• Snapshot of deployment in 50 cities around the world shows that wealthier urban areas are generally transforming faster, although many have low public awareness and usage of the applications they have implemented. Asian megacities, with their young populations of digital natives and big urban problems to solve, are achieving exceptionally high adoption. Measured against what is possible today, even the global leaders have more work to do in building the technology base, rolling out the full range of possible applications, and boosting adoption and user satisfaction. Many cities have not yet implemented some of the applications that could have the biggest potential impact. Since technology never stands still, the bar will only get higher.

• The public sector would be the natural owner of 70 percent of the applications examined. But 60 percent of the initial investment required to implement the full range of applications could come from private actors. Furthermore, more than half of the initial investment made by the public sector could generate a positive return, whether in direct savings or opportunities to produce revenue.

• The technologies analysed in this report can help cities make moderate or significant progress toward 70 percent of the Sustainable Development Goals. Yet becoming a smart city is less effective as an economic development strategy for job creation.

• Smart cities may disrupt some industries even as they present substantial market opportunities. Customer needs will force a re-evaluation of current products and services to meet higher expectations of quality, cost, and efficiency in everything from mobility to healthcare. Smart city solutions will shift value across the landscape of cities and throughout value chains. Companies looking to enter smart city markets will need different skill sets, creative financing models, and a sharper focus on civic engagement. Becoming a smart city is not a goal but a means to an end. The entire point is to respond more effectively and dynamically to the needs and desires of residents. Technology is simply a tool to optimize the infrastructure, resources, and spaces they share. Few cities want to lag behind, but it is critical not to get caught up in technology for its own sake. Smart cities need to focus on improving outcomes for residents and enlisting their active participation in shaping the places they call home.

Smart Homes for Elderly Healthcare—Recent Advances and Research Challenges[109] states that advancements in medical science and technology, medicine and public health coupled with increased consciousness about nutrition and environmental and personal hygiene have paved the way for the dramatic increase in life expectancy globally in the past several decades. However, increased life expectancy has given rise to an increasing aging population, thus jeopardizing the socio-economic structure of many countries in terms of costs associated with elderly healthcare and wellbeing. In order to cope with the growing need for elderly healthcare services, it is essential to develop affordable, unobtrusive and easy-to-use healthcare solutions. Smart homes, which incorporate environmental and wearable medical sensors, actuators, and modern communication and information technologies, can enable continuous and remote monitoring of elderly health and wellbeing at a low cost. Smart homes may allow the elderly to stay in their comfortable home environments instead of expensive and limited healthcare facilities. Healthcare personnel can also keep track of the overall health condition of the elderly in real-time and provide feedback and support from distant facilities.

The Covenant on Demographic Change[110] aims at gathering all local, regional and national authorities, and other stakeholders, that commit to cooperate and implement evidence-based solutions to support active and healthy ageing as a comprehensive answer to Europe's demographic challenge. It supports local and regional authorities, and other stakeholders, in developing environments for active and healthy ageing that:

- improve healthy life year expectancy (HLY);
- enhance opportunities for independent living of older people and;
- support a society for all ages.

Universal Guide for addressing accessibility in standards[111] The purpose of this Guide is to assist developers of standards (e.g., technical committees or working groups) to address accessibility in standards that focus, whether directly or indirectly, on any type of system that people use. It provides guidance for developing and writing appropriate accessibility requirements and recommendations in standards.

Since 2018 a new network is active in Europe. Social care provider Cáritas Diocesana de Coimbra (Portugal) and SME AFEdemy, Academy on age-friendly environments BV (The Netherlands), took the initiative to launch a proposal on the instalment of a European Thematic Network on Smart Healthy Age-Friendly Environments (SHAFE). The proposal was approved by the European Commission (DG SANTE). In November 2018 the network presented the Joint Statement to the European Commission at the Health Policy Platform meeting in Brussels. The main aim of SHAFE is to better align the construction and ICT sector to improve the implementation and upscaling of smart living environments in Europe. This aim can only be achieved when both construction and ICT co-create smart living solutions with end-users (older people, caregivers, etc.)[112].

Example 5 - Existing policies on accessibility in Spain (see in Attachment1). See also Attachments 2 and 3

# <sup>6</sup> Conclusions

- It is important that the elderly maintain physical activity and capacity through training and exercise. In addition, their living environment should be adapted with safety precautions, telecommunication, virtual reality training, and indoor design that inspire to move. These are important tools to achieve that older individual can live at home for as long as possible.

- Safety, accessibility, independence and social participation are vital for the elderly to remain active to maintain their abilities and capacities.

- An ageing population with increased longevity as a result of improved living conditions and healthcare leads to an urgent need for new and innovative forms of support and health care for older people wishing to stay at home.

- Older people are more sensitive to pollution exposures than younger adults, which should be taken into account in process of planning and maintaining buildings. It should be taken into account thermo sensation , adequate ventilation, to ensure good indoor quality.

- To estimate savings in health- and social care it is important to consider: a) BIA (budget impact analysis) method that suggests a process (model) for the formulation of the impact of expenses connected to the aging population on the public expenses, mainly in healthcare and social care; b) use the access to other professionals (due to the membership in European Cooperation) that consider ICT solutions for elderly people to suggest a model which formulates the potential savings (when supported by evidence) using the specific ICT solutions in health- and social care while considering future changes of the elderly population; c) the creation of other types of expense models for specific disabilities of elderly people, which ICT should find solutions to – and the impact on health- and social care.

- The existing smart ageing approaches have not addressed various dimensions of QoL of elderly, nor have they included different smart ageing determinants. In other words, they are not multidimensional and comprehensive, i.e., not in line with QoL nature.

- The majority of solutions have not been developed for direct usage by the elderly but for people around them. Most solutions were never properly tested and validated by the elderly, so one cannot say that they have been beneficial to them.

- The increased availability and use of the technology is hindered by legislative aspects and greatly challenged by the diversity of governmental approaches in individual countries. The use of wireless devices and storage of information on the internet also leads to potential security concerns.

- It is recommended that practitioners, policy makers, care insurers, and care providers work together with technology developers and researchers to prepare strategies for the implementation of assisting technologies in different care settings.

# 7 Future work

During the first and second periods of the COST Action CA 16226 "Indoor living improvement: Smart Habitat for the Elderly" (**Sheld-on**)the scoping review of the technologies for the elderly with Alzheimer's Disease[113], definition of smart furniture (currently – paper under review), overview on specific physical and psychological needs of elderly people (WG3 report), current practices from different countries (WG3 report and Attachment 1 and 2), and examples of projects(WG3 report and Attachment 3) was mapped out.

In the future work the theoretical background will be combined with specific IT solutions from other working groups to analyse and describe new solutions and their market potential as well as legislative and social impacts.

# 8 Examples of practices from WG3 members' institutions and countries

## Example 1 Interventions for elderly people in Norway[1,2]

**Safety:**
"safety alarm", telecommunication, personal assistant, adapting the environment: streets, shops, public places possibility to admit yourself into a "Safety department" in community-based care

**Accessibility:**
home adaptations: lighting, furniture supports, etc.

**Independence-mobility:**
care from the municipality, assistive devices

**Social participation:**
group activities, senior centres, independent organisations,
Primary care: health interventions: medical and when in need of special support provided by private services or municipality services
Living at home after acute illness

**Specialist services:** acute

**Municipality services:** rehabilitation, short-term care and, if needed, long-term care
"Hverdagsrehabilitering" = "everyday rehabilitation"[3]

**Living in an institution**
"Your home away from home" [4, 5]
* Physical activity: maintenance of function
* Wellbeing: the facilities – buildings, one's own room, design of interior spaces – that inspire to move (for long enough but not excessively)
* Health: teeth / nutrition
* Mobility: falls

**Environment:**
- inside: micro climate, comfort, wellbeing, quality of life
- outside: physical accessibility, mobility, adaptations
- person-environment interaction ("connectedness")
- *climate: inside (comfort) / outside (mortality)

## Example 2 Achievements and findings on Indoor Air Quality from Holzforschung, Austria

Holzforschung Austria has a long tradition of accredited testing, inspection and certification, active standardization, as well as research and development with respect to emissions, especially those coming from wood and timber products. Throughout centuries, formaldehyde has been the main issue of concern. For more than a century issues related to VOC-emissions (volatile organic compounds) have been studied. This report focuses especially on the major research activities within the last century concerning indoor air quality.

**VOC - emissions from timber and toxicological aspects**

The project ran from 2008-2010 and was financially supported by FFG (Österreichische Forschungsförderungsgesellschaft, FFG number 818628) - the Austrian research promotion agency, and by the Austrian Association of the wood working industry (Fachverband der Holzindustrie Österreichs) and their members.

ACHIEVEMENT
Step-in-stone for VOC measurement based on thermodesorption gaschromatography coupled with mass spectrometry (TD-GC/MS), substance specific qualitative and quantitative determination of VOCs (calibrated ever since approximately based on 75 typical individual substances) and generation of the base for accredited testing according to relevant standards.

CONTENT AND FINDINGS
Generation of a wide matrix for typical VOC emissions of building products, furniture products, flooring products, wood species, wood qualities while considering the impact of origin, as well as thermally modified wood. Despite that, a set of typical accompanying non-wood products were evaluated.
Support of Karl Dobianer (independent researcher) within the development of an in-dependent toxicological assessment system (TIAC: tolerable indoor air concentration) based on published and reliable toxicological thresholds, optionally also including odour. Evaluation of the overall toxicological profile based on the hazard index HI, where values up to 1 are safe for all humans, even for young children, elderly people or persons suffering from specific illnesses (e.g., upper respiratory defects). Values between 1 and 1.5 can be seen as an unspecific border zone whereas with values above, individuals might start showing specific reactions of the contamination. At a range of approximately 4, alterations are usually recommended. The system was since used in a different project, and Karl Dobianer frequently adapts it based on the current state of the art. The TIAC system is also accepted at Austrian cords based on expert opinions.

Most assessed products turned out to be safe with respect to their VOC emissions. Some common substances were identified that might lead to customer complaints. In rare occasions, individual products failed, for example, due to the emission of CMR substances (carcinogenic, mutagenic, or reproduction toxic) such as furfural (e.g., due to specific conditions of thermal treatments).

## HFA Timber - Indoor Air Quality

The project ran from 2009-2014 and was financially supported by FFG (Österreichische Forschungsförderungsgesellschaft, FFG number 820501) - Austrian research promotion agency, and by a selection of individual industry partners from wood and wood-related industry.

ACHIEVEMENT
Within this project, the first two 30 m² model rooms, according to the CEN TC 351 specifications of a theoretical model room (now covered by EN 16516) were constructed as real stand-alone buildings fabricated of conventional building products. One is still in use and can be equipped with any kind of structural components or furnishings, such as floorings, walls, ceilings, or furniture.

CONTENT AND FINDINGS
The project gained the first dynamic profiles of VOC-decline in the built environment under controlled climatic conditions. Different building materials were accessed in labs and selected for the construction of the model rooms. The interior walls were designed in OSB (oriented strand board), plaster board, or a combination of both with or without water vapour barrier in between. As expected, natural building materials such as OSB give higher VOC emissions compared to almost inert materials. However, within reasonable time after construction, these emissions fall under any relevant threshold. By introduction of water vapour barriers, these emissions can he held back, but typically protrude through after a certain period of time leading to delayed emission peaks.

### Wood Comet cooperation project with Wood K plus

The project ran from 2012-2014 and was financially supported by FFG (Österreichische Forschungsförderungsgesellschaft) - the Austrian research promotion agency, and by the Austrian Association of the wood working industry (Fachverband der HolzindustrieÖsterreichs) and their members. Holzforschung Austria was a subcontractor of Wood K plus.

ACHIEVEMENT
The inter-laboratory comparison between the involved partners clearly demonstrated the problem of applying Toluene-d8-equalents for the characterisation of VOCs in comparison to substance specific evaluation based on calibration. The Td8-resultes deviated almost up to a factor of ten from the reference.

CONTENT AND FINDINGS
Different building products used for the creation of 30 m² model rooms were assessed under controlled conditions. The VOC concentrations within two different model rooms were measured following each individual finishing stage until final applications. Typical daily activities, such as cooking, smoking, cleaning etc., were performed. These activities most commonly lead to the exceedance of threshold values for specific substances within the indoor air (e.g., limonene when peeling oranges). The construction products and furniture could be seen as only a baseline for indoor VOCs compared to the constant impact of human activity.

### BIGConAir

The project ran from 2012-2015 and was financially supported by FFG (Österreichische Forschungsförderungsgesellschaft, FFG number 836468) - the Austrian research promotion agency, and by a selection of individual industry partners from wood and wood-related industry. Holzforschung Austria was subcontractor of University of Innsbruck.

ACHIEVEMENT
Based on the data generated within this project, the first mathematical model for the long-term development of VOC-concentrations in the indoor air was applied. It was possible to fit the model used within EN 717-1 for formaldehyde emissions from a single product under controlled testing conditions for up to 28 days.

VOC-measurements (single substances and sum-parameter) derived from real environments and were measured for a much longer time.

CONTENT AND FINDINGS
Long-term indoor air quality was measured in two different office containers made of wood with different constructions. To enable comparison, the same was done in a steel container. Later, the steel container was equipped with loam plastering. All containers were equipped with typical and comparable office furniture and used on daily basis. In order to see the additional contribution from the furniture, the latter was frequently removed and brought back before and after the measurement. While the steel container showed almost no VOC-emissions, the wooden containers showed a typical decline throughout several months. In all cases, the furniture gave a certain additional contribution. However, in all cases, VOC contamination of indoor air was acceptable or low. In contrast, the subjective sensation was, that the indoor air quality within the steel container was worse compared to the wooden one. Loam plastering improved the subjective sensation of indoor air quality within the steel container.

**Wood2New**

The project ran from 2014-2017 and was financially supported within the WoodWisdom Net+ program (number 101005), and by CEI-Bois - the European Confederation of Woodworking Industries as well as a selection of individual industrial partners from wood and wood-related industry. Holzforschung Austria was partner within a consortium led by Aalto-University.

ACHIEVEMENT
The development of the ISO 16000-3 method for the measurement of carbonyl substances such as formaldehyde.
First known long-term indoor air quality assessment within a real environment of residential housing.

CONTENT AND FINDINGS
Thirteen building objects got monitored during the phase of building, move in, and operation within the first months. Six objects were timber frame constructions, six were made of solid wood, and one built of concrete served as a reference. In addition to VOCs and VVOCs, other indoor air quality parameters, such as particular mater, airborne microorganisms (mould, yeast), temperature, relative humidity and ventilation rate, were assessed. Furthermore, human health parameters were assessed: blink rate, pulmonary function, pulse and blood pressure, self-assessments of well-being.
Although VOC emissions were high during the building phase and most likely increased at the move in, they showed a specific decline within the first months. In general, good indoor air quality was reached within reasonable time in all objects. However, toxicologically relevant substances could be found in detached occasions, and their origin was found. Controlled ventilation proved to lead towards better indoor air quality compared to manual ventilation. The self-assessment of well-being of residents was generally on a very high level ("excellent", "outstanding"), even throughout times of elevated emissions.
In addition, also the influence of different wood modification and surface treatment methods on the emission profile, depending on variable climatic conditions from different wood species, was studied. Specific alterations of sorption isotherms could be seen.

**Brand Wasser Schaden (FireWaterDamage)**

The project started in 2016, ending 2019, and is financially supported by FFG (Österreichische Forschungsförderungsgesellschaft, FFG number 850936522) being the Austrian research promotion agency, and by a selection of companies from building refurbishment industry.

ACHIEVEMENT
The development of a straight-forward measurement method for odour-relevant VOCs after fire. Assessment of typical odour-relevant VOCs after fire depending on different construction- and interior products.

CONTENT AND FINDINGS
Specimens from selected building and interior products were exposed to smoke under controlled conditions and their typical emission profile was compared to the emission profile after different refurbishment techniques (e.g., ozone or enzyme treatment). Currently, experiments are being performed after a controlled fire in a specifically designed research building.

**IASca – Indoor Air Scavenger**

The project started in 2017, ending 2020, and is financially supported by FFG (Österreichische Forschungsförderungsgesellschaft, FFG number 860587) - the Austrian research promotion agency, and by the Austrian Association of the wood working industry (Fachverband der Holzindustrie Österreichs) and their members.

ACHIEVEMENT
The development of a novel dynamic measurement method for the evaluation of sorption and desorption of VOC and VVOC on loose materials that may be used for indoor air cleaning.

CONTENT AND FINDINGS
The indoor air quality of existing timber- and mineral based buildings was assessed within the phase of long-term use (for several years). Some of these objects were even assessed within earlier projects, which helped in interpreting newly acquired data. The goal is to understand and model real long-term emission developments within the built environment. This will help with the selection of optimum sorption media in different stages of object use and thus improve indoor air quality.

**Public contributions with respect to the above-mentioned projects**

**Example 3** Introduction to the topic "person- environment interrelation"

Homo sapiens has lived in buildings for only thousands of years, but 6 million years ago he lived in wild nature surrounded by all sorts of natural dangers that forced him to look for supportive natural, and later built environments for survival. The nervous system was developed according to theses condition and it has remained almost the same also in contemporary human beings[1].

Why did Homo sapiens prefer certain places and avoid others? The earliest human beings needed food, water and protection to survive, but also new challenges to develop further.

These basic human preferences in space are expressed by the five archetypal characteristics of a supportive environment by the American architect Grant Hildebrand in 1999, based on knowledge from evolutionary biology and his own research, and they are: prospect refuge, attraction, peril and complex order[2].

As contemporary "cultural animals", human have inborn instincts, emotions and certain behavioral patterns and preferences, that can be overcome with consciousness. Human beings can perform in an "autopilot" regime thanks to cultural stereotypes and habituation/ somatization.

Instinctive behavior comes into conflict with western cultural habits. "Our behavior, outlook, general wellbeing and everyday social interactions are directly tied to the natural and built environment. How spaces and their structures control our everyday use have consequences which are far-reaching - potentially impacting our long-term mental and physical health. Built environments for long term stays should be supportive and stimulating in an appropriate way, preventing environmental and social stress - helping us to be in state of complex comfort or well-being. We set the assumption that the body-conscious designing or design for well-being applied consciously in public and private spaces for long term stay can help not only improve public health, but also foster better relations to social and natural environments.

We can state that at least 3rd age older adults of the upcoming digital future will want and have to work (due to the prolonging working age and leaving for being retired) with computers and other digital equipment used for work and entertainment thus we have to speak also about the supportive working built-environment for older adults. In the 4th age more the health care and social life issues are relevant.

For staying healthy and active up to 3rd age of humans the prevention of civilization diseases is extremely important. By exploring those in the research project within BCDlab.

Thus to handle the topic of ageing properly it is necessary to speak about prevention of civilization diseases already in youth and middle age. By exploring this topic we have found out two most important issues related to the most of civilization diseases – a lot of environmental stress and lack of movement. Thus dynamization of environment and reduction of environmental stress can lead to the prevention and healing of civilization diseases. Bringing into living, working and public space appropriate measure of neural physical activity and freedom of choosing body positions and reduction of environmental stimuli in the long term stay spaces can contribute to public health. In the projects of BCDlab there were elaborated also recommendations for preventing and healing the single diseases[3].

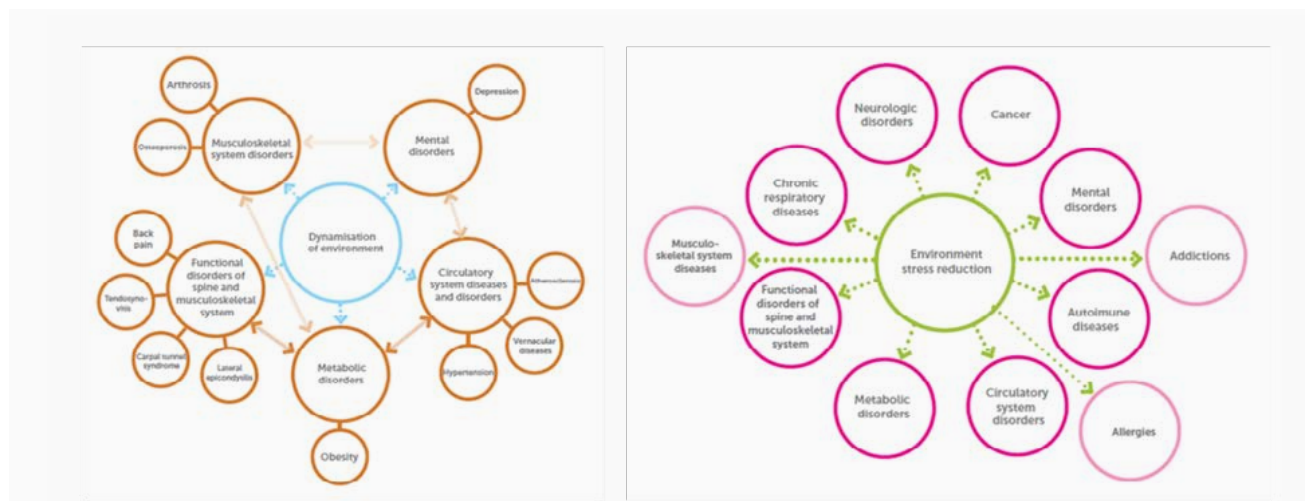

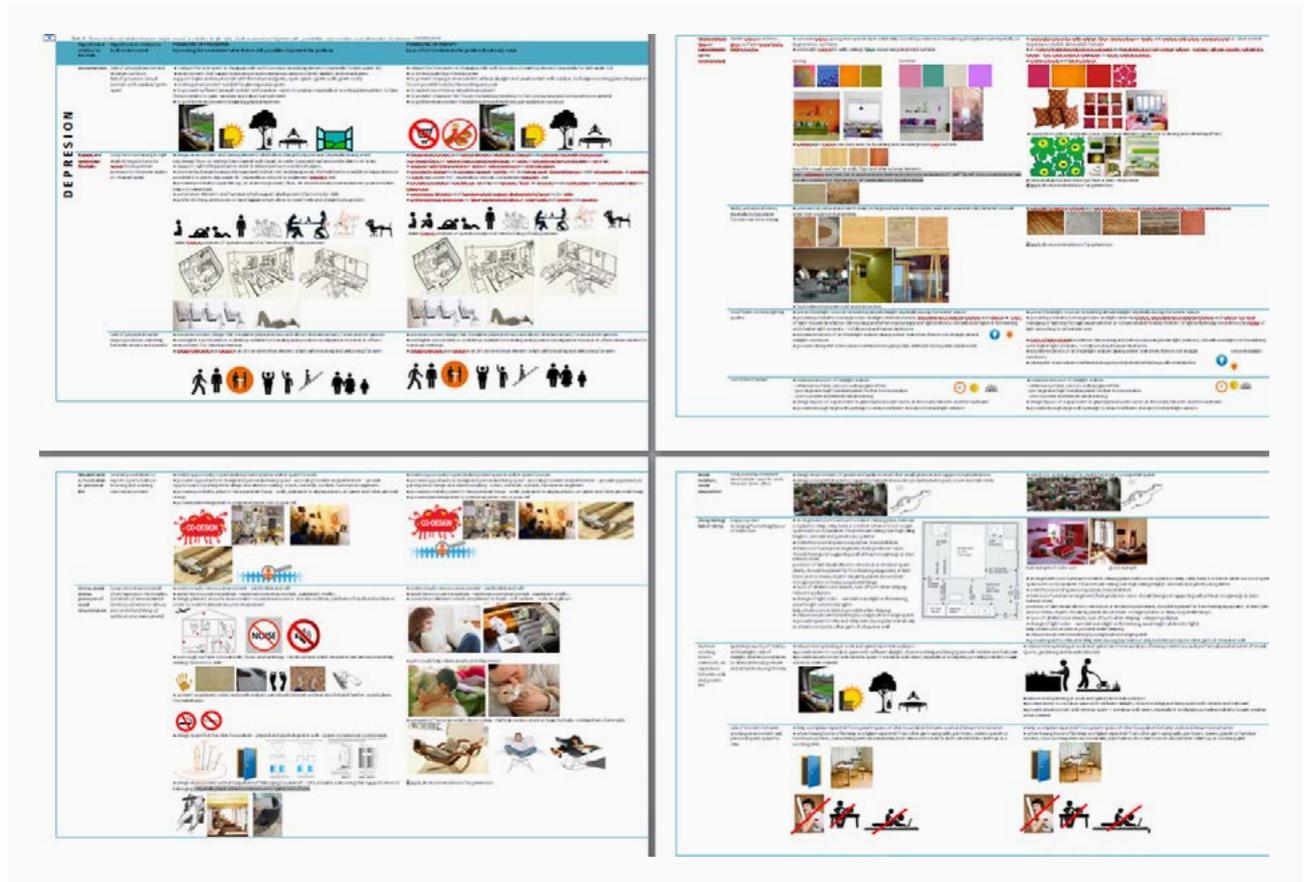

Understanding of comfort – well-being in the western culture is connected with many stereotypes that we are questioning and re-evaluating with our research studies. Ergonomics/ human factors are in our research understood in a broader context and relation and it is enriched by knowledge from other sciences dealing with the human body and behavior in relation to the environment. So the summarized knowledge from studies from environmental, cognitive, social and cultural psychology, neuroergonomy, environmental ergonomic, physiology and own experimental research have led to the formulation of 11 principles of spatial design for well-being. These principles are now being specially reconsidered from older adult's needs point of view.

**Features of supportive environment / principles of spatial design for well-being**

FEELING OF SAFETY

A feeling of safety is a basic condition for a feeling of well-being. In term of safety in the macroenvironment, it means e.g. being away from endangerment from nature elements or being in a safe quarter of a city with low crime. In terms of the microenvironment, it means being safe at home – with less endangering of privacy and territory, or feeling safe from any kind of mechanical or technical /static endangering – like instable objects. Even though in western cultures, vision is the dominant sense especially by entering a space, safety is matter of all senses. Our visual culture precludes stimuli from other senses, but nonetheless they are more connected to human unconsciousness, where instincts initiate an intuition that can give us "warnings" in risky situations which lead us to change our behavior.

This topic is related to our ability to be oriented in a space while disorientation can cause people to feel stressed up and unsafe, that can lead to potential aggressive or psychotic behavior.

The feeling of safety is supported by these features:

- the ability to protect one's own private space,
- the presence of a physical barrier behind one's back- e.g. in the form of a wall, screen or piece of furniture that unconsciously grant protection from attack from the back,
- overview about entrance to the space and control over territory,
- horizontal structuring of space,
- lower ceiling,
- panoramatic outlook over surroundings,
- presence of limits and barriers that according to the situation are flexible and adaptable.

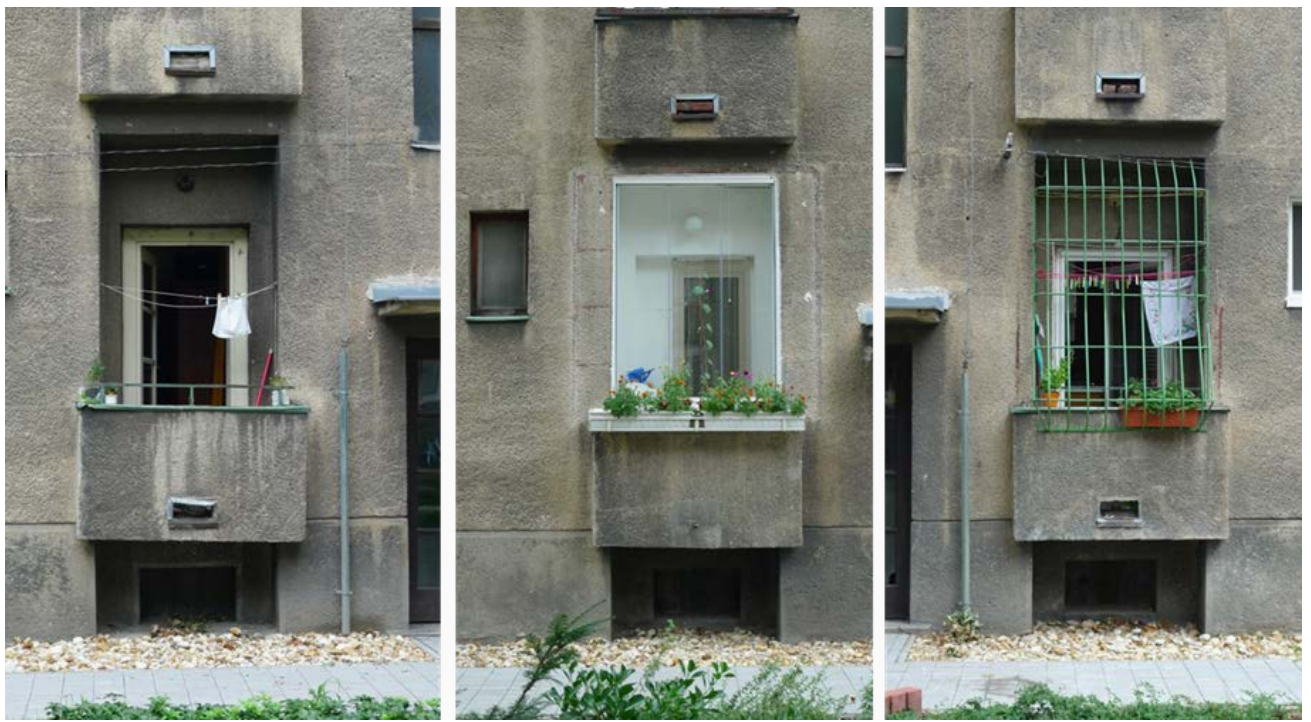

*Figure 8.1: Different arrangements of risk management in one house in Bratislava, foto: Veronika Kotradyová*

In this relation it is important to have the competence to manage risks and attractions where it is possible.

APPROPRIATE RATIO OF PROSPECT AND REFUGE

The possibility to combine prospect / perspective and refuge is a very basic need of Homo sapiens. It is connected with an instinctive search for situations and environmental settings where a person has a protected back and at the same time an overview of the actions and events in the space. It is about the possibility to see and be seen that is controlled while being culturally conditioned. It is essential for a person to alternate between staying in a safe controlled place (with socialization possibilities) and staying on "hunting grounds" overlooking other perspec-

tives - alternating between privacy, socialization and challenges. Our goal of behaving in space is therefore to take the position with the best possible outlook and to have control over our territory - that is, to control not only our personal space but also the space that we can control by sight. For example, if somebody is choosing a place to sit in a café or restaurant, he usually chooses a place where he can have a wall behind his back and can overlook the door and often view through the window. It's the setting that most evokes a sense of security (the back wall hides and the view from the door and the windows look out).

Important is:

- to have a possibility to combine prospect and refuge
- to support searching for situations in a space where human has protected back and view in the same time

- to have a possibility to control the situation when we see other but they do not see us and opposite, it depends on cultural background.

SUFFICIENT VISUAL AND SOMATIC CONTACT WITH OUTDOORS

For a person - a cultural creature suffering from civilization diseases – staying in nature is actually a recovery environment and the prosperity of contact with the outdoor is there evident. The everyday view from the window can determine / influence the whole life perspective. It is definitely better to have any kind of view in our day-to-day work and housing. It is significant not only because of the need for daylight from a light-technical point of view, but to keep our natural biorhythm tuned to the period of the day, the weather and the season.

Moreover the possibility of viewing out to the distance allows the eye to alternate between concentrated viewing on what is near (such as reading, writing and working on the computer) and looking out into the distance, thereby supporting - Saccadic eye movement. The eye will relax in this alternation, so we instinctively search for the setting in the environment. We do not have to talk about the extremes of uncivilized prisons, it is enough to go to production facilities or supermarkets with all-day artificial lighting. People exposed to long-term deprivation of contact with the exterior suffer more from civilization diseases, not to mention the frequent loss of life perspectives leading to passivity and life resignation.

As for the administrative workplace, contact with the exterior is equally important. Having a desk / desk at a window and not in the middle of a room is often a matter of organizational status and a certain privilege - the privilege of quality visual contact with the outdoors. Another very natural thing that is very often reduced in modern intelligent buildings is the possibility of opening a window and having a direct physical contact with the exterior. Providing employees with greenery and the possibility of resting on benches or platforms could be motivation to spend one's break even without smoking somewhere at a service entrance. Daylight exposure is the minimum that can be done for employees, but the standard should also be direct visual contact with the exterior. Sufficient daylight from a light-technical point of view is not enough; direct visual contact with the exterior is also important for a healthy biorhythm. If at least basic contact with the exterior is secured through windows, we can talk about quality contact.

What is a nice view to the outdoors? Views on mountain peaks in national parks are incomparable with the outlook on most of a city's real estate. However, placing the building and disposition of apartments in a way to achieve the most pleasant view in the given situation while preserving all the other requirements for the construction, is a great challenge. It speaks about the architect's quality and the character and precision of the developer. A nice view (garden, park, forest, mountain range, water area, harmonized or at least sensitively placed in the surroundings, etc.) is undeniably an added value of each property. External visual contact, important for all life functions, biorhythm, well-being, and eye relax, especially when one is working with a computer. But sufficient physical stay right in nature is obviously irreplaceable. Therefore, it is also natural to create green zones in the immediate vicinity of dwellings and workplaces, particularly industrial ones. Even the smallest area of greenery makes it possible to interact with nature, which brings a general slowdown, recognition and digestion, fascination, respect for nature, regeneration and overall relaxation. Going out to nature should be an accepted a part of the life style and culture of an organization.

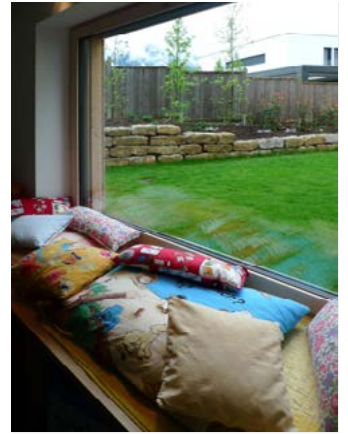

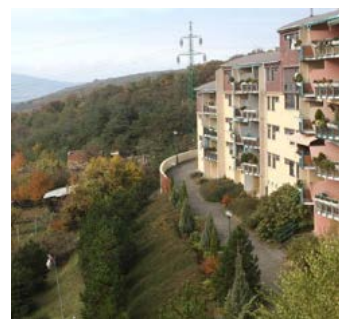

*Figure 8.2: Visual contact with outdoor within normal housing unit*

By creating housing concepts and public places it is very helpful to offer potential of nature watching. Why nature watching? It is great occasion not only for regeneration, relax and slow down, but also to be fascinated, to experience, to enjoy respect to nature which offer a deep contemplation and feeling of belonging.

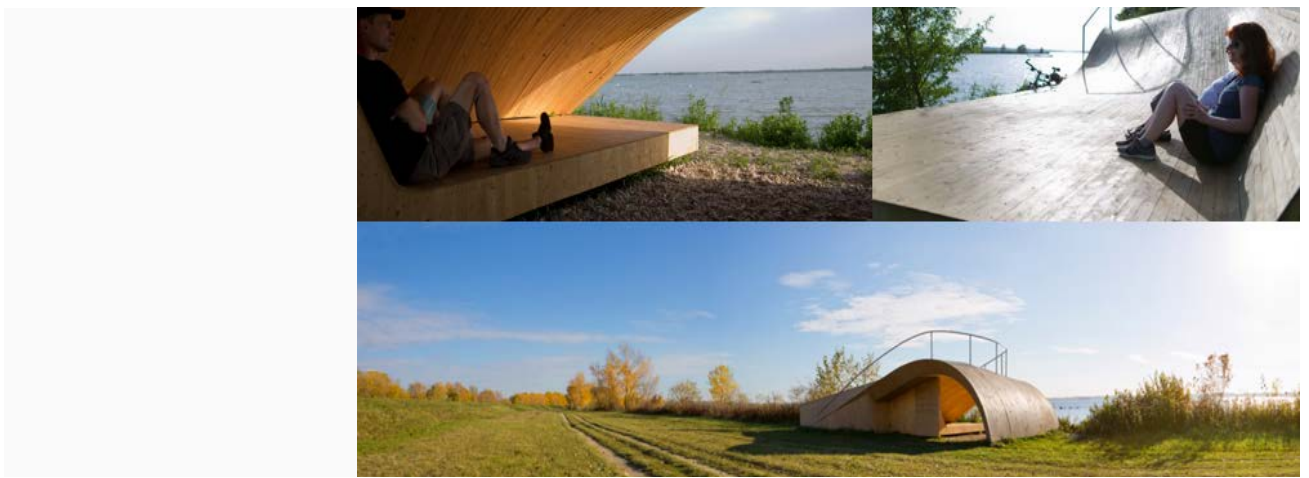

*Figure 8.3 DUNA bird watching near Bratislava is a great occasion for observing burds and other nature, project of Faculty of architecture and Bergen school of Architecture, www.watchamber.com*

## ACCURATE PERSONAL SPACE

The sense of security is closely related to the feeling of preserving the personal space. On this occasion, it is also necessary to mention the science dealing with the distance between persons and furnishings in space, i.e. proxemics and sociometry. The concept of proxemics was introduced in 1959 by modern American anthropologist Edward T. Hall in modern anthropology. Essential for a sense of security is personal distance (personal space, zone of intimacy). It serves as an invisible personal bubble that surrounds the organism of any non-contact animal. Hall, in his book from 1969 - Hidden Dimension, also gave precise numerical expressions of acceptable distances for certain activities and cultures. In negotiations, for example, it is 120 to 360 cm. The opposite is, according to anthropologists, an intimate distance, which is 0 to 40 cm – where the human will not let anyone else go. In western culture for example if someone gets too close to somebody else sitting on a bench in the park – it is felt as a disturbance of privacy, an infringement of the most private zone around the body. Even sitting very close to some strange person in transport or a waiting area is kind of disturbance in cases where there is a choice and a free capacity of places.

Edward T. Hall even worked out a detailed zoning chart, dividing the zones according to the reach of senses where the visual zone is the largest, followed by the acoustic, the olfactory and the smallest and the most intuitive, the tactile. We suffer more when sitting close to someone else, for example in transport means or in the waiting room at the doctor's. There is a greater likelihood of talking to a stranger at a safe social distance of about 120-240 cm than someone who interferes with us in the intimate zone, e. g in the waiting areas or transportation[4]. According to the influential sociologist John Fruin, the shape of the touch zone is variable according to human activities[5]. The perception of the disturbance of the territory is cultural and the great difference is in the personal distance between Western and Eastern cultures.

Distances between people defined by the interior of the premises also significantly affect the non-verbal communication.

That's why, for a stay at home and at the workplace, the interior and exterior furnishings allow users to protect, occupy, adapt and control the size and shape of their personal space, their own territory, through its size and arrangement.

Another important topic is the density of space. We can talk about density at all levels of the environment, from the microenvironment to the scale of human settlements. Density interferes with the private zone, which causes discomfort in the long-term interaction, but is very strongly culturally justified. In western cultures it increases the risk of increased stress and aggression in both private and public areas.

The problem of territoriality occurs, in particular, in large-scale offices with little personal space and, on the contrary, too much space in large-scale dwellings, where it is then difficult to experience the feeling of intimacy, cosines and a sense of home.

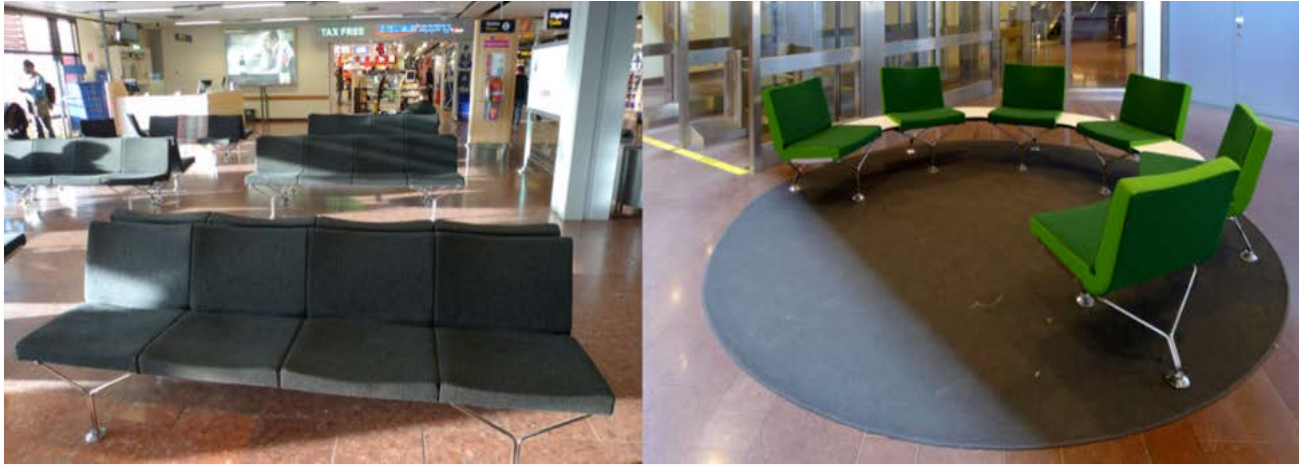

*Figure 8.4 Sociofugal that reduce or even discourage communication and the right picture is an example of sociopetal that is encouraging communication by placing elements "face to face", Stockholm Arlanda airport in Sweden, foto: Veronika Kotradyova*

INTIMACY VERSUS SOCIALIZATION – SOCIAL INCLUSION

We can talk about intimacy and occasions for communication- socialization at several levels of environment, but the three of them (own body, family or other group and society- community) are the most significant whereas at all levels the main issue), is to have the possibility to switch between privacy and socialization, to switch between encouraging and preventing communication. Essential here is the choice and possibility to switch between privacy and socialization as needed, which is helpful in arranging space to encourage or discourage communication and simultaneously to have control over privacy -intimacy and own territory.

Body

First one is perceived on the intimate body level in interaction with direct immediate microenvironment, Lower level – the personal intimacy with occasion for rich socialization means the maintaining of the aforementioned personal distance, which is necessary not only for the feeling of security, but also for a feeling of intimacy.

The need for an intimate sphere is defined from childhood along with the formation of personality and its extroversion / introversion, which are individual to each child. The older a child is, the more he needs intimacy. And with increasing age, the need to have a territory that is representative of friends also grows. We can hypothesize that a living space opened to children or the absence of children's room on the one hand promotes extroversion, socialization and interaction in children; on the other hand, the child does not have the ability to tighten and process stimuli from the environment and from his personal everyday life, which may negatively affect his personal development.

Family or other small group

At another level, intimacy is perceived on a broader scale in a "populated" space through the possibility of retiring or tightening into privacy, into a separate, clear or at least symbolic space, to reach a state where a person is not seen by the others, or to see the others but not to be seen by them).

The need for intimacy while doing private activities or resting within own territory is an important part of the feeling of complex comfort by living and working in a group. However, this is also true for public spaces where work or other activity is done for longer times in one place (receptionists in a foyer, etc.).The feeling of intimate privacy cannot be achieved with the physical presence of another strange person or the fact that we are in their view because of the inappropriate layout of the space in terms of its building elements or furniture. Intimacy is the key issue, especially when performing hygiene, sleeping and other intimate activities, and it is directly related to the feeling of security. Here the physical limits and barriers are indispensable, whether there is a complete barrier such as a wall or a wall with doors that can be closed / locked or a wall with a non-closing opening or incomplete partition wall or just a hint - such as glass partitions.

For large open-space offices, the territory defined by partitions is an indicator of status and hierarchy in an organization. It can work in business, family, or any community sharing common spaces. In the case of glass partitions, however, privacy is provided only in a very suggestive way or it does not provide it at all. But it is an attribute of status and the hierarchy within the organization. The location of doors, their presence or ability to be closed, is also strategic when planning a layout, with its associated operations. For example the best place in a business meeting is that which faces the door and the back wall. Opened doors, stair entrances and any other communications encourage movement or departure, so it is not appropriate to have them in the visual field when remaining, concentration on work or complete retiring are intended. When talking about intimacy at the workplace, the issue of the size of a direct work team arises, directly related to the need for privacy and overall comfort when carrying out a particular job.

Many studies show that the size of a working group has a direct impact on labour productivity, more than lighting and noise. Harold Wilensky's early study on interpersonal relationships in industry, where he observed that a bigger size of an immediate working group was negatively related to labour productivity or job satisfaction, regular attendance, or an industrial peace agreement – assuming the other factors were the same. According to him, the partial cause of people feeling better in small workgroups is that, in general, the majority of workers are satisfied with more primary relationships (relations that are more intimate, personal, inclusive, and spontaneous), and these will more easily develop in small ones than in large work collectives. That is why we have to deal with the size of the workgroup, which is often determined by the space. The workgroup is mostly space-bound, whether it is by a lightweight partition, such as partitions, cabinets or just the tables themselves, or in a better case by walls or walls that form a complete room that is partially or completely separated from other working groups.

Society - community

At this level, social interaction, inclusiveness into mixed age multi-generation society - being valid part of mixed age society and appropriate occasions for socializing within community is crucial especially for older adults, in 3rd and 4th age.

It is very helpful to support heterogeneous community and multi-generation housing and to prevent homogeneity and concentration of elderly at one place.

Huge potential by creating inclusive housing and public spaces has an interface between indoor and outdoor. This topic is also connected with the issue of contact with outdoor. To create settings where users are feeling protected by own home but in the same time having benefits from being outside and having social and health benefits.

In many cultures we can observe very interesting phenomena by older adults – preferring staying in area of interface like e. g. phenomena of bench by front door by traditional rural housing concepts.

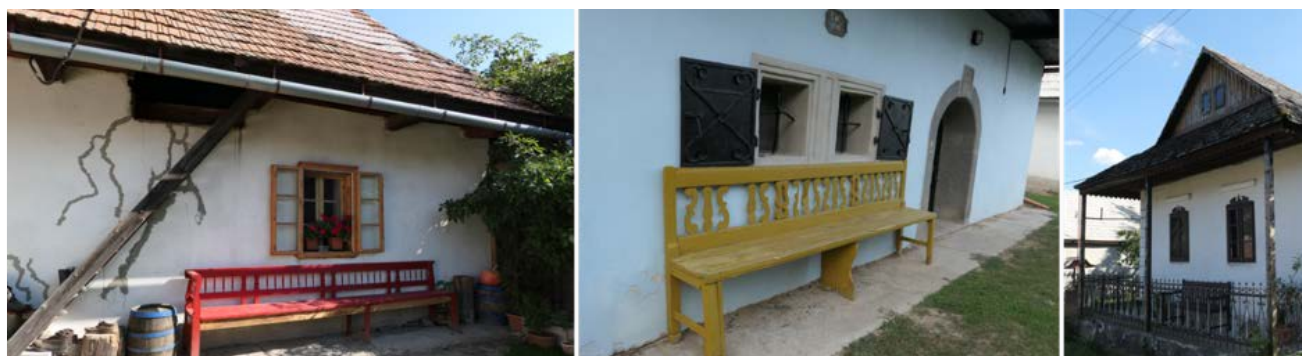

*Figure 8.5 Phenomena of bench in front of entrance to traditional houses in south Slovakia.*

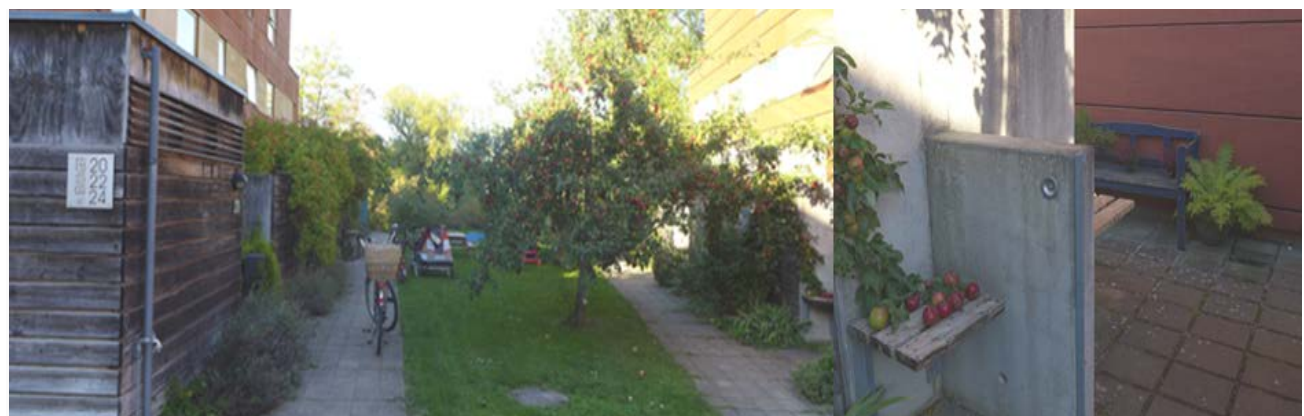

*Figure 8.6 Housing settlement in Bamberg, Germany that was built on previous brown filed, consist of more different sorts of wood structure houses with different housing programs and typology that creates setting with common green yards, by entrance into the housing units there is high quality interface between outdoor and indoor supported by possibility to sit on the wooden bench and support socialization with neighbors. With its concept it remains lining in village and provides more occasions to spend free time together*

Supporting models of more-generation housing without physical barriers housing units with mixed facilities program, occupying the social housing units with mixed age population – like a rough orientation e.g. 2/3 of occupants up to age 60 years and 1/3 over age 60 years possibility to use and adapt common interior and exterior spaces.

Intimacy is a basic need of human beings and it is basic condition for comfort and personal development. It is also about tolerance and empathy to being different, respect to personal limits and needs. And to provide appropriate environment that is flexible and adaptable and enables to occupants freedom of choice.

HUMAN SCALE AND MEASURE

The need for an adequate human scale and balanced proportions applies to all levels of the environment, from the microenvironment to the macro- and semi-environment. When designing products and built environments, it should be taken into account that each space and material has its own radiating expression that needs to be recognized and respected in forming. Together with colour, shape is the most important artistic category. In order for any object to be mastered on the aesthetic side, and thus to extend its moral life, all the simpering means must be mastered in it: rhythm, proportion, scale and dimension, equilibrium, unity, style and character. Most of them are derived from nature and the human body.

In macro and semi-environment, cities are now overwhelmed by "oversize" administrative and commercial buildings, more users prefer smaller, more adequate housing such as family houses or lower-storey dwelling units. That is why residential units similar to arrangement of the village or the medieval town with squares are still very welcomed[6].Overly large and vast spaces that the human nervous system cannot handle and process in order to be able to recover is one extreme. The opposite extreme is too small and densified spaces that stress users especially during long-term stays. But here it has to be said that the scale of the space and well-being are influenced by the cultural background.

Especially with long-term residence in the premises, each extreme creates immediate stress for our nervous system, but also deforms perceptions of the self and of life perspectives. The scale of every investment, design, and architectural achievement should be the human being - both body and mind[7][8].

TO ALLOW ATTACHMENT

Human being built the "Self" through the extending, mirroring, processing and getting feedback and self-identification through the extensions[4] that is why we need the possibility to be attached to places, environmental settings and things.

Important is to have the possibility (or competence) to be attached to a place or products, to have the competence to adapt them, personalize, to mirror and extend the Self /ego into the occupied space where a human lives, and thus gain a state of self-identification. This feature is important for the human as a cultural creature, but also as an animal that needs its own marked habitat. This need is extremely intensified with aging.

According to Robert Gifford, one of the main experts from the field of environmental psychology, place attachment represents a deep experience of feeling part of a place. It is related to the richness of meaning and sense that is developed out of acquaintance with a place and, subsequently, when the place gets to be familiar. This attachment can be to our homes, properties, communities or local nature sceneries and settings. Where the attachment rises, the intensity and meaning of the place and the meaning of Self become affiliated[9]. Then the meaning of the place can become so strong that ´s self-identity starts to be restricted by the place. On the smaller scale, many people are identified with their neighborhood, quarter, village, farms, houses and rooms.

Attachment to a place has some serious implications. Its close relative - the identity of the place / place identity, is an important dimension of the personality of the individual. The ability to adhere to spaces and things, to mirror them, to build the Self upon them, to identify oneself with them, are very strong human needs. Each individual has a different intensity of projection of unprocessed unconscious issues into their immediate living space, into the things they own, depending on their life strategy and the evolutionary stage of personality development. With elderly people the process can be very intensive. Meaning of place can be so intensive that somebody´s Self-identity is bordered /limited by the place.

From research into this issue in different social sciences we can summarize these points:

- our environment and its compounds are extensions of the Self and a screen for projecting unconscious unprocessed issues, whereas nesting and homemaking are major means towards self-expression and development of the personality (anthropology and psychoanalytic psychology),
- "the Joneses" - others and their opinions are very important for us according to studies of cultural anthropologist Daniel Miller[10],
- with identify with a place and build our own personal identity (environmental psychology),
- changes inside of a person are reflected on the desire and action to change its surrounding environment, and vice versa. To have the choice and ability to change anything (on one's own body or in the he immediate space around the person) means creating an opportunity to continuously grow (psychosomatic medicine, analytical psychology, Jungian psychoanalysis).

That´s why it is important to leave users the power and to give them the choice to personalize their housing and working space, to adapt it according to actual personal needs and preferences, to extend and to represent a Self that can move towards self-reflection, self-identification, acceptance and communication. Attachment to objects and spaces is painfully affected by the divorce of any kind of partnership; property that was common before now has to be separated too. Innovative concepts of administration buildings office spaces, such as desk-sharing systems etc. promote full or partly loosening of the fixed workplace, transforming into semi-fixed or completely loose, often with compulsory booking of a working space for the next day. This concept works against the attachment and the only place to leave own personal belongings and potential objects for personalisation is a box with a lock. This system can work very well for workers spending most of their time outside the office - like auditors, or and for very modern jobs without compulsory physical work places, working from home. Will this concept be suiting to the workers of 3rd age?

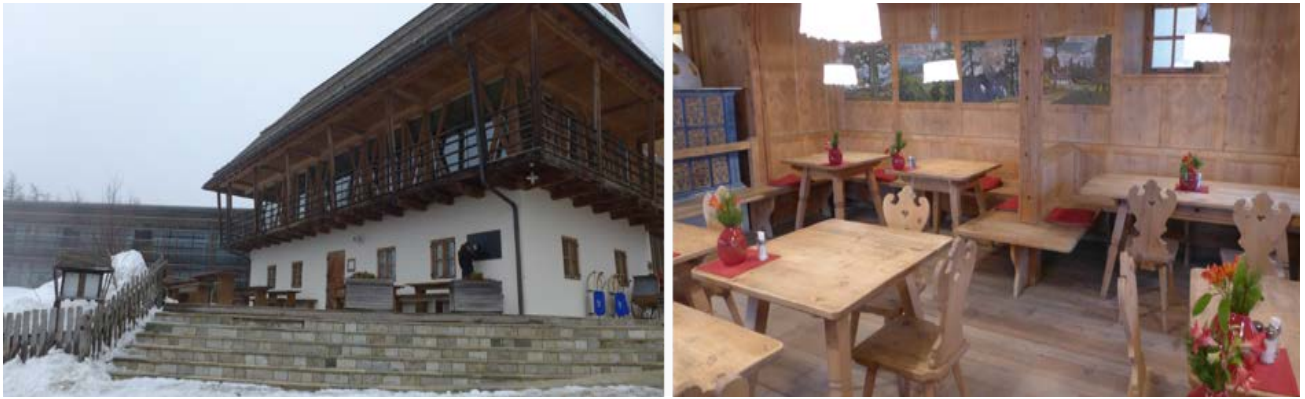

*Figure 8.7: Example of bringing local identity into the built –environment, Vigilius resort, architect Mattheo Thun, Lana, South Tirol, photo: Veronika Kotradyová*

Maintaining the cultural sustainability through giving local identity to built environment and to life style - this means to bring back local materials, principles, concepts, stories into material culture – architecture, housing, life style, product design.

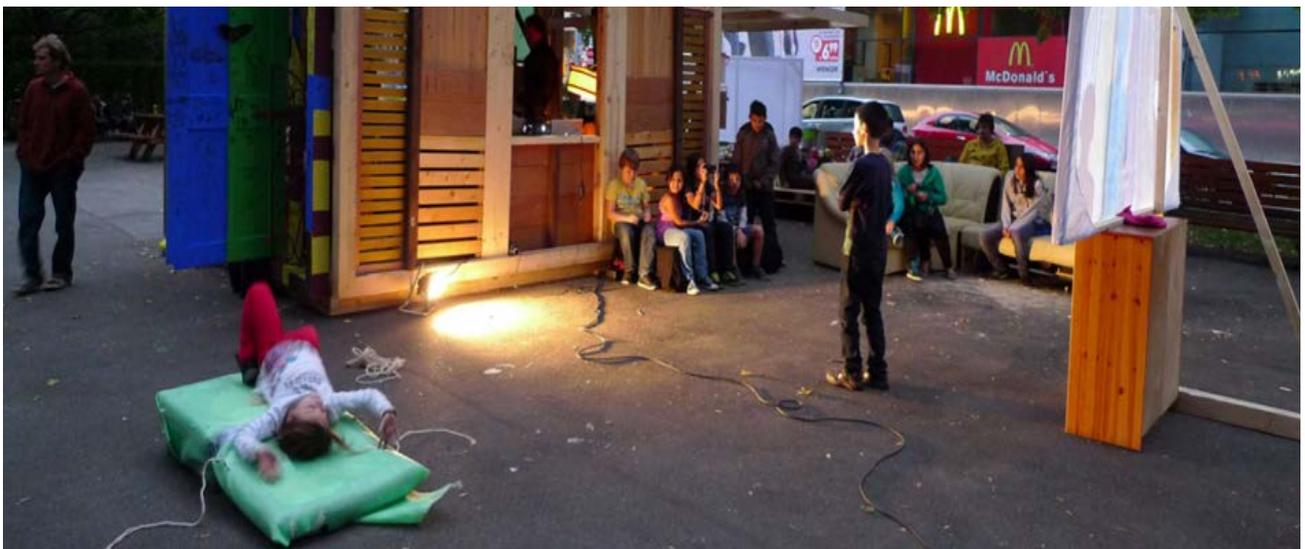

*Figure 8: BAECH in the city, voluntary built temporary community center in Sandleiten, Vienna, workshop with participants from Austria, New Zeeland and Slovakia, 2012, photo: Veronika Kotradyová*

For older adults it is highly appreciated when being in environment that reminds and supports local culture and place of origin. If it is not possible to do on general level, it is at least possible within own microenvironment as it was mentioned in the previous chapter.

Another need and necessity is being valid part of participatory and activities and projects in the local community. Older adults can share their wisdom, experiences and deeper cultural values especially with younger generations and thus to support socio-cultural sustainability. This can secure identity in private and public spaces and preventing loss of local identity thanks to globalization.

## ACCORDING TO HUMAN BODY

In industry, human factors are known mainly as one-sided stress and physical loads, but in working, school and health care facilities and households especially the problem is long-term sitting; physical passivity that can turn to dependence. The lack of physical activity in ordinary activities is a serious and complex problem of contemporary western civilization. Technological achievements physically facilitate survival, resulting in the "laziness" of civilized people. The result is a sedentary culture that has many health consequences.

A serious fact here is that a long-term static convoluted sitting with a C-shaped spin causes not only problems with the spine and the motion apparatus, but also puts pressure on the internal organs, which can result in problems with digestion, fertility and mental health.

Many scientific and commercial studies have already been written lobbying for dynamic / active seating. For dynamic sitting an unstable, springy or tilted seat is common. The constant movement of the human back and legs relaxes the whole apparatus and in addition prevents insufficient nutrition of the intervertebral disc. This concept also includes sitting on a fit ball, which should be optimally sized for our body in order to function properly. Just simply sloping the whole body forward and back as needed not only helps prevent spinal deformation, but also improves blood circulation in the brain and the whole body. It is interesting how many of us still in adulthood are springing or sloping on our stool, thus quietening our need to change our body position and calming down at the same time.

Getting dynamic sitting is easy nowadays, but it's more appropriate to alternate it with regular static sitting and objects for other body postures. Today, active chairs are available on the market, which can be hampered by springing. It thus has the advantage of long-lasting micro-movements which, however, can tire the entire nervous system. One of the solutions is to create multi-positional elements that allow for a whole range of body positions, such as the iconic Capisco chair by Peter Opsvik. We have also embraced this multi-position concept in the Horse Office concept, where it is possible to sit at work, reading or drinking, but also taking up a few resting positions or even exercise. Whether universal and relaxing passive or dynamic sitting or a perching, it is advisable to have available support for the feet to unhook the legs. Having the ability to "move" our legs even during sitting has a lot of psychic and physical benefits. When coming to school, children have the problem of sitting serenely and fixing their attention for long periods. Allowing children to move and maintain discipline may seem impossible, but functional demonstrations from alternative education systems point to better pupil's concentration if they have some freedom of movement and body choice during a lesson.

Work outside standard sitting and the ability to change body positions can have the following effects:

- reducing the pressure on the intervertebral discs common from regular sitting prevents many functional diseases of the spine and reduces the degree of monotonic / unilateral body loading, improving blood circulation, reducing pressure on the internal organs (pressure on the internal organs results in impaired digestion and also a negative influence on the mental state)
- enhancing work style, promoting creativity, interactivity, and thus interpersonal workplace relationships.

From this working perspective, we can divide body positions into formal, semi-formal and informal. Many progressive organizations have already implemented informal spaces and space-freer working styles into their administrative and training facilities to allow people to stay in semiformal and informal working constellations and to alternate with formal ones. Freedom, relaxation and informality in physical positions can bring non-standard and progressive ideas and problem solving. This freedom must be embedded not only in the products used in the working environment themselves, but also into the whole working environment and organization culture.

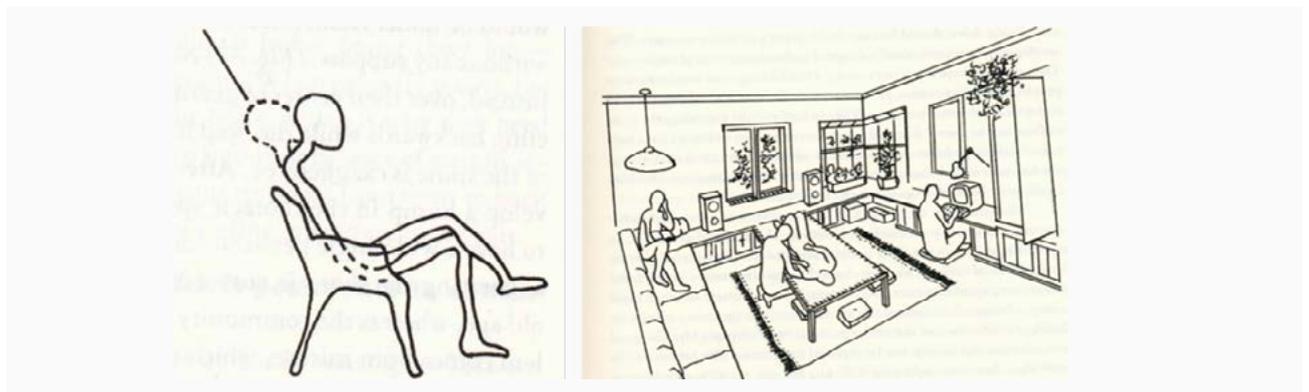

*Figure 8.9a Chair sitting actively creates a hump in the upper spine (left), concept of living room according to GalnCranz where all possible body postures are possible*

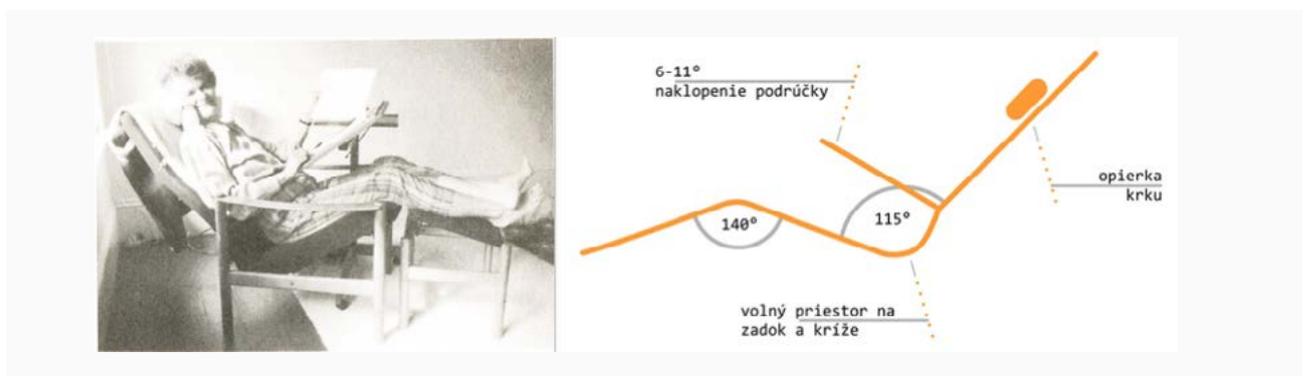

*Figure 8.9b: Relaxing recling positions according Galen Cranz (left) and optimized standard for recling according to BCDlab*

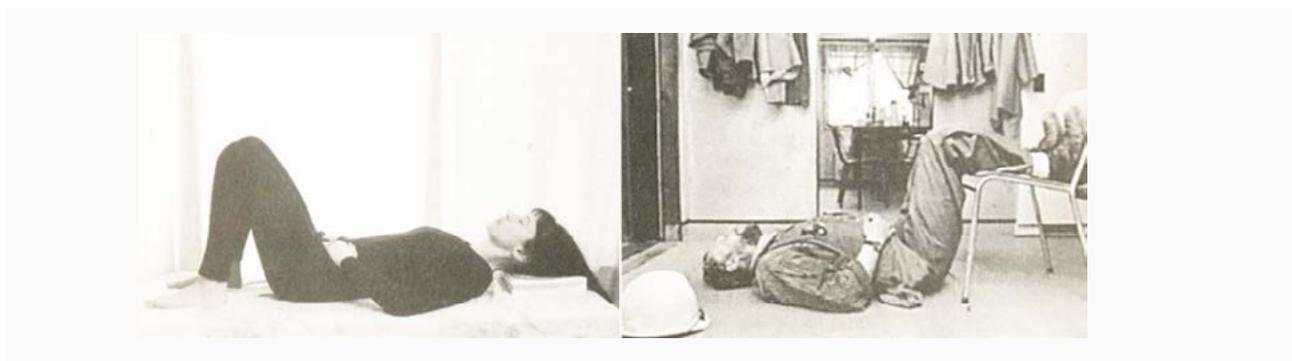

*Figure 8.9c :Construcitve relaxing postion according to Galen Cranz[11]*

It can be very difficult to incorporate these concepts without initial discomfort or reactions from conservative milieu. In the past, in our research we have also been intensively engaged in an acceptable degree of added movement within the residential interior, but we have realized that increased movement or deliberate discomfort in the home environment, in order to get more movement, can be rather counterproductive. Nonetheless the intimate environment stimulates laziness, which is here the right measure in the right place. Increased movement is likely to get more into school facilities, workplaces and public spaces. Even physical laziness can be addictive. A simple solution for stimulating users to physical activity in administrative workplaces is to place stairs and ramps in more accessible and visible places in public buildings, and to motivate people to use them even more because of their visuals and performative experience.

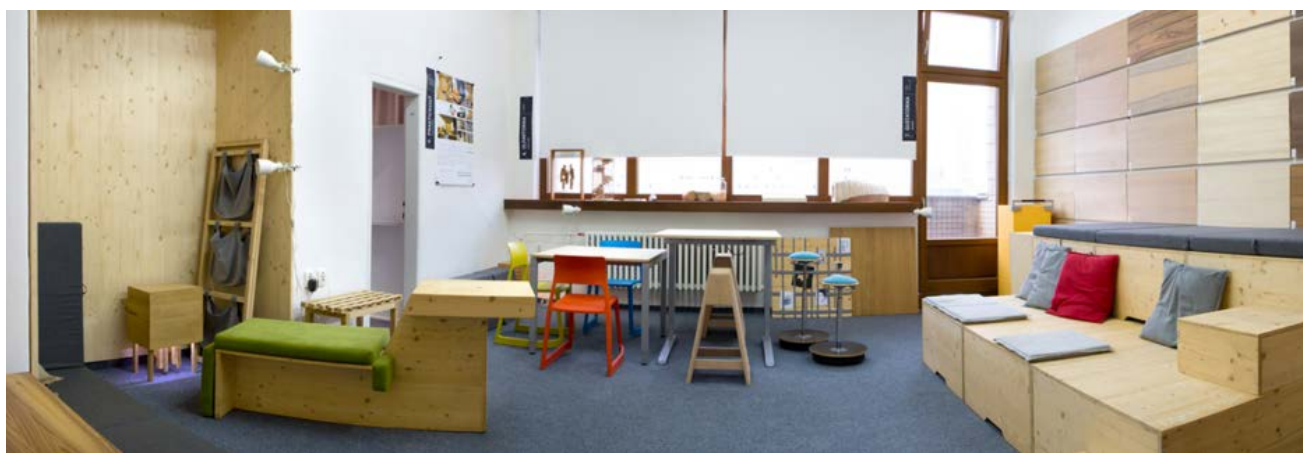

*Figure 8.10: premises of BCDlab at Faculty of Architecture, STU in Bratislava, where all options of body position, including active sitting, saddle sitting and perching is possible during working meeting and teaching, showing universal podiums, saddle chair Soma and multipositional workplace Horse office oto: Noro Knap*

After several decades of development of various alternative ergonomic concepts for working and housing spaces, it is obvious that no other concept can easily fully replace standard sitting, where we are able to withstand the most prolonged time in the waking state. Sitting on standard chairs is a strong part of our western culture and getting it out of this pedestal would need several generations and the elimination of sitting as the only possibility for schools and workplaces. But the point is to equalize it with other body positions suitable for everyday activities. To achieve better physiological responses and the most natural body posture, as the most suitable is considered to be saddle sitting[11]. Developing new products with this philosophy therefore makes sense, and on Figure 8.10 are couple of BCDlab products – saddle chair SOMA. However, if such a product is not fitted with upholstery, there is the risk that contact discomfort will soon arise from prolonged leaning, although the chair is made of softwood with good contact comfort. It is equally beneficial to equip an interior with floors and elevated surfaces such as podiums, benches or deep window parapets that are comfortable to the touch, along with adding a lot of scattered upholstery (pillows, matts, loose light mattresses, etc.). This concept has the greatest potential to create a versatile and flexible environment for taking up any physical position in the interior, as evidenced by the interior of one of the designs by BCDlab (Figure 8.10).

How is it with active sitting or saddle sitting by older adults? As the physical activity including anaerobic training, the active sitting and saddle sitting and perching are more suitable in the phase of prevention, in 3rd age, in the 4th age it can be already dangerous.

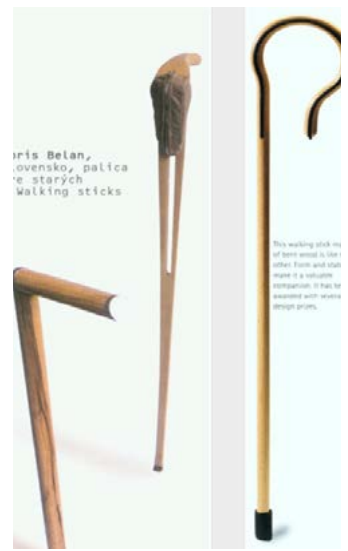

Special issue is a support of positive tactile stimulation considering interrelation of object volume and surface and human hand. It is caused by the fact that evolution of human skills is connected with improvements of the other senses, especially vision, hearing and development of their relation. This sensuality of architecture is deeply explored in studies of finish architect Juhani Pallasmaa[13].

The impact was that the measure of a human hand is the measure of beauty. Most valuable for humans are things that match perfectly into hand and can be managed easily. By older adults the pleasant feeling of safety and manageability regarding satisfying tactile experience and form matching to human fingers is especially important for wellbeing.

*Figure 8.11:Examples of well designed supporting sticks with added value, giving supportive tactile stimuli*

FEWER ENVIRONMENTAL STIMULI

The nervous system of the contemporary modern human being is overwhelmed by stimuli from the environment, especially from public spaces, workplaces, digital and virtual environments. It is therefore better to "save" and reduce them and to prefer pleasant and moderate sensory stimuli, especially in long-term stay environments. It is a way to prevent further environmental stress and to reduce that which is already part of the working process, housing or some other component of the way of life.

Moderate environmental stimuli relate to proportions, scales, outlines, shapes, materials - their surfaces and colours. Simplicity and clarity of the solution, easy orientation, all make it easier for the nervous system to function. Architect Mies van der Rohe's "less is more" as well as a strong Zen approach to design have their merits here, and not only due to the minimalism trend.

We therefore recommend natural or neutral surfaces on large areas in long-term stay interiors, distinctive colours to a lesser degree than accents - supplements that can be easily changed and personalized. Strong stimuli are good in public spaces, with the intention to challenge, to experience, to feel emotions, understand the context, etc., such as museums, galleries, bars, showrooms, playgrounds, etc...

USING OF MORE AUTHENTIC NATURAL MATERIALS

In general, it is possible to declare that the more natural materials in their authentic form and more nature-evoking solutions present in the immediate surrounding environment of a human being during his long-term stay, the better. The explanation comes from evolutionary biology and psychology. Natural materials are beneficial to the nervous system, which responds to something well known to it, which does not need to be constantly "scanned". A nervous system that is directly linked to our unconscious and genetic memory responds to natural materials, is gently stimulated and can easily be recovered. Natural materials are part of our archetypes in our traditional material culture. Their use significantly contributes to the creation of a supportive environment and an environment for well-being.

However, in order to have an ecological meaning, it is necessary to apply natural materials from renewable raw materials from sustainable forestry or agriculture.

First we examined wood as the most used natural, domestic and renewable material for interior design. In the project APVV 0594-12 Interaction of man and wood, we examined wood itself and its application as a phenomenon of supportive environment and a "healthy interior". Its positive properties for the comfort and health of the human, for the internal environment and its micro-climates can be formulated as follows:

- enhances and softens the overall atmosphere of the space, supports cosines,
- is aesthetically attractive - its shape and surface properties - the colour of the wood is warm to earthy, but not "hot" like red or orange; its colour, texture, structure, tone can also soften and "warm" otherwise cold lighting in space,
- excellent haptic properties - contact comfort, especially with softwoods,
- wood-based materials are well formable and adaptable to the shape of the human body, thanks to the comfort features, they allow good control over the body's positions and the possibilities for their variation,
- maintenance is possible through renewability - e.g. weaker wet-wise maintenance can be compensated, for example, by surface sanding,
- has antibacterial properties, especially pine, larch and oak,
- produces a pleasant sound in human interaction, and wood-based materials can enhance acoustic interaction when used on acoustic panels,
- regulates air humidity and emissions,
- smell (mainly coniferous species) and "taste", in direct and indirect interaction –e.g. when food is served on it,
- is part of our culture, collective unconscious, and is in genetic memory.

Most of these strengths of wood are effective when the wood is free from any surface treatment.

The weakest aspect of natural materials, just in connection with its surface treatment, is its maintenance by wet methods (water and strong detergents).

Application of wood significantly supports regenerative effect of built-environment for the nervous system of human, and can contribute to recovery in health care facilities.

Anyway, there are many stereotypes that are again using natural materials in their authentic form in the health care environment because of their maintenance.

But there are many examples e.g. in Austria or Swiss where natural authentic wood or clay are used at least in less exposed zones of health care facilities. (Figure 8.12 8.13)

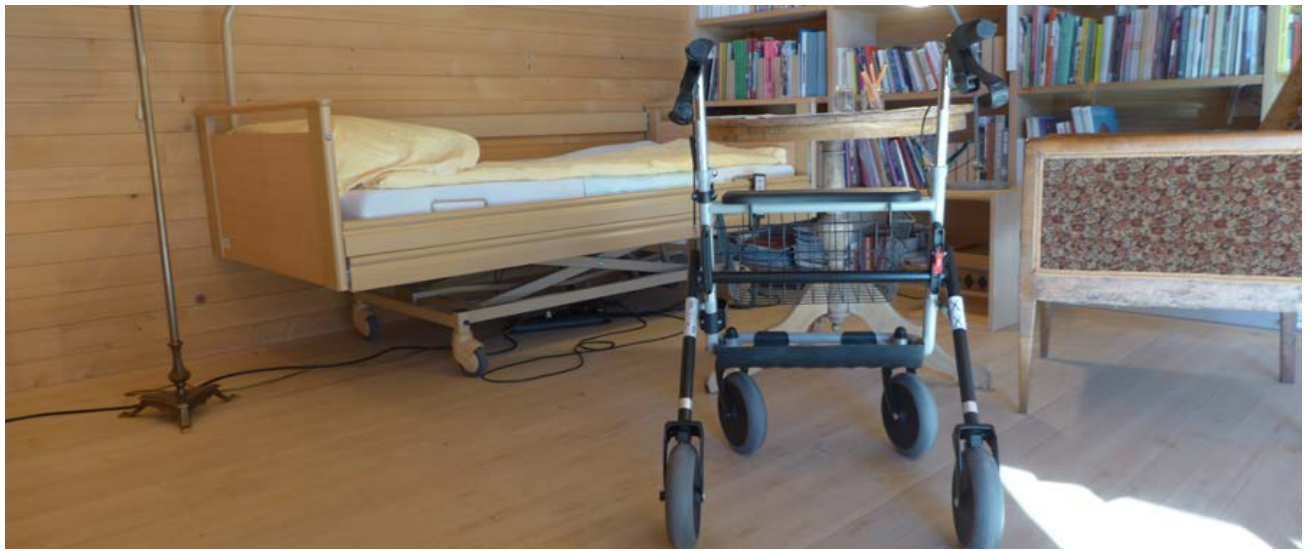

*Figure 8.12: Exposition about Home health care in Frauenmuseum in Hitisau, Vorarlberg, photo: Veronika Kotradyová*

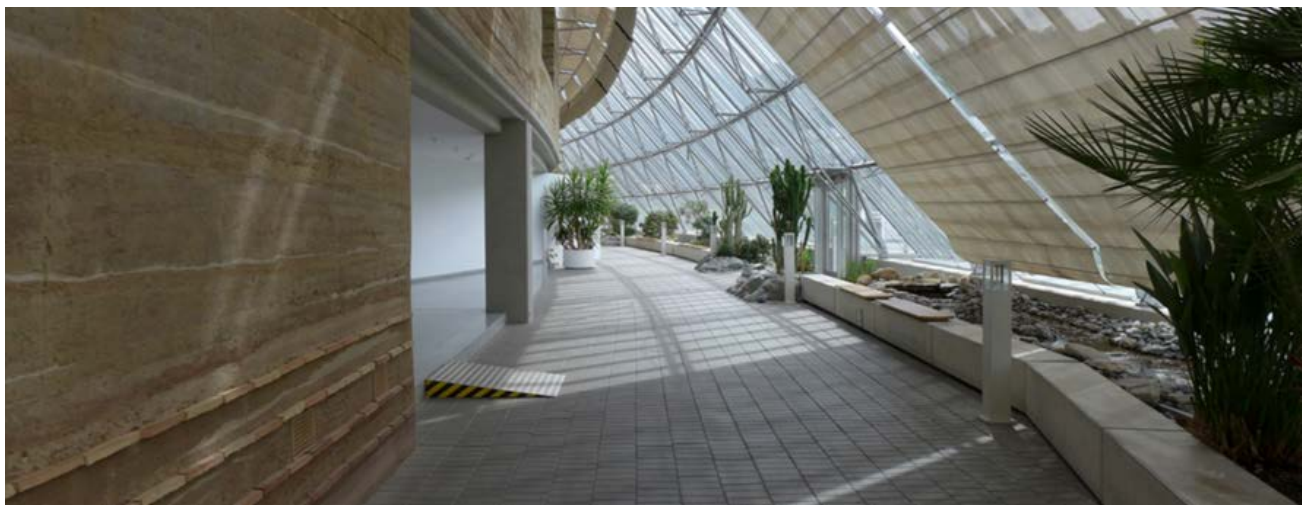

*Figure 8.13: Hospital in Feldkirch, Vorarlberg, whre clay is used for main bearing wall., photo: Veronika Kotradyová*

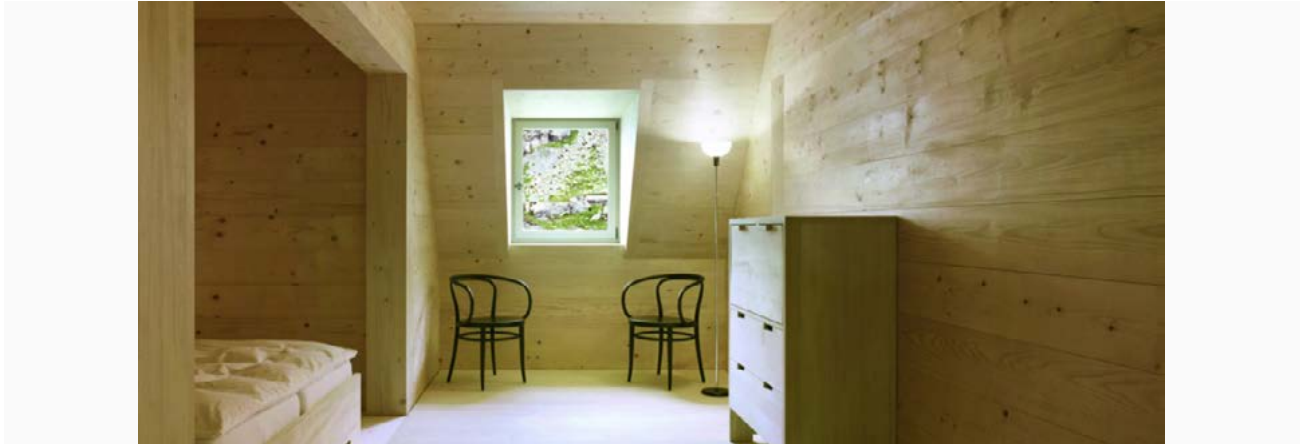

*Figure 8.14.: Addition of Hospic in St. Gotthard pass in Swiss, where the atuthenticswiss pine is used for interior furnishings*

These properties were tested in the case study in the National Oncological institute in Bratislava, by the emplacement of larch benches and pine wood wall and ceiling panelling with very simple archetypal design during reconstruction of part of the foyer (Figure 8.15). There were executed further microbial tests of wooden surfaces and air in the space prior to reconstruction and 3 weeks, 4 and 7 months after reconstruction. Only the heavily used parts of sitting benches lost most of their natural antimicrobial effects, which is why it was necessary to find a solution for intensive cleaning. We tested and found appropriate way of cleaning by maintaining all positive features of wood - cleaning with ethanol. As added value to the study there were executed also tests of VOC´s emission to support the sink-effect of wood hypothesis. Physiological reactions of respondents were further tested by entering this space to see the reactions of nervous system[16].

In 2017 another case study in a health care facility – revitalization of room for mothers for preparation of mother milk at the Neonatology Clinic in the Faculty hospital in Košice, was executed (Figure 8.16).

Similarly, wood works with other renewable raw materials (flax, hemp, straw, cork, clay, sheep wool, etc.) and natural materials made from them. There is a need in this field to educate both the professional and the general public.

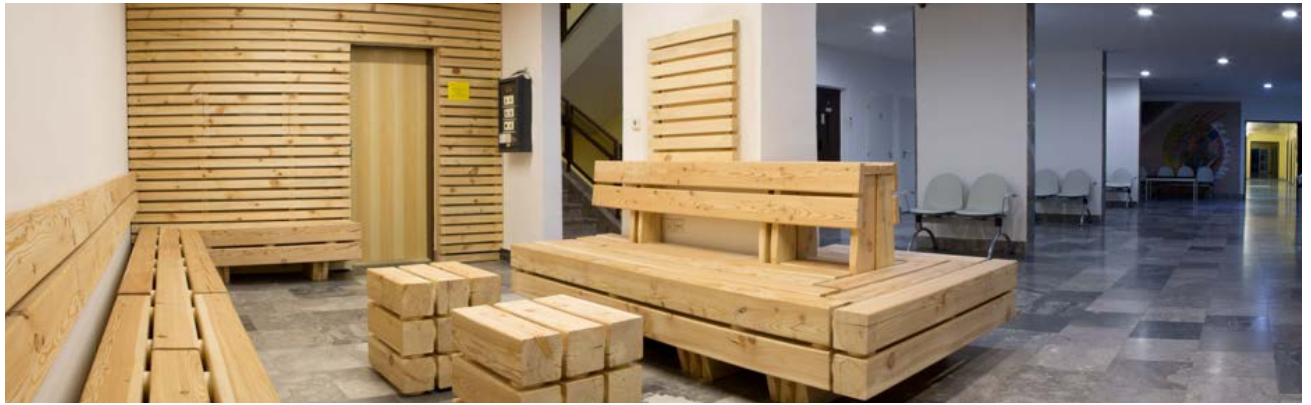

*Figure 8.15: Reconstructed waiting zone in foyer of pavilion M in National Onlogical Institute in Bratislava made for demonstration and testing the application of wood in health care facilities, there is used pine and larch without finishing, design: Veronika Kotradyová, Martin Boleš, photo: Noro Knap*

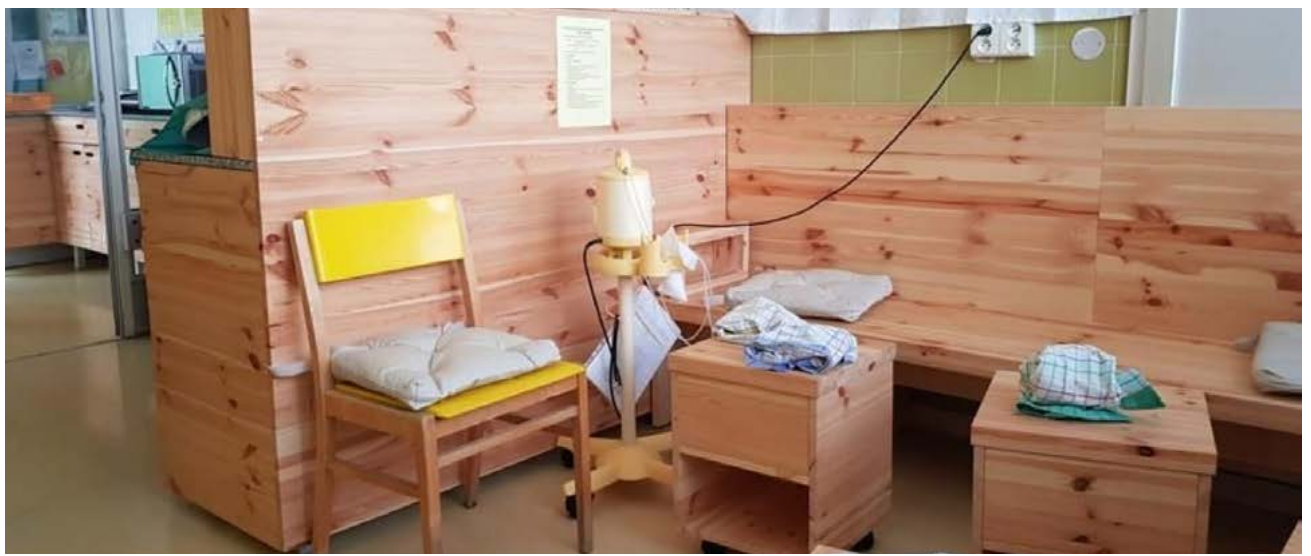

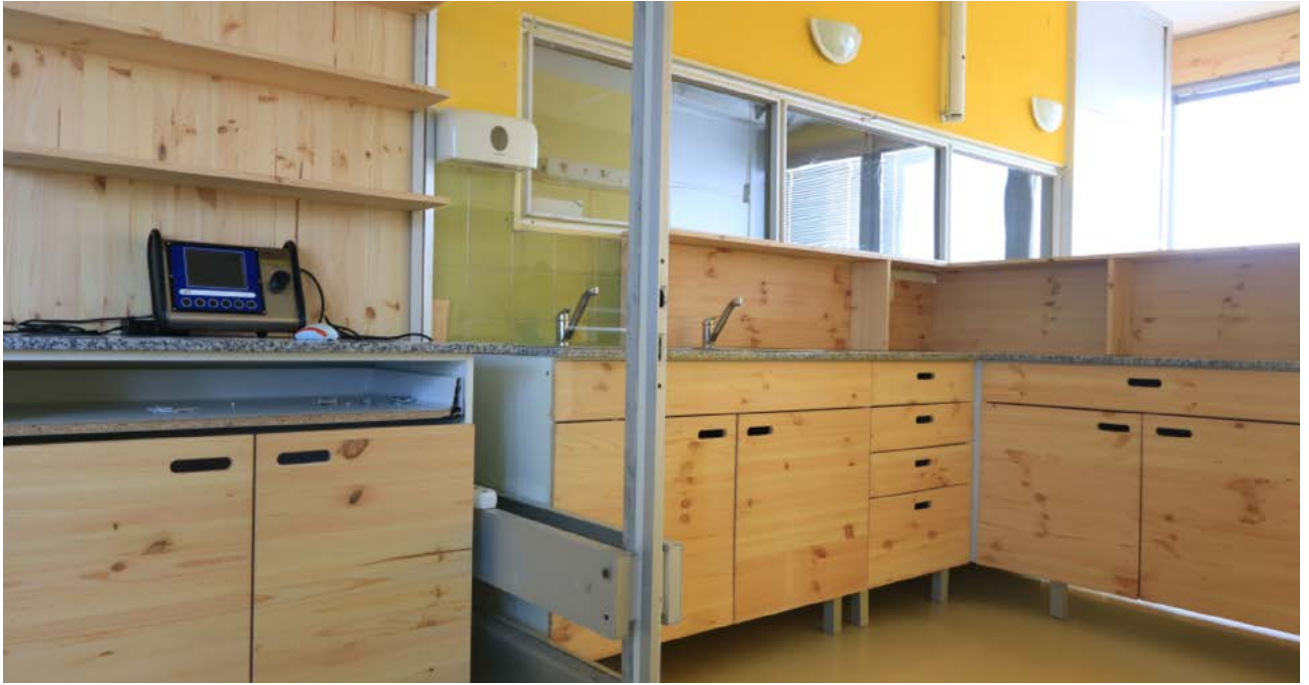

*Figure 8.16: Reconstructed room for soaking mother milks at the department of neonatologiy, University hospital in Košice, all furnishings are made of solid pine wood with flax oil finishing, Slovakia, Deisgn: Veronika Kotradyova, photo: Petr Krcho*

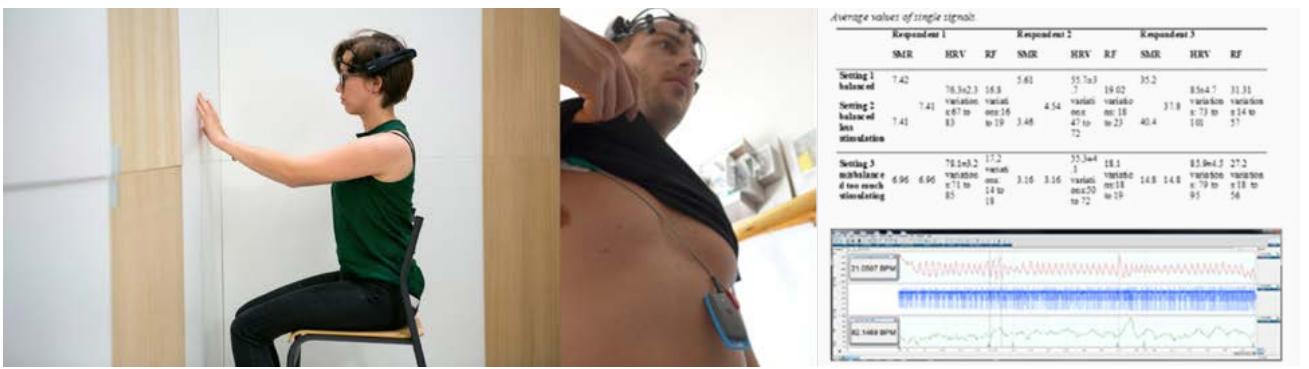

*Figure 8.17:.Measuring of EEG responses to various surfaces, using of EEG hotlers and EPOC plus and evaluation with lab chart and mat lab, by visual and tactile contact in BCDlab[15]*

## COMFORT OR WELL BEING BY EXPERIENCING BEING CONSCIOUS ABOUT NATURE PROTEC-TION AND CONTRIBUTION TO SUSTAINABILITY

Conscious users in this area can experience the advanced well-being by experiencing using of products, services, concepts, environmental settings that have some ecological/green con-cept and can contribute to save planet. The awareness in field of sustainability, protection of natural environment and social accesability are part of wellbeing in a broader context.

Conclusions

To explore the problematic of well-being or complex comfort in built environments in general and for older adults particularly requires a holistic approach. The 11 principles presented in this paper are a supportive skeleton for further research. Each principle is a topic to itself and setting some hierarchy among them is hard to do. One important principle has not yet been added in this paper- social affordability, because it transcends well-being and has s more broad meaning. One positive finding is that environmental settings have an effect on human behaviour that it is possible to document and evaluate, and a physiology which is measurable. From our pilot research studies we have found out that it is possible to set a multidisciplinary methodology and also to cooperate with scientists from different related branches of social sciences, medicine, physiology and engineering. It has to be further deepened and tested on a larger amount of respondents. In the premises of research design center BCDlab there is a testing space where the respondents are more isolated and the responses can be measured in a more effective way. This complexity should also be considered in the education of future architects and designers and informing the public on a broader level.

## Acknowledgement

The paper was supported by project APVV 016-0567 IDENTITY.SK – common platform of design, architecture and the social sciences.

# Example 4: Social inclusion from the perspective of older adults in Lithuania and Portugal

## Use of technologies in maintaining autonomy of frail older persons

Although Lithuanian population is ageing, it has one of the shortest expectancies of autonomous life in Europe. An increasing number of the elderly results not only in ageing society, but in increasing morbidity as well. Reality of life is encouraging healthcare systems to consider the needs of older patients and to assess them comprehensively, which is not limited to a physical assessment but includes an evaluation of functional state, cognition, socioeconomic status, and home environment, identification of geriatric syndromes and vulnerability factors, and frailty. The term "frailty" is used to describe the status of the elderly person who can usually perform basic daily tasks, but due to the decline in various functions, changes, or disorders cannot quickly restore the strength. Concomitant diseases, disorders, disability, and frailty were considered as synonyms, but the researchers have recently agreed that these terms do not mean the same. The social component of supportive environments is composed of people (family, friends, and professional caregivers) who provide help. The physical component includes technologies that make living easier and more autonomous. Gerontechnologies in vulnerable and frail elderly people can compensate for their impaired orientation and memory, widen their communication options, exchange information, and move. With the help of gerontechnologies, safe home environment and health monitoring can be ensured. In this article, algorithm for maintaining autonomy in older persons (created by the authors of this text) is presented[119].

## Needs of geriatric patients living at home and meeting them by technical means

Studies show that people with disabilities and older people are reluctant to use services of healthcare institutions. Similarly, families and other informal carers prefer home care. The purpose of this review is to examine the needs of geriatric patients living at home that can be fulfilled by technical means, and the options of remote care for the elderly. Needs of geriatric patients that can be met by technical means, are mobility, security, communication, recreation, vital functions monitoring, chronic disease control assurance, and assistance in memory. There is a large number of tele-homecare technologies that can monitor chronic illnesses and vital functions. The main factors that determine the development of these technologies are an aging society and rational health care policies. The purpose of remote home care is to improve quality of life and promote independence by rationally using the resources and providing services at home. Telemedicine devices make it possible to monitor patients' body temperature, heart rate, respiratory rate, blood pressure, blood glucose, prothrombin time, and oxygen saturation[120].

**Levels and characteristics of the digital divide: a case study of Lithuania**

After the sudden growth of popularity of IT and the internet, scholars have noticed emerging differences between usage or non-usage of the internet among different individuals and groups. These differences were defined as the digital divide. At first, more attention was given to the differences of access to IT and the internet, but today, scholars are focusing on another level of the digital divide – one determined by skills, perceived value, and motivation. In this article, we examine the characteristics of the different levels of the digital divide by analyzing various statistical indicators. We seek to explain which groups of Lithuanian society feel the social and digital divide the most; in addition, we explain the features and changes of every level of the digital divide[121].

**Use of technologies in maintaining autonomy and QoLof older persons**[122]

Cáritas Coimbra has strategically been focusing on promoting a more active and healthy ageing. This focus has been translated into the adoption and availability of technologies to its users, since they have the potential to contribute to improving the older people's quality of life, facilitating their daily routines and improving their levels of social involvement. In this sense, a package of activities has been implemented using innovative technologies, through practical and informal ICT training, with the permanent provision of tablets and its use in workshops at Cáritas Coimbra centres; through the participation in the different GrowMeUp project initiatives (funded by the H2020 programme under the Grant Agreement no. 643647, led by the University of Coimbra); Usability tests by older people using the CaMeLi technology in Portugal; and the intervention of the pilot-project primer COG using the online platform for cognitive stimulation. In this cross-institutional initiative, the older people of the daycare centres, home support services and residential structures, have been involved, as well as multidisciplinary teams of professionals from these social responses, the innovation department and the partner institutions.

## Objectives

- Increasing the confidence of older people in their capabilities, reinforcing their value in society;
- Improving their cognitive conditions;
- Improving digital literacy;
- Preventing isolation and loneliness;

- Increasing self-esteem and well-being;
- Promoting alternatives that enable older people to stay longer in their homes;
- Introducing the voice of the users in the technology development process.

## Outcomes

This package of activities is in phase of implementation in several social and care services, being constantly updated due to the opportunities that arise. It is not yet a closed process, but the routine usage of new tools and the existence of a permanent space in the programs of activities specifically aimed at ICT and innovation. Although it is being continually matured and worked according to the characteristics, needs and requirements of the local population, this package has already three years of experimentation and phased implementation, which allows some maturation and stability. A first global evaluation is being performed in 2019 with the MAFEIP[123] tool.

# **Example 5**: Existing policies on accessibility in Spain

### Criterios Dalco 

Accessibility refers to the different dimensions of human activity: moving, communicating, reaching, understanding, using, and manipulating are some of the basic forms of human activity. Ensuring accessibility means ensuring that these activities can be accomplished by any user without encountering any type of barrier.

These activities are summarized in four major groups: Ambulation, Apprehension, Localization, and Communication (DALCO).

normas@aenor.com

### Estudio de la Fundación Vodafone 

The study of the Vodafone Foundation "The elderly before the Tic" explains why the reduction of the digital gap has to be a social priority, especially in the older adults where the digital gap creates greater impact, due to the problems of the elderly related to communication loneliness and care, among others.

## Sistema de la dependencia en España y Tecnologias Implementadas 

The system of dependency is the set of services and economic resources intended to promote personal autonomy as well as care and protection of the dependant people, through accredited public and private services, and contributes to the improvement of the living conditions of the citizens. System is supported by the most important Law 39/2006, of December 14, on the Promotion of Personal Autonomy and Care for people in situations of dependency.

## The only successful technological service implemented in Spain is the Andalusian Tele-Assistance Service (SAT) of the Ministry of Health and Social Welfare of the Andalusian 

The Andalusian Telecare Service manages more than 17,100 calls per day. This service has served more than 35 million claims since its creation in 2002. The Andalusian Tele-Assistance Service (SAT) of the Ministry of Health and Social Welfare of the Andalusian Government has attended more than 17,100 calls per day in the first half of 2013, which translates into a total of 3,101,914 calls since January to June, which has involved more than 104,000 hours of conversation. Since the start of this service in 2002, more than 35 million calls have been answered.

In addition to the management of the calls from the SAT, in the first semester of 2013 more than 4,700 home monitoring visits were carried out in order to increase the effectiveness of the service and the perception of safety in the users. Initially collected data and the usage of tele-assistance devices were reviewed, and new useful information in managing the service was obtained.

## LIVING LAB-participation of end users 

Living lab has up-to-date didactic methods which are used for research in everyday environment. The lab offers a comprehensive and direct service providing customized solutions for research projects and in the development and innovation of products and services related to the Ageing challenge.

Living Lab is a tool enabling the end user to be actively involved, from their own everyday environment, in the design process, redesign of projects, products and services, as well as in the pilot stage.

Some examples could be found in the following links:   whose objectives are:

- Testing, validating, and manufacturing prototypes and refining complex solutions in real life - constantly evolving environments where multiple unknown factors in traditional research interfere.
- Allowing the customization of products and services to the real needs of the end user.
- Studying the older people in their real-life environment, with a frequent observation.
- This methodology is endorsed by the European Network of Living Lab.

## 9 Existing practices regarding health care in relation to smart living spaces for older persons

### Products

**HoneyCo** 

Based in Nashville USA

HoneyCo delivers peace of mind to families concerned about a loved one living independently. The user-friendly software collects information from smart devices throughout the home to provide resident activity summaries, wellness trends, and proactive alerts so that you are prepared to take action should a concerning event arise.

In addition, with HoneyCo, gain insights into care visits to ensure your loved one is receiving the level of care they need and deserve. The mission is to improve the level of care and safety of aging adults. This is done by designing easy-to-use technology. Specifically, for the elder care industry and families of older adults. Aging in place can be safer and more comfortable with the help of these technology platforms.

These products check and report on professional care visits and daily in-home safety. With HoneyCo, home care providers and families can work together.

**QORVO** 

Based in USA

The Senior Lifestyle System, developed by Qorvo® and Sensara®, is designed to help seniors feel safe and live independently at home. The system learns seniors' everyday behavior patterns using unobtrusive wireless sensors located around the home, and detects changes in behavior that may cause concern. It then alerts family members and/or caregivers via messages. The system uniquely uses proven technology that has been successfully tested in residential care facilities for several years, and is now available to service providers, retailers, and smart-home system integrators for home installation. Qorvo's wireless Internet of Things technology makes the system robust, virtually maintenance free, low cost and easy to install.

SENIOR BEHAVIOR PATTERN RECOGNITION

When does Grandma get up in the morning, eat breakfast, or watch TV? When does she leave and come back? Within a few weeks, the application learns what is "normal" vs. unexpected and sends an alert when something is wrong or irregular.

ALERTS ON EXCEPTIONS

When the system identifies irregular behavior or exceptional situations, it notifies family or caregivers via an alert on their smartphone or messages on social media platforms like WhatsApp or QQ. The system always keeps family members informed wherever they are.

LONG-TERM TRENDS

The Senior Lifestyle Service also provides information about longer-term behavioral trends, such as sleep patterns, eating less frequently or starting to move slower over time.

UNOBTRUSIVE AND RESPECTS PRIVACY

The system does not use cameras or require people to wear devices.

**Lively** 

Based in USA

Lively is a new smart home system, which is specifically designed for senior citizens. It gives children the ability to monitor their parent's every move in a less invasive manner. The system is capable of monitoring a wide array of products and items: televisions, pill bottles, doors, electricity usage, flood monitoring, stoves, etc.

How it Work The Lively System includes several different components, such as a smart watch. The watch, which is worn by the elderly, is capable of monitoring their moves. It can also help to provide them with daily reminders. If they're required to consume pills each and every day, the watch will ensure that they do so. For sensors, which can be placed almost anywhere, accelerometers are used. These devices are capable of determining, whether or not an item has been moved. If pill bottles have not been touched, you will receive a notification. This will give you the ability to visit your parent's residence or place a call to the authorities. All of these activities can be monitored from your smart phone.

Lively allows: to customize what to watch (with the use of the sensors, you have total control over what exactly to monitor. If you're worried about your mother and father using their oven, you can place a sensor on it. You can even place a sensor on their television, which will let you know, when they go to sleep at night), monitoring activities of daily living and monitoring blood pressure.

**Research papers** 

Medical and Home automation Sensor Networks for Senior Citizens Telehomecare.

The research paper presents the research about a home Tele-healthcare system for the senior citizens. The system uses home automation sensors to detect activity level of the persons. It is equipped also with other technologies to monitor them and detect any abnormal state in their health situation, like as bed and chair sensors, a mini PC connected to a TV, medical sensors, a wireless camera network. The system detects health abnormalities at an early stage through the frequent monitoring of physiological data. The system is designed for the elderly patients who wish to spend their old age in their own home, because of its potentials to increase independence and quality of life for seniors who prefer to live in their own homes and to realize cost savings for the health care system. By using this system, MEDeTIC (www.medetic.com) , a non-profit organization, offers a new concept of building smart homes by using telemedicine and home automation, named in French Maisons Vill'Âge. The first housing schemes are in building with implements of the system's components.

# ¹⁰List of identified projects on national and European level

 Smart Gerontechnology for Healthy Aging

*Lithuania, National research programme „Healthy Ageing": SEN-16048. Habil. dr. Vita Lesauskaitė Lietuvos sveikatos mokslų universitetas (2016-2018)*

 Effects of air pollution on the lungs: monitoring indicators and regulation by phytochemicals

*Lithuania, National research programme „Healthy Ageing": SEN-16074. Dr. Rūta Aldonytė Valstybinis mokslinių tyrimų institutas Inovatyvios medicinos centras (2016-2018)*

 Transformations in the older people care sector: the need for services, labor fource and the quality of employment

*Lithuania, National Research Programme „Welfare Society": GER-15056 Habil. dr. Laimutė Žalimienė Vilniaus universitetas (2015-2017)*

 Older people living alone: trends, profiles and challenges for the integration of generations

*Lithuania, National Research Programme „Welfare Society": GER-17001. Dr. Sarmitė Mikulionienė Lietuvos socialinių tyrimų centras (2017-2018)*

 Air quality management in low energy buildings

The selection of a ventilation strategy and air distribution method, as well as the air change rate, may affect both building energy consumption and indoor environmental health. The aim of this project was to examine the effect of various air supply strategies on the dispersion of aerosol particles in a full-scale test chamber by utilising highly time and size resolved measurements of aerosol concentration. The effectiveness of certain ventilation strategies were evaluated by several calculated indicators, such as the age of air and the particle removal efficiency.

*Lithuania, European Social Fund under Global Grant Scheme Project.*
*Kaunas University of technology (2013 – 2015)*
*Dainius Martuzevicius dainius.martuzevicius@ktu.lt*
*Andrius Jurelionis Andrius.jurelionis@ktu.lt*
*Lina Seduikyte lina.seduikyte@ktu.lt*

## Improving Energy Efficiency of Housing Stock: Impacts on Indoor Environmental Quality and Public Health in Europe (INSULATE) 🔗

The INSULATE project focused on the assessment of national programmes to improve the energy performance of existing housing stock, including cost-effective and proven measures such as government-supported improvements in thermal insulation. The project's specific objectives included developing a common protocol for assessing the impacts of a building's energy performance on indoor environmental quality and health; establishing an integrated approach for the assessment of environmental and health information, including demonstrating the use of relevant environmental and health indicators; demonstrating the effects (both positive and negative) of energy efficiency on Indoor Environment Quality (IEQ) and health in up to three different European countries; developing guidelines to support the implementation of related policies; and facilitating transnational networking and the dissemination of information.

## High-level R & D (SMART) the EU Structural Funds Investment Program for 2014-2020, measure 01.2.2-LMT-K-718

## Hybrid air ventilation unit with superior functionality

## Creation of technology for wood-modifying, which is environmentally friendly and giving added value to products

*Lithuania. 01.2.2-LMT-K-718-01-0021. Marius Aleinikovas, Lietuvos agrarinių ir miškų mokslų centras*

## Complex research on complementary reality for socially disadvantaged persons (blind and visually impaired) with superior functionality

*Lithuania.01.2.2-LMT-K-718-01-0060. Darius Plikynas, Vilniaus Gedimino technikos universitetas*

## Citizens' opportunities to provide theirselves with housing and measures to increase the availability of housing

*Lithuania. P-REP-18-41. Lietuvos energetikos institutas (2018-2019)*

## GERIA - -Geriatric study in Portugal on Health Effects of Air Quality in Elderly Care Centers (2011-15)

*Portugal. PTDC/SAU-SAP/116563/2010.*
*National Institute of Health Dr. Ricardo Jorge, Porto, Portugal*
*Institute of Public Health, University of Porto, Portugal*
*Ana Mendes | asestevao@gmail.com |a.sofia.mendes@insa.min-saude.pt*

## ConTerMa - Analysis of thermal comfort in nursing homes for the elderly in the cross-border cooperation area of Spain-Portugal (2019-20)

Funded by 'Fondo Europeo de Desarrollo Regional en el marco del Programa de Cooperación Interreg V-A España – Portugal, (POCTEP) 2014-2020. Expediente: 6/2018_CIE_6'.

*Portugal. 6/2018_CIE_6*
*National Institute of Health Dr. Ricardo Jorge, Porto, Portugal*
*Institute of Public Health, University f Porto, Portugal*
*Ana Mendes | asestevao@gmail.com |a.sofia.mendes@insa.min-saude.pt*
*Universitat Politécnica de Catalunya, Barcelona, Spain*
*Nuria Forcada | nuria.forcada@upc.edu*

BioFrail - An Exposome Approach
to Frailty in Older Adults (2016-20)

*National Institute of Health Dr. Ricardo Jorge, Porto, Portugal*
*Institute of Public Health, University f Porto, Portugal*
*Solange Costa | solange.costa2@gmail.com | solange.costa@insa.min-saude.pt*

Reducing Old-Age Social Exclusion: Collaborations in Research and
Policy (ROSEnet) (2016-20)

Reducing the number of people at risk of social exclusion is a headline target of the Europe 2020 strategy. Population ageing and low economic growth pose major challenges to meeting this target, emphasising the necessity to tackle old-age exclusion. While risks of exclusion of older people are widening and deepening, damaging gaps in understanding old-age exclusion exist across Europe. Existing knowledge is poorly developed, lacks synthesis and is spread across highly disparate disciplines. This Action aims to overcome fragmentation and critical gaps in conceptual innovation on old-age exclusion across the life course, in order to address the research-policy disconnect and tackle social exclusion amongst older people in Europe. The action will engage with researchers and policy stakeholders to develop shared under-standings and to direct the development of new policy and practice interventions, that can be practically and effectively implemented, for reducing exclusion in diverse European ageing societies. The Action will establish an innovative participatory, interdisciplinary and cross-European collaboration that will: (1) synthesise existing knowledge; (2) critically investigate the construction of life-course old-age exclusion (3) assess the implications of old-age exclusion across the life course; (4) Develop new conceptual frameworks on old-age exclusion; and (5) identify innovative, and implementable, policy and practice for reducing old-age exclusion. The Action focuses on economic, social, service, civic rights, and community/spatial exclusion. With deliverables that include conferences, workshop-policy events, briefing papers, early-career investigator development, and a repository of innovative practice and policy, the Action will forge much-needed new links between research and policy, enhancing evidence-based and effective innovation.

*CA15122*
*Dr Kieran WALSH Chair +35391495460 kieran.walsh@nuigalway.ie*
*https://www.cost.eu/actions/CA15122/#tabs|Name:overview*

Accessibility for all to services and terminals for
next generation networks (2003-07)

The main objective of the Action is to increase the accessibility of next generation telecommu-nication network services and equipment to elderly people and people with disabilities by design or, alternatively, by adaptation when required. In cases where this cannot be achieved, the Action will aim at promoting the establishment of appropriate supplementary assistive services and equipment. Taking always into account the "Design for All" concept in telecommunications and teleinformatics, especially in the mobile field, the objectives of the Action can be specified in operational terms as follows: Information Collaction. Extend the existing cost 219 database and the knowledge required for designers on consumers and their requirements, so that many more disabled and elderly people can be catered for in mainstream design, · Support the exchange of Inclusion and Accessibility issues so that these can be freely explored with developers, researchers and representatives of the telecommunications industries and service providers, so that · disabled or elderly people are enabled to share in the benefits of new communication systems as discriminating consumers from the onset, but not being discriminated against.

*219ter*
*Mr Patrick ROE Chair +41 76 329 47 36 patrick.roe@epfl.ch*
*https://www.cost.eu/actions/219ter/#tabs|Name:overview*

## Reducing Old-Age Social Exclusion: Collaborations in Research and Policy (ROSEnet) (2016-20)

Reducing the number of people at risk of social exclusion is a headline target of the Europe 2020 strategy. Population ageing and low economic growth pose major challenges to meeting this target, emphasising the necessity to tackle old-age exclusion. While risks of exclusion of older people are widening and deepening, damaging gaps in understanding old-age exclusion exist across Europe. Existing knowledge is poorly developed, lacks synthesis and is spread across highly disparate disciplines. This Action aims to overcome fragmentation and critical gaps in conceptual innovation on old-age exclusion across the life course, in order to address the research-policy disconnect and tackle social exclusion amongst older people in Europe. The action will engage with researchers and policy stakeholders to develop shared understandings and to direct the development of new policy and practice interventions, that can be practically and effectively implemented, for reducing exclusion in diverse European ageing societies. The Action will establish an innovative participatory, interdisciplinary and cross-European collaboration that will: (1) synthesise existing knowledge; (2) critically investigate the construction of life-course old-age exclusion (3) assess the implications of old-age exclusion across the life course; (4) Develop new conceptual frameworks on old-age exclusion; and (5) identify innovative, and implementable, policy and practice for reducing old-age exclusion. The Action focuses on economic, social, service, civic rights, and community/spatial exclusion. With deliverables that include conferences, workshop-policy events, briefing papers, early-career investigator development, and a repository of innovative practice and policy, the Action will forge much-needed new links between research and policy, enhancing evidence-based and effective innovation.

*CA15122*
*Dr Kieran WALSH Chair +35391495460 kieran.walsh@nuigalway.ie*
*https://www.cost.eu/actions/CA15122/#tabs|Name:overview*

## European Network on New Sensing Technologies for Air-Pollution Control and Environmental Sustainability – EuNetAir (2012-16)

This Action will focus on a new detection paradigm based on sensing technologies at low cost for Air Quality Control (AQC) and set up an interdisciplinary top-level coordinated network to define innovative approaches in sensor nanomaterials, gas sensors and devices, wireless sensor-systems, distributed computing, methods, models, standards and protocols for environmental sustainability within the European Research Area (ERA). The State of the Art showed that research on innovative sensing technologies for AQC based on advanced chemical sensors and sensor-systems at low-cost, including functional materials and nanotechnologies for eco-sustainability applications, the outdoor/indoor environment control, olfactometry, air-quality modelling, chemical weather forecasting, and related standardisation methods is performed already at the international level, but still needs serious efforts for coordination to boost new sensing paradigms for research and innovation. Only a close multidisciplinary cooperation will ensure cleaner air in Europe and reduced negative effects on human health for future generations in smart cities, efficient management of green buildings at low CO2 emissions, and sustainable economic development. The objective of the Action is to create a cooperative network to explore new sensing technologies for low-cost air-pollution control through field studies and laboratory experiments to transfer the results into preventive real-time control practises and global sustainability for monitoring climate changes and outdoor/indoor energy efficiency. Establishment of such a European network, involving Non-COST key-experts, will enable EU to develop world capabilities in urban sensor technology based on cost-effective nanomaterials and contribute to form a critical mass of researchers suitable for cooperation in science and technology, including training and education, to coordinate outstanding R&D and promote innovation towards industry, and support policy-makers.

*TD1105*
*Dr Michele PENZA Chair +390831201422 michele.penza@enea.it*
*https://www.cost.eu/actions/TD1105/#tabs|Name:overview*

## Indoor Air Pollution Network (2018-22)

In developed countries, we spend 80-90% of our time indoors, where we receive most of our exposure to air pollution. However, regulation for air pollution focuses mainly on outdoors and the indoor environment is much less well characterised. The concentrations of many air pollutants can be higher indoors than out, particularly following activities such as cleaning and cooking. With increasing climate change impacts, related energy efficiency measures are making buildings considerably more airtight. Such measures can increase indoor pollutant concentrations even further. Therefore, to reduce our exposure to air pollution, we must consider both the indoor and outdoor environments and the role of ventilation, in order to mitigate through appropriate building operation, use and design.

INDAIRPOLLNET (INDoor AIR POLLution NETwork) will improve our understanding of the cause of high concentrations of indoor air pollutants. It will assemble experts in laboratory and chamber experiments, modelling studies and measurements of relevance to indoor air quality (IAQ), including outdoor air chemists. Our network includes experts in chemistry, biology, standardisation, particulate matter characterisation, toxicology, exposure assessment, building materials (including those manufactured specifically to improve IAQ such as green materials), building physics and engineering (including ventilation and energy) and building design. This Action aims to significantly advance the field of indoor air pollution science, to highlight future research areas and to bridge the gap between research and business to identify appropriate mitigation strategies that optimise IAQ. The findings will be disseminated to relevant stakeholders such as architects, building engineers and instrument manufacturers.

*CA17136*
*Dr Nicola CARSLAW Chair +441904324777 nicola.carslaw@york.ac.uk*
*https://www.cost.eu/actions/CA17136/#tabs|Name:overview*

## European Network for Environmental Citizenship (2017-21)

European Network for Environmental Citizenship (ENEC) aims to improve understanding and assessment of environmental citizenship in European societies and participating countries. Environmental Citizenship is a key factor in EU's growth strategy (Europe 2020) and its vision for Sustainable Development, Green and Cycle economy and Low-carbon society (EU-roadmap 2050). The Integrated Network of the Action will diminish the barriers between human, economic, social, political and environmental sciences multiplying the knowledge, expertise, research and insights of different stakeholders (researchers, scholars, teachers, practitioners, policy officials, NGOs, etc.) related in Environmental Citizenship. The different macro- and micro- level dimensions of formal and non-formal education that could lead to Environmental Citizenship will be focused. By developing National, European and International collaborations ENEC will enhance the scientific knowledge and attention to Environmental Citizenship. Expected deliverables include: a) the creation of a web-site, b) a repository database of scientific measures and evidence based interventions that target Environmental Citizenship, c) the facilitation of scientific training schools, short term scientific missions, conferences and d) the dissemination of collaborative working papers, scientific reports, proceedings, academic publications, policy and recommendation papers and an edited book on Environmental Citizenship. The Action will conceptualize and frame the Environmental Citizenship and will develop new research paradigms and metrics for assessing the Environmental Citizenship. Good examples and best educational practices leading to pro-environmental attitudes, behaviour and values will be highlighted and promoted. Policy measures and recommendations will be proposed. The Action will serve as a vehicle to defragment the knowledge and expertise in Environmental Citizenship.

*CA16229*
*Dr Andreas HADJICHAMBIS Chair*
*+35799477309 a.chadjihambi@cytanet.com.cy*
*https://www.cost.eu/actions/CA16229/#tabs|Name:overview*

## Health monitoring and sOcial integration environMEnt for Supporting WidE ExTension of independent life at HOME (HOME SWEET HOME)  (2010-14)

HOME SWEET HOME brings together a set of services which, combined, allow extending the independent life of elderly people. HOME SWEET HOME (HSH) is trialling a new, economically sustainable home assistance service which extends elders independent living. HSH intends to achieve this by providing a comprehensive set of services which support elders in their daily activities and allows carers to remotely assess their ability to stay independent. HSH privileges features which the elders themselves can use and limits the need for other people to interfere with their private life, unless the system detects a clear need. The project measures the impact of monitoring, cognitive training and e-Inclusion services on the quality of life of the elderly, on the cost of social and healthcare delivered to them, and on a number of social indicators.

*https://cordis.europa.eu/project/rcn/191712_en.html*
*https://cordis.europa.eu/docs/projects/cnect/9/250449/080/reports/001-HomeSweetHomePublishableSummary.pdf*

## Integrated Cognitive Assistive and Domotic Companion Robotic Systems for Ability and Security (2008-12)

The main unique selling point of the CompanionAble project lies in the synergetic combination of the strengths of a mobile robotic companion with the advantages of a stationary smart home, since neither of those approaches alone can accomplish the demanding tasks to be solved. Positive effects of both individual solutions shall be combined to demonstrate how the synergies between a stationary smart home solution and an embodied mobile robot companion can make the care and the care person's interaction with her assistive system significantly better.

Starting with a profound requirement engineering for ICT-supported care and therapy management for the care persons, basic technologies for multimodal user observation and human-machine interaction will provide the fundamentals for the development of a stationary smart home assistive system and a mobile robot assistant, building the cornerstones of the overall system integrating the promising solutions of both parts. Substantial support comes from the research activities focusing on an architectural framework, allowing such a complex care scenario solution be achievable. After the realization of the respective scenarios, long lasting field experiments will be carried out to evaluate and test the system, and both scenarios can be evaluated to show their strength and weaknesses. This will initiate the development of an overall, integrated care scenario (smart home with embedded robot companion).

*https://cordis.europa.eu/project/rcn/85553_en.html*

# ❞ Full bibliography

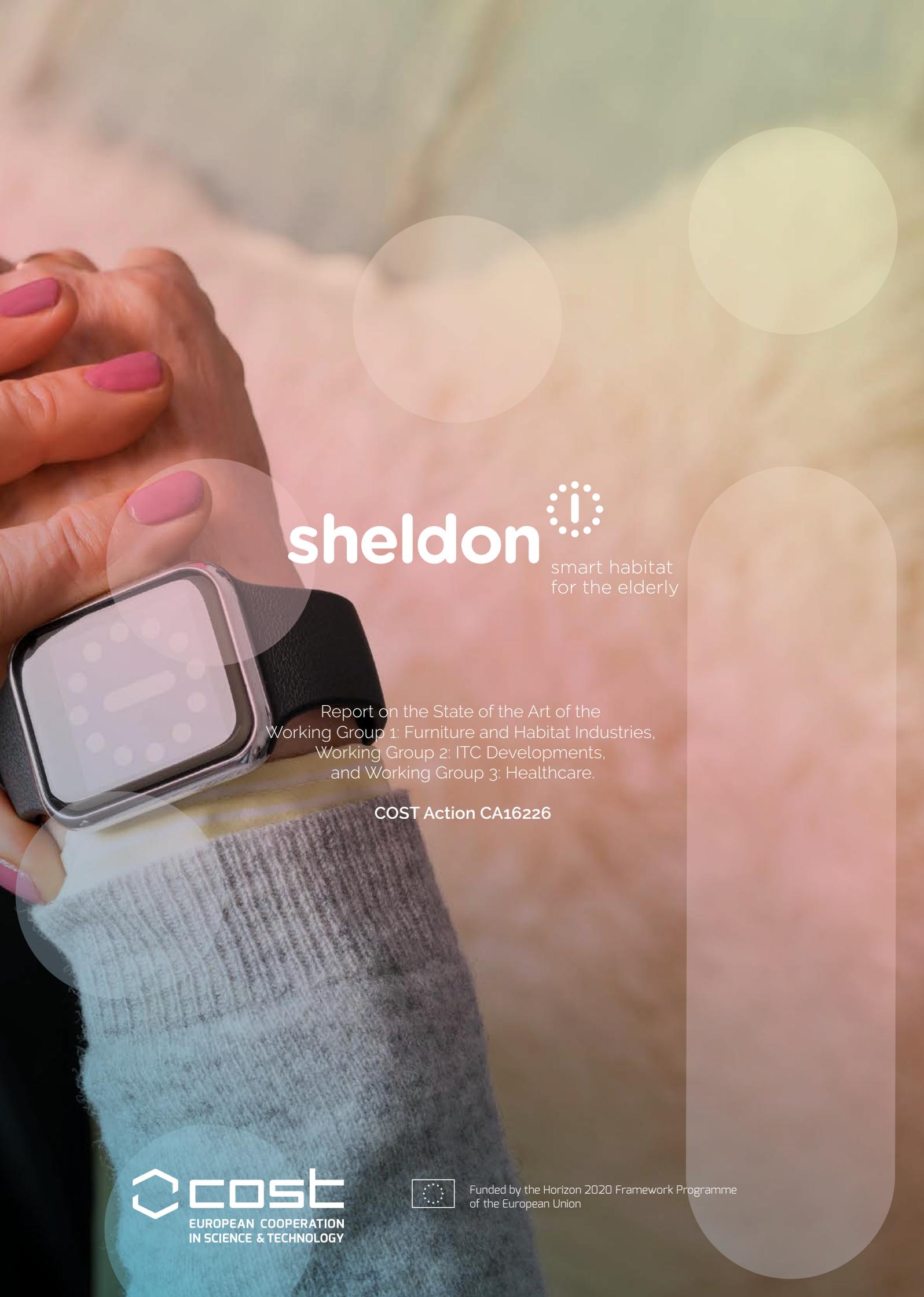

# sheldon

smart habitat
for the elderly

Report on the State of the Art of the
Working Group 1: Furniture and Habitat Industries,
Working Group 2: ITC Developments,
and Working Group 3: Healthcare.

**COST Action CA16226**

cost
EUROPEAN COOPERATION
IN SCIENCE & TECHNOLOGY

Funded by the Horizon 2020 Framework Programme
of the European Union